\documentclass[12pt,oneside]{book}
\usepackage{amssymb}
\usepackage{a4}
\usepackage{amsmath}
\usepackage{latexsym}
\usepackage{amsfonts}
\usepackage{multibox}
\usepackage{amsmath,amsfonts,amsbsy}
\usepackage[dvips]{graphicx,epsfig}
\usepackage{graphics}
\usepackage{graphicx}
\usepackage{epic}
\usepackage{eepic}

\begin{document}
\oddsidemargin 6pt\evensidemargin 6pt\marginparwidth
48pt\marginparsep 10pt

\renewcommand{\thefootnote}{\fnsymbol{footnote}}
\renewcommand{\theequation}{\thesection.\arabic{equation}}
\pagenumbering{roman}
\thispagestyle{empty}


\noindent \vskip3.3cm

\bigskip \bigskip

\begin{center}
\LARGE {YEREVAN PHYSICS INSTITUTE}
\end{center}

\bigskip\bigskip\bigskip\bigskip \bigskip

\begin{center}
{DEPARTMENT OF THEORETICAL PHYSICS}
\end{center}

\bigskip\bigskip\bigskip\bigskip \bigskip\bigskip \bigskip\bigskip

\begin{center}
{\Large\bf Higher Spin Interacting Quantum Field Theory
and Higher Order Conformal Invariant Lagrangians}
\end{center}

\bigskip\bigskip\bigskip\bigskip \bigskip\bigskip \bigskip\bigskip

\begin{center}
{\large
Karapet Mkrtchyan}\\
\medskip
{\small\it Yerevan Physics Institute\\ Alikhanian Br.
Str.
2, 0036 Yerevan, Armenia}\\
\medskip
{\small\tt e-mail karapet@yerphi.am}

\bigskip \bigskip \bigskip \bigskip \bigskip \bigskip\bigskip

{\large
PhD Thesis}\\

\bigskip \bigskip\bigskip \bigskip

Advisor: Prof. Ruben Manvelyan \\

\bigskip \bigskip

\end{center}

\bigskip \bigskip \bigskip\bigskip

\vspace*{22cm}

\begin{center}
{\sc Abstract}
\end{center}

\quad This thesis includes several original results. All of them are already published or submitted for publication.
The thesis is based on articles \cite{MMM}, \cite{Manvelyan:2008ks}, \cite{Manvelyan:2009tf}, \cite{Mkrtchyan}, \cite{Manvelyan:2009vy}, \cite{Manvelyan:2010wp}, \cite{Manvelyan:2010jr}, \cite{Manvelyan:2010je} and reproduces the results of \cite{MT} for completeness.

I present here the short summary of main results:

The ultraviolet singular structure of the
bulk-to-bulk propagators for higher spin gauge fields in $AdS_{4}$
space is analyzed in details. One loop mass renormalization is studied on a simple example.

The conformal invariant Lagrangian with the $k$-th power of Laplacian for the
 hierarchy of conformally coupled scalars  with increasing scaling dimensions connected
with the $k$-th Euler density is rederived using the Fefferman-Graham ambient space approach.
The corresponding \emph{gauged} ambient metric, Fefferman-Graham expansion  and extended
Penrose-Brown-Henneaux transformations are proposed and analyzed.

Linearized gauge invariant interactions of scalar and general higher even spin fields in the $AdS_{D}$ space are obtained.
A generalized Weyl transformation is proposed and the corresponding Weyl invariant action for cubic coupling of a scalar
to a spin $\ell$ field is constructed.

Using Noether's procedure several cubic interactions between different HS gauge fields are derived, including cubic selfinteraction of even spin gauge fields in a flat background. Then the main result - \emph{the complete off-shell gauge invariant Lagrangian for the trilinear interactions of Higher Spin Fields with arbitrary spins $s_{1}, s_{2}, s_{3}$ in a flat background} is presented. All possibilities with different numbers of derivatives are discussed. Restrictions on the number of derivatives are obtained. For any possible number of derivatives this interaction is uniquely fixed by gauge invariance up to partial integration and field redefinition.

Finally an off-shell generating function for all cubic interactions of Higher Spin gauge fields is presented. It is written in a compact way, and turns out to have a remarkable structure.

\newpage

\begin{flushright}
{\large {\em "The good lord is subtle, but he is not malicious."\\
    Albert Einstein}}
\end{flushright}

\newpage
{\LARGE {\textbf{Acknowledgments}}}

\vspace{1.5cm}

This Thesis contains almost all the work I have done 
during my PhD course in Yerevan Physics Institute.

First of all I would like to thank my advisor Ruben Manvelyan, who got me acquainted with the fascinating field of High Energy Physics. Who helped and supported my research and has taught me almost all that I know in High Energy Physics and particularly in the theory of Higher Spin Fields. His patience and willingness to answer any question, his readiness for discussions at any time of the day made this work possible. He has been also a very good friend whom I talked to with great interest not only about science. I would like to thank him for the opportunities he gave me to get in touch with other researchers and to work at the cutting edge of modern High Energy Physics.

I also want to thank my collaborators Werner R\"uhl and Ruben Mkrtchyan with whom it was a pleasure to work with. I would like to thank them for everything I have learned from them, for all the discussions we had, for being very kind to me, for readiness to help always and for their warm attention towards me and my work. I also want to thank Werner R\"uhl for inviting me to the Technical University of Kaiserslautern several times, where big part of this work was done.

I have been traveling frequently during my PhD years, participated in several conferences and schools, got acquainted with the leading researchers
in the field of Higher Spin gauge field theory and High Energy Physics in general. I would like to thank Mikhail Vasiliev and Igor Klebanov with
whom it is a pleasure to talk about science and any other topic. They were very pleasant and didn't leave any question unanswered.
My gratitude goes also to Ruben Poghosyan, Tigran Tchrakyan, Evgeny Ivanov, George Savvidy, Arthur Lipstein, Xi Yin,
Eugene Skvortsov, Vyacheslav Didenko, Maxim Grigorev, Igor Samsonov, Marine Samsonyan, 
Peter Koroteev, Alexander Monin, Alexander Zhiboedov, Dmitry Krotov, Viktor Alexandrov, Adrian Mertens, Michael Kay,
Kentaro Hanaki, Aram Saharyan, Alexei Isaev, Armen Nersisyan, Tigran Hakobyan, Vadim Ohanyan, Vahagn Eghikyan, Armen Tumasyan and Murad Tovmasyan for thoughtful discussions. I would like to thank also Augusto Sagnotti and Massimo Bianchi for inviting me to Italy to participate
in the workshop "New Perspectives in String Theory", to present my work and also participate in the conference "Strings 2009". I would like also to thank Eugene Skvortsov, Vyachaslav Didenko, Nicholas Boulanger, Per Sundell, Alexei Isaev and Armen Nersisyan for their kind invitations to present my work in their institutions. I also want to thank Evgeny Ivanov and Ruben Poghosyan for carefully refereeing this Thesis.

I would like to thank also the Institute for Advanced Study in Princeton, the Joint Institute for Nuclear Research in Dubna and of course the Technical University of Kaiserslautern for their hospitality.

I am indebted to all my colleagues from the Department of Theoretical Physics in Yerevan Physics Institute.

I would like to thank also my friends who are always near to me and always ready to help: Khachatur, Artsrun, Harutyun, Vardan, Armen, Vahe, David, Anna, Arthur, Khachatur, Mamikon, Marine, Anna, Gor, Karen, Bagrat, Vazgen, Vahan, Arshak, Murad and many others.

And finally my gratitude goes to my parents for their constant encouragement and support, for always being understanding and helping.

This work is dedicated to them.

\pagebreak

\tableofcontents


\chapter{Introduction}

\pagenumbering{arabic} \setcounter{page}{1}

\quad Higher Spin gauge field theory is one of the most important and puzzling problems in modern quantum field theory. The first attempts to deal with a quantum theory of high spin particles date back to late 1930-s and 1940-s, with the works of Dirac \cite{Dirac:1936tg}, Fierz and Pauli \cite{Fierz:1939ix}, Rarita and Schwinger \cite{Rarita:mf} and others. One of the most important concepts in quantum field theory is Poincar\'e symmetry. In the classical papers \cite{Wigner:1939cj} and \cite{Bargmann:1948ck} the irreducible representations of Poincar\'e group were classified. These irreducible representations are characterized by two quantities - mass and spin. After discovery of nonabelian gauge theory by Yang and Mills \cite{Yang:1954ek} the role of gauge symmetries in quantum field theory was acknowledged. It became clear that the invariance of the Lagrangian with respect to \emph{local} symmetry is the cornerstone of any field theory.

There are higher rank tensor representations of the Poincar\'e group \cite{Wigner:1939cj,Bargmann:1948ck}, which are not yet associated with any physical field theory. It is clear that the higher rank tensor representations of the Poincar\'e group, the Higher Spin fields, are gauge fields that one has to introduce gauging higher derivative symmetries of the action.

It is natural to assume that Higher Spin gauge field interactions are much weaker than gravitational ones, that's why we don't see any evidence for these interactions. They should be important only in a very high energy regime. An interesting speculative application of Higher Spin gauge fields might be also it's connection to Dark Matter and/or Dark Energy.

Nowadays the best candidate for quantum gravity is String Theory. In the spectrum of the String Theory there are excitations with any high spin, therefore String Theory gives another motivation for investigations of Higher Spin gauge fields to take place.

A new motivation for investigating Higher Spin gauge field theories arose during the last decade after discovering the holographic duality between the $O(N)$ sigma model in $d=3$ space and Higher Spin gauge field theory living in the space $AdS_4$ \cite{Klebanov}, which is an interesting special case of AdS/CFT conjecture \cite{Maldacena:1997re}. This case of holography is especially important  by the existence of two conformal points of the boundary theory and the possibility to describe them by the same HSF gauge theory with the help of \emph{spontaneous breaking of higher spin gauge symmetry and mass generation by a corresponding Higgs mechanism} (\cite{Ruehl}-\cite{Sezgin:2002rt} and references therein).

After the nonabelian gauge theory of vector fields \cite{Yang:1954ek} appeared, there were numerous attempts to construct a gauge theory with a gauge group which mixes space-time and internal symmetries in a nontrivial way. These attempts resulted in several no-go theorems. The most general result was obtained by Coleman and Mandula in \cite{Coleman:ad}. They have proved a theorem on the impossibility of combining space-time and internal symmetries in any but a trivial way, which holds not only for Lie groups but is also applicable to infinite-parameter groups. The moral of this theorem is that \emph{if the assumptions of the theorem hold, there can't be Higher Spin charges}. Therefore Higher Spin fields, if existing, don't participate in interactions. In order to have an interacting theory of Higher Spin fields one has to loosen some of the assumptions of the Coleman-Mandula theorem.
Then it was shown that it is possible to overcome this theorem introducing graded Lie algebras \cite{Golfand:1971iw,Volkov:1973ix,Wess:1974tw}, which give rise to the supersymmetric theories. It was shown in \cite{Haag:1974qh} that the only possible algebras that mix space-time and internal symmetries are graded Lie algebras, which in addition to the standard generators of Poincar\'e algebra include also supersymmetry generators with spin one-half.

Two other no-go theorems were formulated by Steven Weinberg and Edward Witten in 1980 \cite{Weinberg:1980kq}. One of them rules out electrically charged fields with spin $s>1/2$, the other theorem forbids theories with a Lorentz covariant energy-momentum tensor which include fields with spin $s>1$. These theorems don't apply to gauge theories though, therefore the search for Higher Spin gauge theories wasn't proved to be meaningless.

Despite all the no-go results (see also \cite{Aragone:1979hx}), using the Lagrangian formulation of Higher Spin theories by Singh and Haagen \cite{Singh:qz,Singh:rc}, the consistent Lagrangian description for \emph{free} Higher Spin gauge fields both in flat space and in constantly curved backgrounds was given by Fronsdal in \cite{Fronsdal:1978rb,Fronsdal:1978vb} for bosonic fields and by Fang and Fronsdal in \cite{Fang:1978wz,Fang:1979hq} for fermionic Higher Spin fields. Fronsdal's theory of Higher Spin gauge fields includes some new features. There are constraints on Higher Spin gauge fields and the gauge parameters. In order to have a gauge invariant field equation of motion that is linear in the Higher Spin gauge field and of second order in the derivatives, the gauge parameter should satisfy the tracelessness condition. This is so-called Fronsdal's first constraint. In order to have a gauge invariant kinetic Lagrangian for free Higher Spin gauge fields, the field itself should be \emph{double traceless}. This is Fronsdal's second constraint. On-shell gauge symmetry allows to gauge away all nonphysical components of the field and to obtain a traceless transversal tensor field with a simple Klein-Gordon-like equation of motion. For the spin $s$ field we get two possible helicities: $\pm s$.
The possible deformation of the gauge algebra of Fronsdal's Lagrangian, which should lead to the Higher Spin interacting gauge theory is considered as a challenge already 30 years. Fronsdal's theory of Higher Spin gauge fields is a natural generalization of linearized gravity, and is called also metric-like formulation of Higher Spin gauge fields.

The generalization of Christoffel symbol and Riemann curvature of linearized gravity for Higher Spin cases was given by deWitt and Freedman in \cite{deWit:pe}. There are $s-1$ Christoffel symbols for Higher Spin gauge fields, with different numbers of derivatives (from 1 to $s-1$). They are all linear in the field and transform under gauge transformations in a simple way. The curvature of the Higher Spin field, also linear in the field, is invariant with respect to gauge transformations. Interesting properties of Higher Spin field curvatures are discussed in \cite{Damour:1987vm}. The full nonlinear form for the deWitt-Freedman curvature and Christoffel symbols (if any) is still unknown.

Despite the fact that consistent equations of motion for Higher Spin gauge fields are known over twenty years \cite{VasilievEqn}, the question of existence of \emph{Lagrangian} for interacting Higher Spin gauge fields is still open. The subject of special interest is a minimal selfinteraction of even spin gauge fields, where one can naively expect the existence of an Einstein-Hilbert type nonlinear action for any single even spin gauge field. Although there are known restrictions on Higher Spin theories in flat space-time, the recent development \cite{Manvelyan:2010jr} has shown that \emph{there is a local higher derivative cubic interaction Lagrangian for gauge fields with any higher spins in flat space-time of any dimensions}. This shifts the no-go theorems to the quartic power of fields in interaction Lagrangians, where one can expect the final battle for the existence of local (or nonlocal) Lagrangians for interacting HS gauge field theory in flat space.

Gauge symmetry, which is a redundancy of non-physical degrees of freedom in the Lagrangian, is the main principle which helps to choose the right Lagrangian for the given theory. To quote C. N. Yang, gauge \emph{symmetry dictates the form of the interaction}\footnote{This quotation along with a very beautiful review of the history of gauge symmetries you can find in \cite{DavidGross}.}. In this thesis we will show that for Higher Spin gauge fields interactions are uniquely determined by gauge symmetry.

The free Lagrangian for Higher Spin gauge fields both in flat space and in constantly curved backgrounds (dS and AdS) are known over thirty years \cite{Fronsdal:1978rb,Fang:1978wz,Fronsdal:1978vb,Fang:1979hq}. In contrast to free theory, attempts to construct Lagrangians for interacting theories haven't been successful yet beyond the cubic vertices. In this thesis we are going to discuss only trilinear interactions of Higher Spin gauge fields.

Our recent results \cite{Manvelyan:2010jr}, \cite{Manvelyan:2004mb}-\cite{Manvelyan:2010je} on Higher Spin gauge field cubic interactions in flat space, which certainly reproduce the flat limit of the famous Fradkin-Vasiliev vertex for higher spin coupling to gravity \cite{Vasiliev}, show that all interactions of higher spin gauge fields with any spins $s_{1}, s_{2}, s_{3}$ both in flat space and in dS or AdS are unique\footnote{The cubic interaction Lagrangian is unique up to partial integration and field redefinition.}. This was already proven for some low spin cases of both the Fradkin-Vasiliev vertex for $2, s, s$ and the nonabelian vertex for $1, s, s$  in \cite{boulanger}.

The first important step towards cubic interactions in Higher Spin gauge field theory in covariant formulation was done in 1984 by Berends, Burgers and van Dam \cite{vanDam}. They constructed a cubic selfinteraction Lagrangian for spin three gauge fields and proved impossibility of extension to higher orders. Their arguments are based on gauge algebra, which does not close for a single spin three nonabelian field. The authors give an optimistic hope that it will be possible to extend this Lagrangian to higher orders if one takes into account corrections from interactions with gauge fields with spins higher than three. A recent discussion on this subject appeared from Bekaert, Boulanger and Leclerq \cite{bek}. They show the impossibility to close this non-abelian (spin 3) algebra taking into account corrections from interactions of other fields with spins higher (or lower) than three.

The first successful result on Higher Spin gravitational interactions was derived in the already mentioned work by Fradkin and Vasiliev \cite{Vasiliev}, where a cubic coupling of Higher Spin gauge fields to linearized gravity was constructed in the constantly curved background. The interesting property of this Lagrangian is it's non-analyticity in the cosmological constant, therefore excluding a flat space limit. However it was shown already  in \cite{boulanger} that after rescaling of Higher Spin gauge fields one can observe a flat limit for the Fradkin-Vasiliev interactions. In our approach the spin $s$ gauge field has scaling dimension $[length]^{s-2}$, and the Fradkin-Vasiliev vertex has a flat limit with $2s-2$ derivatives (minimal possible number) in the $2-s-s$ interaction which has the same scaling dimension as the Einstein-Hilbert Lagrangian terms. As it was shown by Metsaev in \cite{Metsaev} using a light cone gauge approach, there are three different couplings to linearized gravity with different numbers of derivatives for any higher spin $s$ field, and in general $min\{s_{1},s_{2},s_{3}\}+1$ different possibilities with different numbers of derivatives for the $s_{1}-s_{2}-s_{3}$ interaction. All these interactions were derived in a covariant off-shell formulation in \cite{Manvelyan:2010jr}, and I am going to discuss them in this Thesis.

For some important results on higher spin cubic interactions see \cite{Vasiliev1}-\cite{Zinoviev} and references therein. For recent reviews see \cite{review}.

The only Section in this Thesis that is not directly connected to Higher Spin theory is Section \ref{conform}, where I give brief introduction independently.

All the Chapters in this Thesis are more or less independent, in some cases there are even differences in conventions, therefore I give notations and conventions independently where needed.

The formalism which I use in this Thesis is developed in Chapter \ref{free}, where I present also well known results in the free theory of Higher Spin fields: the Fronsdal Lagrangian, the deWitt-Freedman curvatures and Christoffel symbols for HS gauge fields and the Bianchi identities that connect them as well as some new connections between these quantities following from their Bianchi identities.

In Chapter \ref{Propagator} the ultraviolet singular structure of the bulk-to-bulk propagators for higher spin gauge fields in $AdS_{4}$ space is analyzed in details. One loop mass renormalization corresponding to interactions with the Higgs scalar are studied. This mass renormalization is finite and connected with the anomalous dimensions of those currents in the corresponding boundary $CFT_{3}$ that cease to be conserved when the interaction is switched on. In particular it is proportional to $\ell-2$ for a spin $\ell$ field.

In Chapter \ref{confL} the hierarchy of conformally coupled scalars  with increasing scaling dimensions $\Delta_{k}=k-d/2$, $k=1,2,3,\dots $ connected with the $k$-th Euler density  in the corresponding space-time dimensions $d\geq 2k$ is proposed. The corresponding conformal invariant Lagrangian with the $k$-th power of Laplacian for the already known cases $k=1,2$ is reviewed, and the subsequent case of $k=3$ is completely constructed and analyzed. The same hierarchy is rederived using the Fefferman-Graham $d+2$ dimensional ambient space approach. The corresponding mysterious "holographic" structure of these operators is clarified. We explore also the $d+2$ dimensional ambient space origin of the Ricci gauging procedure proposed by A.~Iorio, L.~O'Raifeartaigh, I.~Sachs and C.~Wiesendanger as another method of constructing the Weyl invariant Lagrangians.
The corresponding \emph{gauged} ambient metric, Fefferman-Graham expansion and extended Penrose-Brown-Henneaux transformations are proposed and analyzed.

Then another generalization of conformal coupling of the scalar to the gravity is considered. The explicit form of linearized gauge invariant interactions of scalar and general higher even spin fields in the $AdS_{D}$ space is obtained. In the case of general spin $\ell$  a generalized 'Weyl' transformation is proposed and the corresponding 'Weyl' invariant action is constructed. In both cases the invariant actions of the interacting higher even spin gauge field and the scalar field include the whole tower of invariant actions for couplings of the same scalar with all gauge fields of smaller even spin.

In section \ref{off} of Chapter \ref{cubic} several trilinear interactions of higher spin fields involving two equal ($s=s_{1}=s_{2}$) and one higher even ($s_{3}>s$) spin are presented. Interactions are constructed on the Lagrangian level using Noether's procedure together with the corresponding next to free level fields of the gauge transformations. In certain cases when the number of derivatives in the transformation is $2s-1$ the interactions lead to the currents constructed from the generalization of the gravitational Bell-Robinson tensors.
In section \ref{self} of Chapter \ref{cubic} using Noether's procedure the complete cubic selfinteraction for the case of spin $s=4$ in a flat background is presented and the cubic selfinteraction for general spin $s$ with $s$ derivatives in the same background is discussed. The leading term of the latter interaction together with the leading gauge transformation of first field order are presented.
In section \ref{general} of Chapter \ref{cubic} the complete solution for the trilinear interactions of arbitrary spins $s_{1},s_{2}, s_{3}$ in a flat background is presented, the possibility to enlarge this construction to higher order interactions in the gauge field is discussed. Finally the expansion of a general spin $s$ gauge transformation into powers of the field and the related closure of the gauge algebra in the general case are discussed.

In the recent paper \cite{ST} by Sagnotti and Taronna the authors proposed an on-shell generating function for the general HS cubic interaction presented in \cite{Manvelyan:2010jr} from a massless limit of String Theory. In the final Chapter \ref{GF} of this thesis I am going to present an off-shell extension of that generating function, which can surprisingly be enhanced with a beautiful Grassmann structure, the string origin of which is not clear yet.

In this Thesis I am not going to address many important directions in Higher Spin gauge field theories like Mixed Symmetry Higher Spin fields (see \cite{Skvortsov:2010nh}-\cite{Aulakh:1986cb} and references therein), Unconstrained Higher Spins ( \cite{Campoleoni:2008jq}, \cite{Campoleoni:2009gs}, \cite{Francia:2007qt} and references therein), Unfolded formulation of Higher Spin dynamics (\cite{Didenko:2009tc}-\cite{Barnich:2005ru} and references therein), Vasiliev equations (\cite{VasilievEqn}, \cite{Iazeolla:2008bp} and references therein), BRST approach to Higher Spins (\cite{Barnich:2005ru}-\cite{Barnich:2005ga}), Light Cone gauge formulation (\cite{Metsaev},\cite{Bengtsson:1983pg}-\cite{Bengtsson:1986kh} and references therein), some AdS/CFT aspects (\cite{Giombi:2010vg}, \cite{Metsaev:2009hp} and references therein), some String-inspired constructions (\cite{Bengtsson:1986ys}-\cite{Antoniadis:2009rd} and references therein), partially massless Higher Spin fields (\cite{Skvortsov:2006at}-\cite{Deser:2001pe} and references therein).

Another weakness of this Thesis is the lack of a group-theoretical description: the study of the non-abelian gauge algebra which stands behind the interactions presented in this Thesis as well as "Higher Spin geometry" interpretations (if any) will follow in the future. The interactions we discuss in this Thesis don't include fermionic half-integer spin fields, which is another important direction to generalize these results.


\chapter{Free Higher Spin Gauge fields}\label{free}

\setcounter{equation}{0}

\section{Technical setup and important relations in free HSF theory}\label{techsetup}

\quad We work with Higher Spin Gauge Fields in Fronsdal (metric-like) formulation \cite{Fronsdal:1978rb,Fronsdal:1978vb}, in which the spin $s$ field is double traceless fully symmetric s-th rank tensor. The most elegant and convenient  way of handling symmetric tensors such as $h^{(s)}_{\mu_1\mu_2...\mu_s}(z)$ is by
contracting it with the $s$'th tensorial power of a vector $a^{\mu}$ of the tangential space at
the base point $z$
\begin{equation}
h^{(s)}(z;a) = \sum_{\mu_{i}}(\prod_{i=1}^{s} a^{\mu_{i}})h^{(s)}_{\mu_1\mu_2...\mu_s}(z) .
\label{2.5}
\end{equation}
In this way we obtain a homogeneous polynomial in the vector $a^{\mu}$ of degree $s$.
In this formalism the symmetrized gradient, trace and divergence are\footnote{To distinguish easily between "a" and "z" spaces we introduce for space-time derivatives $\frac{\partial}{\partial z^{\mu}}$ the notation $\nabla_{\mu}$.}
\begin{eqnarray}
&&Grad:h^{(s)}(z;a)\Rightarrow Gradh^{(s+1)}(z;a) = (a\nabla)h^{(s)}(z;a) , \\
&&Tr:h^{(s)}(z;a)\Rightarrow Trh^{(s-2)}(z;a) = \frac{1}{s(s-1)}\Box_{a}h^{(s)}(z;a) ,\\
&&Div:h^{(s)}(z;a)\Rightarrow Divh^{(s-1)}(z;a) = \frac{1}{s}(\nabla\partial_{a})h^{(s)}(z;a) .
\end{eqnarray}
Next we introduce the notation $*_a, *_b$ for a contraction in the symmetric spaces of indices $a$ or $b$
\begin{eqnarray}
  *_{a}&=&\frac{1}{(s!)^{2}} \prod^{s}_{i=1}\overleftarrow{\partial}^{\mu_{i}}_{a}\overrightarrow{\partial}_{\mu_{i}}^{a} .
   \label{star}
\end{eqnarray}
To manipulate reshuffling of different sets of indices we employ other differentials with respect to $a$ and $b$, e.g. $(a\partial_{b})$ or $(b\partial_{a})$.
Then we see that operators $(a\partial_{b}), a^{2}, b^{2}$ are dual (or adjoint) to $(b\partial_{a}),\Box_{a},\Box_{b}$ with respect to the "star" product of tensors with two sets of symmetrized indices  (\ref{star})
\begin{eqnarray}
    \frac{1}{n}(a\partial_{b})f^{(m-1,n)}(a,b)*_{a,b} g^{(m,n-1)}(a,b)&=& f^{(m-1,n)}(a,b)*_{a,b} \frac{1}{m}(b\partial_{a})g^{(m,n-1)}(a,b),\qquad\quad \label{4.13}\\
    a^{2}f^{(m-2,n)}(a,b)*_{a,b} g^{(m,n)}(a,b)&=&f^{(m-2,n)}(a,b)*_{a,b} \frac{1}{m(m-1)}\Box_{a} g^{(m,n)}(a,b) \nonumber\\
\textnormal{(an analogous equation for $b^2$)}
\label{4.14}
\end{eqnarray}
In the same fashion gradients and divergences are dual with respect to the full scalar product in the space $(z;a,b)$
\begin{eqnarray}
  (a\nabla)f^{(m-1,n)}(z;a,b)*_{a,b} g^{(m,n)}(z;a,b) &=& -f^{(m-1,n)}(z;a,b)*_{a,b}\frac{1}{m}(\nabla\partial_{a})g^{(m,n)}(z;a,b) \nonumber\\
\textnormal{(an analogous equation for $(b\nabla)$)}
   \label{dual}
  \end{eqnarray}
We will use the deWit-Freedman curvature and Christoffel symbols \cite{deWit:pe, MR6}.
The n-th deWit-Freedman-Christoffel symbol is
\begin{eqnarray}
\Gamma_{(n)}^{(s)}(z;b,a)\equiv \Gamma^{(s)}_{(n)\rho_{1}...\rho_{n},\mu_{1}...\mu_{\ell}}b^{\rho_{1}}...b^{\rho_{n}}
a^{\mu_{1}}...a^{\mu_{\ell}}=[(b\nabla)-\frac{1}{n}(a\nabla)(b\partial_{a})]\Gamma_{(n-1)}^{(s)}(z;b,a),\quad\quad
\end{eqnarray}
or in another way
\begin{equation}
\Gamma_{(n)}^{(s)}(z;b,a)=(\prod_{k=1}^{s}[(b\nabla)-\frac{1}{k}(a\nabla)(b\partial_{a})])h^{(s)}(z;a) .
\end{equation}
We contracted them with the degree $s$
tensorial power of one tangential vector $a^{\mu}$ in the first set of s indices  and with a similar tensorial power of another tangential vector $b^{\nu}$ in its second set. The deWit-Freedman curvature and n-th Christoffel symbol  are then written as
\begin{eqnarray}
&&\Gamma^{(s)}(z;b,a):\qquad\Gamma^{(s)}(z; b,\lambda a) = \Gamma^{(s)}(z;\lambda b, a)= \lambda^{s}\Gamma^{(s)}(z;b, a) ,\\
&&\Gamma^{(s)}_{(n)}(z;b,a):\qquad\Gamma^{(s)}_{(n)}(z;b,\lambda a) = \lambda^{s}\Gamma^{(s)}_{(n)}(z;b, a) ,\\
&&\quad\quad\quad\quad\quad\quad\quad\,\,\,\,\Gamma^{(s)}_{(n)}(z; \lambda b, a)=\lambda^{n}\Gamma^{(s)}_{(n)}(z;b, a) ,\\
&&\Gamma^{(s)}(z;b,a)=\Gamma^{(s)}_{(n)}(z;b,a)|_{n=s} .
\label{2.6}
\end{eqnarray}

Now one can prove that \cite{deWit:pe, MR5}:
\begin{equation}
(a\partial_{b})\Gamma^{(s)}(z;a,b) = (b\partial_{a})\Gamma^{(s)}(z;a,b) = 0 .
\label{Bianchi}
\end{equation}
These "primary Bianchi identities" are manifestations of the hidden antisymmetry.

Using the following commutation relations
\begin{eqnarray}
&&[(b\partial_{a})^{n},(a\nabla)^{m}]=\sum_{k=1}^{min\{n,m\}}k!\binom{n}{k}\binom{m}{k}(b\nabla)^{k}(a\nabla)^{m-k}(b\partial_{a})^{n-k},\label{commutator}\\
&&\Box_{b}(b\nabla)^{i}=i(i-1)(b\nabla)^{i-2}\Box,\\
&&\partial^{b}_{\mu}(b\nabla)^{i}\partial_{b}^{\mu}(b\partial_{a})^{j}=ij(b\nabla)^{i-1}(b\partial_{a})^{j-1}(\nabla\partial_{a}),\\
&&\Box_{b}(b\partial_{a})^{j}=j(j-1)(b\partial_{a})^{j-2}\Box_{a} ,\label{4.26}
\end{eqnarray}
and mathematical induction we can prove that
\begin{eqnarray}
\Gamma_{(n)}^{(s)}(z;b,a)=\sum_{k=0}^{n}\frac{(-1)^{k}}{k!}(b\nabla)^{n-k}(a\nabla)^{k}(b\partial_{a})^{k}h^{(s)}(z;a) .\label{gumar}
\end{eqnarray}
The gauge variation of a spin $s$ field is
\begin{eqnarray}\label{variation}
\delta h^{(s)}(z;a)=s (a\nabla)\epsilon^{(s-1)}(z;a) ,
\end{eqnarray}
with traceless gauge parameter
\begin{eqnarray}
\Box_{a}\epsilon^{(s-1)}(z;a)=0 ,\label{tracelessnes}
\end{eqnarray}
for the double traceless gauge field
\begin{eqnarray}
\Box_{a}^{2}h^{(s)}(z;a)=0 .
\end{eqnarray}
The gauge variation of n-th Christoffel symbol is
\begin{eqnarray}
&&\delta \Gamma_{(n)}^{(s)}(z;b,a)=\frac{(-1)^{n}}{n!}(a\nabla)^{n+1}(b\partial_{a})^{n}\epsilon^{(s-1)}(z;a) ,
\end{eqnarray}
putting here $n=s$ we obtain gauge invariance for the curvature
\begin{equation}\label{curvinv}
\delta \Gamma_{(s)}^{(s)}(z;b,a)=0 .
\end{equation}
Tracelessness of the gauge parameter (\ref{tracelessnes})
implies that b-traces of all Christoffel symbols are gauge invariant
\begin{eqnarray}
&& \Box_{b}\delta \Gamma_{(n)}^{(s)}(z;b,a)=\frac{(-1)^{n}}{(n-2)!}(a\nabla)^{n+1}(b\partial_{a})^{n-2}\Box_{a}\epsilon^{(s-1)}(z;a)=0 .
\end{eqnarray}
Thus for the second order gauge invariant field equation we can use the trace of the second Christoffel symbol,
the so called Fronsdal tensor:
\begin{eqnarray}
\mathcal{F}^{(s)}(z;a)&=&\frac{1}{2}\Box_{b}\Gamma_{(2)}^{(s)}(z;b,a)\nonumber\\
&=&\Box h^{(s)}(z;a)-(a\nabla)(\nabla\partial_{a})h^{(s)}(z;a)
+\frac{1}{2}(a\nabla)^{2}\Box_{a}h^{(s)}(z;a) .\quad\quad\label{4.32}
\end{eqnarray}
Using equation (\ref{gumar}) for Christoffel symbols and
after long calculations we obtain the following expression
\begin{eqnarray}
\Box_{b}\Gamma_{(n)}^{(s)}(z;b,a)=\sum_{k=0}^{n-2}\frac{(-1)^{k}}{k!}(n-k)(n-k-1)
(b\nabla)^{n-k-2}(a\nabla)^{k}(b\partial_{a})^{k}\mathcal{F}^{(s)}(z;a) .\quad\quad
\end{eqnarray}
We have expressed the b-trace of any $\Gamma_{(n)}^{(s)}$ through the Fronsdal tensor\footnote{Which means that all b-traces of all Christoffel symbols are zero On-shell when Fronsdal equation $\mathcal{F}^{(s)}(z;a)=0$ holds.} (or the b-trace of the second Christoffel symbol, which is the same (\ref{4.32})), which means that there are only two nontrivial gauge invariant objects--Fronsdal tensor and deWit-Freedman curvature of higher spin gauge field. But this is not the whole story. Using mathematical induction and (\ref{commutator})-(\ref{4.26}) again we can show that
\begin{eqnarray}
&&\sum_{k=0}^{n-2}\frac{(-1)^{k}}{k!}(n-k)(n-k-1)(b\nabla)^{n-k-2}(a\nabla)^{k}(b\partial_{a})^{k}\mathcal{F}^{(s)}(z;a)\nonumber\\
&&\quad\quad\quad\quad=n(n-1)(\prod^{n}_{k=3}[(b\nabla)-\frac{1}{k}(a\nabla)(b\partial_{a})])\mathcal{F}^{(s)}(z;a) .
\end{eqnarray}
In particular for the trace of the curvature we can write
\begin{eqnarray}
&&\Box_{b}\Gamma^{(s)}(z;b,a)=s(s-1)\mathcal{U}(a,b,3,s)\mathcal{F}^{(s)}(z;a) ,\label{4.35}
\end{eqnarray}
where we introduced an operator mapping the Fronsdal tensor on the trace of the curvature
\begin{equation}\label{4.36}
   \mathcal{U}(a,b,3,s)=\prod^{s}_{k=3}[(b\nabla)-\frac{1}{k}(a\nabla)(b\partial_{a})] .
\end{equation}

Now let us consider this curvature in more detail. First we have the symmetry under exchange of $a$ and $b$
\begin{equation}
\Gamma^{(s)}(z;a,b) = (-1)^{s}\Gamma^{(s)}(z;b,a) .
\end{equation}
Therefore the operation "$a$-trace" can be defined by (\ref{4.35}) with exchange of $a$ and $b$ at the end.
The mixed trace of the curvature can be expressed through the $a$ or $b$ traces using "primary Bianchi identities" (\ref{Bianchi})
\begin{equation}\label{4.38}
(\partial_{a}\partial_{b})\Gamma^{(s)}(z;b,a)=-\frac{1}{2}(b\partial_{a})\Box_{b}\Gamma^{(s)}(z;b,a)=
-\frac{1}{2}(a\partial_{b})\Box_{a}\Gamma^{(s)}(z;b,a) .
\end{equation}

The next interesting properties of the higher spin curvature and corresponding Ricci tensors are so called generalized secondary or differential Bianchi identities. We can formulate  these identities  in our notation in the following compressed form ($[\dots]$ denotes antisymmetrization )
\begin{equation}\label{4.39}
\frac{\partial}{\partial a^{[\mu}}\frac{\partial}{\partial b^{\nu}}\nabla_{\lambda]}\Gamma^{(s)}(z;a,b)= 0 .
\end{equation}
This relation can be checked directly from representation (\ref{gumar}). Then contracting with $a^{\mu}$ and $b^{\nu}$ we get a symmetrized form of (\ref{4.39})
\begin{equation}\label{4.40}
    s \nabla_{\mu}\Gamma^{(s)}(z;a,b)=(a\nabla)\partial^{a}_{\mu}\Gamma^{(s)}(z;a,b)+(b\nabla)\partial^{b}_{\mu}\Gamma^{(s)}(z;a,b) .
\end{equation}
Now we can contract (\ref{4.40}) with a $\partial^{\mu}_{b}$ and using (\ref{4.38}) obtain a connection between the divergence and the trace of the curvature
\begin{equation}\label{4.41}
    (s-1)(\nabla\partial_{b})\Gamma^{(s)}(z;a,b)=[(b\nabla)-\frac{1}{2}(a\nabla)(b\partial_{a})]\Box_{b}\Gamma^{(s)}(z;a,b) .
\end{equation}
These two identities with a similar identity for the Fronsdal tensor
\begin{equation}\label{BianchiF}
    (\nabla\partial_{a})\mathcal{F}^{(s)}(z;a)=\frac{1}{2}(a\nabla)\Box_{a}\mathcal{F}^{(s)}(z;a) ,
\end{equation}
play an important role for the construction of the interaction Lagrangian. To complete the free field information we present here Fronsdal's Lagrangian in terms of our quantities:

\begin{equation}\label{4.42}
 \mathcal{L}_{0}(h^{(s)}(a))=-\frac{1}{2}h^{(s)}(a)*_{a}\mathcal{F}^{(s)}(a)
    +\frac{1}{8s(s-1)}\Box_{a}h^{(s)}(a)*_{a}\Box_{a}\mathcal{F}^{(s)}(a) .
\end{equation}
To obtain the equation of motion we vary (\ref{4.42}) and obtain
\begin{equation}\label{lagrangianvariation}
     \delta\mathcal{L}_{0}(h^{(s)}(a))=-(\mathcal{F}^{(s)}(a)-\frac{a^{2}}{4}\Box_{a}\mathcal{F}^{(s)}(a))*_{a}\delta h^{(s)}(a) .
\end{equation}
Zero order gauge invariance can be  checked easily by substitution of (\ref{variation}) into this variation and use of the duality relation (\ref{dual}) and identity (\ref{BianchiF}) taking into account tracelessness  of the gauge parameter (\ref{tracelessnes}).



\chapter{Propagator of HSF and One-Loop Diagram}\label{Propagator}

\section{General setup for higher spin propagators}
\setcounter{equation}{0}
\quad Here we would like to give all the conventions about $AdS_{d+1}$ metric\footnote{We will always try to keep general $d$ in all possible formulas admitting of course that for $AdS_{4}$ theory it should be set to $3$ at the end.}.
We work in Euclidian $AdS_{d+1}$ with the following metric,
curvature and covariant derivatives:
\begin{eqnarray}
&&ds^{2}=g_{\mu \nu }(z)dz^{\mu }dz^{\nu
}=\frac{L^{2}}{(z^{0})^{2}}\delta _{\mu \nu }dz^{\mu }dz^{\nu
},\quad \sqrt{g}=\frac{L^{d+1}}{(z^{0})^{d+1}}\;,
\notag  \label{ads} \\
&&\left[ \nabla _{\mu },\,\nabla _{\nu }\right] V_{\lambda }^{\rho }=R_{\mu
\nu \lambda }^{\quad \,\,\sigma }V_{\sigma }^{\rho }-R_{\mu \nu \sigma
}^{\quad \,\,\rho }V_{\lambda }^{\sigma }\;,  \notag \\
&&R_{\mu \nu \lambda }^{\quad \,\,\rho
}=-\frac{1}{(z^{0})^{2}}\left( \delta _{\mu \lambda }\delta _{\nu
}^{\rho }-\delta _{\nu \lambda }\delta _{\mu }^{\rho }\right)
=-\frac{1}{L^{2}}\left( g_{\mu \lambda }(z)\delta _{\nu
}^{\rho }-g_{\nu \lambda }(z)\delta _{\mu }^{\rho }\right) \;,  \notag \\
&&R_{\mu \nu }=-\frac{d}{(z^{0})^{2}}\delta _{\mu \nu }=-\frac{d}{L^{2}}%
g_{\mu \nu }(z)\quad ,\quad R=-\frac{d(d+1)}{L^{2}}\;.  \notag
\end{eqnarray}%
For simplicity we will from now on put $L=1$ during all calculations keeping in mind that we can always restore the $AdS$ radius from dimensional consideration.

Then we can write the starting point of the investigation of higher spin gauge field propagators, namely Fronsdal's equation of motion (we introduce here the ``geometric AdS mass'' $\mu^{2}_{\ell}$)
\cite{Fronsdal:1978vb} for the double traceless spin $\ell$ field, $\Box _{a}\Box _{a}h^{(\ell)}=0$:
\begin{eqnarray}
\mathcal{F}(h^{(\ell )}(z;a))&=&[\Box-\mu^{2}_{\ell}] h^{(\ell )}(z;a)-a^{2}\Box
_{a}h^{(\ell)}(z;a)\quad   \nonumber  \\
&-&(a\nabla )\Big[\nabla ^{\mu
}\frac{\partial }{\partial a^{\mu }}h^{(\ell )}-\frac{1}{2}(a\nabla
)\Box _{a}h^{(\ell )}(z;a)\Big]=0 ,\label{FronsAdS}\\
\mu^{2}_{\ell}&=&\left( \ell ^{2}+\ell (d-5)-2(d-2)\right) ,  \label{AdSmass}
\end{eqnarray}
The most important property of this equation is higher spin gauge invariance with the
traceless parameter $\epsilon ^{(\ell -1)}(z;a),$
\begin{eqnarray}
\delta h^{(\ell )}(z;a)&=&(a\nabla )\epsilon ^{(\ell -1)}(z;a),\label{1.7}\\
\Box_{a}\epsilon ^{(\ell -1)}(z;a)&=&0,\label{1.8}\\
\delta \mathcal{F}(h^{(\ell)}(z;a))&=&0.  \label{1.9}
\end{eqnarray}%
The most natural gauge fixing condition for Fronsdal's equation is the so called \emph{traceless} de Donder gauge
\begin{eqnarray}
&&\mathcal{D}^{(\ell -1)}(h^{(\ell )})=\nabla ^{\mu }\frac{\partial }{%
\partial a^{\mu }}h^{(\ell )}-\frac{1}{2}(a\nabla )\Box _{a}h^{(\ell )}=0,
\label{1.11}
\end{eqnarray}
In this gauge Fronsdal's equation simplifies to
\begin{eqnarray}
&&\mathcal{F}^{dD}(h^{(\ell )})=[\Box-\mu^{2}_{\ell}] h^{(\ell )}-a^{2}\Box _{a}h^{(\ell)}=0.\quad \quad \label{FronsdD}
\end{eqnarray}
Note that any deviation in (\ref{FronsdD}) of $\mu^{2}_{\ell}$ from (\ref{AdSmass}) leads to a massive higher spin field.

We can write now Fronsdal's  gauge invariant action in the concise form
\begin{eqnarray}
  \mathcal{S^{\ell}} = \frac{1}{2} \int\sqrt{g}d^{d+1}z\left\{h^{(\ell )}(z;a)*_{a}\mathcal{F}(h^{(\ell )}(z;a))-\frac{1}{4s(s-1)}\Box_{a}h^{(\ell )}(z;a)*_{a}\Box_{a}\mathcal{F}(h^{(\ell )}(z;a))\right\}\quad\label{1.16}
\end{eqnarray}
This action is gauge invariant due to the "Bianchi" identity (\ref{BianchiF}).

Next we can write our double traceless field $h^{(\ell)}(z;a)$ as a set of the two traceless spin $\ell$ and $\ell-2$
fields $\psi^{(\ell)}(z;a)$ and $\theta^{(\ell-2)}(z;a)$
\begin{eqnarray}
&& h^{(\ell)}(z;a)= \psi^{(\ell)}+\frac{a^{2}}{2\alpha_{0}}
\theta^{(\ell-2)}(z;a)\quad,  \label{1.17} \\
&&\Box_{a}h^{(\ell)}=\theta^{(\ell-2)}\quad,\quad
\Box_{a}\psi^{(\ell)}=\Box_{a}\theta^{(\ell-2)}=0 ,\label{1.18p}\\
&&\alpha_{0}=d+2\ell-3 .\label{1.19p}
\end{eqnarray}
Applying the de Donder gauge condition we see that the fields $\psi^{(\ell)}(z;a)$ and $\theta^{(\ell-2)}(z;a)$ completely separate in the action (\ref{1.16})
\begin{eqnarray}
  \mathcal{S^{\ell}} &=& \frac{1}{2} \int\sqrt{g}d^{d+1}z\Big\{\psi^{(\ell )}(z;a)*[\Box-\mu^{2}_{\ell}]\psi^{(\ell )}(z;a) \nonumber\\
  &-&\frac{\alpha_{0}-2}{4\alpha_{0}}\theta^{(\ell-2)}(z;a)*[\Box-\mu^{2}_{\theta^{(\ell-2)}}]\theta^{(\ell-2)}(z;a)\Big\} ,\label{1.20p}\\
  \mu^{2}_{\theta^{(\ell-2)}}&=& \mu^{2}_{\ell}+2\alpha_{0}=\ell(\ell+d-1)-2 ,\label{1.21p}
\end{eqnarray}
with the following diagonal field equations and de Donder gauge condition connecting  $\psi^{(\ell)}$ and $\theta^{(\ell-2)}$
\begin{eqnarray}
&& \nabla^{\mu}\frac{\partial}{\partial a^{\mu}}\psi^{(\ell)}=
\frac{\alpha_{0}-2}{2\alpha_{0}}(a\nabla)\theta^{(\ell-2)}-\frac{a^{2}}{%
2\alpha_{0}} \nabla^{\mu}\frac{\partial}{\partial
a^{\mu}}\theta^{(\ell-2)},\label{dD}\\
&&\left(\Box+\ell\right)\psi^{(\ell)}=\Delta_{\ell}(\Delta_{\ell}-d)
\psi^{(\ell)} ,\qquad  \label{1.23p} \\
&&\left(\Box+\ell-2\right)\theta^{(\ell-2)}=\Delta_{\theta}(\Delta_{%
\theta}-d)\theta^{(\ell-2)} ,
\label{1.24} \\
&&\Delta_{\ell}=d+\ell-2\quad,\quad \Delta_{\theta}=d/2+1/2\sqrt{(\alpha_{0}-1)(\alpha_{0}+7)} .\label{1.25}
\end{eqnarray}
So we realize that only in the de Donder gauge we have a diagonal
equation of motion for the physical $\psi^{(\ell)}$ components but
this component is not transversal due to the presence of
$\theta^{(\ell-2)}$ . This is the
most convenient gauge for the quantization and construction of
bulk-to-bulk propagators and for the investigation of
$AdS_{4}/CFT_{3}$ correspondence in the case of the critical
conformal $O(N)$ boundary sigma model. We also mentioned that in the boundary limit only the traceless mode $\psi^{(\ell)}$ survives but the nonphysical trace mode $\theta^{(\ell-2)}$ can create a Goldstone mode and enters the bulk tree dynamics and the loops.

The negative sign of the $\theta$ part in the action (\ref{1.16}) suggests to quantize this
higher spin field with a formalism of Gupta-Bleuler type, so that a state with $n$ quanta of $\theta$ has norm squared of signature $(-1)^{n}$ yielding a pseudo Hilbert space. Applying de Donder's constraint (\ref{dD}) on field operators, the ``physical'' Hilbert space is the kernel of the annihilation operator part of this constraint inside the pseudo Hilbert space. Finally zero norm states are projected out. In the context of this work it is only
relevant that the two-point function of $\theta$ satisfies
\begin{equation}
\big<\theta^{(\ell-2)}(z_1;a),\theta^{(\ell-2)}(z_2;c)\big> \quad \leq \quad 0
\label{1.26}
\end{equation}
as a distribution.

\section{Propagators in de Donder's gauge}
\setcounter{equation}{0}
\quad
On AdS space which is a constant curvature space the geodesic distance $\eta$ enters all invariant expressions of the relative distance of two points. The standard variable $\zeta = \cosh \eta$ can be expressed by Poincar\'e coordinates
as
\begin{equation}
        \zeta(z_{1},z_{2})=\frac{(z^{0}_{1})^{2}+(z^{0}_{2})^{2}+(\vec{z}_{1}-\vec{z}_{2})^{2}}
        {2z^{0}_{1}z^{0}_{2}}=1+\frac{(z_{1}-z_{2})^{\mu}(z_{1}-z_{2})^{\nu}
        \delta_{\mu\nu}}{2z^{0}_{1}z^{0}_{2}} .
 \end{equation}
The propagators are bitensorial quantities which are presented in the algebraic basis
of homogeneous functions of $I_1, I_2, I_3, I_4$
\begin{eqnarray}
       && I_{1}(a,c):=(a\partial_{1})(c\partial_{2})\zeta(z_{1},z_{2}) , \\
       && I_{2}(a,c):=(a\partial_{1})\zeta(z_{1},z_{2})(c\partial_{2})\zeta(z_{1},z_{2}),\\
       && I_{3}(a,c):=a^{2}_{1}I^{2}_{2c}+c^{2}_{2}I^{2}_{1a} , \\
       && I_{4}:=a^{2}_{1}c^{2}_{2} ,\\
       && I_{1a}:=(a\partial_{1})\zeta(z_{1},z_{2})\quad ,
       \quad I_{2c}:=(c\partial_{2})\zeta(z_{1},z_{2}) , \\
       &&(a\partial_{1})=a^{\mu}\frac{\partial}{\partial
       z_{1}^{\mu}} ,\quad (c\partial_{2})=c^{\mu}\frac{\partial}{\partial
       z_{2}^{\mu}} ,\\&& a^{2}_{1}=g_{\mu\nu}(z_{1})a^{\mu}a^{\nu} ,
       \quad c^{2}_{2}=g_{\mu\nu}(z_{2})c^{\mu} c^{\nu} .
\end{eqnarray}
of degree $\ell$, the spin of the field. All important formulas for this "advanced technology" of working with higher spin field theory in $AdS$ space one can find in Appendix A.
We are interested only in that part of the propagator expansion which neglects traces. So it is a map from a space of $\ell+1$ functions $\left\{F_{k}(\zeta)\right\}_{k=0}^{\ell}$ to a space of bitensors  parameterized by $I_{1}$ and $I_{2}$ only, namely
\begin{eqnarray}
 \Psi^{(\ell)}[F_{k}]& =& \sum_{k=0}^{\ell} F_{k}(\zeta) I_{1}^{\ell-k} I_{2}^{k} ,\label{2.0}\\
 \left(\Box+\ell\right)\Psi^{(\ell)}[F_{k}]&=&\Delta_{\ell}(\Delta_{\ell}-d)\Psi^{(\ell)}[F_{k}] +O(a^{2}_{1};c^{2}_{2}) .
\end{eqnarray}

In the variable $\zeta$ the
analytic properties of QFT n-point functions are conveniently described.
In particular the two-point functions or propagators are analytic in the $\zeta$ plane with singularities at $\zeta = \pm 1$ and at $\zeta = \infty$, which in most cases are
logarithmic branch points. Analyticity is therefore meant in general on infinite covering planes. All $AdS$ field theories are symmetric under the exchange $\zeta$ against $-\zeta$.

Another variable used often is $u = \zeta -1$, the ``chordal distance'', more precisely one half the square of the chordal distance. The series expansions for two-point functions in $u$ converge in a radius $2$,
whereas the series expansions in powers of $\zeta^{-1}$ converge for $\mid\zeta\mid > 1$.
These analytic properties remind us of Legendre functions. Indeed if propagator functions can be identified as Gaussian hypergeometric functions, these are Legendre functions and the "quadratic transformations" can be applied.
Using formulas from Appendix A we can show that
in de Donder's gauge the propagator satisfy the following set of differential equations
for the functions $F_{k}(\zeta)$ or correspondingly $\Phi_{k}(u)$ of (\ref{2.0}) following from equation (\ref{1.23p})
\begin{eqnarray}
(\zeta^{2} -1)F_{k}'' +(d+1+4k)\zeta F_{k}' +X_{k}F_{k} +2\zeta(k+1)^{2}F_{k+1}
+2(\ell-k+1)F_{k-1}' = 0 ,\quad\quad\label{2.01}\\
X_{k} = k(d+2\ell-k) +2l - (\ell-2)(\ell+d-2) .\qquad\qquad\qquad
\label{2.03}
\end{eqnarray}
The "dimension" of the higher spin field $\Delta_{\ell} = \ell+d-2$
has been inserted. Moreover we use $F_{-1} = F_{\ell+1} = 0$. The dimension of the
AdS space is $d+1$, we interpolate analytically in $d$ if this is technically required.
Our issue is to solve these equations by expansion in powers of $\zeta^{-1}$ or $u$.
This leads to matrix recursion equations which necessitate some linear algebra operations.

As an ansatz for the series expansion of $F_{k}(\zeta)$ at $\zeta = \infty$ we use
\begin{equation}
F_{k}(\zeta) = \zeta^{-\alpha -k}\sum_{n=0}^{\infty} c_{kn} \zeta^{-2n} .
\label{2.1}
\end{equation}
Denote $\xi = \alpha +2n $. Then a two term recursion of the form
\begin{eqnarray}
D_{n} \left(\begin{array}{ccc}  c_{0n}\\c_{1n}\\ \vdots \\c_{\ell,n}
\end{array}\right) = C_{n-1} \left( \begin{array}{ccc} c_{0,n-1}\\ c_{1,n-1}\\
 \vdots \\ c_{\ell,n-1} \end{array} \right) ,
\label{2.2}
\end{eqnarray}
results with the two matrices
\begin{equation}
C_{n-1} = diag\{(\xi -1)(\xi -2),\xi (\xi-1), \ldots (\xi+\ell-1)(\xi +\ell-2)\} ,
\label{2.3}
\end{equation}
and the entries of the matrix $D_{n}$
\begin{eqnarray}
(D_{n})_{k,k-1} &=& -2(\ell-k+1)(\xi+k-1) ,\label{2.4}\\
(D_{n})_{k,k}  &=& \xi^{2} -\xi (d+2k) -4k^{2}+2\ell(k+1) -(l-2)(\ell+d-2)\ \ \ \label{2.5p} \\
(D_{n})_{k,k+1} &=& 2(k+1)^{2} .\label{2.6p}
\end{eqnarray}
The determinant of $D_{0}$ is a polynomial of degree $2(\ell+1)$ of the variable $\alpha$
with roots which we identify with the "roots" of the differential equation system.
For arbitrary $\ell$ we have
\begin{eqnarray}
det D_{0} = [(\alpha +\ell-2)(\alpha +2-\ell-d)][(\alpha +\ell-2)(\alpha -\ell-d)] \nonumber\\ \times \prod_{n=0}^{\ell-2}[\alpha^{2} -(d+4+2n)\alpha -((\ell-2)d + (\ell+n)^{2}-(n+2)(3n+4))] .
\label{2.7}
\end{eqnarray}
Each square bracket represents one eigenvalue of $D_{0}$ and contributes two roots. The quadratic factors lead in almost all cases to two irrational roots that are neither degenerate among themselves nor with the other roots, but there are exceptions which have two integer roots e.g. for $d=3: (\ell,n)\in \{(4,1), (6,4), (9,2),\\ (9,5), (11,8), (15,8)... \}$. Two roots are said to be degenerate, if their difference is an integer. For the case of expansions in powers of $\zeta^{2}$ as in (\ref{2.1}), this integer must be even. In such case the solution with the bigger root enters the other one with a $log\zeta$ factor.

The following roots are of particular (physical) importance
\begin{eqnarray}
\alpha_{p} &=& \ell+d-2 ,\label{2.8}\\
\alpha_{c} &=& \ell+d .
\label{2.9}
\end{eqnarray}
We call the first root $\alpha_{p}$ "principal"
because it has the value of the dimension $\Delta$ of the field which enters the field
equation in the form $\Delta(\Delta-d)$. The second root is a "companion" of it, since they appear for all $\ell$ as such pair (see (\ref{2.7})). It is degenerate with the principal root and the solution of it
enters the principal solution with a $log \zeta$ factor on the next to leading power in the expansion. The bigger ones of the two roots in the exceptional cases quoted above are also bigger than the principal root $\ell +1$ (for the same $\ell$) but their distance to it are odd numbers except for the case $(\ell,n) = (15,8)$, where the distance to $\ell+1$ is sixteen
and the $log\zeta$ term appears at a very high power.

For the principal root the equation for the eigenvector of $D_{0}$
\begin {eqnarray}
D_{0}(\alpha_{p}) \left (\begin{array}{ccc}  c_{00}^{(\alpha_{p})}\\ c_{10}^{(\alpha_{p})}\\ \vdots \\ c_{l0}^{(\alpha_{p})} \end{array} \right) = 0 ,
\label{2.10}
\end{eqnarray}
can be solved for each $\ell$. We find
\begin{equation}
c_{k,0}^{(\alpha_{p})} = (-1)^{k} \binom{\ell}{k} ,
\label{2.11p}
\end{equation}
which is easy to prove by using the general expression for the rows of the matrix $D_{n}$
as given in (\ref{2.4}) - (\ref{2.6p}). The consequence of this result is that the leading term
of $\Psi^{(\ell)}[F_{k}(\alpha_{p})]$ at $\zeta = \infty$ is the well known expression $\zeta^{-\Delta}(I_{1} - \zeta^{-1}I_{2})^{\ell}$. Already at next order in $\zeta^{-2}$ log-terms appear.

For the companion root $\alpha_{c}$ the eigenvector for $D_{0}$ can be derived by
a little bit more algebra for any $\ell$
\begin{equation}
c_{k,0}^{(\alpha_{c})} = (-1)^{k}\left( \binom{\ell}{k} +(d+2\ell-2) \binom{\ell-1}{k-1} \right) .
\label{2.13}
\end{equation}
The actual construction of a solution for the pair of roots starts with the bigger one, $\alpha_{c}$. Its solution takes the form
\begin{equation}
F_{k}(\zeta; \alpha_{c}) = \zeta^{-\Delta -2} \sum_{n=0}^{\infty} \zeta^{-2n}\sum_{s=0}^{\ell}
\Pi_{n}(\alpha_{c})_{k,s}c_{s,0}^{(\alpha_{c})} ,
\label{2.14}
\end{equation}
where we used
\begin{eqnarray}
H_{n}(\alpha_{c}) &=& D_{n}(\alpha_{c})^{-1} C_{n-1}(\alpha_{c})\nonumber\\
                  &=& H_{1}(\alpha_{c} +2(n-1)) ,\label{2.15}\\
\Pi_{n}(\alpha_{c}) &=& \Pi_{r=0}^{n-1,\leftarrow}H_{1}(\alpha_{c} +2r) .
\label{2.16}
\end{eqnarray}
and the left arrow denotes ordering of the product with increasing $r$ from right to left.
In this context we note that if a nonsingular matrix $S(\alpha)$ would exist, so that $H_{1}$ could be diagonalized by
\begin{equation}
H_{1}(\alpha) = S^{-1}(\alpha +2) \Delta(\alpha) S(\alpha) ,
\label{2.17}
\end{equation}
then $F_{k}(\zeta;\alpha)$ would be a generalized hypergeometric function.

Having constructed the solution for the companion root we turn to the principal root.
We recognize that $D_{n}(\alpha_{p})$ can be spectrally decomposed in the following fashion
\begin{eqnarray}
D_{n}\chi_{i} &=& \lambda_{i}\chi_{i} ,\label{2.18}\\
D_{n}^{T}\psi_{i} &=& \lambda_{i}\psi_{i} ,\label{2.19}\\
D_{n} &=& \sum_{i=0}^{l} \lambda_{i} \chi_{i} \otimes \psi_{i}^{T} ,\label{2.20}\\
 \psi_{i}^{T} \chi_{j} &=&\delta_{ij} \label{2.21}
\end{eqnarray}
Denote further
\begin{equation}
\rho ^{T} = \psi^{T} C_{n-1} .
\label{2.22}
\end{equation}
All these quantities can be determined as functions of $\xi$, and it is easily verified that
(\ref{2.17}) is not fulfilled.

One of the eigenvalues of $D_{1}(\alpha_{p})$ vanishes, we denote it $\lambda_{0}$,
so that $D_{1}(\alpha_{p})$ cannot be inverted. We perform a deformation
of our differential equation system replacing $\alpha_{p}$ only in $\lambda_{0}$
and in the prefactor $\zeta^{-\alpha_{p}}$ by $\alpha_{p} + \epsilon$. All other eigenvalues and the eigenvectors remain unchanged. Then we continue the whole procedure known from the
companion root, all $H_{n}$ will remain singularity free. At the end we subtract a certain multiple $\gamma$ of
$(\epsilon^{-1}+\mu)\Psi^{(\ell)}[F_{k}(\alpha_{c})]$ so that the limit $\epsilon \rightarrow 0$
can be performed and the log-terms appearing are $-\gamma log\zeta \Psi^{(l)}[F_{k}(\alpha_{c})]$. The additional parameter $\mu$ is in principle arbitrary
showing that the principal solution containing a log factor is a coset with respect to adding the companion solution. This parameter can, however, be normalized in a standard fashion by requiring that the $(l+1)$-tupel of coefficients $c_{k,n}^{(\alpha_{p})}$ where at level $n$ the log term appears first, is orthogonal to the eigenvector $\psi_{0}$ belonging to the deformed eigenvalue.
We close this discussion with the remark that on the boundary of AdS space i.e. $\zeta = \infty$ any linear combination
\begin{equation}
\Psi^{(\ell)}[F_{k}(\alpha_{p})] + A \Psi^{(\ell)}[F_{k}(\alpha_{c})]
\label{2.23}
\end{equation}
is indistinguishable from the pure principal solution. Thus the boundary constraint
fixes only the whole coset and not any representative of it.

In order to render the expansions of $F_{k}$  around $\zeta =1(u=0)$ a visually different expression, we shall denote them $\Phi_{k}$. The expansions are
\begin{equation}
\Phi_{k}(u) = u^{\alpha}\sum_{n=0}^{\infty}a_{k,n}u^{n} .
\label{3.1p}
\end{equation}
Again we obtain matrix recursion relations
\begin{eqnarray}
A_{n}\left( \begin{array} {ccc}a_{0,n}\\ a_{1,n}\\ \vdots\\ a_{\ell,n}\end{array} \right) +
B_{n-1} \left( \begin{array}{ccc} a_{0,n-1}\\ a_{1,n-1}\\
\vdots\\ a_{\ell,n-1}\end{array} \right) + E \left( \begin{array}{ccc}a_{0,n-2}\\a_{1,n-2}\\\vdots\\ a_{\ell,n-2}\end{array} \right)
= 0 .
\label{3.2p}
\end{eqnarray}
We define
\begin{equation}
\xi = \alpha + n ,
\label{3.3p}
\end{equation}
and obtain the matrices
\begin{eqnarray}
(A_{n})_{k,k} &=& \xi(2\xi +d +4k -1) ,\label{3.4}\\
(A_{n})_{k,k-1} &=& 2\xi(\ell-k+1) ,\label{3.5p}\\
(B_{n-1})_{k,k} &=& (\xi-1)(\xi + d +4k -1) + X_{k} ,\label{3.6}\\
(B_{n-1})_{k,k+1} &=& 2(k+1)^{2}
                  = (E)_{k,k+1} \label{3.7} .
\end{eqnarray}
Here we used the shorthand (see (\ref{2.03}))
\begin{equation}
X_{k}(\lambda) = k(2\lambda +2\ell -k +1) +2\ell -(\ell-2)(2\lambda +\ell -1) ,
\label{3.8p}
\end{equation}
and $ d=2\lambda+1 $ has been introduced. Therefore $A_{n}$ is of lower triangular shape with eigenvalues $\xi(2\xi +d +4k-1)$. The root
system is therefore
\begin{itemize}
\item  $\ell+1$ times the root zero;
\item  the $\ell+1$ roots $\alpha_{m} =-\lambda -2m, \qquad 0 \leq m \leq \ell$.
\end{itemize}

Both sets are degenerate among themselves, and if $d$ is odd, the second set is degenerate with respect to the first one. The first set produces regular solutions, the second set
produces poles if $d$ is odd, which it is in the case of present interest. Nevertheless we will regard $d$ as a free real parameter in order to handle the degeneracy with the regular cases. The solution for $\alpha_{0}$ in combination with any regular solution has the appropriate singular behaviour at $u=0$ needed for a propagator, namely applying Fronsdal's differential operator the correct delta function is created.

Any solution is obtained by requiring
\begin{eqnarray}
A_{0} \left( \begin{array}{ccc} a_{0,0} \\a_{1,0}\\ \vdots\\ a_{\ell,0}\end{array} \right) = 0 .
\label{3.9p}
\end{eqnarray}
This requirement is solved for the regular solutions $\Phi_{k}^{(r)}(u)$ (for which $A_{0} = 0$ and the solution is trivial) by
\begin{equation}
a_{k,0}^{(r)} = \delta_{k,r} .
\label{3.10p}
\end{equation}
For any such solution $r$ we obtain next
\begin{eqnarray}
a_{k,1}^{(r)} &=& -(A_{1}^{-1}B_{0})_{k,r} \nonumber \\
              &=& -(A_{1}^{-1})_{k,r}(B_{0})_{rr} -(A_{1}^{-1})_{k,r-1}(B_{0})_{r-1,r} ,
\label{3.11}
\end{eqnarray}
where we insert
\begin{eqnarray}
(A_{1})_{r,r} &=& d+4r+1 ,\label{3.12p}\\
(A_{1})_{r,r-1} &=& 2(\ell-r+1) ,
\label{3.13p}\\
(B_{0})_{r,r} &=& X_{r} ,
\label{3.14p}\\
(B_{0})_{r-1,r} &=& 2r^{2} ,
\label{3.15p}
\end{eqnarray}
and obtain
\begin{eqnarray}
(A_{1}^{-1})_{k,r} &=& (-2)^{k-r}\prod_{s=r+1}^{k} (\ell-s+1)\quad [\prod_{s=r}^{k}(d+4s+1)]^{-1}\nonumber\\
                        &&\quad \quad \ \ (\textnormal{for}\quad k > r) ,\label{3.16}\\
(A_{1}^{-1})_{r,r} &=& (d+4r+1)^{-1}\label{3.17} ,\\
(A_{1}^{-1})_{k,r} &=& 0 \quad \quad (\textnormal{for} \quad k<r).
\label{3.18}
\end{eqnarray}
There is no sign of any singularity caused by the degeneracy. Finally we get
\begin{equation}
a_{k,1}^{(r)} = -X_{r}(A_{1}^{-1})_{k,r} -2r^{2}(A_{1}^{-1})_{k,r-1} ,
\label{3.19}
\end{equation}
which vanishes for $r>k+1$.

We turn now to the nonanalytic solutions $\Phi_{k}(u,\alpha_{m})$ with roots $\alpha_{m} = -\lambda -2m$ and concentrate on the case $m=0$ because this is the perturbative Green function for the Fronsdal differential
operator. At the beginning we assume $\lambda \notin \textbf Z$ in order to avoid the degeneracy with the regular solutions. In this case we have
\begin{eqnarray}
(A_{0})_{k,k} &=&-4\lambda k ,\label{3.20p}\\
(A_{0})_{k,k-1} &=&-2\lambda (\ell-k+1) ,
\label{3.21p}
\end{eqnarray}
and the equation
\begin{equation}
\sum_{r}(A_{0})_{k,r} c_{r,0}^{(\alpha_{0})} = 0
\label{3.22p}
\end{equation}
is solved by
\begin{equation}
c_{k,0}^{(\alpha_{0})} = \left(-\frac{1}{2}\right)^{k} {\ell\choose k} .
\label{3.23p}
\end{equation}
Next we treat the $A_{1}$ matrix
\begin{eqnarray}
(A_{1})_{k,k} &=& 2(1-\lambda)N_{k},\quad N_{k} = 2k+1 ,\label{3.24p}\\
(A_{1})_{k,k-1} &=& 2(1-\lambda)(\ell-k+1) ,\label{3.25p}\\
(A_{1}^{-1})_{k,r} &=& [2(1-\lambda)]^{-1} \beta_{k,r}, \textnormal{for}\quad k\geq r
\quad \textnormal{and zero else} ,\label{3.26}\\
\beta_{k,r} &=&(-\ell)_{k-r}[\prod_{s = r}^{k} N_{s}]^{-1} .
\label{3.27}
\end{eqnarray}
The $B_{0}$ matrix is
\begin{eqnarray}
(B_{0})_{k,k} &=& -\lambda(\lambda +4k+1) + X_{k} :=Z_{k}(\lambda) ,\label{3.28}\\
(B_{0})_{k,k+1} &=& 2(k+1)^{2} .
\label{3.29}
\end{eqnarray}
The matrix $E$ is still not needed for $n=1$.

We define the matrix
\begin{equation}
(H_{1})_{k,r} = - (A_{1}^{-1}B_{0})_{k,r} =  [2(\lambda-1)]^{-1}\{ \beta_{k,r}(B_{0})_{r,r}
+ \beta_{k,r-1}(B_{0})_{r-1,r} \} ,\label{3.30p}
\end{equation}
and obtain for the coefficients $ c_{k,1}^{(\alpha_{0})}$
\begin{equation}
c_{k,1}^{(\alpha_{0})} = \sum_{r=0}^{k+1} (H_{1})_{k,r} \left(-\frac{1}{2}\right)^{r} {\ell \choose r} .
\label{3.31}
\end{equation}
All these coefficients inherit a pole in $\lambda$ at the value $1$.

This pole does not appear in one eigenvalue only as in the $\zeta = \infty$ case. This is due to
the fact that for $\lambda =1$ there exist $ \ell+1$ degenerate regular solutions and therefore the pole appears in all $\ell+1$ eigenvalues simultaneously. It is straightforward to calculate the residues of all matrix elements of $H_{1}$ and to derive the expressions
\begin{equation}
\rho_{k} = \sum_{r=0}^{k+1} res (H_{1})_{k,r} \left( -\frac{1}{2}\right)^{r}{\ell\choose r} .
\label{3.32}
\end{equation}
Then we subtract from this solution at $n=1$ the regular solution
\begin{equation}
(\lambda -1)^{-1} [\sum_{r=0}^{\ell} \rho_{r} \Phi^{(r)}(u)] ,
\label{3.33}
\end{equation}
obtaining in the limit the log term of $\Psi^{(\ell)}[\Phi_{k}(u,\alpha_{0})]$
\begin{equation}
-\log u \quad[\sum_{r=0}^{\ell} \rho_{r} \Phi^{(r)}(u)] .
\label{3.34}
\end{equation}
We mention that the leading term of $\Psi^{(\ell)}[\Phi_{k}(u,\alpha_{0})]$ is
\begin{equation}
u^{-1}(I_{1} -\frac{1}{2} I_{2})^{\ell} .
\label{3.35}
\end{equation}

The situation with the Green function type solution is the same as with the solution which
is constrained by the AdS boundary condition: The UV constraint is satisfied by a coset, namely any linear combination of regular solutions can be added to the solution $\Psi^{(\ell)}[\Phi(\alpha_{0})]$. In turn this may also be used to normalize the
solutions $\Phi_{k}(\alpha_{0})$. We can namely require that on the level $n=1$ on which $\log{u}$ appears first, all the coefficients $c_{k,1}^{(\alpha_{0})}$ are made to vanish by appropriate subtraction of regular solutions.

\section{Propagators in Feynman's gauge}
\setcounter{equation}{0}
In this section we consider the higher spin gauge field propagators analyzed in the previous section and in \cite{Manvelyan:2005fp}, \cite{Leonhardt:2003qu}, \cite{Leonhardt:2003sn}, in an approach developed originally for the spin $\ell=0,1,2$ only in \cite{Scal}, \cite{AllenJ}, \cite{Freed}, but now generalized for all $\ell$ with a slight modification of arguments. Namely we consider our propagator working directly in the space of conserved currents
\begin{equation}\label{4.1}
h^{(\ell )}(z_{1};a)= \int\sqrt{g}d^{d+1}z_{2}K^{(\ell)}(z_{1},a;z_{2},c)*_{c}J^{(\ell)}(z_{2},c) ,
\end{equation}
where
\begin{equation}\label{4.2}
  K^{(\ell)}(z_{1},a;z_{2},c)=\Psi^{(\ell)}[F_{k}(u(z_{1};z_{2}))] + \textnormal{traces} .
\end{equation}
Taking into account the conservation properties of the current $J^{(\ell)}(z_{2},c)$ we can formulate the ansatz following from (\ref{1.23p})
\begin{eqnarray}
    [\Box_{1} +\ell-\Delta_{\ell}(\Delta_{\ell}-d)]\Psi^{(\ell)}[F_{k}(u(z_{1};z_{2}))]&=&-I^{\ell}_{1}\delta_{d+1}(z_{1};z_{2})+ \textnormal{traces}  \nonumber\\ &+&(c\nabla_{2})\left(I_{1a}\Psi^{(\ell-1)}[\Lambda_{k}(u(z_{1};z_{2}))]\right) .\qquad\quad\label{4.3}
\end{eqnarray}
This means that applying the gauge fixed equation of motion at the first argument of the bilocal propagator we get zero (or more precisely a delta function in the coincident points) due to a gauge transformation at the second argument.

Here we should make some comments on the delta function in curved $AdS$ space. Our notation in (\ref{4.3}) means
\begin{equation}\label{4.4}
   \delta_{(d+1)}(z_{1};z_{2})= \frac{\delta_{(d+1)}(z_{1}-z_{2})}{\sqrt{g(z)}} ,\quad\quad\quad\quad \int\delta_{(d+1)}(z_{1}-z_{2})d^{d+1}z_{1}=1 .
\end{equation}
In the polar coordinate system defined in Appendix A the invariant
measure (for $d=3$) is
\begin{equation}\label{4.5p}
\sqrt{g}d^{4}z=u(u+2)dud\Omega_{3} .
\end{equation}
Therefore we can define
\begin{eqnarray}
  && \frac{\delta_{(4)}(z-z_{pole})}{\sqrt{g(z)}}=\frac{\delta(u)}{u(u+2)\Omega_{3}}
  =-\frac{\delta^{(1)}(u)}{(u+2)\Omega_{3}} ,\label{4.6p}\\
  && u\delta^{(1)}(u)=-\delta(u)\nonumber .
\end{eqnarray}
This $u$- dependence of the measure leads to the idea that short distance singularities in $D=d+1=4$ dimensional $AdS$ space should start from
$\frac{1}{u^{2}}$ not from $\frac{1}{u}$.

Then using the gradient map (\ref{grad1}), (\ref{grad2}) we can derive
\begin{eqnarray}\label{4.7}
(c\nabla_{2})\left(I_{1a}\Psi^{(\ell-1)}[\Lambda_{k}(u)]\right)=\Psi^{(\ell)}[\Lambda'_{k-1}(u)+(k+1)\Lambda_{k}(u)], \quad\Lambda_{\ell}=0
\end{eqnarray}
Combining this with the Laplacian map (\ref{lm1})-(\ref{lm3}) and (\ref{4.1}) we obtain the following set of $\ell+1$ equations for $z_{1}\neq z_{2}$ (unlike the case (\ref{2.01}) we do not insert the value of $\Delta_{\ell}$ here)
\begin{eqnarray}
  && u(u +2)F_{k}'' +(d+1+4k)(u+1)F_{k}'+2(\ell-k+1)F_{k-1}'+2(u+1)(k+1)^{2}F_{k+1}\nonumber
\\&&+[2\ell+k(d+2\ell-k)]F_{k}-\Delta_{\ell}(\Delta_{\ell}-d)F_{k} = \Lambda'_{k-1}+(k+1)\Lambda_{k} .\label{4.8}
\end{eqnarray}
To analyze this system we write the $k=0,1$ and $\ell-1, \ell$ cases explicitly
\begin{eqnarray}
&& u(u +2)F_{0}''+(d+1)(u+1)F_{0}'-\Delta_{\ell}(\Delta_{\ell}-d)F_{0}+2(u+1)F_{1}\nonumber\\
&&\hspace{7.5cm}+2\ell F_{0} = \Lambda_{0} ,\label{4.9}\\
&&\hspace{3.5cm} O(F_{1}'',F_{1}', F_{1}, F_{2})+2\ell F_{0}'= \Lambda'_{0}+2\Lambda_{1} ,\label{4.10}\\
&&\hspace{3.5cm} \vdots  \nonumber\\
&&\hspace{3.5cm} O(F_{\ell-1}'',F_{\ell-1}', F_{\ell-1}, F_{\ell}, F_{\ell-2}') = \Lambda'_{\ell-2}+\ell\Lambda_{\ell-1} ,\quad\label{4.11}\\
&& u(u +2)F_{\ell}''+(d+1+4\ell)(u+1)F_{\ell}'+[\ell^{2}+\ell(d+2) -\Delta_{\ell}(\Delta_{\ell}-d)]F_{\ell}\nonumber\\
&&\hspace{7.5cm}+2F'_{\ell-1} = \Lambda'_{\ell-1} ,\label{4.12}
\end{eqnarray}
and we see that this system for $2\ell+1$ functions is separable. One solution is obtained if we put
\begin{eqnarray}
  F_{k} &=& 0 , \quad k=1,2,\dots \ell ,\label{4.13p}\\
  \Lambda_{k}&=& 0 , \quad k=1,2,\dots \ell-1 ,\label{4.14p}
\end{eqnarray}
and submit $F_{0}(u)$ to the Gaussian hypergeometric equation
 \begin{equation}\label{4.15}
   u(u +2)F_{0}''(u) +(d+1)(u+1)F_{0}'(u)-\Delta_{\ell}(\Delta_{\ell}-d)F_{0}(u)=0 ,
 \end{equation}
supplemented with a noncontradictory solution for the remaining gauge parameter $\Lambda_{0}(u)$
\begin{equation}\label{4.16}
    \Lambda_{0}(u)=2\ell F_{0}(u) .
\end{equation}

So we prove that with an appropriate choice of the gauge freedom we can obtain the propagator in Feynman's gauge in the form
\begin{equation}\label{4.17}
    K^{(\ell)}(z_{1},a;z_{2},c)= I^{\ell}_{1}F_{0}(u)+ \textnormal{traces} ,
\end{equation}
where $F_{0}(u)$ is the solution of the equation for the scalar field with dimension $\Delta_{\ell}$ (\ref{4.15}) \cite{Scal}. The solution of this equation is well known and can be written in two different forms \cite{Freed, FFF}. The first form is ($\zeta=u+1$)
\begin{equation}\label{4.18}
    F_{0}(\zeta)=C(\ell,d) 2^{\Delta_{\ell}} \zeta^{-\Delta_{\ell}}{}_{2}F_{1}\left(\frac{\Delta_{\ell}}{2},
    \frac{\Delta_{\ell}+1}{2},\Delta_{\ell}-\frac{d}{2}+1;\frac{1}{\zeta^{2}}\right) .
\end{equation}
This form is suitable for an investigation of the infrared behaviour. We see immediately that near the boundary limit we have
\begin{equation}\label{4.19}
     F_{0}(\zeta) \sim \zeta^{-\Delta_{\ell}}|_{d=3}=\zeta^{-(\ell+1)},   \quad\textnormal{if} \quad\zeta \rightarrow \infty ,
\end{equation}
which is just wanted for AdS/CFT correspondence. Indeed comparing  $\Delta_{\ell}$ and $\Delta_{\theta}$ in (\ref{1.23p})-(\ref{1.25})  we see that the propagator of the nonphysical mode $\theta$ falls off in the boundary limit faster than the propagator for the physical mode $\psi$, as it should be.

But for us  the second form of this expression obtained after a quadratic transformation of the hypergeometric function listed in the Appendix B (\ref{B1}) is more interesting
\begin{equation}\label{4.20}
     F_{0}(u)=C(\ell,d) \left(\frac{2}{u}\right)^{\Delta_{\ell}}{}_{2}F_{1}\left(\Delta_{\ell},
   \Delta_{\ell} -\frac{d}{2}+\frac{1}{2},2\Delta_{\ell}-d+1;-\frac{2}{u}\right) .
\end{equation}
The normalization constant $C(\ell,d)$ is chosen to obtain the $\delta$ function on the right hand side of (\ref{4.3})
\begin{equation}\label{4.21}
    C(\ell,d)=\frac{\Gamma(\Delta_{\ell})\Gamma(\Delta_{\ell}
    -\frac{d}{2}+\frac{1}{2})}{(4\pi)^{\frac{(d+1)}{2}}\Gamma(2\Delta_{\ell}-d+1)}|_{d=3}=\frac{\ell!(\ell-1)!}{16\pi^{2}(2\ell-1)!} .
\end{equation}

To investigate the ultraviolet limit of (\ref{4.20}) we have to use the second formula  (\ref{B2}) of Appendix B and take carefully the limit $d\rightarrow 3$ to obtain
\begin{eqnarray}
  &&\left(\frac{2}{u}\right)^{\Delta_{\ell}}{}_{2}F_{1}\left(\Delta_{\ell},
   \Delta_{\ell} -\frac{d}{2}+\frac{1}{2},2\Delta_{\ell}-d+1;-\frac{2}{u}\right)|_{d\rightarrow 3}=\frac{(2\ell-1)!}{(\ell-1)!}\left\{\frac{2}{\ell! u}\right.\quad\quad\quad\quad\nonumber\\
  &&\left.+\frac{1}{(\ell-2)!}\sum^{\ell-2}_{n=0}\frac{(\ell+1)_{n}(2-\ell)_{n}}{n!(n+1)!}
  \left[\Upsilon_{\ell,n}+\log{\frac{u}{2}}\right]\left(-\frac{u}{2}\right)^{n}\right\} ,\label{4.22}
\end{eqnarray}
where the rational number $\Upsilon_{\ell,n}$ is expressed by the $\psi$ functions
\begin{equation}
   \Upsilon_{\ell,n}=\psi(\ell+n+1)+\psi(\ell-n-1)-\psi(n+1)-\psi(n+2) .
\end{equation}

So we see now that in the ultraviolet limit we get
\begin{equation}\label{4.23}
    F_{0}(u)|_{d=3}\cong\frac{1}{8\pi^{2}}\frac{1}{u} + O(1,u,\log{u},u\log{u}, \dots) .
\end{equation}
This main singular term in the propagator of the scalar field with dimension $\Delta_{\ell}$ does not depend on the field dimension and behaves always like $\frac{1}{8\pi^{2}u}$. For example we have the same singularity in the propagator of the conformally coupled scalar in $AdS_{4}$
(see \cite{Manvelyan:2006zy})
\begin{eqnarray}
  \Sigma[u(z_{1},z_{2})] &=&\frac{1}{8\pi^{2}}\left(\frac{1}{u}\pm \frac{1}{u+2}\right)  ,\label{4.24}\\
  (\Box+2)\Sigma[u(z_{1},z_{2})]&=& -\delta_{(4)}(z_{1};z_{2}) .\label{4.25}
\end{eqnarray}
So we observe some universality in  the UV behaviour of higher spin propagators in Feynman's gauge:

\emph{For any spin $\ell$ the main term of the propagator has the form $I^{\ell}_{1}\frac{1}{8\pi^{2}u}$.}

Comparing with (\ref{3.35}) we deduce that in de Donder gauge we have the same picture because
\begin{itemize}
  \item $I_{1}(a,c;u)\rightarrow a^{\mu}c_{\nu}$
  if $u\rightarrow 0$ .
  \item $I_{2}(a,c;u)=I_{3}(a,c;u) \rightarrow 0$ if $u\rightarrow 0$ .
  \item $I_{4}(a,c;u)\rightarrow a^{2}c^{2}$ if $u\rightarrow 0$ .
\end{itemize}
So finally we can formulate the following statement:

\emph{The higher spin propagator in Feynman's gauge  is simplest and most convenient for the calculation of any Feynman diagram. Just we have to couple it with conserved currents to make sure that we preserve gauge invariance. The UV-behaviour of the propagator is universal and described by (\ref{4.23})}.

\section{Spin $\ell$, $\ell-2$ and scalar interaction and mass renormalization}
\setcounter{equation}{0}

In this section we will discuss some interaction between two neighboring higher spin  gauge fields and a scalar containing two derivatives. On this linearized level of understanding the higher spin gauge invariance it is possible to construct an interaction of the gauge field contracted with the conserved current formed from gauge fields of the nearest different spin ($\ell\pm 2$), comformally coupled in the $AdS_{d+1}$ background with the scalar $\sigma(z)$
\begin{equation}\label{5.1p}
   \Box \sigma(z)+\frac{d^{2}-1}{4}\sigma(z)=0 ,
\end{equation}
and two derivatives
\begin{equation}\label{5.2p}
    S_{int}=\frac{g_{\ell}}{\sqrt{N}}\int \sqrt{g}d^{4}z h^{(\ell)}(z;a)*_{a}J^{(\ell)}[h^{(\ell\pm 2)}(z;a),\sigma(z)] .
\end{equation}
Here we introduce an unknown coupling parameter $g_{\ell}$ normalized as $O(\frac{1}{\sqrt{N}})$ as it follows from $AdS_4$/$CFT_{3}$ correspondence.
The conservation condition following from the gauge transformation for $h^{(\ell)}(z)$ (\ref{1.7}) with the \emph{traceless} parameter $\epsilon^{(\ell-1)}$ is
\begin{equation}\label{5.3p}
    \nabla^{\mu}\frac{\partial}{\partial a_{\mu}}J^{(\ell)}[h^{(\ell\pm 2)}(z;a),\sigma(z)]= O(a^{2}) .
\end{equation}
This equation  could be used to construct all possible currents with  properties mentioned above.

Returning to the equation (\ref{5.3p}) we note first that the operator
\begin{equation}\label{5.4p}
\mathcal{\widetilde{D}}^{(+1)}=(a\nabla)-\frac{1}{2}a^{2}\nabla ^{\mu }\frac{\partial }{\partial a^{\mu }}
\end{equation}
is dual  to the de Donder gauge operator
\begin{eqnarray}
&&\mathcal{D}^{(-1)}=\nabla ^{\mu }\frac{\partial }{\partial a^{\mu }}-\frac{1}{2}(a\nabla )\Box _{a},
\label{5.5p}
\end{eqnarray}
with respect to the full scalar product,
and second that this operator commutes with the divergence in the following way (see (\ref{A.7}) and (\ref{A.8}))
 \begin{equation}\label{5.6p}
    \nabla ^{\mu }\frac{\partial }{\partial a^{\mu }}\mathcal{\widetilde{D}}^{(+1)}j^{(\ell-1)}(z,a)=[\Box-(\ell-1)(d+\ell-2)]j^{(\ell-1)}(z,a) +O(a^{2}) .
 \end{equation}
Then taking into account (\ref{A.9}) we can see that with the natural choice $j^{(\ell-1)}(z,a)=(a\nabla)[h^{(\ell-2)}(z;a)\sigma(z)]$
one can obtain ($\mu^{2}_{\theta^{(\ell-2)}}$ is the $AdS$ mass of the trace part of $h^{(\ell)}$ (\ref{1.21p}))
\begin{equation}\label{5.7p}
    \nabla ^{\mu }\frac{\partial }{\partial a^{\mu }}\mathcal{\widetilde{D}}^{(+1)}(a\nabla)[h^{(\ell-2)}(z;a)\sigma(z)]
    =(a\nabla)[\Box-\mu^{2}_{\theta^{(\ell-2)}}](h^{(\ell-2)}(z;a)\sigma(z))+O(a^{2}) .
\end{equation}
This can be integrated easily and we restore conserved current from \cite{Manvelyan:2004ii} in the following form
\begin{equation}\label{5.8p}
    J^{(\ell)}[h^{(\ell-2)},\sigma]=\mathcal{\widetilde{D}}^{(+1)}(a\nabla)[h^{(\ell-2)}(z;a)\sigma(z)]
    -\frac{a^{2}}{2}[\Box-\mu^{2}_{\theta^{(\ell-2)}}](h^{(\ell-2)}(z;a)\sigma(z))+O(a^{4}) .
\end{equation}
Note that all $O(a^{4})$ terms are unimportant due to the double tracelessness of $h^{(\ell)}(z;a)$.
At this point we will apply for simplicity de Donder's gauge condition to all types of gauge fields.
Then using free equations of motion only for the fields $h^{(\ell-2)}$ and $\sigma$ that form the conserved current, and neglecting the first part due to de Donder's gauge condition for the gauge field $h^{(\ell)}$, we obtain the following effective current
\begin{equation}\label{5.9p}
    J^{(\ell)}[h^{(\ell-2)}, \sigma]=-\frac{a^{2}}{2}\left[2\nabla^{\mu}(\nabla_{\mu}h^{(\ell-2)}\sigma)
    +\left(\frac{1-d^{2}}{4}-\mu^{2}_{(\ell-2)}-\mu^{2}_{\theta^{(\ell-2)}}\right)h^{(\ell-2)}\sigma\right] .
\end{equation}

Note that this interaction vanishes if we require a free equation of motion for  the field $h^{(\ell)}$ coupled to the conserved current.

The next step of our consideration is the construction of the conserved current $ J^{(\ell)}[h^{(\ell+2)},\sigma]$ which is dual to the former one, where the gauge field inside the current has a spin higher than the gauge field coupled with the current.
Exploring in a similar way the conservation condition (\ref{5.3p}) and using divergence instead of gradient on stage (\ref{4.7}) and formula (\ref{A.6}) instead of (\ref{A.9}) we obtain the following solution
\begin{equation}\label{5.10p}
 J^{(\ell)}[h^{(\ell+2)},\sigma]=\mathcal{\widetilde{D}}^{(+1)}(\nabla \partial_a )[\theta^{(\ell)}(z;a)\sigma(z)]
    -[\Box-\mu^{2}_{(\ell)}](\theta^{(\ell)}(z;a)\sigma(z)) ,
\end{equation}
where
\begin{equation}\label{5.11p}
    \theta^{(\ell)}(z;a)=\Box_{a}h^{(\ell+2)}(z;a) .
\end{equation}
Inserting in (\ref{5.10p}) $\ell-2$ instead of $\ell$ and using the equation of motion for the fields inside the current and de Donder' gauge condition for the external gauge field, we obtain the effective current
\begin{equation}\label{5.12p}
     J^{(\ell-2)}[h^{(\ell)},\sigma]= -\left[2\nabla^{\mu}(\nabla_{\mu}\theta^{(\ell-2)}\sigma)
    +\left(\frac{1-d^{2}}{4}-\mu^{2}_{(\ell-2)}-\mu^{2}_{\theta^{(\ell-2)}}\right)\theta^{(\ell-2)}\sigma\right] .
\end{equation}
Comparing with (\ref{5.9p}) we see that in both cases we have the same effective interaction between the physical mode $\psi^{(\ell-2)}$, the trace mode $\theta^{(\ell-2)}$ and the scalar, and the conserved current (\ref{5.12p}) up to an overall normalization is dual to the conserved current (\ref{5.9p}) due to the equations of motion for the fields forming the currents in each cases.
Thus we prove that one can use Feynman's gauge for propagators coupled to these two currents and can turn now to investigate some loop diagram for a study of mass renormalization or quantum mass generation phenomena. Actually we considered all interactions of two neighbouring higher spin fields with a Higgs scalar that are minimal with respect to the number of derivatives, and which can generate mass in a loop. Though from (\ref{5.6p}) follows that we can introduce in principle many other $j^{(\ell-1)}(z;a)$, all of them will contain more derivatives  in front of the quantized fields and will generate counterterms that are not suitable for finite mass renormalization.

So we see that only one reasonable one loop  diagram can be constructed from the interactions considered in this section. It is a loop formed by the scalar $\sigma$ and the nonphysical trace mode $\theta^{(\ell)}$ and with physical but off-shell external lines $\psi^{(\ell)}$. Actually we would like to calculate the following quadratic part of the effective action
\begin{equation}\label{5.13p}
    \frac{g^{2}_{\ell}}{N}\int \sqrt{g}d^{4}z_{1}\int \sqrt{g}d^{4}z_{2} h^{(\ell)}(z_{1};a)*_{a}\big< J^{(\ell)}[h^{(\ell+2)},\sigma;z_{1};a],J^{(\ell)}[h^{(\ell+2)},\sigma;z_{2};c)]\big>*_{c}h^{(\ell)}(z_{2};c) ,
\end{equation}
where $J^{(\ell)}[h^{(\ell+2)},\sigma;z_{2};c)]$ is presented in (\ref{5.10p}). Performing a partial integration and taking into account tracelessness of $\theta^{(\ell)}$ we get the following expression
\begin{equation}\label{5.14p}
     \frac{g^{2}_{\ell}}{N}\int \sqrt{g}d^{4}z_{1}\int \sqrt{g}d^{4}z_{2}\mathcal{F}(\psi^{(\ell )}(z_{1};a))*_{a}\Sigma[u(z_{1},z_{2})]\Theta^{(\ell)}[u(z_{1},z_{2});a,c]*_{c}\mathcal{F}(\psi^{(\ell )}(z_{2};c)) .
\end{equation}
Here $\mathcal{F}(\psi^{(\ell )})$ is the traceless part of Fronsdal's operator, $\Sigma[u]$ is the scalar propagator (\ref{4.24}) and
\begin{equation}\label{5.15}
    \Theta^{(\ell)}[u(z_{1},z_{2});a,c]=\big<\theta^{(\ell)}(z_1;a),\theta^{(\ell)}(z_2;c)\big>
\end{equation}
is a trace part of the $h^{(\ell+2)}$ propagator. We want to understand the singular part of this loop.

>From now on we follow a technique developed in \cite{Manvelyan:2006zy} and \cite{Manvelyan:2005ew}
where the trace anomaly of the scalar mode in external higher spin field was successfully calculated from a one loop diagram.
First we can use an $AdS$ transformation to fix the point $z_{1}$ as a
pole for the coordinate system $z_{2}$. Then the integration measure can be expressed through the chordal distance $u$ as it is explained in the Appendix A. The singularity of the product of the scalar and the higher spin propagators is relevant if it is at least $1/u^{2}$ because one power of $u$ is compensated by the integration measure (see (\ref{invv}) and explanation hereafter). Then from the relative coefficient between the $\psi$ and $\theta$ modes in (\ref{1.20p}) evaluated for spin $\ell+2$ and $d=3$, from (\ref{4.23}) and an additional sign from the indefinite metric (\ref{1.26}) we deduce
 \begin{equation}\label{5.16}
   \big< \theta^{(\ell)}(z_1;a),\theta^{(\ell)}(z_2;c)\big>=-\frac{4(\ell+2)}{\ell+1}I^{\ell}_{1}\frac{1}{8\pi^{2}u} + O(u, \log{u},\dots) ,
 \end{equation}
Multiplying hereafter with the scalar propagator we get the unique singular term of the loop function
 \begin{equation}\label{5.17}
   \left\{\Sigma[u]\Theta^{(\ell)}[u]\right\}_{sing}=-\frac{(\ell+2)}{16\pi^{4}(\ell+1)}I^{\ell}_{1}\frac{1}{u^{2}} .
 \end{equation}
 Using a standard formula of analytic dimensional regularization in $AdS$ space (see \cite{Manvelyan:2006zy} and \cite{Manvelyan:2005ew})
 \begin{equation}\label{5.18}
    \left[\frac{1}{u^{n-\epsilon}}\right]_{sing}=-
    \frac{1}{\epsilon}\frac{(-1)^{n-1}}{(n-1)!}\delta^{(n-1)}(u) .
 \end{equation}
for our distribution with $n=2$ and the definition of the delta function (\ref{4.6p}), we obtain
\begin{equation}\label{5.19}
  \left\{\Sigma[u]\Theta^{(\ell)}[u]\right\}_{sing}=-\frac{\Omega_{3}(\ell+2)}{8\pi^{4}(\ell+1)}(a^{\mu}c_{\mu})^{\ell}\frac{1}{\epsilon} \delta_{(4)}(z_{1};z_{2}) .
\end{equation}

Before inserting this expression in (\ref{5.14p}) we have to be sure that we preserved gauge invariance during regularization.
At this stage it means that we have to preserve the conservation condition (\ref{5.3p}) for the current as a Ward identity for the correlator in (\ref{5.13p}). Then taking into account that after partial integration we got gauge invariant Fronsdal's operators instead of external lines we deduce that we just should write these external lines as a gauge invariant object during dimensional regularization or in other words for  $d=3-\epsilon$. From the formula for the geometric $AdS$ mass $\mu^{2}_{\ell}$  (\ref{AdSmass}) we see that
\begin{equation}\label{5.20}
    \mathcal{F}^{d=3-\epsilon}(h^{(\ell)}(z;a))=\mathcal{F}^{d=3}(h^{(\ell)}(z;a))+\epsilon(\ell-2)h^{(\ell)}(z;a) .
\end{equation}
Then inserting this  and (\ref{5.19}) in (\ref{5.14p}) we obtain immediately as local singularity of our diagram
\begin{equation}\label{5.21}
    -\frac{1}{\epsilon}\frac{g^{2}_{\ell}\Omega_{3}(\ell+2)}{8N\pi^{4}(\ell+1)}\int \sqrt{g}d^{4}z\mathcal{F}(\psi^{(\ell )}(z;a))*_{a}\mathcal{F}(\psi^{(\ell )}(z;a)) ,
\end{equation}
supplemented with the additional finite local term
\begin{equation}\label{5.22}
    -\frac{g^{2}_{\ell}\Omega_{3}(\ell-2)(\ell+2)}{4N\pi^{4}(\ell+1)}\int \sqrt{g}d^{4}z\psi^{(\ell )}(z;a)*_{a}\mathcal{F}(\psi^{(\ell )}(z;a)) .
\end{equation}

The first singular term can be dropped  adding the same  singular local and gauge invariant counterterm to the effective action as (\ref{5.21}) but with opposite sign. The second finite local part is not gauge invariant itself and  cannot be absorbed by adding the local finite invariant counterterm but can be absorbed by finite renormalization of the mass term. Indeed let us add an additional finite local counterterm proportional to
\begin{equation}\label{5.23}
    \int \sqrt{g}d^{4}z[\mathcal{F}(\psi^{(\ell )})-\delta m^{2}_{\ell} \psi^{(\ell )} ]*_{a}[\mathcal{F}(\psi^{(\ell )})-\delta m^{2}_{\ell} \psi^{(\ell )}] .
\end{equation}
Then we see that if
\begin{equation}\label{5.24}
    \delta m^{2}_{\ell} =\frac{g^{2}_{\ell}\Omega_{3}(\ell-2)(\ell+2)}{8N\pi^{4}(\ell+1)}=\frac{g^{2}_{\ell}(\ell-2)(\ell+2)}{4N\pi^{2}(\ell+1)} ,
\end{equation}
we will cancel (\ref{5.22}) without any additional term in the given order of perturbation theory ($O(1/N)$), absorbing the additional $O(1)$ finite local $\mathcal{F}^{2}$ term in the infinite singular \emph{gauge invariant} counterterm, since an additional finite renormalization supplementing the infinite one fixes the renormalization scheme. We see that our mass renormalization implies a soft symmetry breaking because we got only a finite mass generation. In other words all our infinite counterterms are gauge invariant.

So we see that we got mass renormalization as it was expected from $AdS_{4}/CFT_{3}$ correspondence and formulated in terms of boundary $CFT$ theory in \cite{Ruehl}. In principle we can compare this mass with the answer obtained in \cite{Ruehl} from anomalous dimensions of higher spin currents in the $O(N)$ sigma model
\begin{equation}
\delta m_{\ell}^{2} = \frac{1}{N}\frac{16(\ell-2)}{3\pi^{2}}.
\end{equation}
We got the same interesting $(\ell-2)$ factor protecting the spin 2 graviton field, that corresponds to the boundary energy momentum tensor, from renormalization and found a prediction for the coupling $g_{\ell}$. But we will not compare them at this stage because it is not the full solution of the problem. We did not include in our consideration the all other possible interactions and the corresponding one loop diagrams. It is also interesting to compare this  UV approach with another IR ansatz including the St\"uckelberg and Goldstone mechanism which was considered in \cite{Manvelyan:2004ii}.


\chapter{Conformal invariant Lagrangians}\label{confL}
\setcounter{equation}{0}
\section{Conformal invariant powers of the laplacian, FG ambient metric and Ricci gauging}\label{conform}

\quad The problem of constructing conformally invariant Lagrangians or differential operators in various dimensions and for various fields has quite a long history. This problem attracts attention primarily because it is always a nontrivial task to construct local conformal or Weyl invariants in higher dimensions \cite{Bon, Erd, Boul}. The $AdS/CFT$ correspondence \cite{Maldacena:1997re} increased interest in this old problem as well as returned the attention to the seminal mathematical paper by Fefferman and Graham (FG) on conformal invariants  \cite{FG}.
In this section we discuss conformal coupling of a scalar field with gravity in different dimensions which has
been a subject of interest in quantum field theory in curved
spacetimes~\cite{BD}. In recent years it has attracted special
attention in the context of new developments in the area of
$AdS/CFT$ \cite{Maldacena:1997re} correspondence, and in investigations of
higher order and higher spin gravitating systems in
general~\cite{OP}. Conformally invariant field theories in higher
dimensions are particularly interesting because they present a
universal tool for investigations of their quantum properties, such
as conformal or trace anomalies~\cite{anom}. Another important
properety of conformally invariant theories in arbitrary dimensions
is, that the method of dimensional regularization can be employed as
a conformally invariant regularization in higher dimensions for the
construction of anomalous effective actions \cite{Schwim}. Note also
that in connection with higher spin gauge field interactions with a
scalar field, this coupling and Weyl invariance itself, can be
generalized (see next section and \cite{Manvelyan:2004mb},\cite{Manvelyan:2009tf}).
Our goal in this section is to establish the connections between different ways of construction of the local conformal invariant Lagrangians or differential operators in $d$ dimensions \cite{MT},\cite{oraf} and the FG $d+2$ dimensional ambient  Ricci flat space method \cite{FG}.

In this work we propose a hierarchy of such couplings of gravity
to scalar fields with increasing scaling dimensions parameterized by
a natural number $k$, and living in all space-time dimensions $d\geq
2k$. Actually this hierarchy corresponds to the conformally
invariant $k$-th power of the Laplacian acting on a scalar field
with conformal dimension $\Delta_{(k)}=k-d/2$, in spacetime
dimensions $d\geq 2k$. From the other hand we propose the connection
between this hierarchy and the $k$-th Euler density $E_{(k)}$ lifted
to spacetime dimensions greater than $2k$. For completeness, we
verify this proposal in the well known text book case of
$k=1$~\cite{BD}. We then turn to the known case in
$d=4$~\cite{Riegert,FT}, and the fourth order conformally covariant
operator in dimension $d\geq 4$ obtained in \cite{pan,es} long ago,
which provides us with a further check of our proposal, now
involving the second Euler density $E_{(2)}$. In the subsection \ref{delta3} we perform the
new calculation of the locally Weyl invariant
third power of the Laplacian in spacetime dimensions $d\geq 6$, or
in another words we construct a conformally invariant action for the
scalar with conformal dimension $3-d/2$ coupled with gravity. In all
three cases we have found the corresponding Euler density $E_{(k)}$
as part of the invariant action, proportional to the first order of
$\Delta_{(k)}$, and without derivatives. Taking into account the
rather technical character of this section we devote a substantial
subsection, subsection {\bf \ref{NC}}, with a more or less complete technical
setup and all the formulas which we have used in our calculations.

The main FG idea consists in the confidence that the lower dimensional diffeomorphisms and local conformal invariants can be obtained from corresponding reparametrization invariant counterparts in the higher dimensional space where $d$ dimensional conformal invariance is realized as a part of $d+2$ dimensional diffeomorphisms (we review the FG method in subsection \ref{FG}). On the other hand the FG expansion is connected with $AdS_{d+1}/CFT_{d}$ correspondence and plays a crucial role in derivation of the holographic anomalies in different dimensions \cite{HS}. This point forced us in subsection \ref{ambient} to derive again, using the FG ambient space method, the
hierarchy of conformally invariant powers of the Laplacian (or invariant Lagrangian) in spacetime dimensions $d\geq 2k$ acting on a scalar field obtained in subsections \ref{hierarchy}, \ref{delta3} by the direct Noether procedure, whose conformal dimension is $\Delta_{(k)}=k-d/2$. This ambient space derivation unveiled the remarkable and mysterious feature of these differential invariants namely the appearance of the $2k$ dimensional holographic anomaly in the $k$-th member of this hierarchy \cite{MT} (recent mathematical development in the holographic  formalism for conformally invariant operators is considered in \cite{Juhl}).

\quad Then we propose also (subsection \ref{PBH}) an extended or gauged FG $d+2$ dimensional space to establish a connection  between  the FG expansion and another interesting method of constructing  the Weyl invariant Lagrangians obtained in \cite{oraf} by A.~Iorio, L.~O'Raifeartaigh, I.~Sachs and C.~Wiesendanger and  named ``Ricci gauging''. The magic and universality of the $d+2$ dimensional FG method is defined by the existence of so-called Penrose-Brown-Henneaux (PBH) diffeomorphisms \cite{PBH} considered in details for usual FG metric in \cite{Schwim} and \cite{OA}. In subsection \ref{PBH} we consider the new PBH transformation for gauged ambient spaces to explore some properties of the FG expansion in the presence of the Weyl gauge field and the holographic origin of the Ricci gauging.

\subsection{Notations and Conventions}\label{NC}
We work in a $d$ dimensional curved space with metric $g_{\mu\nu}$ and use the following
conventions for covariant derivatives and curvatures:
\begin{eqnarray}
  \nabla_{\mu}V^{\rho}_{\lambda}&=& \partial_{\mu}V^{\rho}_{\lambda}+
  \Gamma^{\rho}_{\mu\sigma}V^{\sigma}_{\lambda}-\Gamma^{\sigma}_{\mu\lambda}V^{\rho}_{\sigma} , \\
  \Gamma^{\rho}_{\mu\nu} &=& \frac{1}{2} g^{\rho\lambda}\left(\partial_{\mu}g_{\nu\lambda}+
  \partial_{\nu}g_{\mu\lambda} - \partial_{\lambda}g_{\mu\nu}\right) , \\
  \left[\nabla_{\mu} , \nabla_{\nu}\right]V^{\rho}_{\lambda} &=&
  R^{\quad\,\,\rho}_{\mu\nu \sigma}V^{\sigma}_{\lambda}
  -R^{\quad\,\,\sigma}_{\mu\nu \lambda}V^{\rho}_{\sigma} ,\\
  R^{\quad\,\,\rho}_{\mu\nu \lambda}&=& \partial_{\mu}\Gamma^{\rho}_{\nu\lambda}
  -\partial_{\nu}\Gamma^{\rho}_{\mu\lambda}+\Gamma^{\rho}_{\mu\sigma}\Gamma^{\sigma}_{\nu\lambda}
  -\Gamma^{\rho}_{\nu\sigma}\Gamma^{\sigma}_{\mu\lambda} ,\\
  R_{\mu\lambda}&=& R^{\quad\,\,\rho}_{\mu\rho\lambda}\quad , \quad
  R=R^{\,\,\mu}_{\mu} .
\end{eqnarray}
The corresponding local conformal transformations (Weyl rescalings)
\begin{eqnarray}
  \delta g_{\mu\nu}&=&2\sigma(x) g_{\mu\nu} , \quad\
  \delta g^{\mu\nu} = -2\sigma(x) g^{\mu\nu} ,\\
   \delta\Gamma^{\lambda}_{\mu\nu}&=& \partial_{\mu}\sigma\delta^{\lambda}_{\nu}
  +\partial_{\nu}\sigma\delta^{\lambda}_{\mu}-
  g_{\mu\nu}\partial^{\lambda}\sigma , \label{christ}\\
   \delta R^{\quad\,\,\rho}_{\mu\nu \lambda}&=&\nabla_{\mu}
  \partial_{\lambda}\sigma\delta^{\rho}_{\nu}-
  \nabla_{\nu}\partial_{\lambda}\sigma\delta^{\rho}_{\mu}+
  g_{\mu\lambda}\nabla_{\nu}\partial^{\rho}\sigma
  - g_{\nu\lambda}\nabla_{\mu}\partial^{\rho}\sigma ,\\
  \delta R_{\mu\lambda}&=&(d-2)\nabla_{\mu}\partial_{\lambda}\sigma +
g_{\mu\lambda}\Box
\sigma ,\\
 \delta R&=& -2\sigma R + 2(d-1)\Box \sigma\,,
\end{eqnarray}
are first order in the infinitesimal local scaling parameter
$\sigma$.

We then introduce the Weyl ($W$) and Schouten ($K$) tensors, as well as
the scalar $J$
\begin{eqnarray}
  R_{\mu \nu } &=& (d - 2)K_{\mu \nu }  + g_{\mu \nu}J , \quad J=\frac{1}{2(d - 1)}R\,\,, \\
  W_{\mu \nu \lambda }^{\quad\,\,\rho}   &=& R_{\mu \nu \lambda } ^\rho   - K_{\mu \lambda }
 \delta _\nu  ^\rho   +
 K_{\nu \lambda } \delta _\mu  ^\rho   - K_\nu  ^\rho  g_{\mu \lambda }  +
 K_\mu  ^\rho  g_{\nu \lambda } \,\,, \label{n12}\\
  \delta K_{\mu \nu } &=& \nabla _\mu  \partial _\nu
 \sigma \,,\,\,\,\, \label{n13}\\ \delta J&=&-2\sigma J + \Box
 \sigma\,,\,\,\,\,\,\label{deltaJ}\\
 \delta W_{\mu \nu \lambda }^{\quad\,\,\rho}&=& 0\,,
\end{eqnarray}
which are more convenient because their conformal transformations
are "diagonal".

To describe the Bianchi identity with these tensors, we have to
introduce the so called Cotton tensor
\begin{eqnarray}
  C_{\mu \nu \lambda } &=& \nabla _\mu  K_{\nu \lambda }  - \nabla _\nu
 K_{\mu \lambda } \,\,,
  \label{bi1}\\ \delta C_{\mu \nu \lambda }
 &=&  - \partial _\alpha  \sigma W_{\mu \nu \lambda
 }^{\quad\,\,\alpha}\,\,\,\,,\,\,\,\,\,  C_{[\mu \nu \lambda ] = 0\,\,.}
\end{eqnarray}
All important properties of these tensors following from the Bianchi
identity can then be listed as
\begin{eqnarray}
  \nabla _{[\alpha } W_{\mu \nu ]\lambda }^{\quad\,\,\,\rho} &=& g_{\lambda [\alpha }
  C_{\mu \nu ]}^{\quad\,\,\rho}
   - \delta _{[\alpha }^\rho  C_{\mu \nu ]\lambda } \,\,,\, \label{n18}\\
 \nabla _\alpha  W_{\mu \nu \lambda }^{\quad\,\,\alpha}&=& \left( {3
- d}\right)C_{\mu \nu \lambda } \,\,\,,\\
 \nabla^{\mu}K_{\mu\nu}&=&\partial_{\nu}J\,,\label{bi2}\\
 C_{\mu \nu }^{\quad\,\nu} &=& 0\,\,,\quad\quad
\nabla^{\lambda}C_{\mu\nu\lambda}=0\,.
\end{eqnarray}
Finally we introduce the last important conformal
object in the above listed hierarchy, namely the symmetric and traceless
Bach tensor
\begin{eqnarray}
B_{\mu\nu}&=&\nabla^{\lambda}C_{\lambda\mu\nu}
  +K^{\lambda}_{\alpha}W_{\lambda\mu \nu  }^{\quad\,\,\alpha} ,
\end{eqnarray}
whose conformal transformation and divergence are expressed in terms
of the Cotton and the  Schouten tensors as follows
\begin{eqnarray}
    \delta B_{\mu\nu}&=&-2\sigma B_{\mu\nu}+(d-4)\nabla^{\lambda}\sigma
  \left( C_{\lambda\mu\nu}+C_{\lambda\nu\mu}\right) ,\label{bachtr}\\
   \nabla^{\mu}B_{\mu\nu}&=&(d-4)C_{\alpha\nu\beta}K^{\alpha\beta} .
  \end{eqnarray}
Note that only in four dimensions is the Bach tensor
conformally covariant and divergenceless.

This basis of $B,C,K,J,W$ tensors we will use in subsections \ref{hierarchy}, \ref{delta3} to construct directly\footnote{This basis of $B,C,K,J,W$ tensors forms  a closed system with respect to local conformal (or Weyl) transformations of the boundary metric $\delta g_{ij}(x)=2\sigma(x)g_{ij}(x)$
\begin{eqnarray}
 && \delta W_{ijk}^{\quad\,\,m}=0 ,\quad  \delta K_{ij}=\nabla_{i}\partial_{j}\sigma ,\quad \delta J=-2\sigma J +\Box \sigma\nonumber\\
  && \delta C_{ijk}= -\partial_{m}\sigma W_{ijk}^{\quad\,\,m} ,\quad \delta B_{ij}= -2\sigma B_{ij} +(d-4)\nabla^{k}\sigma \left(C_{kij}+C_{kji}\right), \nonumber
\end{eqnarray} and it is all one needs to construct any conformally invariant object in arbitrary dimensions  \cite{MT},\cite{Erd}.}
a hierarchy of
conformally invariant Lagrangians or differential operators originating from  powers of the Laplacian in spacetime dimensions $d\geq
2k$, describing the nonminimal coupling of gravity with a scalar
field whose conformal dimension is $\Delta_{(k)}=k-d/2$.
 Finally for any
scalar $f^{\Delta}(x)$ with arbitrary scaling dimension $\Delta$ we
can easily derive the following important relations
\begin{eqnarray}
   \delta\left(\nabla_{\mu}\partial_{\nu}f^{\Delta}\right)
  &=&\Delta\sigma \nabla_{\mu}\partial_{\nu}f^{\Delta}
  +\Delta f^{\Delta} \nabla_{\mu}\partial_{\nu}\sigma
  +(\Delta-1)\partial_{(\mu}\sigma\partial_{\nu)}
  f^{\Delta}+g_{\mu\nu}\partial^{\lambda}\sigma\partial_{\lambda}f^{\Delta} \quad\quad\quad\quad  \\
   \delta\left(\Box f^{\Delta}\right)&=&(\Delta-2)\sigma\Box
  f^{\Delta}+\Delta f^{\Delta}\Box \sigma +(d+2\Delta-2)
  \partial^{\lambda}\sigma\partial_{\lambda}f^{\Delta} \label{lap}
\end{eqnarray}
by using the
transformation (\ref{christ}) for Christoffel symbols.

\subsection{Ambient metric and Fefferman-Graham expansion}\label{FG}
In this section we review the FG ambient space method for constructing local conformal invariants \cite{FG}.
We define the $d+2$ dimensional ambient space with the set of coordinates
$\{x^{\mu}\}= \{t,\rho,x^{i}; i=1,2,\dots d \}$ and the following Ricci flat
 metric
\begin{equation}\label{1}
ds^{2}_{A}=g^{A}_{\mu\nu}(t,\rho,x^{i})dx^{\mu}dx^{\nu}=
\frac{t^{2}}{\ell^{2}}h_{ij}(x,\rho)dx^{i}dx^{j}-\rho dt^{2}-tdtd\rho\,,
\end{equation}
where
\begin{equation}\label{2}
    h_{ij}(x,\rho)=g_{ij}(x)+\rho h^{(1)}_{ij}(x)+\rho^{2}h^{(2)}_{ij}(x)+\dots\dots
\end{equation}
is the well known FG expansion with an arbitrary boundary value of the metric $g_{ij}(x)=h_{ij}(x,\rho)|_{\rho=0}$ and  a set of  the higher $\rho$ derivatives $n!h^{(n)}_{ij}(x)=\frac{\partial^{n}}{\partial\rho^{n}}h_{ij}(x,\rho)|_{\rho=0}$ fixed by the Ricci flatness condition in ambient space
\begin{equation}\label{3}
    R^{A}_{\mu\nu}=0\,.
\end{equation}
This condition produces the following set of equations
\begin{eqnarray}
  && R^{A}_{it} = R^{A}_{\rho t}= R^{A}_{tt}\equiv 0 ,\label{4}\\
  &&R^{A}_{\rho\rho} =\frac{1}{2}\left[h^{kl}h''_{kl}
  -\frac{1}{2}h^{ij}h'_{jk}h^{kl}h'_{li}\right]=0 \,,\label{5}\\
  && R^{A}_{i\rho}=\frac{1}{2}h^{kl}\left[\nabla^{(h)}_{i}
  h'_{kl}-\nabla^{(h)}_{k}h'_{il}\right]=0\,,\label{6}\\
  &&\ell^{2}R^{A}_{ij}=\ell^{2}R_{ij}[h]-(d-2)h'_{ij} - h^{kl}h'_{kl}h_{ij}\nonumber\\&&+
  \rho\left[2h''_{ij}-2h'_{il}h^{lm}h'_{mj}
  +h^{kl}h'_{kl}h'_{ij}\right]=0\,,\label{7}
\end{eqnarray}
where $\dots'=\partial_{\rho}\dots$ and $\nabla^{(h)}_{i}, R_{ij}[h]$ are covariant derivative and Ricci tensor of the metric
$h_{ij}(x,\rho)$, respectively. It was shown in \cite{FG} that this system of equations is equivalent to the
$d+1$ dimensional Einstein's equations with negative cosmological constant (see \cite{OA} for details). This can be easily seen from the following consideration
\begin{itemize}
  \item The $AdS_{d+1}$ bulk can be found in $d+2$ dimensional ambient space as a $d+1$ dimensional surface defined as
      \begin{equation}\label{8}
        t^{2}\rho=\ell^{2} ,\quad \rho >0.
      \end{equation}
      on which the metric (\ref{1}) induces the standard Poincar\'{e} metric for coordinates $\{x^{a}\}=\{\rho, x^{i}\}$
      \begin{equation}\label{9}
      ds^{2}_{Bulk}=g^{Bulk}_{ab}(x,\rho)dx^{a}dx^{b}
      =\frac{\ell^{2}}{4\rho^{2}}d\rho^{2}+\frac{1}{\rho}h_{ij}(x,\rho)dx^{i}dx^{j}\,.
\end{equation}
    \item The corresponding bulk Ricci tensor is related to the nonzero components of the ambient Ricci tensor in the way
        \begin{equation}\label{10}
            R^{A}_{ab}=R^{Bulk}_{ab}+\frac{d}{\ell^{2}}g^{Bulk}_{ab}\,,
        \end{equation}
       and condition (\ref{3}) leads to the negative constant curvature
       \begin{equation}\label{11}
         R^{A}_{ab}=0 \Rightarrow R^{Bulk}=R^{Bulk}_{ab}g^{ab}_{Bulk}=-\frac{d(d+1)}{\ell^{2}}\,.
       \end{equation}
       \end{itemize}
 Therefore  (\ref{3}) leads, as in the case of $AdS_{d+1}/CFT_{d}$ correspondence \cite{HS}, to the same solutions for $h^{(n)}_{ij}(x)$ in the FG expansion (\ref{2}) in terms of covariant objects constructed from the boundary value $g_{ij}(x)$
 \begin{eqnarray}
   && h^{(1)}_{ij}(x)=\ell^{2}K_{ij} ,\quad
   h^{(1)}=g^{ij}(x)h^{(1)}_{ij}(x)=\ell^{2}J \,,\label{12}\\
   &&h^{(2)}_{ij}(x)=\frac{\ell^{4}}{4}\left\{\frac{B_{ij}}{d-4}+K^{m}_{i}K_{mj}\right\}, \quad  h^{(2)}=g^{ij}(x)h^{(2)}_{ij}(x)=\frac{\ell^{4}}{4}K^{ij}K_{ij},\label{13}\quad \quad \quad \\
   &&  h^{(3)}=g^{ij}(x)h^{(3)}_{ij}(x)=\frac{\ell^{6}}{6(d-4)}K^{ij}B_{ij} \,,\label{14}
 \end{eqnarray}

The connection of the Fefferman-Graham ambient metric expansion and\\ $AdS/CFT$ correspondence was investigated and developed by many authors. We do not pretend here to present an exhaustive list of citations in this field and just quote a number of  articles important for us in this area  \cite{Schwim}, \cite{HS}, \cite{OA}. For us the most important result of \cite{FG} is the elegant method of constructing conformal invariants (covariants) in $d$ dimensions from reparametrization invariant (covariant) combinations of the curvature and it's covariant derivatives in $d+2$ dimensional ambient space equipped with a Ricci flat metric (\ref{1}) by truncation to the $d$ dimensional boundary  at $\rho=0$ and $t=const$. In the simplest case of a Riemannian  curvature tensor this prescription gives for nonvanishing  components  (see \cite{OA} for detailed derivation)
\begin{eqnarray}
  && R^{A\quad l}_{ijk}|_{\rho=0}=W_{ijk}^{\quad l} \,,\label{18}\\
  && R^{A\quad t}_{ijk}|_{\rho=0}=t C_{ijk} \label{19}\,,\\
  && R^{A\quad t}_{\rho ij}|_{\rho=0}=\frac{t \ell^{2}}{2}\frac{B_{ij}}{d-4}\,. \label{20}
\end{eqnarray}
Using this the authors derived in \cite{FG} the first nontrivial invariant obtained from $(\nabla^{A}_{m}R^{A}_{ijkl})^{2}$ and discussed in details in \cite{Erd}. In the same article Fefferman and Graham predicted that usual Laplacian in ambient $d+2$ dimensional space should produce conformal invariant second order differential operator in dimension $d$, which is the first representative in the hierarchy of conformal operators for scalar fields constructed here in subsections \ref{hierarchy}, \ref{delta3}.

\subsection{Hierarchies of conformal scalars and Euler densities}\label{hierarchy}

\quad In this section we introduce the hierarchy of scalar fields
$\varphi_{(k)}$, where $k=1,2,3,\dots$ with the corresponding
scaling dimensions and infinitesimal conformal transformations
\begin{eqnarray}
   \Delta_{(k)}&=&k-d/2\,,\\
    \delta \varphi_{(k)}:&=&\Delta_{(k)}\sigma\varphi_{(k)} .\label{trans}
\end{eqnarray}
Each of these exist in the spacetime dimensions $d\geq 2k$, and with
the minimal dimension vanishing, $\Delta_{(k)}=0$ when $d=2k$.

Let us now introduce the hierarchy of the Euler densities~\footnote{Note
that the usual Einstein--Hilbert Lagrangian in $d$ dimensions is
the $k=1$ member of this hierarchy of gravitational Lagrangians.}
\begin{eqnarray}
  && E_{(k)}:=\frac{1}{2k (d-2k)!}\delta^{\alpha_{1}
  \dots\alpha_{d-2k}\mu_{1}\mu_{2}\dots\mu_{2k-1}\mu_{2k}}_{\alpha_{1}
  \dots\alpha_{d-2k}\nu_{1}\nu_{2}\dots\nu_{2k-1}\nu_{2k}}
  R^{\nu_{1}\nu_{2}}_{\mu_{1}\mu_{2}}\dots R^{\nu_{2k-1}\nu_{2k}}_{\mu_{2k-1}
  \mu_{2k}} .\label{Ek}
\end{eqnarray}
This set of objects exist as Lagrangians in space time dimensions
$d\geq 2k$, but for the minimal case $d=2k$, $E_{k}$ is a total
divergenece such that its integral is a
topological invariant, the Euler characteristic. In these dimensions
$E_{k}$ trivialize as Lagrangians but describe the topological
part of the trace anomaly in the corresponding even space-time
dimension $2k$.

The idea of this section is the following observation: \emph{There should
be a one to one correspondence between the conformally coupled scalars
$\varphi_{(k)}$ and the Euler densities $E_{(k)}$}.

Our first step in proving this is to start from the action of the well
known non minimal conformally coupled scalar in the space-time dimension
$d$ and with conformal dimension $\Delta_{1}=1-d/2$
\begin{eqnarray}
  && S_{(1)}=\frac{1}{2}\int d^{d}x\sqrt{g}\left\{
  g^{\mu\nu}\partial_{\mu}\varphi_{(1)}\partial_{\nu}
  \varphi_{(1)}-\frac{d-2}{4(d-1)}R\varphi^{2}_{(1)}\right\} .\label{act1}
\end{eqnarray}
We first see that the second term without derivatives and proportional
to the scaling dimension can be written in the form
$-\frac{d-2}{4(d-1)}R=\Delta_{(1)}J$. After that the proof of the
conformal invariance of the action (\ref{act1}) becomes trivial: We
write (\ref{lap}) for $\Delta=\Delta_{(1)}$ and use
(\ref{deltaJ}), from which it follows that
$\delta\left[\sqrt{g}\varphi_{(1)}\left(\Box-\Delta_{(1)}J\right)\varphi_{(1)}\right]=0$. We next evaluate (\ref{Ek}) for $k=1$
\begin{equation}\label{e1}
E_{(1)}=2(d-1)J \, .
\end{equation}
Finally we see that (\ref{act1}) can be rewritten in the form
\begin{eqnarray}
  && S_{(1)}=\frac{1}{2}\int d^{d}x\sqrt{g}
  \left\{-\varphi_{(1)}\Box\varphi_{(1)}
  +\frac{\Delta_{(1)}}{2(d-1)}E_{(1)}\varphi^{2}_{(1)}\right\}\,. \label{act11}
\end{eqnarray}
We now see that \emph{derivative independent part of the conformally
invariant action is proportional to the scaling dimension times the first
Euler density}. Note again that both objects degenerate in minimal
dimension $d=2$ where the Laplacian itself is conformally invariant
and the Euler density describes the topological invariant, which is the
two dimensional trace anomaly.

The next step in our considerations is the $k=2$ case. Again this
higher derivative action (or 4-th order conformal invariant
operator) is known since many years~\cite{Riegert,FT} for dimension
$4$ as well as for general $d$~\cite{pan,es}. All this is presented
in \cite{Erd} where many of the invariant objects are considered. In
our work, we rederived this Lagrangian just applying the Noether
procedure to the local conformal variation of the following suitable
object
\begin{equation}\label{k21}
    S_{(2)}^{0}=\frac{1}{2}\int d^{d}x\sqrt{g}\left(\widehat{D}_{(2)}\varphi_{(2)}\right)^{2}
    ,
\end{equation}
whose Weyl transformation includes only the first derivatives of
the parameter. In (\ref{k21}) and
thereafter, we use the notation
\begin{eqnarray}\label{dk}
    \widehat{D}_{(k)}&:=&\Box - \Delta_{(k)}J , \quad k=1,2,3,\dots \,\,,\\
    \delta\left(\widehat{D}_{(k)}\varphi_{(k)}\right)&=&(\Delta_{(k)}-2)\widehat{D}_{(k)}
    \varphi_{(k)} +
    2(k-1)\partial^{\mu}\sigma\partial_{\mu}\varphi_{(k)} ,\\
    \widehat{D}^{(k)}_{\mu\nu}&:=&\nabla_{\mu}\partial_{\nu}-
    \Delta_{(k)}K_{\mu\nu}\,,\quad
    g^{\mu\nu}\widehat{D}^{(k)}_{\mu\nu}=\widehat{D}_{(k)}\,,\\
    \delta\left(\widehat{D}^{(k)}_{\mu\nu}\varphi_{(k)}\right)&=&
    \Delta_{(k)}\sigma\widehat{D}^{(k)}_{\mu\nu}\varphi_{(k)}+(\Delta_{(k)}-1)\partial_{(\mu}\sigma\partial_{\nu)}
  \varphi_{(k)}+g_{\mu\nu}\partial^{\lambda}\sigma\partial_{\lambda}\varphi_{(k)}\quad\quad\quad
\end{eqnarray}
Performing the functional integration of the Weyl variation of the
(\ref{k21}) is now just a matter of some partial integration,
elimination of the second derivatives of $\sigma$ using
(\ref{n13}),(\ref{deltaJ}) and cancelation of terms linear in
$\partial\sigma$ using the Bianchi identity (\ref{bi2}). It should
be noted here that all these types of calculations could instead be
performed using the powerful method proposed in \cite{oraf}.  Here
we presented only the direct Noether procedure because that will be
more suitable for us in the next section. After all these
manipulations we arrive at the following action
\begin{eqnarray}
  S_{(2)}^{1}&=&\frac{1}{2}\int d^{d}x\sqrt{g}\Big\{\left(\widehat{D}_{(2)}
  \varphi_{(2)}\right)^{2} +4K^{\mu\nu}\partial_{\mu} \varphi_{(2)}
  \partial_{\nu}\varphi_{(2)} - 2 J\partial^{\mu}
  \varphi_{(2)}\partial_{\mu} \varphi_{(2)}\nonumber\\
  & & +2\Delta_{(2)}\left(K^{2}-J^{2}\right)\varphi^{2}_{(2)}\Big\}\label{s2c}
\end{eqnarray}
Then after some work we can evaluate $E_{(2)}$ using (\ref{Ek}) and
(\ref{n12}) to be
\begin{eqnarray}
  E_{(2)}&=& W^{2}-4(d-3)(d-2)\left(K^{2}-J^{2}\right)\,. \label{E2}
\end{eqnarray}
We see that the $\varphi^{2}_{(2)}$ term in
(\ref{s2c}) which is linear in $\Delta_{(2)}$, is proportional to the
Weyl tensor independent part of the Euler
density. The other term without derivatives is proportional to
$\Delta_{(2)}^{2}$. This noninvariant part of the four dimensional
trace anomaly arises in $AdS/CFT$ \cite{HS} and carries the name
"holographic", and corresponds to the maximally supersymmetric gauge
theory on the boundary of $AdS_{4}$.

The combination
\begin{equation}\label{w2c}
    -\frac{1}{2}\int
    d^{d}x\sqrt{g}\left\{\frac{\Delta_{2}}{2(d-3)(d-2)}W^{2}\varphi^{2}_{(2)}\right\}
    ,
\end{equation}
on the other hand is also conformally invariant and can be added to (\ref{s2c}) at no cost. This leads us to our final result
\begin{eqnarray}
  S_{(2)}^{E}&=&\frac{1}{2}\int d^{d}x\sqrt{g}
  \left\{\varphi_{(2)}\Box^{2}\varphi_{(2)}
 -2\Delta_{(2)}J\varphi_{(2)}\Box\varphi_{(2)}
  + \Delta_{(2)}^{2}J^{2}\varphi^{2}_{(2)} \nonumber\right.\\
 &-& \left. 2 J\partial^{\mu}
  \varphi_{(2)}\partial_{\mu} \varphi_{(2)}
  +4K^{\mu\nu}\partial_{\mu} \varphi_{(2)} \partial_{\nu}\varphi_{(2)}
   -\frac{\Delta_{(2)}}{2(d-3)(d-2)}E_{(2)}\varphi^{2}_{(2)}
   \right\},\quad\quad\quad\quad\label{se}
\end{eqnarray}
confirming our main observation in the $k=2$ case.


\subsection{The $\Delta_{3}=3-d/2$ case}\label{delta3}
To confirm our main observation, verified for $k=1,2$ above, and present
it as an assertion for general $k$, we need to carry out this verification
in the next nontrivial case of $k=3$. This is the content of the present
subsection, which consists of the explicit calculation of the conformally
invariant action analogous to (\ref{act11}) and (\ref{se}) for $k=1,2$.
In this case we will follow again the same strategy.

Taking into account that $\widehat{D}_{(3)}\varphi_{(3)}$ scales as an
object with the dimension $\Delta_{(1)}=\Delta_{(3)}-2$ we start
from the following initial Lagrangian
\begin{equation}\label{ks0}
    S_{(3)}^{0}=-\frac{1}{2}\int
    d^{d}x\sqrt{g}\left\{\widehat{D}_{(3)}\varphi_{(3)}
    \left(\widehat{D}_{(3)}+2J\right)\widehat{D}_{(3)}\varphi_{(3)}\right\}
    ,
\end{equation}
with the more or less simple Weyl variation
\begin{eqnarray}
  &&\delta S_{(3)}^{0}=-\int d^{d}x\sqrt{g}
  \left\{4\widehat{D}_{(3)}\varphi_{(3)}\left(\Delta_{(3)}
  \varphi_{(3)}\partial^{\lambda}\sigma\partial_{\lambda}J +4(\Delta_{(3)}-2)
  K^{\mu\nu}\partial_{\mu}\sigma\partial_{\nu}\varphi_{(3)}\right)\right.\quad\nonumber\\
  &&\left. -2\widehat{D}_{(3)}\varphi_{(3)}\left(\widehat{D}_{(3)}\varphi_{(3)}
  \delta J -4\widehat{D}^{(3)}_{\mu\nu}
  \varphi_{(3)}\delta K^{\mu\nu} -
  2\partial_{\lambda}\varphi_{(3)}\partial^{\lambda}\delta
  J -2
  \Delta_{(3)}\delta(K^{2})\varphi_{(3)}\right)\right\}.\quad\quad\quad\quad\label{1st}
 \end{eqnarray}
The second line in (\ref{1st}) can be integrated adding to the
$S_{(3)}^{0}$ the following terms
\begin{eqnarray}
  S_{(3)}^{1}&=&-\int d^{d}x\sqrt{g}\left\{2(\widehat{D}_{(3)}\varphi_{(3)})^{2}
  J
  -8\widehat{D}_{(3)}\varphi_{(3)}\widehat{D}^{(3)}_{\mu\nu}
  \varphi_{(3)}K^{\mu\nu}\right.\nonumber\\
  & &\left.-
  4\widehat{D}_{(3)}\varphi_{(3)}\partial_{\lambda}\varphi_{(3)}\partial^{\lambda}
  J -4\Delta_{(3)}
  \widehat{D}_{(3)}\varphi_{(3)}K^{2}\varphi_{(3)}\right\} .\label{n44}
\end{eqnarray}
Writing the variation of the $ S_{(3)}^{0+1}$ is rather more
complicated. First we should separate the Laplacians from
$\Delta_{(3)}J$ in the terms with $\widehat{D}_{(3)}\varphi_{(3)}$,
then, performing some partial integrations we redistribute
derivatives and separate the terms
$\partial_{\mu}\varphi_{(3)}\partial_{\nu}\varphi_{(3)}$,\,
$\partial_{\mu}\varphi_{(3)}\partial^{\mu}\varphi_{(3)}$ and
$\varphi_{(3)}^{2}$, that are irreducible under partial integration. After some manipulations, using (\ref{bi1}) and Bianchi
identities, we obtain
 \begin{eqnarray}
   && \delta S_{(3)}^{0}+\delta S_{(3)}^{1}=-\delta S_{(3)}^{2}-\delta S_{(3)}^{\Delta_{(3)}}\nonumber\\
   && +
   \int d^{d}x\sqrt{g}\left\{16C^{\lambda\mu\nu}\partial_{\lambda}
   \sigma\partial_{\mu}\varphi_{(3)}\partial_{\nu}
   \varphi_{(3)}+24\Delta_{(3)}C^{\lambda\mu\nu}\partial_{\lambda}
   \sigma K_{\mu\nu}\varphi_{(3)}^{2}\right\},\quad\label{cotton}
 \end{eqnarray}
where
\begin{eqnarray}
  &&S_{(3)}^{2}= \int d^{d}x\sqrt{g}\left\{24K^{2\mu\nu}-16JK^{\mu\nu}
  -4K^{2}g^{\mu\nu}\right\}\partial_{\mu}\varphi_{(3)}
  \partial_{\nu}\varphi_{(3)} , \label{n46}\\
  && S_{(3)}^{\Delta_{(3)}}=4\Delta_{(3)}\int
  d^{d}x\sqrt{g}\left\{J^{3}-3K^{2}J +
  2K^{3}\right\}\varphi_{(3)}^{2} .
\end{eqnarray}
Now to cancel the second line in (\ref{cotton}) with the Cotton tensor
we have to turn to the Bach tensor transformation (\ref{bachtr}). It
is easy to see that the following combination
\begin{equation}\label{bpart}
    S_{(3)}^{B}=-\frac{8}{d-4}\int d^{d}x\sqrt{g}\left\{B^{\mu\nu}
    \partial_{\mu}\varphi_{(3)}\partial_{\nu}\varphi_{(3)}
    +\Delta_{(3)}B^{\mu\nu}K_{\mu\nu}\varphi_{(3)}^{2}\right\} ,
\end{equation}
make our action completely locally conformal invariant. It follows
that the required locally Weyl invariant action for the $k=3$ case is
\begin{equation}\label{final}
    S_{(3)}=\sum^{2}_{i=0}S_{(3)}^{i}+S_{(3)}^{\Delta_{(3)}} +
    S_{(3)}^{B} .
\end{equation}
Now we analyze the linear on $\Delta_{(3)}\varphi_{(3)}^{2}$ part of
(\ref{final}):
\begin{equation}\label{hol}
   4\Delta_{(3)}\int
  d^{d}x\sqrt{g}\left\{J^{3}-3K^{2}J + 2K^{3}-\frac{2}{d-4}
  B^{\mu\nu}K_{\mu\nu}\right\}\varphi_{(3)}^{2} .
\end{equation}
We see again that this part coincides with the so-called
"holographic" anomaly \cite{HS} in 6 dimensions written in
general spacetime dimension $d$ ( see  also \cite{OA} for the role
of the Bach tensor in holography ). The main property of the holographic
anomaly is that it is a special combination of the Euler density
with the other three Weyl invariants \cite{Deser:1996na} which
reduce the topological part of the anomaly to the expression (\ref{hol})
(see \cite{Bastianelli:2000hi} for the correct separation), which is
zero for the Ricci flat metric.

But this is for the anomaly itself in $d=6$. Here we are concerned
with the invariant Lagrangian and presence of the scalar field and the
integral make our considerations easier. To get the invariant action with
the whole third Euler density, we have to perform some more work,
and find that there is another invariant action with the maximum of
four derivatives.
This action can be obtained, using the same Noether
procedure, to render the following initial term
\begin{equation}\label{Cpart}
    S_{W}^{0}=\frac{8}{(d-3)(d-4)}\int d^{d}x\sqrt{g}
    W^{\mu\alpha\nu\beta}\widehat{D}^{(3)}_{\mu\nu}
    \varphi_{(3)}\widehat{D}^{(3)}_{\alpha\beta}\varphi_{(3)}
\end{equation}
invariant. After some lengthy but straightforward calculation we arrive at
the following locally conformal invariant action.
\begin{equation}\label{w2}
    S_{W}=S_{(3)}^{B}-S_{W}^{0}-S_{W}^{1}-S_{W}^{\Delta_{(3)}}, \quad \delta
    S_{W}=0 ,
\end{equation}
where
\begin{eqnarray}
   S_{W}^{1}&=&\int d^{d}x\sqrt{g}\left\{\frac{16W^{\mu\alpha\nu\beta}
  K_{\alpha\beta}}{(d-3)}
  + \frac{3W^{2}g^{\mu\nu}-12W^{2\mu\nu}}{(d-3)(d-4)}
  \right\} \partial_{\mu}\varphi_{(3)}\partial_{\nu}\varphi_{(3)},\quad\quad\quad\label{n53}\\
  S_{W}^{\Delta_{(3)}}&=&\Delta_{(3)}\int d^{d}x\sqrt{g}\left\{\frac{12W^{\mu\alpha\nu\beta}
  K_{\mu\nu}K_{\alpha\beta}}{(d-3)}
  + \frac{3W^{2}J-12W^{2\mu\nu}K_{\mu\nu}}{(d-3)(d-4)}
  \right\}\varphi_{(3)}^{2}.\quad\quad\quad\quad\label{n54}
\end{eqnarray}
To derive this we used the Bianchi identity (\ref{n18}) contracted
with the Weyl tensor. This leads to the following relation
\begin{equation}\label{bi3}
    \frac{1}{2}\partial_{\alpha}W^{2}-2\nabla_{\mu}W^{2\mu}_{\alpha}
    =2(d-4)C^{\quad\nu}_{\lambda\rho}W_{\nu\alpha}^{\quad\lambda\rho}\,,
\end{equation}
which generates the terms quadratic in the Weyl tensor in
(\ref{w2})-(\ref{n54}). Therefore the existence of the invariant
(\ref{w2}) allows us to replace the Bach tensor
dependent term $S^{B}_{(3)}$ in (\ref{final}) with $W$ dependent terms and obtain
\begin{equation}\label{finalw}
    S_{(3)}^{\mathcal{A}}=\sum^{2}_{i=0}S^{i}_{(3)}
    +S^{0}_{W}+S^{1}_{W}+S^{\Delta_{(3)}}_{(3)}+S^{\Delta_{(3)}}_{W}
    .
\end{equation}
Then we see that all terms proportional to
$\Delta_{3}\varphi_{(3)}^{2}$ are accumulated in the last two terms
of (\ref{finalw})
\begin{equation}\label{anom}
    S^{\Delta_{(3)}}_{(3)}+S^{\Delta_{(3)}}_{W}=\frac{3\Delta_{3}}{(d-5)(d-4)(d-3)}\int
    d^{d}x\sqrt{g}\mathcal{A}\varphi_{(3)}^{2} ,
\end{equation}
where
\begin{eqnarray}
  \mathcal{A}&=&(d-5)
  [W^{2}J-4W^{2\mu\nu}K_{\mu\nu}]
  +4(d-5)(d-4)W^{\mu\alpha\nu\beta}K_{\mu\nu}
  K_{\alpha\beta}\nonumber\\ &&+\frac{4}{3}(d-5)(d-4)(d-3)
  [J^{3}-3K^{2}J + 2K^{3}] .
\end{eqnarray}
We can now insert (\ref{n12}) in (\ref{Ek}) for $k=3$ and get
\begin{eqnarray}
   E_{(3)}&=&\frac{16}{3} W^{3}+\frac{32}{3} W^{\tilde{3}}+ 8\mathcal{A} ,\\
   W^{3}&=&W^{\alpha\beta}_{\mu\nu}W^{\mu\nu}_{\lambda\rho}
   W^{\lambda\rho}_{\alpha\beta}\quad\,,
   W^{\tilde{3}}=W_{\alpha\mu\nu\beta}W^{\lambda\mu\nu\rho}
   W^{\alpha\,\,\,\,\,\beta}_{\,\,\,\,\lambda\rho} .
\end{eqnarray}
In the same way as in the $k=2$ case we can add
these two additional invariant actions with the appropriate
coefficients:
\begin{eqnarray}
  && S_{W^{3}}=\frac{1}{2}\int
    d^{d}x\sqrt{g}\frac{4\Delta_{3}(W^{3}+2 W^{\tilde{3}})}{(d-5)(d-4)(d-3)}
    \varphi_{(3)}^{2} ,
\end{eqnarray}
and restore the Euler density containing Lagrangian
\begin{eqnarray}\label{eds}
   S_{(3)}^{E_{(3)}}&=& S_{(3)}^{\mathcal{A}}+ S_{W^{3}}\nonumber\\
   &=&\frac{1}{2}\int d^{d}x\sqrt{g}
  \left\{-\varphi_{(2)}\Box^{3}\varphi_{(2)}+\,\,\dots\,\,
  +\frac{3\Delta_{3}}{4(d-5)(d-4)(d-3)}E_{(3)}\varphi_{(3)}^{2}\right\},\quad\quad\quad\quad
\end{eqnarray}
where we put $\dots$ instead of the other terms with derivatives, or
terms proportional to $\Delta^{2}_{(3)}$ and $\Delta^{3}_{(3)}$.
These terms can be readily read off (\ref{ks0}), (\ref{n44}),
(\ref{Cpart}) and (\ref{n53}).

We have proved our assertion concerning the connection between the hierarchy
of conformally coupled scalars with the dimensions $\Delta_{k}$  and
Euler densities $E_{(k)}$ for the $k=1,2,3$, and have constructed the
conformal coupling of the third scalar with gravity in dimensions
$d\geq 6$. This action in spacetime dimension $d=6$ or
equivalently for $\Delta_{(3)}=0$ degenerates to a conformal invariant
operator for dimension $0$ scalars obtained in \cite{KMM,AKMM} from
cohomological considerations of the effective action.


\subsection{Hierarchies of conformal invariant powers of Laplacian from ambient space }\label{ambient}

\quad In subsection \ref{hierarchy} we introduced the hierarchy of scalar fields
$\varphi_{(k)}$, where $k=1,2,3,\dots$ with the corresponding
scaling dimensions $\Delta_{(k)}=k-d/2$ and infinitesimal conformal transformations (\ref{trans}).
Each of these exists in the spacetime dimensions $d\geq 2k$, and with
the minimal vanishing dimension, $\Delta_{(k)}=0$ when $d=2k$ and couples with gravity in the conformally invariant way through the hierarchy of the conformally invariant $k$-th power of the Laplacian
\begin{equation}\label{23}
    \hat{\mathcal{L}}_{(k)}=\Box^{k}+\dots+\Delta_{(k)}\mathbf{a}_{{(k)}} .
\end{equation}
The interesting point of this consideration was the appearance \cite{MT} of the so-called holographic anomaly $\mathbf{a}_{{(k)}}$ \cite{HS} namely
\emph{the derivative independent part of the conformally
invariant $k$-th power of Laplacian is the scaling dimension times the holographic anomaly in dimension $d=2k$ written in general spacetime dimension $d$}.

In this subsection we will explain this remarkable property of the above hierarchy, namely that one obtains
conformal invariant operators from the $k$-th power of the Laplace-Beltrami operator constructed from the ambient metric which acts on the $d+2$ dimensional scalar field and from using the FG holographic expansion (\ref{2}). So we concentrate on \footnote{We use the notation  $\Box_{A}$ for the Laplacian in ambient space. The $\Box_{h}$ is the Laplacian constructed from $h_{ij}(x,\rho)$ and  a simple $\Box$ corresponds to the boundary metric $g_{ij}(x)$ .}
\begin{equation}\label{24}
    \left(\Box_{A}\right)^{k}f(x,t,\rho)\,,
\end{equation}
where
\begin{equation}\label{25}
    \Box_{A}=\frac{\ell^{2}}{t^{2}}\Box_{h}
    +\frac{4\rho}{t^{2}}\partial^{2}_{\rho}
    -\frac{4}{t}\partial_{t}\partial_{\rho}
    +h^{ij}h'_{ij}\left(\frac{2\rho}{t^{2}}\partial_{\rho}
    -\frac{1}{t}\partial_{t}\right)-\frac{2(d-2)}{t^{2}}\partial_{\rho}\,.
    \end{equation}
For doing that first of all we have to understand the right truncation for the $d+2$ dimensional scalar $f(x,t,\rho)$ to the $d$ dimensional scalar $\varphi_{k}(x)$. Taking into account that we do not want to consider $AdS/CFT$ behaviour for the scalar field we can  take it $\rho$ independent. Then from simple scaling arguments we arrive at the following ansatz
\begin{equation}\label{26}
    f(x,t,\rho) = t^{\Delta_{(k)}}\varphi_{(k)}(x)\,.
\end{equation}
Then we see that (\ref{25}) reduces to
\begin{eqnarray}
  \Box_{A}\left[t^{\Delta_{(k)}}\varphi_{(k)}(x)\right]
  &=&\ell^{2}t^{\Delta_{(k)}-2}\left[\Box_{h} \varphi_{(k)}(x)-\frac{\Delta_{(k)}}{\ell^{2}}h^{ij}h'_{ij}\varphi_{(k)}(x)\right], \label{27}
\end{eqnarray}
so that inserting $k=1$ and using (\ref{12}) we obtain
\begin{equation}\label{28}
     \Box_{A}\left[t^{\Delta_{(1)}}\varphi_{(1)}(x)\right]|_{\rho=0}=\ell^{2}t^{-d/2}\left(\Box -\Delta_{(1)}J\right)\varphi_{(1)}(x) ,
\end{equation}
where we recognize in the brackets the well known conformal Laplacian
\begin{equation}\label{29}
   \hat{\mathcal{L}}_{(1)}= \Box -\Delta_{(1)}J = \Box+\frac{(d-2)}{4(d-1)}R\,.
\end{equation}

The next step in our ambient space considerations is the $k=2$ case.
First we rewrite the last term in (\ref{25}) in the $\Delta_{(k)}$ dependent form
\begin{equation}\label{30}
    -2\frac{d-2}{t^{2}}\partial_{\rho}=\frac{4\Delta_{(k)}-4(k-1)}{t^{2}}\partial_{\rho}\,.
\end{equation}
Inserting (\ref{27}) in (\ref{25}) and expanding in $\rho$ we obtain
\begin{eqnarray}
   &&\Box^{2}_{A}\left[t^{\Delta_{(k)}}\varphi_{(k)}(x)\right]
  =\ell^{4}t^{\Delta_{{(k)}}-4}f_{(k)}(\rho,x)=\ell^{4}t^{\Delta_{{(k)}}-4}\Big\{\left(\Box- \frac{\Delta_{(k)}}{\ell^{2}}h^{(1)}\right)^{2}+\frac{2}{\ell^2}h^{(1)}\Box \nonumber\\
  && -\frac{4(3-k)}{\ell^{2}}\left[h^{(1)ij}\nabla_{i}\partial_{j}+\frac{1}{2}(\nabla^{n}h^{(1)})\partial_{n}\right]
  +\frac{2}{\ell^{4}}\Delta_{(k)}\left[(3-k)h^{(1)ij}h^{(1)}_{ij}-h^{(1)^{2}}\right] \nonumber\\
  &&+\frac{\rho\Delta_{(k)}}{\ell^{4}}\left( 8h^{(1)ij}h^{(2)}_{ij}
  -4h^{(1)ij}h^{(1)}_{jn}h^{(1)n}_{i}+ 3h^{(1)}
  h^{(1)ij}h^{(1)}_{ij}\right)\nonumber\\
   && + \rho O(\nabla)+\rho O(3-k)+ \rho O(\Delta_{(k)}^{2})+O(\rho^{2})\Big\}\varphi_{(k)}(x) ,\,\,\,\,\label{31}
\end{eqnarray}
where we use the  following relations
\begin{eqnarray}
  && \nabla_{j}h^{(1)j}_{i}=\nabla_{i}h^{(1)} ,\quad h^{(2)}=\frac{1}{4}h^{(1)ij}h^{(1)}_{ij}\,, \label{32}\\
  && \nabla_{j}h^{(2)j}_{i}+\frac{1}{2}\nabla_{i}h^{(2)}
  =\frac{1}{2}h^{(1)jn}\nabla_{j}h^{(1)}_{ni}+\frac{1}{4}h^{(1)}_{ij}\nabla^{j}h^{(1)} \,, \label{33}\\
  &&h^{(3)}=\frac{2}{3}h^{(1)ij}h^{(2)}_{ij}-\frac{1}{6}h^{(1)ij}h^{(1)}_{jn}h^{(1)n}_{i} \,, \label{34}
\end{eqnarray}
obtained from $\rho$ expansion of  (\ref{5}) and (\ref{6}).
Now inserting in (\ref{31}) $k=2$ and $\rho=0$  and using (\ref{12}) we obtain
\begin{eqnarray}
  &&\Box^{2}_{A}\left[t^{\Delta_{(2)}}\varphi_{(k)}(x)\right]|_{\rho=0}
  =\ell^{4}t^{\Delta_{{(2)}}-4}\hat{\mathcal{L}}_{(2)}\varphi_{(k)}(x) \,,\label{35}\\
  && \hat{\mathcal{L}}_{(2)}=\left(\Box-\Delta_{(2)}J\right)^{2}
  -4\nabla_{i}K^{ij}\partial_{j}+2\nabla^{i}J\partial_{i}
  +2\Delta_{(2)}\left(K^{2}-J^{2}\right).\quad \label{36}
\end{eqnarray}
Again this fourth order
higher derivative  conformal invariant
operator is known since many years~\cite{Riegert,FT} for dimension
$4$ as well as for general $d$~\cite{pan,es}. This operator was rederived in \cite{MT} and here in subsection \ref{hierarchy} as a kinetic operator for the second  Lagrangian of the hierarchy of conformally coupled scalars by simply applying the Noether procedure.

Now we can evaluate the general expression for Euler densities
\begin{eqnarray}
  && E_{(k)}:=\frac{1}{2k (d-2k)!}\delta^{i_{1}
  \dots i_{d-2k}j_{1}j_{2}\dots j_{2k-1}j_{2k}}_{i_{1}
  \dots i_{d-2k}k_{1}k_{2}\dots k_{2k-1}k_{2k}}
  R^{k_{1}k_{2}}_{j_{1}j_{2}}\dots R^{k_{2k-1}k_{2k}}_{j_{2k-1}
  j_{2k}} \,.\label{37}
\end{eqnarray}
for $k=2$ and obtain
\begin{eqnarray}
  2\Delta_{(2)}\left(K^{2}-J^{2}\right)=-\frac{\Delta_{(2)}}{2(d-3)(d-2)}\left(E_{(2)}- W^{2}\right)\,. \label{38}
\end{eqnarray}
So we see that the last term in (\ref{36}), which is linear in $\Delta_{(2)}$, is proportional to the
Weyl tensor independent part of the Euler
density.  Thus we recognize as $\mathbf{a}_{(k)}$ of (\ref{23}) for both the $k=1,2$ cases (\ref{28}), (\ref{35})
\begin{eqnarray}
  && \mathbf{a}_{(1)}=-\frac{1}{\ell^{2}}h^{(1)}=-\frac{1}{2(d-1)}E_{(1)} \,,\label{39}\\
  && \mathbf{a}_{(2)}=2(h^{(1)ij}h^{(1)}_{ij}-h^{(1)2})
  =-\frac{1}{2(d-3)(d-2)}\left(E_{(2)}- W^{2}\right)\,.\label{40}
\end{eqnarray}

The "holographic" trace anomaly arises in $AdS/CFT$ \cite{HS}
and corresponds to the maximally supersymmetric gauge
theories on the boundary of $AdS_{3}$ and $AdS_{5}$.
To check our statement as an assertion for general $k$, we need to carry out this verification
in the next nontrivial case of $k=3$ obtained in subsection \ref{delta3} by the Noether procedure \cite{MT}(the sixth order conformally invariant operator in $d=6$ was obtained in \cite{KMM} from cohomological consideration).
We performed the full calculation  inserting (\ref{31}) with   $k=3$ in (\ref{25}) and have found full agreement with the formula (\ref{finalw}) ((56) of \cite{MT}). Here, to avoid cumbersome formulas, we will trace only the derivative independent term linear in $\Delta_{(3)}$.
First of all we see from (\ref{25}) and (\ref{30}) the relation
\begin{eqnarray}\label{41}
    \Box_{A}\ell^{4}t^{\Delta_{(k)}-4}f_{(k)}(\rho,x)&=&\ell^{6}t^{\Delta_{(k)}-6}
    \left[\Box+(4-\Delta_{(k)})h^{(1)}\right.\nonumber\\&+&\left.4(5-k)\partial_{\rho} +O(\rho)\right]f_{(k)}(\rho,x)\,.
\end{eqnarray}
Then it is easy to see that the relevant terms in (\ref{31}) are only two derivative free expressions with the $\ell^{-4}$ in front. Now because both derivative free terms in (\ref{31}) are already with a $\Delta_{(k)}$ factor, the operator (\ref{41}) contributes only as $4h^{(1)}+8\partial_{\rho}$ if $k=3$ and we have to just multiply the derivative free part of the second line in (\ref{31}) (it is just $-\frac{2\Delta_{(3)}}{\ell^{4}}h^{(1)2}$ for k=3) by $4h^{(1)}$ and add it to the third line of (\ref{31}) with factor $8$ instead of the $\rho$. So finally we have
\begin{eqnarray}
  && \Box^{3}_{A}\left[t^{\Delta_{(3)}}\varphi_{(k)}(x)\right]|_{\rho=0}
  = \ell^{6}t^{\Delta_{(3)}-6}\hat{\mathcal{L}}_{(3)}\varphi_{(3)}(x)=\ell^{6}
  t^{\Delta_{(3)}-6}\Big\{\Box^{3}+ \dots\dots \nonumber\\ && +\frac{8\Delta_{(3)}}{\ell^{6}}\left[8h^{(1)ij}h^{(2)}_{ij}
  -4h^{(1)ij}h^{(1)}_{jk}h^{(1)k}_{i}+3 h^{(1)}
  h^{(1)ij}h^{(1)}_{ij}-h^{(1)3}\right]\Big\}\varphi_{(3)}(x).\quad\quad\quad\quad\label{42}
\end{eqnarray}
Now using again (\ref{12}) and (\ref{13}) we see that
\begin{eqnarray}
  && \mathbf{a}_{(3)}=-8\left[J^{3}-3K^{ij}K_{ij}J+
  2K^{ij}K_{jn}K^{n}_{j}-\frac{2}{d-4}K^{ij}B_{ij}\right].\label{43}
\end{eqnarray}
We see again that this part coincides with the so called
"holographic" anomaly \cite{HS} in 6 dimensions written in
general spacetime dimension $d$ ( see  also \cite{OA}). The important property of the holographic anomaly is that it is a special combination of the Euler density
with three other  Weyl invariants \cite{Deser:1996na},\cite{Bastianelli:2000hi}  which
reduce the topological part of the anomaly to the expression (\ref{43}), which is zero for the Ricci flat metric (see  \cite{Boul2} for recent results on purely algebraic considerations of the general structure of the Weyl anomaly in arbitrary $d$ ).


\subsection{The ambient space, PBH diffeomorphisms  and Ricci gauging}\label{PBH}

\quad In this subsection we consider an ambient space origin of another method of construction of $d$ dimensional local conformal invariants. This is the so-called Ricci gauging proposed by A.~Iorio, L.~O'Raifeartaigh, I.~Sachs and C.~Wiesendanger in \cite{oraf}. Ricci gauging is very effective when we start from a scale invariant matter field Lagrangian and want to generalize it to a local Weyl or conformal invariant Lagrangian.
The prescription developed in \cite{oraf} consists of two steps
\begin{enumerate}
\item First of all we have to perform Weyl gauging by introduction of the corresponding Weyl gauge field $A_{i}(x)$. For the scalar field it looks like
\begin{eqnarray}
&&\partial_{i}\varphi_{(k)}(x)\rightarrow D_{i}\varphi_{(k)}(x)= (\partial_{i}-\Delta_{(k)}A_{i}(x))\varphi_{(k)}(x) ,\label{44}\\
&&\delta A_{i}(x)=\partial_{i}\sigma(x) ,\quad \delta D_{i}\varphi_{(k)}(x)=\Delta_{(k)}D_{i}\varphi_{(k)}(x),\,\label{45}
\end{eqnarray}
   with the additional "pure gauge" conditions $\nabla_{i}A_{j}=\nabla_{j}A_{i}$ for elimination of the  self invariant combinations of $A_{i}$ constructed from the field strength $F_{ij}=\partial_{[i}A_{j]}$ .
\item After Weyl gauging the actions with a conformally invariant flat space limit (scale invariant) contain  the field $A_{i}$ only in the combinations
\end{enumerate}
\begin{eqnarray}
 && \Omega_{ij}[A]=\nabla_{i}A_{j}(x)+A_{i}A_{j}-\frac{g_{ij}}{2}g^{kl}A_{k}A_{l} ,\quad \delta \Omega_{ij}[A]=\nabla_{i}\partial_{j}\sigma(x), \quad\quad\label{46}\\
 && \Omega[A]=g^{ik}\nabla_{i}A_{k}(x)+\frac{d-2}{2}g^{kl}A_{k}A_{l} ,\quad \delta \Omega[A]=\Box \sigma(x)  \,,\label{47}
\end{eqnarray}

and therefore can be replaced by
\begin{equation}\label{48}
 K_{ij}=\Omega_{ij}[A]\quad \textnormal{and}\quad  J=\Omega[A]\,.
\end{equation}
 The authors of \cite{oraf} called  this procedure \emph{Ricci gauging}.

To understand this Ricci gauging on the level of $d+2$ dimensional \emph{gauged} ambient space of Fefferman and Graham we turn first to the idea of PBH diffeomorphisms \cite{PBH} of the higher dimensional spaces, which reduce to conformal transformations on the lower dimensional boundary or embedded subspace. Actually the PBH transformations can be defined as higher dimensional diffeomorphisms which leave the form of the higher dimensional metric invariant.
The PBH transformations for the bulk metric (\ref{9}) are constructed and analyzed in \cite{Schwim} and \cite{S}. For the $d+2$ dimensional ambient metric (\ref{1}) PBH diffeomorphisms are considered in \cite{OA}. The existence of such a transformations is another reason why the reparametrization invariant powers of the Laplacian in ambient space reduce to the Weyl invariant
operators in $d$ dimensional space as considered in the previous section.
Following \cite{OA} we define PBH transformations of (\ref{1}) as diffeomorphisms (Lie derivative along the vector $\zeta^{\mu}(t,\rho,x)$)
\begin{eqnarray}
    \delta g^{A}_{\mu\nu}(x^{\mu})&=&\mathcal{L}_{\zeta(t,\rho,x)}g^{A}_{\mu\nu}(t,\rho,x)
    =\zeta^{\lambda}(t,\rho,x)\partial_{\lambda}g^{A}_{\mu\nu}(t,\rho,x)\nonumber\\
    &+&g^{A}_{\mu\lambda}(t,\rho,x)\partial_{\nu}\zeta^{\lambda}(t,\rho,x)
    +g^{A}_{\nu\lambda}(t,\rho,x)\partial_{\mu}\zeta^{\lambda}(t,\rho,x) ,\label{49}
\end{eqnarray}
satisfying the conditions
\begin{equation}\label{50}
    \delta g^{A}_{tt}(t,\rho,x)=\delta g^{A}_{t\rho}(t,\rho,x)
    =\delta g^{A}_{\rho\rho}(t,\rho,x)=\delta g^{A}_{ti}(t,\rho,x)=\delta g^{A}_{\rho i}(t,\rho,x)=0\,.
\end{equation}
The corresponding infinitesimal PBH transformations are \cite{Schwim},\cite{OA}
\begin{eqnarray}
  && \zeta^{t}(t,\rho,x)=t\sigma(x) \,,\label{51}\\
  && \zeta^{\rho}(t,\rho,x)=-2\rho\sigma(x)\,,\label{52}\\
  && \zeta^{i}(t,\rho,x)=\zeta^{i}(x,\rho) \,,\quad h_{ij}(\rho ,x)\partial_{\rho}\zeta^{i}(\rho,x)=\frac{\ell^{2}}{2} \partial_{i}\sigma(x)\,,\label{53}\\
  && \delta h_{ij}(\rho,x)=2\sigma(x)(1-\rho\partial_{\rho})h_{ij}(\rho,x)
  +\mathcal{L}_{\zeta(\rho,x)}h_{ij}(\rho,x) \,.\label{54}
\end{eqnarray}
We see that PBH transformations depend on two free parameters $\sigma(x)$ and $\zeta^{i}(x)=\zeta^{i}(0,x)$ corresponding to the local Weyl and local diffeomorphisms of the boundary metric $g_{ij}(x)=h_{ij}(0,x)$. All other terms $n!\zeta^{(n)i}(x)=\frac{\partial^{n}}{\partial_{\rho}^{n}}\zeta(\rho,x)|_{\rho=0}$ of the $\rho$ expansion of the $\zeta^{i}(\rho,x)$ are expressed through $\sigma(x)$ according to the relation (\ref{53}). This dependence fixes the special unhomogeneous  forms of the Weyl transformations of the FG coefficients,  which is in full agreement with the  direct solution (\ref{11})-(\ref{13}) of the corresponding equations (\ref{3}) or (\ref{11}) (see \cite{Schwim} for details).

To include the Weyl gauge field $A_{i}(x)$ in this game and find an ambient space description of the Ricci gauging we introduce a generalized $d+2$ dimensional gauged ambient space with the following metric
\begin{equation}\label{55}
    ds^{2}_{GA}=\frac{t^{2}}{\ell^{2}}\left[h_{ij}(\rho,x)+\rho \ell^{2}A_{i}(x)A_{j}(x)\right]dx^{i}dx^{j}-\rho dt^{2} -t\left[dt+tA_{i}(x)dx^{i}\right]d\rho\,.
\end{equation}
Then we consider corresponding $d+2$ dimensional diffeomorphisms conserving the form of (\ref{55})
\begin{equation}\label{56}
  \delta g^{GA}_{tt}(t,\rho,x)=\delta g^{GA}_{t\rho}(t,\rho,x)
    =\delta g^{GA}_{\rho\rho}(t,\rho,x)=\delta g^{GA}_{ti}(t,\rho,x)=0\,,
\end{equation}
and giving for $A_{i}(x)$  a gauge transformation with the Weyl parameter $\sigma(x)$ (\ref{45}). The corresponding solution gives for new PBH transformations
\begin{eqnarray}
  && \zeta^{t}(t,\rho,x)=t\sigma(x) \,,\label{57}\\
  && \zeta^{\rho}(t,\rho,x)=-2\rho\sigma(x)\,,\label{58}\\
  && \zeta^{i}(t,\rho,x)=\zeta^{i}(x) \,,\label{59}\\
  && \delta h_{ij}(\rho,x)=2\sigma(x)(1-\rho\partial_{\rho})h_{ij}(\rho,x)
  +\mathcal{L}_{\zeta(x)}h_{ij}(\rho,x) \,,\label{60}\\
  &&\delta A_{i}(x)=\partial_{i}\sigma(x) +\mathcal{L}_{\zeta(x)}A_{i}(x)\,.\label{61}
\end{eqnarray}
Comparing with (\ref{51})-(\ref{54}) we see that we were lucky with the ansatz (\ref{55}) to restore the Weyl part of the PBH transformation with the proper gauge transformation for $A_{i}(x)$.
The only difference that we have here is the $\rho$-independence of the bulk diffeomorphisms $\zeta^{i}(x)$ and correspondingly the absence of the  condition (\ref{53}).
It is a price for the additional gauge field transformation (\ref{61}). However, this difference is very essential for the FG expansion. Putting $\zeta^{i}(x)=0$ we get from (\ref{60}) for pure Weyl transformations of the FG coefficients $n!h^{(n)}_{ij}(x)$ only the homogeneous parts
\begin{eqnarray}
  && \delta g_{ij}(x)=2\sigma(x)g_{ij}(x) \,,\label{62}\\
  && \delta h_{ij}^{(1)}(x)=0 \,,\label{63}\\
  && \delta h_{ij}^{(2)}(x)=-2\sigma(x)h^{(2)}_{ij}(x)\,.\label{64}
\end{eqnarray}
 So it seems really as a Weyl gauged version of the FG expansion.
For making the final check of this assertion we turn now to the Ricci flatness condition for the gauged ambient metric (\ref{55}). Inverting the metric (\ref{55}) we obtain
\begin{eqnarray}
&&\left(
  \begin{array}{ccc}
   \ell^{2}A^{2} & -\frac{2\gamma}{t} & -\frac{\ell^{2}}{t}A^{j}\\
    -\frac{2\gamma}{t} & \frac{4\rho\gamma}{t^{2}} & \frac{2\rho\ell^{2}}{t^{2}} A^{j}\\
     -\frac{\ell^{2}}{t}A^{i} & \frac{2\rho\ell^{2}}{t^{2}} A^{i} & \frac{\ell^{2}}{t^{2}}h^{ij} \\
  \end{array}
\right)\,,\label{65}\end{eqnarray}
where
 \begin{eqnarray}
   && \gamma=1+\rho\ell^{2}A^{2} , \quad A^{2}(\rho,x)=h^{nm}(\rho,x)A_{n}(x)A_{m}(x)\,,\label{66}\\
   && A^{i}(\rho,x)=h^{ik}(\rho,x)A_{k}(x)\,.
 \label{67}
 \end{eqnarray}
 Then  the calculation of the Christoffel symbols and Ricci tensor became straightforward if we admit the condition $F_{ij}=0$ .
After a long calculation we see that the first four equations
\begin{eqnarray}
  && R^{GA}_{it} = R^{GA}_{\rho t}= R^{GA}_{tt}\equiv 0 \,,\label{68}\\
  &&R^{GA}_{\rho\rho} =\frac{1}{2}\left[h^{kl}h''_{kl}
  -\frac{1}{2}h^{ij}h'_{jk}h^{kl}h'_{li}\right]=0 \,,\label{69}
\end{eqnarray}
are the same as in the usual ambient space. But the last two undergo a change
\begin{eqnarray}
   R^{GA}_{i\rho}&=&\frac{1}{2}h^{kl}\left[\nabla^{(h)}_{i}
  h'_{kl}-\nabla^{(h)}_{k}h'_{il}\right] + \frac{1}{2}h^{kl}h'_{kl}A_{i}+\frac{d-2}{2}h'_{ik}h^{kl}A_{l}-\rho h''_{ik}h^{kl}A_{l}=0\nonumber\\&&\hspace{8cm}\,,\label{70}\\
  \ell^{2}R^{GA}_{ij}&=&\ell^{2}R_{ij}[h]-(d-2)h'_{ij} - \gamma h^{kl}h'_{kl}h_{ij}+
  \rho\gamma\left[2h''_{ij}-2h'_{il}h^{lm}h'_{mj}
  +h^{kl}h'_{kl}h'_{ij}\right]\nonumber\\
  &-&(d-2)(\nabla^{(h)}_{i}A_{j}+A_{i}A_{j}-A^{2}h_{ij})-h_{ij}\nabla^{(h)}_{k}A^{k}\nonumber\\
  &+&\rho[h^{kl}h'_{kl}\nabla^{(h)}_{i}A_{j}-(d-4)A^{2}h'_{ij}
  -2A^{k}(h'_{ik}A_{j}+h'_{jk}A_{i})\nonumber\\
  &-& h^{kl}(h'_{ki}\nabla^{(h)}_{l}A_{j}+h'_{kj}\nabla^{(h)}_{l}A_{i})+\nabla^{(h)}_{k}(h'_{ij}A^{k})+2\rho h'_{ik}A^{k}h'_{jl}A^{l}]=0.\quad\quad\quad\quad\label{71}
\end{eqnarray}
Then inserting in (\ref{71}) $\rho=0$ we obtain instead of (\ref{12}) the following solution for the first coefficient of the FG expansion
\begin{eqnarray}
  \frac{1}{\ell^{2}}h^{(1)}_{ij}(x)&=& K_{ij}-\nabla_{i}A_{j}-A_{i}A_{j}
  +\frac{1}{2}g_{ij}A_{k}A_{l}g^{kl}\nonumber\\
  &=& K_{ij}-\Omega_{ij}[A] \,,\label{72}\\
 \frac{1}{\ell^{2}}h^{(1)}(x)&=& J-\Omega[A] \,.\label{73}
\end{eqnarray}
So we see that (\ref{72}) is Weyl invariant which is in agreement with the PBH transformation (\ref{63}). On the other hand we see that Ricci gauging  leads to a \emph{trivialization of the Fefferman-Graham expansion}. Indeed the Ricci gauging condition (\ref{48}) means
\begin{equation}\label{74}
    h^{(1)}_{ij}\equiv 0 .
\end{equation}
Moreover because equations (\ref{69})-(\ref{71}) express recursively each next $h^{(n)}_{ij}$  through the nonzero powers of previous ones  we can conclude that all higher $h^{(n)}_{ij}$ coefficients of the FG expansion are trivialized after imposing the Ricci gauging condition.
The final conclusion which we can make now is the following:
\emph{The FG expansion for a gauged ambient metric (\ref{55}) can be obtained from the usual expansion for (\ref{1}) by  the Weyl gauging}.
For example we can easily guess the next coefficient
\begin{equation}\label{75}
    h^{(2)}_{ij}(x)=\frac{\ell^{4}}{4}\left\{\frac{\tilde{B}_{ij}}{d-4}+(K^{m}_{i}
    -\Omega^{m}_{i}[A])(K_{mj}-\Omega_{mj}[A])\right\} ,
\end{equation}
where
\begin{equation}\label{76}
    \tilde{B}_{ij}=B_{ij}-(d-4)A^{k}(C_{kij}+C_{kji})-(d-4)A^{k}A_{l}W^{\quad l}_{kij}
\end{equation}
is the Weyl gauged Bach tensor.


\section{Conformal invariant interaction  of a scalar field with the higher spin field in
$AdS_{D}$}\label{scalar}

\setcounter{equation}{0}
\subsection{The cases of spin two and spin four}

\quad We work in Euclidian $AdS_{D}$ with the following metric,
curvature and covariant derivatives:
\begin{eqnarray}
&&ds^{2}=g_{\mu \nu }(z)dz^{\mu }dz^{\nu
}=\frac{L^{2}}{(z^{0})^{2}}\delta _{\mu \nu }dz^{\mu }dz^{\nu
},\quad \sqrt{-g}=\frac{L^{D}}{(z^{0})^{D}}\;,
\notag  \\
&&\left[ \nabla _{\mu },\,\nabla _{\nu }\right] V_{\lambda }^{\rho }=R_{\mu
\nu \lambda }^{\quad \,\,\sigma }V_{\sigma }^{\rho }-R_{\mu \nu \sigma
}^{\quad \,\,\rho }V_{\lambda }^{\sigma }\;,  \notag \\
&&R_{\mu \nu \lambda }^{\quad \,\,\rho
}=-\frac{1}{(z^{0})^{2}}\left( \delta _{\mu \lambda }\delta _{\nu
}^{\rho }-\delta _{\nu \lambda }\delta _{\mu }^{\rho }\right)
=-\frac{1}{L^{2}}\left( g_{\mu \lambda }(z)\delta _{\nu
}^{\rho }-g_{\nu \lambda }(z)\delta _{\mu }^{\rho }\right) \;,  \notag \\
&&R_{\mu \nu }=-\frac{D-1}{(z^{0})^{2}}\delta _{\mu \nu }=-\frac{D-1}{L^{2}}%
g_{\mu \nu }(z)\quad ,\quad R=-\frac{(D-1)D}{L^{2}}\;.  \notag
\end{eqnarray}%

In \cite{Manvelyan:2004mb} the authors constructed gauge and generalized Weyl invariant actions for spin two and four gauge fields
interacting with a scalar field. Here we review these results in the form suitable for a generalization to arbitrary higher even spin fields. We work with double traceless higher spin fields in Fronsdal's formulation \cite{Fronsdal:1978rb},\cite{Fronsdal:1978vb} where the free field equation of motion for the higher spin $\ell$ field $h_{\mu_{1}...\mu_{s}}$ reads

\begin{eqnarray}
\mathcal{F}_{\mu_{1}...\mu_{\ell}}=\Box h_{\mu_{1}...\mu_{\ell}}
-\ell \nabla_{(\mu_{1}}\nabla^{\rho}h_{\mu_{2}...\mu_{\ell})\rho}
+\frac{\ell(\ell-1)}{2}\nabla_{(\mu_{1}}\nabla_{\mu_{2}}h^{\ \ \ \ \ \ \ \ \rho}_{\mu_{3}...\mu_{\ell})\rho}\nonumber\\
+\frac{\ell^{2}+\ell(D-6)-2(D-3)}{L^{2}}h_{\mu_{1}...\mu_{\ell}}
+\frac{\ell(\ell-1)}{L^{2}}g_{(\mu_{1}\mu_{2}}h^{\ \ \ \ \ \ \ \ \rho}_{\mu_{3}...\mu_{\ell})\rho}=0
\end{eqnarray}

This equation is invariant under gauge transformation\footnote{We denote symmetrization of indices by
round brackets.}

\begin{eqnarray}
\delta h_{\mu_{1}...\mu_{\ell}}=\ell\nabla_{(\mu_{1}}\epsilon_{\mu_{2}...\mu_{\ell})}=\nabla_{\mu_{1}}\epsilon_{\mu_{2}...\mu_{\ell}}+c.p.
\end{eqnarray}
where

\begin{eqnarray}
h^{\ \ \ \ \ \ \ \ \ \ \ \rho\sigma}_{\mu_{1}...\mu_{\ell-4}\rho\sigma}=0,\\
\epsilon^{\ \ \ \ \ \ \ \ \ \ \rho}_{\mu_{1}...\mu_{\ell-3}\rho}=0.
\end{eqnarray}
The trace of Fronsdal's tensor reads as

\begin{eqnarray}
r^{(\ell)\mu_{1}...\mu_{\ell-2}}=-\frac{1}{2}Tr\mathcal{F}(h^{\ell})=\nabla_{\alpha}\nabla_{\beta}h^{(\ell)\alpha\beta\mu_{1}...\mu_{\ell-2}}-\Box h_{\alpha}^{(\ell)\alpha\mu_{1}...\mu_{\ell-2}}\nonumber\\
-\frac{\ell-2}{2}\nabla^{(\mu_{1}}\nabla_{\alpha}h_{\beta}^{(\ell)\mu_{2}...\mu_{\ell-2})\alpha\beta}
-\frac{(\ell-1)(D+\ell-3)}{L^{2}}h_{\alpha}^{(\ell)\alpha\mu_{1}...\mu_{\ell-2}}.\label{frdtr}
\end{eqnarray}

For the case $\ell=2$  one can see \cite{Manvelyan:2004mb} that a Weyl invariant action is
\begin{eqnarray}
&&S^{WI}(\phi,h^{(2)})=S_{0}(\phi)+S_{1}^{\Psi^{(2)}}(\phi,h^{(2)})+S_{1}^{r^{(2)}}(\phi,h^{(2)}). \label{spin2}
\end{eqnarray}
where
\begin{eqnarray}
S_{0}(\phi)&=&\frac{1}{2}\int d^{D}z\sqrt{-g}[\nabla_{\mu}\phi\nabla^{\mu}\phi+\frac{D(D-2)}{4L^{2}}\phi^{2}], \label{S0}\\
S_{1}^{\Psi^{(2)}}(\phi,h^{(2)})&=&\frac{1}{2}\int d^{D}z\sqrt{-g}h^{(2)\mu\nu}\Psi^{(2)}_{\mu\nu}(\phi)\label{psi12}\\\Psi^{(2)}_{\mu\nu}(\phi)&=&-\nabla_{\mu}\phi\nabla_{\nu}\phi
+\frac{g_{\mu\nu}}{2}(\nabla_{\lambda}\phi\nabla^{\lambda}\phi+\frac{D(D-2)}{4L^{2}}\phi^{2}), \quad\label{psi2}\\
S_{1}^{r^{(2)}}(\phi,h^{(2)})&=&\frac{1}{8}\frac{D-2}{D-1}\int d^{D}z\sqrt{-g}r^{(2)}(h^{(2)})\phi^{2}, \label{Sr2}\\r^{(2)}(h^{(2)})&=&\nabla_{\mu}\nabla_{\nu}h^{(2)\mu\nu}-\Box h_{\mu}^{(2)\mu}
-\frac{D-1}{L^{2}}h_{\mu}^{(2)\mu}\label{Sr12}
\end{eqnarray}
which is of course the linearized form of (\ref{act1}) and is invariant with respect to the gauge and Weyl transformations
\footnote{$\Delta$ is so-called conformal weight of the scalar and gets fixed by conformal invariance condition}
\begin{eqnarray}
\delta_{\varepsilon}^{1}\phi&=&\varepsilon^{\mu}(z)\nabla_{\mu}\phi\label{phig2},\quad
\delta_{\varepsilon}^{0}h_{\mu\nu}^{(2)}=2\nabla_{(\mu}\varepsilon_{\nu)}\label{eps2};\\
\delta_{\sigma}^{1}\phi(z)&=&\Delta\sigma(z)\phi(z)\label{phiw2},\quad
\delta_{\sigma}^{0}h_{\mu\nu}^{(2)}=2\sigma(z)g_{\mu\nu}\label{hw2}.\\
\Delta&=&1-\frac{D}{2}
\end{eqnarray}

Now we turn to the case $\ell=4$. In \cite{Manvelyan:2004mb} the authors started from the action (\ref{S0}) and applied Noether's procedure using the following higher spin 'reparametrization' of the scalar field with a traceless third rank symmetric tensor parameter
\begin{eqnarray}
\delta_{\epsilon}^{1}\phi(z)=\epsilon^{\mu\nu\lambda}(z)\nabla_{\mu}\nabla_{\nu}\nabla_{\lambda}\phi(z),\quad
\epsilon^{\alpha}_{\ \alpha\mu}=0.\label{phig4}
\end{eqnarray}
The variation of (\ref{S0}) is\footnote{From now on we will never make a difference between a variation of the Lagrangians or the actions discarding all total derivative terms and admitting partial integration if necessary. For compactness  we introduce shortened notations for divergences of the tensorial symmetry parameters
\begin{eqnarray}
\epsilon_{(1)}^{\mu\nu\dots}=\nabla_{\lambda}\epsilon^{\lambda\mu\nu\dots},\quad \epsilon_{(2)}^{\mu\dots}=\nabla_{\nu}\nabla_{\lambda}\epsilon^{\nu\lambda\mu\dots}, \label{conv}\quad\dots
\end{eqnarray}}
\begin{eqnarray}
&&\delta_{\epsilon}^{1}S_{0}(\phi)=\int d^{D}z\sqrt{-g}\{-\nabla^{(\alpha}\epsilon^{\mu\nu\lambda)}\nabla_{\mu}
\nabla_{\alpha}\phi\nabla_{\nu}\nabla_{\lambda}\phi+\epsilon_{(1)}^{\mu\nu}[\frac{1}{2}\nabla_{\mu}
\nabla_{\alpha}\phi\nabla_{\nu}\nabla^{\alpha}\phi\nonumber\\
&&+\frac{D(D+2)}{8L^{2}}\nabla_{\mu}
\phi\nabla_{\nu}\phi]
-\nabla^{(\mu}\epsilon_{(2)}^{\nu)}
[-\nabla_{\mu}\phi\nabla_{\nu}\phi+\frac{g_{\mu\nu}}{2}
(\nabla_{\lambda}\phi\nabla^{\lambda}\phi+\frac{D(D-2)}{4L^{2}}\phi^{2})]\nonumber\\&&+[\nabla^{2}\phi
-\frac{D(D-2)}{4L^{2}}\phi]\nabla_{\mu}(\epsilon_{(1)}^{\mu\nu}\nabla_{\nu}\phi)
 \}.\label{var4}
\end{eqnarray}
We see immediately  that the first two lines of (\ref{var4}) produce interactions with the spin four and two currents. From the other hand the last line in (\ref{var4}) is proportional to the equation of motion following from $S_{0}(\phi)$ and  therefore  can be absorbed after gauging by the trace of the spin four gauge field  ($2\epsilon_{(1)}^{\mu\nu}\rightarrow h_{\alpha}^{(4)\alpha\mu\nu}$) performing the following field redefinition of $\phi$
\begin{equation}\label{frd}
    \phi\rightarrow \phi +\frac{1}{2}\nabla_{\mu}(h_{\alpha}^{(4)\alpha\mu\nu}\nabla_{\nu}\phi)
\end{equation}
Such a type of field redefinition is a standard correction of Noether's procedure and means that we always can drop from the cubic part of the action terms proportional to the equation of motion following from the quadratic part of the initial action.

So finally we see that the action
\begin{eqnarray}
S^{GI}(\phi,h^{(2)},h^{(4)})&=&S_{0}(\phi)+S_{1}^{\Psi^{(2)}}(\phi,h^{(2)})+S_{1}^{\Psi^{(4)}}(\phi,h^{(4)})\label{GI4},
\end{eqnarray}
where $S_{0}(\phi)$ , $S_{1}^{\Psi^{(2)}}(\phi,h^{(2)})$ are defined in (\ref{S0})-(\ref{psi2}) and
\begin{eqnarray}
&&S_{1}^{\Psi^{(4)}}(\phi,h^{(4)})=\frac{1}{4}\int d^{D}z\sqrt{-g}h^{(4)\mu\nu\alpha\beta}\Psi^{(4)}_{\mu\nu\alpha\beta}(\phi)\\
&&\Psi^{(4)}_{\mu\nu\alpha\beta}(\phi)=\nabla_{(\mu}
\nabla_{\nu}\phi\nabla_{\alpha}\nabla_{\beta)}\phi-g_{(\mu\nu}[\nabla_{\alpha}
\nabla^{\gamma}\phi\nabla_{\beta)}\nabla_{\gamma}\phi
+\frac{D(D+2)}{4L^{2}}\nabla_{\alpha}
\phi\nabla_{\beta)}\phi]\label{psi4},\quad\quad\quad\quad
\end{eqnarray}
is invariant with respect to the gauge transformations of the spin four field with  an additional  spin two field gauge transformation inspired by the second divergence of the spin four gauge parameter\footnote{Note that the spin two part of our action continues to be invariant in respect of usual linearized  reparametrization (\ref{eps2})}
\begin{eqnarray}
&&\delta_{\epsilon}^{1}\phi(z)=\epsilon^{\mu\nu\lambda}(z)\nabla_{\mu}\nabla_{\nu}\nabla_{\lambda}\phi(z)
,\label{phig4n}\\
&&\delta_{\epsilon}^{0}h^{(4)\mu\nu\alpha\beta}=4\nabla^{(\mu}\epsilon^{\nu\alpha\beta)},\quad
\delta_{\epsilon}^{0}h_{\alpha}^{(4)\alpha\mu\nu}=2\epsilon_{(1)}^{\mu\nu}\label{eps4},\\
&&\delta_{\epsilon}^{0}h^{(2)\mu\nu}=2\nabla^{(\mu}\epsilon_{(2)}^{\nu)}.\label{eps24}
\end{eqnarray}
 Thus we introduced a gauge invariant interaction of the scalar with the spin four gauge field $h^{(4)}_{\mu\nu\alpha\beta}$ in the minimal way. The next step is the spin four Weyl invariant interaction.

We write the generalized Weyl transformation law for the spin four case  as in the \cite{Manvelyan:2004mb}
\begin{eqnarray}
\delta_{\sigma}^{0}h^{(4)\mu\nu\alpha\beta}(z)=12\sigma^{(\mu\nu}(z)g^{\alpha\beta)},\ \ \
\delta_{\sigma}^{1}\phi(z)=\Delta_{4}\sigma^{\alpha\beta}\nabla_{\alpha}\nabla_{\beta}\phi,\label{sig4}
\end{eqnarray}
where we introduced a generalized "conformal" weight $\Delta_{4}$ for the scalar field.
Then following \cite{Manvelyan:2004mb} one can make (\ref{GI4}) Weyl invariant  introducing the following terms
\begin{eqnarray}
S_{1}^{r^{(4)}}=\frac{1}{2}\xi_{4}^{1}\int d^{D}z\sqrt{-g}r^{(4)\mu\nu}\nabla_{\mu}\phi\nabla_{\nu}\phi+
\frac{1}{2}\xi_{4}^{0}\int d^{D}z\sqrt{-g}\nabla_{\mu}\nabla_{\nu}r^{(4)\mu\nu}\phi^{2},\label{Sr4}
\end{eqnarray}
where\footnote{We have to mention that our $\Delta_{4}$ here differs from $\widetilde{\Delta}$ in \cite{Manvelyan:2004mb} because of field redefinition (\ref{frd}) which is the reason why $S_{1}^{\Psi^{(4)}}$ from \cite{Manvelyan:2004mb} turned into (\ref{psi4}). When we make field redefinition, we add to the Lagrangian terms which are not Weyl invariant, and in order to restore Weyl invariance we have to change the coefficient $\Delta_{4}$.}
\begin{eqnarray}
&&r^{(4)\mu\nu}=\nabla_{\alpha}\nabla_{\beta}h^{(4)\alpha\beta\mu\nu}-\Box h_{\alpha}^{(4)\alpha\mu\nu}-\nabla^{(\mu}\nabla_{\beta}h_{\alpha}^{(4)\nu)\beta\alpha}
-\frac{3(D+1)}{L^{2}}h_{\alpha}^{(4)\alpha\mu\nu},\quad\quad\quad\quad\\
&&\delta_{\epsilon}^{1}r^{(4)\mu\nu}=0,\ \ \ \ r_{\mu}^{(4)\mu}=0,\\
&&\xi_{4}^{1}=-\frac{1}{4}\frac{D}{D+3},\ \ \ \ \xi_{4}^{0}=\frac{1}{32}\frac{D(D-2)}{(D+1)(D+3)},\ \ \ \ \ \Delta_{4}=\Delta=1-\frac{D}{2}.
\end{eqnarray}
Thus the linearized action for a scalar field interacting with the spin two and four fields in a conformally invariant way is
\begin{eqnarray}
S^{WI}(\phi,h^{(2)},h^{(4)})=S^{WI}(\phi,h^{(2)})+S_{1}^{\Psi^{(4)}}(\phi,h^{(4)})+S_{1}^{r^{(4)}}(\phi,h^{(4)}),\label{spin4}
\end{eqnarray}
which is invariant with respect to gauge and generalized Weyl transformations
\begin{eqnarray}
&&\delta^{1}\phi=\varepsilon^{\mu}\nabla_{\mu}\phi+\epsilon^{\mu\nu\lambda}\nabla_{\mu}\nabla_{\nu}\nabla_{\lambda}\phi
+\Delta\sigma\phi+\Delta\sigma^{\mu\nu}\nabla_{\mu}\nabla_{\nu}\phi,\\
&&\delta^{0}h^{(2)\mu\nu}=2\nabla^{(\mu}\varepsilon^{\nu)}+2\nabla^{(\mu}\epsilon_{(2)}^{\nu)}
+2(1-\Delta-4D\xi_{4}^{1})\nabla^{(\mu}\sigma_{(1)}^{\nu)}\nonumber\\
&&+2\sigma g^{\mu\nu}+2\xi_{4}^{1}\sigma_{(2)}g^{\mu\nu}\label{rgp}\\
&&\delta^{0}h^{(4)\mu\nu\alpha\beta}=4\nabla^{(\mu}\epsilon^{\nu\alpha\beta)}+12\sigma^{(\mu\nu}g^{\alpha\beta)}.
\end{eqnarray}

\subsection{Gauge invariant interaction for the spin $\ell$ case}

\quad Here we generalize our construction to the general spin  $\ell$ case.
Again following \cite{Manvelyan:2004mb} we apply the following gauge transformation
\begin{eqnarray}
&&\delta_{\epsilon}^{1}\phi(z)=\epsilon^{\mu_{1}\mu_{2}...\mu_{\ell-1}}(z)\nabla_{\mu_{1}}\nabla_{\mu_{2}}...
\nabla_{\mu_{\ell-1}}\phi(z),\label{phigl}\\
&&\delta_{\epsilon}^{0}h^{(\ell)\mu_{1}...\mu_{\ell}}=l\nabla^{(\mu_{\ell}}\epsilon^{\mu_{1}\mu_{2}...\mu_{\ell-1})},\quad
\delta_{\epsilon}^{0}h_{\alpha}^{(\ell)\alpha\mu_{1}...\mu_{\ell-2}}=2\epsilon_{(1)}^{\mu_{1}...\mu_{\ell-2}}\label{tepsl},\\  &&\epsilon^{\alpha}_{\alpha\mu_{3}...\mu_{\ell-1}}=0
\end{eqnarray}
to the action (\ref{S0}) and obtain  the  following starting variation for Noether's procedure
\begin{eqnarray}
&&\delta_{\epsilon}^{1}S_{0}(\phi)=
\int d^{D}z\sqrt{-g}
\{
\nabla^{\alpha}\epsilon^{\mu_{1}...\mu_{\ell-1}}\nabla_{\alpha}\phi\nabla_{\mu_{1}}...\nabla_{\mu_{\ell-1}}\phi+\nonumber\\
&&\epsilon^{\mu_{1}...\mu_{\ell-1}}\nabla_{\alpha}\phi\nabla^{\alpha}\nabla_{\mu_{1}}...\nabla_{\mu_{\ell-1}}\phi+
\frac{D(D-2)}{4L^{2}}\epsilon^{\mu_{1}...\mu_{\ell-1}}\phi\nabla_{\mu_{1}}...\nabla_{\mu_{\ell-1}}\phi
\}.\quad\quad\label{varl}
\end{eqnarray}
Using the following notations
\begin{eqnarray}
T(n,k)&=&\nabla^{\alpha}\epsilon_{(\ell-n)}^{\mu_{1}...\mu_{n-1}}\nabla_{\mu_{1}}...\nabla_{\mu_{k-1}}\nabla_{\alpha}\phi
                                                               \nabla_{\mu_{k}}...\nabla_{\mu_{n-1}}\phi,\\
M(n,k)&=&\epsilon_{(\ell-n-1)}^{\mu_{1}...\mu_{n}}\nabla_{\mu_{1}}...\nabla_{\mu_{k}}\nabla_{\alpha}\phi
                                              \nabla_{\mu_{k+1}}...\nabla_{\mu_{n}}\nabla^{\alpha}\phi,\\
N(n,k)&=&\epsilon_{(\ell-n-1)}^{\mu_{1}...\mu_{n}}\nabla_{\mu_{1}}...\nabla_{\mu_{k}}\phi
                                              \nabla_{\mu_{k+1}}...\nabla_{\mu_{n}}\phi.
\end{eqnarray}
and commutation relation (D.1) from Appendix D we rewrite (\ref{varl}) in the form
\begin{eqnarray}
&&\delta_{\epsilon}^{1}S_{0}(\phi)=
\int d^{D}z\sqrt{-g}\{T(\ell,1)+M(\ell-1,0)+\nonumber\\
&&+\frac{(\ell-1)(\ell-2)}{2L^{2}}N(\ell-1,1)+\frac{D(D-2)}{4L^{2}}N(\ell-1,0)
\}.\label{varl1}
\end{eqnarray}
Then using relations between $T(m,n)$, $M(m,n)$ and $N(m,n)$ from Appendix C and after some algebra we 'diagonalize' (\ref{varl1})
\begin{eqnarray}
&&\delta_{\epsilon}^{1}S_{0}(\phi)=\sum_{m=1}^{\frac{\ell}{2}}(-1)^{m}\binom{\ell-m-1}{m-1}
\int d^{D}z\sqrt{-g}\{
-T(2m,m)+\frac{1}{2}M(2m-2,m-1)\nonumber\\
&&+\frac{(D+2m-2)(D+2m-4)}{8L^{2}}N(2m-2,m-1)\nonumber\\
&&-\frac{m-1}{\ell-2m+1}\epsilon_{(\ell-2m+1)}^{\mu_{1}...\mu_{2m-2}}
(\nabla_{\mu_{1}}...\nabla_{\mu_{m-1}}[\nabla^{2}\phi-\frac{D(D-2)}{4L^{2}}\phi]
\nabla_{\mu_{m}}...\nabla_{\mu_{2m-2}}\phi)\}\label{epslS}
\end{eqnarray}
Further performing a final symmetrization in (\ref{epslS}), we obtain the following elegant expression
\begin{eqnarray}
&&\delta_{\epsilon}^{1}S_{0}(\phi)=\int d^{D}z\sqrt{-g}\Big\{\sum_{m=1}^{\frac{\ell}{2}}\binom{\ell-m-1}{m-1}
[-\nabla^{(\mu_{2m}}\epsilon_{(\ell-2m)}^{\mu_{1}...\mu_{2m-1})}\Psi^{(2m)}_{\mu_{1}...\mu_{2m}}]\nonumber\\
&&+[\nabla^{2}\phi-\frac{D(D-2)}{4L^{2}}\phi]\sum_{m=2}^{\frac{\ell}{2}}\binom{\ell-m-1}{m-2}
\nabla_{\mu_{1}}...\nabla_{\mu_{m-1}}(\epsilon_{(\ell-2m+1)}^{\mu_{1}...\mu_{2m-2}}
\nabla_{\mu_{m}}...\nabla_{\mu_{2m-2}}\phi)
\Big\},\quad\quad\quad\quad\label{varlf}
\end{eqnarray}
where
\begin{eqnarray}
&&\Psi^{(2m)}_{\mu_{1}...\mu_{2m}}=(-1)^{m}\{\nabla_{\mu_{1}}...\nabla_{\mu_{m}}\phi
\nabla_{\mu_{m+1}}...\nabla_{\mu_{2m}}\phi\nonumber\\
&&-\frac{m}{2}g_{\mu_{2m-1}\mu_{2m}}g^{\alpha\beta}
\nabla_{(\mu_{1}}...\nabla_{\mu_{m-1}}\nabla_{\alpha)}\phi
\nabla_{(\mu_{m}}...\nabla_{\mu_{2m-2}}\nabla_{\beta)}\phi\nonumber\\
&&-\frac{m(D+2m-2)(D+2m-4)}{8L^{2}}g_{\mu_{2m-1}\mu_{2m}}\nabla_{\mu_{1}}...\nabla_{\mu_{m-1}}\phi
\nabla_{\mu_{m}}...\nabla_{\mu_{2m-2}}\phi\}\quad\quad\quad\quad \label{psil}
\end{eqnarray}
and we admitted symmetrization for the set $\mu_{1},\dots \mu_{2m}$ of indices.
So we see that miraculously the coefficients in (\ref{psil}) don't depend on $\ell$ ! All $\ell$- dependence is concentrated in the second line of (\ref{varlf}) proportional to the equation of motion for the action (\ref{S0}). This part like in the spin four case can be removed by an appropriate field redefinition (see (\ref{tepsm}), (\ref{vareps}), (D.6))
\begin{eqnarray}
    \phi\rightarrow \phi +\sum_{m=2}^{\frac{\ell}{2}}\frac{m-1}{2(\ell-2m+1)}\nabla_{\mu_{1}}...\nabla_{\mu_{m-1}}(h_{\alpha}^{(2m)\alpha\mu_{1}...\mu_{2m-2}}
                                                                                        \nabla_{\mu_{m}}...\nabla_{\mu_{2m-2}}\phi)\
\end{eqnarray}
and we can drop these terms from our consideration.
Thus we arrive at the following spin $\ell$ gauge invariant action
\begin{eqnarray}
&&S^{GI}(\phi,h^{(2)},h^{(4)},...,h^{(\ell)})=S_{0}(\phi)+\sum_{m=1}^{\frac{\ell}{2}}S_{1}^{\Psi^{(2m)}}(\phi,h^{(2m)})\label{l}
\end{eqnarray}
where
\begin{eqnarray}
&&S_{1}^{\Psi^{(2m)}}(\phi,h^{(2m)})=\frac{1}{2m}\int d^{D}z\sqrt{-g}h^{(2m)\mu_{1}...\mu_{2m}}\Psi^{(2m)}_{\mu_{1}...\mu_{2m}}\nonumber\\
&&=\frac{(-1)^{m}}{2m}\int d^{D}z\sqrt{-g}\{h^{(2m)\mu_{1}...\mu_{2m}}\nabla_{\mu_{1}}...\nabla_{\mu_{m}}\phi\nabla_{\mu_{m+1}}...\nabla_{\mu_{2m}}\phi\nonumber\\
&&-\frac{m}{2}h_{\alpha\mu_{m}...\mu_{2m-2}}^{(2m)\alpha\mu_{1}...\mu_{m-1}}\nabla_{(\mu_{1}}...\nabla_{\mu_{m-1}}\nabla_{\mu)}\phi
\nabla^{(\mu_{m}}...\nabla^{\mu_{2m-2}}\nabla^{\mu)}\phi\nonumber\\
&&-\frac{m(D+2m-2)(D+2m-4)}{8L^{2}}
h_{\alpha}^{(2m)\alpha\mu_{1}...\mu_{2m-2}}\nabla_{\mu_{1}}...\nabla_{\mu_{m-1}}\phi
\nabla_{\mu_{m}}...\nabla_{\mu_{2m-2}}\phi
\},\nonumber\\\label{psi2m}
\end{eqnarray}
and the final form of the improved gauge transformations
\begin{eqnarray}
&&\delta_{\epsilon}^{1}\phi(z)=\epsilon^{\mu_{1}\mu_{2}...\mu_{\ell-1}}(z)\nabla_{\mu_{1}}\nabla_{\mu_{2}}...
\nabla_{\mu_{\ell-1}}\phi(z),\label{phigln}\\
&&\delta_{\epsilon}^{0}h^{(2m)\mu_{1}...\mu_{2m}}=2m\nabla^{(\mu_{2m}}\varepsilon^{(2m)\mu_{1}...\mu_{2m-1})},\quad
\delta_{\epsilon}^{0}h_{\alpha}^{(2m)\alpha\mu_{1}...\mu_{2m-2}}=2\varepsilon_{(1)}^{(2m)\mu_{1}...\mu_{2m-2}},\quad\quad\quad\quad\label{tepsm}\\
&&\varepsilon^{(2m)\mu_{1}...\mu_{2m-1}}=\binom{\ell-m-1}{m-1}\epsilon_{(\ell-2m)}^{\mu_{1}...\mu_{2m-1}}.\label{vareps}
\end{eqnarray}
Now we can insert $m=\frac{\ell}{2}$ into (\ref{psil}) and compare our general expression for $S_{1}^{\Psi^{(\ell)}}(\phi,h^{(\ell)})$ with the already known cases of spin two (the energy momentum tensor for the scalar field) (\ref{psi2}) and spin four (\ref{psi4}).
We can easily see that for these cases $S_{1}^{\Psi^{(\ell=2,4)}}(\phi,h^{(\ell)})$ exactly reproduces (\ref{psi2}) and (\ref{psi4}) respectively.
So we found the gauge invariant action for a general spin $l$ gauge field coupled to a scalar and this action has the following property:

\emph{The gauge invariant action $S^{GI}(\phi,h^{(2)},h^{(4)},...,h^{(\ell)})$ for a spin $\ell$ gauge field coupled to a scalar includes gauge invariant actions of the tower of all smaller even spin gauge fields coupled to the same scalar in an analogous way.}

Note that this statement holds true only if we think of an even number of divergencies applied to the gauge parameter as a possible redefinition of gauge parameter of smaller even spin gauge fields, in that case this amazing hierarchy of all smaller even spin currents appear. Another possibility is to regard divergencies of the gauge parameter as gauge transformation for divergencies of the trace of the spin $\ell$ field and make  an appropriate field redefinition. In that case we don't need to introduce smaller spin currents, but the field redefinition will be of another form. The current of spin $\ell$ is the same in both approaches, it is unique, and in the flat space limit reproduces currents constructed in  \cite{Anselmi:1999bb}, \cite{vanDam} and \cite{Petkou} applying a partial integration and field redefinition. The interesting point is that this symmetric form of currents is unique, and the natural generalization of the energy-momentum tensor of the scalar field (\ref{psi2}).

\subsection{Weyl invariant action for a higher spin field coupled to a scalar}

\quad
In this section we introduce  generalized Weyl transformations for higher spin fields and derive a Weyl invariant action for a higher spin field coupled to a scalar field. Following \cite{Manvelyan:2004mb} we write the generalized Weyl transformation for the even spin $l$ field in the form
\begin{eqnarray}
&&\delta_{\sigma}^{0}h^{(\ell)\mu_{1}...\mu_{\ell}}=\ell(\ell-1)\sigma^{(\mu_{1}...\mu_{\ell-2}}g^{\mu_{\ell-1}\mu_{\ell})},\label{hwl}\\
&&\delta_{\sigma}^{0}h_{\alpha}^{(\ell)\alpha\mu_{1}...\mu_{\ell-2}}=2(D+2\ell-4)\sigma^{\mu_{1}...\mu_{\ell-2}},\\
&&\delta_{\sigma}^{1}\phi=\Delta_{\ell}\sigma^{\mu_{1}...\mu_{\ell-2}}\nabla_{\mu_{1}}...\nabla_{\mu_{\ell-2}}\phi.\label{phiwl}
\end{eqnarray}
Then we assume that the Weyl invariant action for a spin $\ell$ field should be accompanied with similar Weyl invariant actions for smaller spin gauge fields and therefore can be constructed from (\ref{l}) adding $\frac{\ell}{2}$ additional terms
\begin{eqnarray}
S^{WI}(\phi,h^{(2)},h^{(4)},...,h^{(\ell)})&=&S^{GI}(\phi,h^{(2)},...,h^{(\ell)})
+\sum_{m=1}^{\ell/2}S_{1}^{r^{(2m)}}(\phi,h^{(2m)}),\quad\quad\quad\label{WIl}
\end{eqnarray}
where each $S_{1}^{r^{(2m)}}$ is gauge invariant itself.
In the case of spin two we had only the linearized Ricci scalar (see (\ref{Sr2})) and for the spin four case we had two terms constructed from the spin four generalization of the Ricci scalar (see (\ref{Sr4})). Now we will see that the generalization of the Ricci scalar for a higher spin field namely the trace of Fronsdal's operator (\ref{frdtr}) (see \cite{Fronsdal:1978vb},\cite{Manvelyan:2004mb}) is the only gauge invariant combination of two derivatives and a higher spin field which we need to construct the Weyl invariant action (\ref{WIl}) starting from (\ref{l}). We will use the following strategy for solving our problem:
We apply transformation (\ref{hwl})-(\ref{phiwl}) to (\ref{l}) and try to compensate it with the variation of
\begin{eqnarray}
  && \sum_{m=1}^{\ell/2}S_{1}^{r^{(2m)}}(\phi,h^{(2m)}),\ where \nonumber\\
  && S_{1}^{r^{(\ell)}}(\phi,h^{(2)},...,h^{(\ell)})=\nonumber\\
  && =\frac{1}{2}\sum_{m=0}^{\frac{\ell}{2}-1}\xi_{\ell}^{m}\int d^{D}z\sqrt{-g}\nabla_{\mu_{2m+1}}...\nabla_{\mu_{\ell-2}}r^{(\ell)\mu_{1}...\mu_{\ell-2}}\nabla_{\mu_{1}}...\nabla_{\mu_{m}}\phi
  \nabla_{\mu_{m+1}}...\nabla_{\mu_{2m}}\phi \quad\quad\quad\quad\label{Rl}
\end{eqnarray}
introducing necessarily gauge and Weyl transformations for lower spin gauge fields
\begin{eqnarray}
&\delta_{\sigma}h^{(2m)\mu_{1}...\mu_{2m}}=2m(2m-1)C_{\ell}^{m}\sigma_{(\ell-2m)}^{(\mu_{1}...\mu_{2m-2}}g^{\mu_{2m-1}\mu_{2m})},\quad
m=1,...,\ell/2,\quad\quad\label{wwh}\\
&C_{\ell}^{\ell/2}=1 .\label{wwh1}
\end{eqnarray}
In other words we solve the equation
\begin{eqnarray}
&&\delta^{1}_{\sigma}S^{WI}(\phi,h^{(2)},...,h^{(\ell)})=\delta_{\sigma}^{1}S_{0}+\sum_{s=1}^{\ell/2}\delta_{\sigma}^{0}S_{1}^{\Psi^{(2s)}}
+\sum_{s=1}^{\ell/2}\delta_{\sigma}^{0}S_{1}^{r^{(2s)}}=0 \label{WW}
\end{eqnarray}
which consists of a system of $\ell+1$ equations for $(\ell/2+1)(\ell/2+2)/2$ dependent variables
 \begin{eqnarray}
   && \triangle_{\ell} ,\label{v1} \\
   && C_{\ell}^{m},\quad m=1,2,\dots,\ell/2, \label{v2}\\
   && \xi_{2s}^{n},\quad n=0,1,\dots s-1;\ s=1,...,\ell/2. \label{v3}
 \end{eqnarray}
but when we find $\xi_{\ell}^{\ell/2-k}$ we also find $\xi_{2s}^{s-k}$ for any $s\geq k$. In other words we find a whole diagonal of this triangle matrix
\begin{eqnarray}
\left(
  \begin{array}{cccccccc}
    C_{\ell}^{1} & C_{\ell}^{2} & . & . & . & C_{\ell}^{\ell/2-1} & C_{\ell}^{\ell/2} & \Delta_{\ell}\\
    \xi_{\ell}^{0} & \xi_{\ell}^{1} & . & . & . & \xi_{\ell}^{\ell/2-2} & \xi_{\ell}^{\ell/2-1} \\
    \xi_{\ell-2}^{0} & \xi_{\ell-2}^{1} & . & . & . & \xi_{\ell-2}^{\ell/2-2} &  \\
    . & . & . & . & . &   \\
    . & . & . & . &   &   \\
    \xi_{4}^{0} & \xi_{4}^{1} &   &   &   &   \\
    \xi_{2}^{0} &   &   &   &   &   \\
  \end{array}\label{matrix}
\right)
\end{eqnarray}
which helps us to solve the whole system. We have two equations for any column of this matrix besides the last, for which we have one equation for
$\Delta$. We start from the last column and go to the left. When we take any column and two equations for that column of variables, we have only two
variables to find if we already solved all columns to the right of that one (it is easy to see that the first two rows of (\ref{matrix}) are all we need to find out. The second row  gives the solution for any spin $\ell$. $\xi$-s in lower rows are just particular case and can be determined by putting concrete spin value in a general solution, which means that the independent variables are only first two rows of the (\ref{matrix}) and the number of variables in these two rows is $\ell+1$, just as much as equations we have. This right-to-left method can be used only due to the fact that we solve system for general spin case. This is a  deductive method which we use. Another approach is an inductive method - one could solve equations for concrete cases of spin 2,4,6... and obtaining all rows lower than second, and therefore whole Weyl invariant Lagrangian for lower spins, solve first two rows. Of course this is impossible for general spin $\ell$).
 That means that our system has a unique solution.
Placing all complicated Weyl variations of (\ref{WW}) into the Appendix E, we present here the resulting system of equations for the unknown variables (\ref{v1})-(\ref{v3}):
\begin{eqnarray}
&&\Delta_{\ell}=1-\frac{D}{2}\label{sys1}\\
&&\frac{(-1)^{\ell/2}}{2}(\Delta_{\ell}-\frac{\ell-2}{2})-(D+2\ell-5)\xi_{\ell}^{\ell/2-1}=0\\
&&(-1)^{m}C_{\ell}^{m}+\sum_{s=m+1}^{\ell/2}mC_{\ell}^{s}\xi_{2s}^{m}=0,\ \ \ (m=1,...,\ell/2-1)\\
&&\frac{(-1)^{m-1}}{2}(m-1)C_{\ell}^{m}-C_{\ell}^{m}(D+4m-5)\xi_{2m}^{m-1}\nonumber\\
&&+\frac{1}{2}\sum_{s=m+1}^{\ell/2}C_{\ell}^{s}[-m(m-1)\xi_{2s}^{m}-(2s-2m+2)(D+2s+2m-5)\xi_{2s}^{m-1}]=0\nonumber\\
&&(m=1,...,\ell/2-1)\label{syslast}
\end{eqnarray}

The solution of this system is universal $\Delta_{\ell}=\Delta=1-\frac{D}{2}$ and
\begin{eqnarray}
\xi_{\ell}^{m}=\frac{(-1)^{m}}{2^{l-2m}(l/2)}\binom{\ell/2}{m}\frac{(\frac{D}{2}+m-1)_{l/2-m}}{(\frac{D+l-1}{2}+m-1)_{l/2-m}}\\
\xi_{\ell}^{m}=\frac{(-1)^{m}}{2^{\ell-2m}(\ell/2)}\binom{\ell/2}{m}\frac{(\frac{D}{2}+m-1)_{\ell/2-m}}{(\frac{D+\ell-1}{2}+m-1)_{\ell/2-m}}\\
C_{\ell}^{m}=\frac{(-1)^{\ell/2-m}}{2^{\ell-2m}}\binom{\ell/2-1}{m-1}\frac{(\frac{D}{2}+m-1)_{\ell/2-m}}{(\frac{D-1}{2}+2m)_{\ell/2-m}}.
\end{eqnarray}
These expressions completely fix (\ref{Rl}) and therefore the full Weyl invariant action (\ref{WIl}), and also determine the transformation law for the whole tower of higher spin gauge fields (\ref{wwh}).\footnote{It is easy to see from formula (E.4) that we get also a redefinition of the gauge parameters for all lower even spin fields which in the spin 4 case coincides with formula (\ref{rgp}).}


\chapter{Cubic Interactions of HSF}\label{cubic}

\section{Off-Shell construction of some Higher Spin gauge field cubic interactions}\label{off}

\subsection{Exercises on spin one field couplings with the higher spin gauge fields}
\quad We start this section constructing the well known interaction of the  electromagnetic field $A_{\mu}$ in flat $D$ dimensional space-time with the linearized spin two field. Hereby we illustrate how Noether's  procedure regulates the relation between gauge symmetries of different spin fields. The standard free Lagrangian of the electromagnetic field is
\begin{eqnarray}
&&\mathcal{L}_{0}=-\frac{1}{4}F_{\mu\nu}F^{\mu\nu}=-\frac{1}{2}\partial_{\mu}A_{\nu}\partial^{\mu}A^{\nu}+\frac{1}{2}(\partial A)^{2} ,\label{L0}\\
&&F_{\mu\nu}=\partial_{\mu}A_{\nu}-\partial_{\nu}A_{\mu} ,\quad \partial A=\partial_{\mu}A^{\mu} .
\end{eqnarray}
To construct the interaction we propose a possible form for the action of the spin two linearized gauge symmetry
\begin{equation}\label{gi}
    \delta_{\varepsilon}^{0}h^{(2)\mu\nu}(x)=2\partial^{(\mu}\varepsilon^{\nu)}(x)=\partial^{\mu}\varepsilon^{\nu}(x)
    +\partial^{\nu}\varepsilon^{\mu}(x) ,
\end{equation}
on the spin one gauge field $A_{\mu}(x)$. Then Noether's procedure fixes this coupling (1-1-2 interaction) of the electromagnetic field with linearized gravity correcting when necessary the proposed transformation.

We  start from the following general ansatz for a gauge variation of $A_{\mu}$ with respect to a spin 2 gauge transformation with vector parameter $\varepsilon^{\rho}$
\begin{eqnarray}
\delta_{\varepsilon}^{1}A_{\mu}=-\varepsilon^{\rho}\partial_{\rho}A_{\mu}+C\varepsilon^{\rho}\partial_{\mu}A_{\rho} \label{delta2A}.
\end{eqnarray}
Then we apply this variation (\ref{delta2A}) to (\ref{L0}) and after some algebra neglecting total derivatives we obtain\footnote{Using the same conventions as in previous Chapter (see (\ref{conv})).}
\begin{eqnarray}
\delta_{\varepsilon}^{1}\mathcal{L}_{0}&=&\partial^{(\mu}\varepsilon^{\nu)}\partial_{\mu}A_{\rho}\partial_{\nu}A^{\rho}
-\frac{1}{2}\varepsilon_{(1)}\partial_{\mu}A_{\nu}\partial^{\mu}A^{\nu}
+\frac{1}{2}\varepsilon_{(1)}(\partial A)^{2}
+C\partial^{(\mu}\varepsilon^{\nu)}\partial_{\rho}A_{\mu}\partial^{\rho}A_{\nu}\nonumber\\
&-&2C\partial^{(\mu}\varepsilon^{\nu)}\partial_{\rho}A_{(\mu}\partial_{\nu)}A^{\rho}
+\frac{C}{2}\varepsilon_{(1)}\partial_{\mu}A_{\nu}\partial^{\nu}A^{\mu}
-\frac{C}{2}\varepsilon_{(1)}(\partial A)^{2}\nonumber\\
&+&(C-1)(\partial A)\partial^{\mu}\varepsilon^{\nu}\partial_{\nu}A_{\mu}.\label{delta1L0}
\end{eqnarray}
Then we have to compensate (or integrate) this variation using the gauge variation of the spin 2 field (\ref{gi}) and its trace $\delta_{\varepsilon}^{0}h_{\mu}^{(2)\mu}=2\varepsilon_{(1)}$ . We see immediately that the last line in (\ref{delta1L0}) is irrelevant but can be dropped by choice of the free  constant $C=1$. With this choice we have instead of (\ref{delta2A})
\begin{eqnarray}
\delta_{\varepsilon}^{1}A_{\mu}=-\varepsilon^{\rho}\partial_{\rho}A_{\mu}+\varepsilon^{\rho}\partial_{\mu}A_{\rho}
=\varepsilon^{\rho}F_{\mu\rho},\label{git}
\end{eqnarray}
so that our spin two transformation now is manifestly gauge invariant with respect to the spin one gauge invariance
\begin{eqnarray}
\delta^{0}_{\sigma}A_{\mu}=\partial_{\mu}\sigma,\label{sigma}
\end{eqnarray}
and our spin one gauge invariant free action (\ref{L0}) keeps this property also after spin two gauge variation. Namely (\ref{delta1L0}) now can be written as
\begin{eqnarray}
\delta_{\varepsilon}^{1}\mathcal{L}_{0}=\partial^{(\mu}\varepsilon^{\nu)}F_{\mu\rho}F_{\nu}^{\ \rho}-\frac{1}{4}\varepsilon_{(1)}F_{\mu\nu}F^{\mu\nu} .\label{211}
\end{eqnarray}
This variation can be compensated introducing the following 2-1-1 interaction
\begin{eqnarray}
\mathcal{L}_{1}(A_{\mu},h^{(2)}_{\mu\nu})=\frac{1}{2}h^{(2)\mu\nu}\Psi^{(2)}_{\mu\nu} ,\label{L1(2)}
\end{eqnarray}
where
\begin{eqnarray}
\Psi^{(2)}_{\mu\nu}=-F_{\mu\rho}F_{\nu}^{\ \rho}+\frac{1}{4}g_{\mu\nu}F_{\rho\sigma}F^{\rho\sigma} ,
\end{eqnarray}
is the well known energy-momentum tensor for the electromagnetic field.

Thus we solved Noether's equation
\begin{equation}\label{Ne}
    \delta_{\varepsilon}^{1}\mathcal{L}_{0}(A_{\mu})+\delta_{\varepsilon}^{0}\mathcal{L}_{1}(A_{\mu},h^{(2)}_{\mu\nu})=0
\end{equation}
in this approximation completely, defining a first order transformation and interaction term at the same time.
Finally note that the corrected Noether's procedure spin two transformation of the spin one field (\ref{git}) can be written as a combination of the usual reparametrization for the contravariant vector $A_{\mu}(x)$ (non invariant with respect to (\ref{sigma})) and spin one gauge transformation with the special field dependent choice of the parameter $\sigma(x)=\varepsilon^{\rho}(x)A_{\rho}(x)$
\begin{eqnarray}
\delta_{\varepsilon}^{1}A_{\mu}=\varepsilon^{\rho}F_{\mu\rho}=-\varepsilon^{\rho}\partial_{\rho}A_{\mu}
-\partial_{\mu}\varepsilon^{\rho}A_{\rho}+\partial_{\mu}\left(\varepsilon^{\rho}(x)A_{\rho}(x)\right),\label{git1}
\end{eqnarray}
A symmetry algebra of these transformations can be understood from commutator
\begin{eqnarray}
  && [\delta_{\eta}^{1},\delta_{\varepsilon}^{1}]A_{\mu}(x)=\delta_{[\eta,\varepsilon]}^{1}A_{\mu}(x)+\partial_{\mu}
  \left(\varepsilon^{\rho}\eta^{\lambda}F_{\rho\lambda}(x)\right) \\
&& [\eta,\varepsilon]^{\lambda}=\eta^{\rho}\partial_{\rho}\varepsilon^{\lambda}-\varepsilon^{\rho}\partial_{\rho}\eta^{\lambda}
\end{eqnarray}
So we see that algebra of transformations (\ref{git1}) close on field dependent gauge transformation (\ref{sigma}).

Now we turn to the first nontrivial case of the vector field interaction with a spin four gauge field with the following zero order spin four gauge variation
\begin{eqnarray}
\delta_{\epsilon}^{0}h^{\mu\rho\lambda\sigma}=4\partial^{(\mu}\epsilon^{\rho\lambda\sigma)},\quad
\delta_{\epsilon}^{0}h_{\rho}^{\ \rho\lambda\sigma}=2\epsilon_{(1)}^{\lambda\sigma}.\label{h4}
\end{eqnarray}
where we have a symmetric and traceless gauge parameter $\epsilon^{\mu\nu\lambda}$ to construct a gauge variation for $A_{\mu}$.
According to the previous lesson we start from a spin one gauge invariant ansatz for the spin four transformation of $A_{\mu}$ field
\begin{eqnarray}
\delta_{\epsilon}^{1}A_{\mu}=\epsilon^{\rho\lambda\sigma}\partial_{\rho}\partial_{\lambda}F_{\mu\sigma} .\label{delta4A}
\end{eqnarray}
Thus we have now the following variation  of $\mathcal{L}_{0}$
\begin{eqnarray}
\delta_{\epsilon}^{1}\mathcal{L}_{0}=\delta_{\epsilon}^{1}(-\frac{1}{4}F_{\mu\nu}F^{\mu\nu})
=(\delta_{\epsilon}^{1}A_{\nu})\partial_{\mu}F^{\mu\nu}
=-\partial_{\mu}(\epsilon^{\rho\lambda\sigma}\partial_{\rho}\partial_{\lambda}F_{\nu\sigma})F^{\mu\nu} .
\end{eqnarray}

After some algebra, again neglecting total derivatives and using the Bianchi identity for $F_{\mu\nu}$
\begin{eqnarray}
\partial_{\mu}F_{\nu\lambda}+\partial_{\nu}F_{\lambda\mu}+\partial_{\lambda}F_{\mu\nu}=0,\label{BianciF}
\end{eqnarray}
and taking into account the important relation
\begin{eqnarray}
-\partial^{\mu}\epsilon^{\rho\lambda\sigma}\partial_{\rho}F_{\mu}^{\ \nu}\partial_{\lambda}F_{\sigma\nu}=
    -\partial^{(\mu}\epsilon^{\rho\lambda\sigma)}\partial_{(\rho}F_{\mu}^{\ \nu}\partial_{\lambda}F_{\sigma)\nu}
    +\frac{1}{4}\epsilon_{(1)}^{\lambda\sigma}\partial^{\nu}F_{\mu\lambda}\partial^{\mu}F_{\nu\sigma}\nonumber\\
    -\frac{1}{2}\partial^{\nu}\epsilon^{\rho\lambda\sigma}\partial_{\lambda}F_{\sigma\nu}\partial^{\mu}F_{\mu\rho}
    -\frac{1}{4}\epsilon_{(1)}^{\lambda\sigma}\partial^{\mu}F_{\mu\rho}\partial^{\nu}F_{\nu\sigma},\label{sym4}
\end{eqnarray}
we arrive at the following form of the variation convenient for our analysis
\begin{eqnarray}
\delta_{\epsilon}^{1}\mathcal{L}_{0}&=&-\partial^{(\mu}\epsilon^{\rho\lambda\sigma)}\partial_{(\rho}F_{\mu}^{\ \nu}\partial_{\lambda}F_{\sigma)\nu}+\frac{1}{4}\epsilon_{(1)}^{\lambda\sigma}\partial^{\nu}F_{\mu\lambda}\partial^{\mu}F_{\nu\sigma}
+\frac{1}{4}\epsilon_{(1)}^{\lambda\sigma}\partial_{\lambda}F_{\mu\nu}\partial_{\sigma}F^{\mu\nu}\nonumber\\
&-&\partial_{\lambda}(\epsilon_{(1)}^{\lambda\sigma}F_{\mu\sigma})\partial_{\nu}F^{\nu\mu}
-\frac{1}{4}\epsilon_{(1)}^{\lambda\sigma}\partial^{\mu}F_{\mu\lambda}\partial^{\nu}F_{\nu\sigma}
-\frac{1}{2}\partial^{\rho}\epsilon^{\nu\lambda\sigma}\partial_{\lambda}F_{\sigma\rho}\partial^{\mu}F_{\mu\nu}\nonumber\\
&+&\partial^{(\mu}\epsilon_{(2)}^{\nu)}F_{\mu\sigma}F_{\nu}^{\ \sigma}
-\frac{1}{4}\epsilon_{(3)}F_{\mu\nu}F^{\mu\nu}.\label{delta4L0sym}
\end{eqnarray}
Returning to the gauge variation of the spin four field (\ref{h4}) we notice that all terms in the first line of (\ref{delta4L0sym}) and the first two terms in the second line can be integrated to the interaction terms.
The last term in the second line is proportional to the free field equations but is not integrable, so we can cancel this term only by changing the initial variation of $A_{\mu}$ (\ref{delta4A}). The modified form of (\ref{delta4A}) is
\begin{eqnarray}
\delta_{\epsilon}^{1}A_{\mu}=\epsilon^{\rho\lambda\sigma}\partial_{\rho}\partial_{\lambda}F_{\mu\sigma}
+\frac{1}{2}\partial_{\rho}\epsilon_{\mu\lambda\sigma}\partial^{\lambda}F^{\sigma\rho}.
\end{eqnarray}
Therefore
\begin{eqnarray}
\mathcal{L}_{1}&=&\frac{1}{4}h^{(4)\mu\rho\lambda\sigma}\partial_{(\rho}F_{\mu}^{\ \nu}\partial_{\lambda}F_{\sigma)\nu}-\frac{1}{8}h_{\rho}^{(4)\rho\lambda\sigma}\partial^{\nu}F_{\mu\lambda}\partial^{\mu}F_{\nu\sigma}
-\frac{1}{8}h_{\rho}^{(4)\rho\lambda\sigma}\partial_{\lambda}F_{\mu\nu}\partial_{\sigma}F^{\mu\nu}\nonumber\\
&+&\partial_{\lambda}(\frac{1}{2}h_{\rho}^{(4)\rho\lambda\sigma}F_{\mu\sigma})\partial_{\nu}F^{\nu\mu}
+\frac{1}{8}h_{\rho}^{(4)\rho\lambda\sigma}\partial^{\mu}F_{\mu\lambda}\partial^{\nu}F_{\nu\sigma}\nonumber\\
&-&\frac{1}{2}h^{(2)\mu\nu}F_{\mu\sigma}F_{\nu}^{\ \sigma}
+\frac{1}{8}h^{(2)\rho}_{\rho}F_{\mu\nu}F^{\mu\nu}.\label{delta4Lsym}
\end{eqnarray}

But the two terms in the second line are proportional to the equation of motion for the initial Lagrangian (\ref{L0}), hence they are not physical and can be removed by the following field redefinition
\begin{eqnarray}
A_{\mu} \rightarrow A_{\mu}-\partial_{\lambda}(h_{\alpha}^{\ \alpha\lambda\sigma}F_{\mu\sigma})
-\frac{1}{4}h_{\ \alpha\mu\sigma}^{\alpha}\partial_{\beta}F^{\beta\sigma} .
\end{eqnarray}

So we can drop the second line of (\ref{delta4Lsym}).

Another novelty in comparison with the previous case is the third line of (\ref{delta4L0sym}). Comparing with (\ref{211}) we see that we can integrate these two terms introducing an additional spin two field coupling and compensate the first and third line introducing the following linearized Lagrangian for the coupling of the electromagnetic field to the spin four and spin two fields
\begin{eqnarray}
\mathcal{L}_{1}(A_{\mu},h^{(2)\mu\nu},h^{(4)\mu\nu\alpha\beta})=\frac{1}{4}h^{(4)\mu\nu\alpha\beta}\Psi^{(4)}_{\mu\nu\alpha\beta}
+\frac{1}{2}h^{(2)\mu\nu}\Psi^{(2)}_{\mu\nu} ,
\end{eqnarray}
where the current $\Psi^{(2)}_{\mu\nu}$ is the same energy-momentum tensor (\ref{L1(2)}) and
\begin{eqnarray}
\Psi^{(4)}_{\mu\nu\alpha\beta}=\partial_{(\alpha}F_{\mu}^{\ \rho}\partial_{\beta}F_{\nu)\rho}
-\frac{1}{2}g_{(\mu\nu}\partial^{\lambda}F_{\alpha\sigma}\partial^{\sigma}F_{\beta)\lambda}
-\frac{1}{2}g_{(\mu\nu}\partial_{\alpha}F^{\sigma\rho}\partial_{\beta)}F_{\sigma\rho} .
\end{eqnarray}
The whole lagrangian
\begin{equation}\label{l01}
    \mathcal{L}_{0}(A_{\mu})+\mathcal{L}_{1}(A_{\mu},h^{(2)\mu\nu},h^{(4)\mu\nu\alpha\beta}) ,
\end{equation}
  is invariant with respect to the spin one gauge transformations and the following higher spin transformations
\begin{eqnarray}
&&\delta^{1}A_{\mu}=\epsilon^{\rho\lambda\sigma}\partial_{\rho}\partial_{\lambda}F_{\mu\sigma}
+\frac{1}{2}\partial_{\rho}\epsilon_{\mu\lambda\sigma}\partial^{\lambda}F^{\sigma\rho},\nonumber\\
&&\delta^{0}h^{(4)\mu\nu\alpha\beta}=4\partial^{(\mu}\epsilon^{\nu\alpha\beta)},\
\delta_{\epsilon}^{0}h_{\mu}^{\ \mu\alpha\beta}=2\epsilon_{(1)}^{\alpha\beta},\nonumber\\
&&\delta^{0}h^{(2)\mu\nu}=2\partial^{(\mu}\epsilon^{\nu)}_{(2)},\ \delta^{0}h_{\mu}^{(2)\mu}=2\epsilon_{(3)} .
\end{eqnarray}
Therefore we proved that like the previously investigated scalar--higher spin coupling case (previous Chapter), the interaction with the spin four gauge field leads to the additional interaction with the lower even spin two field.
Following \cite{Mkrtchyan} we review here also vector field coupling to the general HS field.
We start from following gauge variation
\begin{eqnarray}
\delta_{\epsilon}^{1}A_{\mu}=\epsilon_{\ell}^{\mu_{1}...\mu_{l-1}}\nabla_{\mu_{1}}...\nabla_{\mu_{l-2}}F_{\mu_{l-1}\mu}.
\label{varA}
\end{eqnarray}

>From very long and tedious calculations we get

\begin{eqnarray}
&&\delta_{\epsilon}^{1}\mathcal{L}_{0}=
\sum_{m=1}^{\ell/2}\binom{\ell-m-1}{m-1}
(-\nabla^{(\mu_{2m}}\epsilon_{\ell(l-2m)}^{\mu_{1}...\mu_{2m-1})}\Psi_{\mu_{1}...\mu_{2m}}(A_{\mu}))\nonumber\\
&&+\sum_{m=1}^{\ell/2}\binom{\ell-m-1}{m-1}\frac{m-1}{m}\nabla_{\mu_{m+1}}...\nabla_{\mu_{2m-2}}
(\nabla_{\nu}\epsilon_{\ell(l-2m)\mu}^{\mu_{1}...\mu_{2m-2}}
\nabla_{\mu_{1}}...\nabla_{\mu_{m-1}}F_{\mu_{m}}^{\ \nu})\nabla_{\alpha}F^{\alpha\mu}\nonumber\\
&&+\sum_{m=1}^{\ell/2}\binom{\ell-m-1}{m-1}\frac{m-1}{2m}\nabla_{\mu_{m}}...\nabla_{\mu_{2m-3}}
(\epsilon_{\ell(l-2m+1)\mu}^{\mu_{1}...\mu_{2m-3}}
\nabla_{\mu_{1}}...\nabla_{\mu_{m-2}}\nabla^{\nu}F_{\nu\mu_{m-1}})\nabla_{\alpha}F^{\alpha\mu}\nonumber\\
&&-\sum_{m=1}^{\ell/2}\binom{\ell-m-1}{m-1}\frac{m-1}{l-2m+1}\nabla_{\mu_{m}}...\nabla_{\mu_{2m-2}}
(\epsilon_{\ell(l-2m+1)}^{\mu_{1}...\mu_{2m-2}}
\nabla_{\mu_{1}}...\nabla_{\mu_{m-2}}F_{\mu_{m-1}\mu})\nabla_{\alpha}F^{\alpha\mu},\quad\quad\quad\quad
\end{eqnarray}

where
\begin{eqnarray}
&&\Psi_{\mu_{1}...\mu_{2m}}(A_{\mu})=(-1)^{m}(-\nabla_{\mu_{1}}...\nabla_{\mu_{m-1}}F_{\mu_{m}}^{\ \nu}
                                            \nabla_{\mu_{m+1}}...\nabla_{\mu_{2m-1}}F_{\mu_{2m}\nu}\nonumber\\
&&+\frac{m-1}{2}g_{\mu_{1}\mu_{2}}\nabla_{\mu_{3}}...\nabla_{\mu_{m}}\nabla^{\alpha}F_{\mu_{m+1}\beta}
                                \nabla_{\mu_{m+2}}...\nabla_{\mu_{2m-1}}\nabla^{\beta}F_{\mu_{2m}\alpha}\nonumber\\
&&+\frac{m}{4}g_{\mu_{1}\mu_{2}}\nabla_{\mu_{3}}...\nabla_{\mu_{m+1}}F^{\rho\sigma}
                                \nabla_{\mu_{m+2}}...\nabla_{\mu_{2m}}F_{\rho\sigma})\label{psi1}
\end{eqnarray}
and we admitted symmetrization for the set $\mu_{1} \dots \mu_{2m}$ of indices.
This means that when we change our initial variation (\ref{varA}) to
\begin{eqnarray}
&&\delta_{\epsilon}^{1}A_{\mu}=\epsilon_{\ell}^{\mu_{1}...\mu_{l-1}}\nabla_{\mu_{1}}...\nabla_{\mu_{l-2}}F_{\mu_{l-1}\mu}
\nonumber\\
&&                             -\sum_{m=1}^{\ell/2}\binom{\ell-m-1}{m-1}\frac{m-1}{m}\nabla_{\mu_{m+1}}...\nabla_{\mu_{2m-2}}
(\nabla_{\nu}\epsilon_{\ell(l-2m)\mu}^{\mu_{1}...\mu_{2m-2}}
\nabla_{\mu_{1}}...\nabla_{\mu_{m-1}}F_{\mu_{m}}^{\ \nu}),\quad\quad\quad\quad
\end{eqnarray}
and also take into account appropriate field redefinition

\begin{eqnarray}
&&A_{\mu} \rightarrow A_{\mu}+\sum_{m=1}^{\ell/2}\binom{\ell-m-1}{m-1}\frac{m-1}{2m}\nabla_{\mu_{m}}...\nabla_{\mu_{2m-3}}
(\epsilon_{\ell(l-2m+1)\mu}^{\mu_{1}...\mu_{2m-3}}
\nabla_{\mu_{1}}...\nabla_{\mu_{m-2}}\nabla^{\nu}F_{\nu\mu_{m-1}})\nabla_{\alpha}F^{\alpha\mu}\nonumber\\
&&-\sum_{m=1}^{\ell/2}\binom{\ell-m-1}{m-1}\frac{m-1}{l-2m+1}\nabla_{\mu_{m}}...\nabla_{\mu_{2m-2}}
(\epsilon_{\ell(l-2m+1)}^{\mu_{1}...\mu_{2m-2}}
\nabla_{\mu_{1}}...\nabla_{\mu_{m-2}}F_{\mu_{m-1}\mu})\nabla_{\alpha}F^{\alpha\mu}
\end{eqnarray}
we can see that the gauge invariant Lagrangian for interaction of electromagnetic field with the higher even spin $\ell$ field is

\begin{eqnarray}
\mathcal{L}_{1}(A_{\mu}, h^{(2)}, h^{(4)}, ..., h^{(\ell)})=
\sum_{m=1}^{\ell/2}\frac{1}{2m}h^{(2m)\mu_{1}...\mu_{2m}}\Psi^{(2m)}_{\mu_{1}...\mu_{2m}}(A_{\mu}).\label{L1}
\end{eqnarray}

This result is similar to the scalar case investigated in the section \ref{scalar}. The same tower of even spin gauge fields appear when we construct gauge invariant interaction with higher spin fields. The generalization to the non-Abelian charged vector (Yang-Mills) fields is trivial. In scalar case we went further and constructed Weyl invariant lagrangian. The Weyl invariance can't be generalized for spin one case. That is the price for spin one manifest gauge invariance (in all interactions vector field is represented by it's curvature $F_{\mu\nu}$).
Here we wanted to mention that $AdS_{D}$ corrections to (\ref{psi1}) have following basic properties. As in the scalar case there are no $1/L^{4}$ or higher corrections. The $1/L^{2}$ term is proportional to $\ell-2$. As a result, for 1-1-2 interaction we don't have any difference between interaction in the flat space and AdS.

\subsection{Generalization to the 2-2-4 and 2-2-6 interactions}

\quad In this section we turn to the spin two field as a lower spin field in the construction of the higher spin gauge invariant interactions with spin 4 and spin 6 gauge potentials. And again we want to keep manifest the lower spin two gauge invariance.

So proceeding similarly as in the previous section we start from the free spin two Pauli-Fierz Lagrangian \cite{Fierz:1939ix}
\begin{equation}\label{3.1}
\mathcal{L}_{0}(h^{(2)}_{\mu\nu})=\frac{1}{2}\partial_{\mu}h^{(2)}_{\alpha\beta}\partial^{\mu}h^{(2)\alpha\beta}
-\partial_{\alpha}h^{(2)\alpha\beta}\partial_{\mu}h^{(2)\mu}_{\beta}+\partial_{\mu}h^{(2)\alpha}_{\alpha}\partial_{\beta}h^{(2)\beta\mu}
-\frac{1}{2}\partial_{\mu}h^{(2)\alpha}_{\alpha}\partial^{\mu}h^{(2)\beta}_{\beta},
\end{equation}
and try to solve the following Noether's equation
\begin{equation}\label{3.2}
    \delta_{\varepsilon}^{1}\mathcal{L}_{0}(h^{(2)}_{\mu\nu})
    +\delta_{\varepsilon}^{0}\mathcal{L}_{1}(h^{(2)}_{\mu\nu},h^{(4)\alpha\beta\lambda\rho} )=0 .
\end{equation}
For this purpose we introduce the following starting ansatz for the spin four transformation of the spin two field
\begin{eqnarray}
\delta^{1}_{\epsilon}h^{(2)}_{\mu\nu}=\epsilon^{\rho\lambda\sigma}\partial_{\rho}\Gamma_{\lambda\sigma,\mu\nu} ,\label{3.3}
\end{eqnarray}
where $\Gamma_{\lambda\sigma,\mu\nu}$ is the spin two gauge invariant symmetrized linearized Riemann curvature
\begin{eqnarray}
&&\Gamma_{\alpha\beta,\mu\nu}=\frac{1}{2}(R_{\alpha\mu,\beta\nu}+R_{\beta\mu,\alpha\nu}) ,\\
&& \Gamma_{(\alpha\beta,\mu)\nu}=0 ,\label{3.5}
\end{eqnarray}
introduced by de Witt and Freedman for higher spin gauge fields together with the higher spin generalization of the
Christoffel symbols \cite{deWit:pe}. This symmetrized curvature is more convenient for the construction of an interaction with symmetric tensors.
The corresponding Ricci tensor (Fronsdal operator for higher spin generalization) and scalar can be defined in the usual manner using traces
\begin{eqnarray}
&&\mathcal{F}_{\mu\nu}=\Gamma_{\mu\nu,\lambda}^{\ \ \ \ \lambda}=\Box h^{(2)}_{\mu\nu}-2\partial_{(\mu}\partial^{\alpha}h^{(2)}_{\nu)\alpha}+\partial_{\mu}\partial_{\nu}h^{(2)\alpha}_{\alpha} , \\
&&\mathcal{F}=\mathcal{F}_{\mu}^{\mu}=2(\Box h^{(2)\mu}_{\mu}-\partial_{\mu}\partial_{\nu}h^{(2)\mu\nu}).
\end{eqnarray}
In terms of these objects the Bianchi identities can be written as
\begin{eqnarray}
  &&\partial_{\lambda}\Gamma_{\mu\nu,\alpha\beta}=\partial_{(\mu}\Gamma_{\nu)\lambda,\alpha\beta}
  +\partial_{(\alpha}\Gamma_{\beta)\lambda,\mu\nu} , \label{3.8}\\
  &&\partial_{\lambda}\mathcal{F}_{\alpha\beta}=\partial^{\mu}\Gamma_{\mu\lambda,\alpha\beta}
  +\partial_{(\alpha}\mathcal{F}_{\beta)\lambda} , \label{3.9}\\
  && \partial^{\lambda}\mathcal{F}_{\lambda\mu}=\frac{1}{2}\partial_{\mu}\mathcal{F}^{\alpha}_{\alpha}.\label{3.10}
\end{eqnarray}
Then a variation of (\ref{3.1}) with respect to (\ref{3.3}) is
\begin{eqnarray}
  && \delta^{1}_{\epsilon}\mathcal{L}_{0}(h^{(2)}_{\mu\nu})=\frac{\delta \mathcal{L}_{0}}{\delta h^{(2)}_{\mu\nu}}\delta^{1}_{\epsilon}h^{(2)}_{\mu\nu}=-(\mathcal{F}^{\mu\nu}-\frac{1}{2}g^{\mu\nu}\mathcal{F})
  \epsilon^{\rho\lambda\sigma}\partial_{\rho}\Gamma_{\lambda\sigma,\mu\nu} .
\end{eqnarray}
To integrate it and solve the equation (\ref{3.2}) we submit to the following strategy:

1) First we perform a partial integration and use the Bianchi identity (\ref{3.9}) to lift the variation to a curvature square term.

2) Then we make a partial integration again and rearrange indices using (\ref{3.5}) and (\ref{3.8}) to extract an integrable part.

3) Symmetrizing expressions in this way we classify terms as
\begin{itemize}
  \item integrable
  \item integrable and subjected to field redefinition (proportional to the free field equation of motion)
  \item non integrable but reducible by deformation of the initial ansatz for the gauge transformation (again proportional to the free field equation of motion)
\end{itemize}

Then if no other terms remain we can construct our interaction together with the corrected first order transformation.
Following this strategy after some fight with formulas we obtain the following expression
\begin{eqnarray}
\delta^{1}_{\epsilon}\mathcal{L}_{0}(h^{(2)}_{\mu\nu})&=&-\partial^{(\alpha}\epsilon^{\beta\mu\nu)}
(\Psi^{(4)}_{(\Gamma)\alpha\beta\mu\nu}-\Psi^{(4)}_{(\mathcal{F})\alpha\beta\mu\nu})
\nonumber\\&-&\epsilon^{\mu\nu}_{(1)}\Gamma_{\mu\nu,\alpha\beta}\frac{\delta \mathcal{L}_{0}}{\delta  h^{(2)}_{\alpha\beta}} +\partial^{\rho}\epsilon^{\,\,\,\mu\nu}_{\alpha}\Gamma_{\beta\rho ,\mu\nu}\frac{\delta \mathcal{L}_{0}}{\delta  h^{(2)}_{\alpha\beta}} , \label{3.12}
\end{eqnarray}
where
\begin{eqnarray}
&&\Psi^{(4)}_{(\Gamma)\alpha\beta\mu\nu}=\Gamma_{(\alpha\beta,}^{\,\,\,\,\,\,\,\,\,\,\,\,\rho\sigma}\Gamma_{\mu\nu),\rho\sigma}
-\frac{2}{3}g_{(\alpha\beta}\Gamma_{\mu}^{\ \rho,\sigma\lambda}\Gamma_{\nu)\rho,\sigma\lambda} ,\label{3.13}\\
&&\Psi^{(4)}_{(\mathcal{F})\alpha\beta\mu\nu}=\mathcal{F}_{(\alpha\beta}\mathcal{F}_{\mu\nu)}
-g_{(\alpha\beta}\mathcal{F}_{\mu}^{\sigma}\mathcal{F}_{\nu)\sigma}=-\frac{\delta \mathcal{L}_{0}}{\delta  h^{(2)(\alpha\beta}}\mathcal{F}_{\mu\nu)}
+g_{(\alpha\beta}\frac{\delta \mathcal{L}_{0}}{\delta  h^{(2)\mu}_{\sigma}}\mathcal{F}_{\nu)\sigma},\quad\quad\quad\quad\label{3.14}\\
&&\frac{\delta \mathcal{L}_{0}}{\delta  h^{(2)\alpha\beta}}=-\mathcal{F}_{\alpha\beta}+\frac{1}{2}g_{\alpha\beta}\mathcal{F} .\label{3.15}
\end{eqnarray}
So we see immediately that in (\ref{3.12}) only the last term of the second line is not integrable but proportional to the equation of motion and can be dropped by the correction to the initial gauge transformation (\ref{3.3}). On the other hand taking into account (\ref{h4}) and (\ref{3.13})-(\ref{3.15}) we can compensate $\Psi^{(4)}_{(\mathcal{F})}$ and the first term in the second line of (\ref{3.12}) by the following field redefinition
\begin{eqnarray}
h^{(2)}_{\mu\nu} \rightarrow h^{(2)}_{\mu\nu} -\frac{1}{2}h^{(4)\alpha\lambda\sigma}_{\alpha}\Gamma_{\lambda\sigma,\mu\nu}
-\frac{1}{4}h^{(4)\alpha\lambda}_{\mu\nu}\mathcal{F}_{\alpha\lambda}
+\frac{1}{4}h^{(4)\alpha\lambda}_{\ \ \ \alpha(\mu}\mathcal{F}_{\nu)\lambda} .
\end{eqnarray}
Thus after field redefinition we arrive at the 4-2-2 gauge invariant interaction
\begin{eqnarray}
&&\mathcal{L}_{1}(h^{(2)}_{\mu\nu},h^{(4)}_{\alpha\beta\mu\nu})=\frac{1}{4}h^{(4)\alpha\beta\mu\nu}
\Psi^{(4)}_{(\Gamma)\alpha\beta\mu\nu}(h^{(2)}_{\mu\nu})\nonumber\\
&&=\frac{1}{4}h^{(4)\alpha\beta\mu\nu}\Gamma_{\alpha\beta,\rho\sigma}\Gamma_{\mu\nu,}^{\ \ \ \rho\sigma}
-\frac{1}{6}h^{(4)\alpha\mu\nu}_{\alpha}\Gamma_{\mu}^{\ \rho,\sigma\lambda}\Gamma_{\nu\rho,\sigma\lambda} ,
\end{eqnarray}
with the following gauge transformations
\begin{eqnarray}
&&\delta_{\epsilon}h^{(2)}_{\mu\nu}=\epsilon^{\rho\lambda\sigma}\partial_{\rho}\Gamma_{\lambda\sigma,\mu\nu}
-\partial_{\rho}\epsilon_{\lambda\sigma(\mu}\Gamma_{\nu)}^{\ \rho,\lambda\sigma} ,\\
&&\delta_{\epsilon}^{0}h^{(4)\mu\rho\lambda\sigma}=4\partial^{(\mu}\epsilon^{\rho\lambda\sigma)},\quad
\delta_{\epsilon}^{0}h_{\rho}^{(4) \rho\lambda\sigma}=2\epsilon_{(1)}^{\lambda\sigma}.
\end{eqnarray}

Now in possession of knowledge about the 2-2-4 interaction we start to construct the most nontrivial interaction in this section between spin 2 and spin 6 gauge fields. We would like to check the appearance of the 2-2-4 coupling during the construction of 2-2-6 which we expect from the analogy with the scalar case considered in \cite{Manvelyan:2004mb, Manvelyan:2009tf} and the 1-1-4 case considered in the previous subsection. To proceed we have to solve the following Noether's equation
\begin{equation}\label{3.20}
    \delta_{\varepsilon}^{1}\mathcal{L}_{0}(h^{(2)}_{\mu\nu})
    +\delta_{\varepsilon}^{0}\mathcal{L}_{1}(h^{(2)}_{\mu\nu},h^{(6)}_{\alpha\beta\lambda\rho\sigma\delta} )=0 ,
\end{equation}
with a starting ansatz for the spin 6 first order gauge transformation for the spin 2 field:
\begin{eqnarray}
\delta^{1}_{\epsilon}h^{(2)}_{\mu\nu}(x)=\epsilon^{\alpha\beta\rho\lambda\sigma}(x)
\partial_{\alpha}\partial_{\beta}\partial_{\rho}\Gamma_{\lambda\sigma,\mu\nu}(x) ,\label{3.21}
\end{eqnarray}
and the standard zero order gauge transformation for the spin 6 gauge field
\begin{eqnarray}
 &&\delta^{0}_{\epsilon}h^{(6)\mu\nu\alpha\beta\sigma\rho}=6 \partial^{(\mu}\epsilon^{\nu\alpha\beta\sigma\rho)}(x) ,\label{3.22}\\
 && \delta^{0}_{\epsilon}h^{(6)\mu\alpha\beta\sigma\rho}_{\mu}=2\epsilon^{\alpha\beta\sigma\rho}_{(1)} .\label{3.23}
\end{eqnarray}
First of all we have to transform the variation
\begin{eqnarray}
  && \delta_{\varepsilon}^{1}\mathcal{L}_{0}(h^{(2)}_{\mu\nu})=-(\mathcal{F}^{\mu\nu}-\frac{1}{2}g^{\mu\nu}\mathcal{F})
  \epsilon^{\alpha\beta\rho\lambda\sigma}
\partial_{\alpha}\partial_{\beta}\partial_{\rho}\Gamma_{\lambda\sigma,\mu\nu} ,\label{3.24}
\end{eqnarray}
into a form convenient for integration. Following the same strategy as before in the 2-2-4 case, using many times partial integration and Bianchi identities (\ref{3.5}), (\ref{3.8})-(\ref{3.10}), we obtain after tedious but straightforward calculations
\begin{eqnarray}
  &&\delta_{\varepsilon}^{1}\mathcal{L}_{0}(h^{(2)}_{\mu\nu})=\partial^{(\alpha}\epsilon^{\beta\mu\nu\lambda\rho)}
\Psi^{(6)}_{(\Gamma)\alpha\beta\mu\nu\lambda\rho}-\partial^{(\alpha}\epsilon^{\beta\mu\nu)}\Psi^{(4)}_{(\Gamma)\alpha\beta\mu\nu} \nonumber\\
  &&+\frac{4}{3}\partial^{\rho}\epsilon^{\,\,\,\,\mu\nu\lambda\sigma}_{\alpha}\partial_{\lambda}\partial_{\sigma}\Gamma_{\beta\rho ,\mu\nu}\frac{\delta \mathcal{L}_{0}}{\delta  h^{(2)}_{\alpha\beta}}-\frac{1}{3}\partial^{\rho}\partial^{\lambda}
  \epsilon^{\,\,\,\,\,\,\,\mu\nu\sigma}_{\alpha\beta}\partial_{\sigma}\Gamma_{\rho\lambda ,\mu\nu}\frac{\delta \mathcal{L}_{0}}{\delta  h^{(2)}_{\alpha\beta}}\nonumber\\
  &&- R_{int}^{\mu\nu}(\Gamma,\mathcal{F})\frac{\delta\mathcal{L}_{0}}{\delta h^{(2)}_{\mu\nu}} ,\label{3.25}
\end{eqnarray}
where
\begin{eqnarray}
 &&  \Psi^{(6)}_{(\Gamma)\alpha\beta\mu\nu\lambda\rho}=
  \partial_{(\alpha}\Gamma_{\beta\mu,}^{\quad\,\,\,\,\,\sigma\delta}\partial_{\nu}\Gamma_{\lambda\rho),\sigma\delta}
-g_{(\alpha\beta}\partial_{\mu}\Gamma_{\nu}^{\ \kappa,\sigma\delta}\partial_{\lambda}\Gamma_{\rho)\kappa,\sigma\delta}\nonumber\\
&&\qquad\qquad \quad -\frac{1}{2}g_{(\alpha\beta}\partial^{\kappa}\Gamma_{\mu\nu ,}^{\quad\sigma\delta}\partial_{\sigma}\Gamma_{\lambda\rho),\kappa\delta} ,\quad\quad\quad\quad\\
   &&\Psi^{(4)}_{(\Gamma)\alpha\beta\mu\nu}=\Gamma_{(\alpha\beta,}^{\quad\quad\rho\sigma}\Gamma_{\mu\nu),\rho\sigma}
-\frac{2}{3}g_{(\alpha\beta}\Gamma_{\mu}^{\ \rho,\sigma\lambda}\Gamma_{\nu)\rho,\sigma\lambda} ,
\end{eqnarray}
and $R_{int}^{\mu\nu}(\Gamma,\mathcal{F})\frac{\delta\mathcal{L}_{0}}{\delta h^{(2)}_{\mu\nu}}$ are remaining integrable terms proportional to the equation of motion. Indeed the symmetric tensor $R_{int}^{\mu\nu}(\Gamma,\mathcal{F})$ is expressed through the only integrable combinations of derivatives of gauge parameter
\begin{eqnarray}
  R_{int}^{\mu\nu}(\Gamma,\mathcal{F},\epsilon)&=&
  \epsilon^{\alpha\beta\lambda\delta}_{(1)}\partial_{\alpha}\partial_{\beta}
  \Gamma_{\lambda\delta,}^{\quad \mu\nu}-\frac{1}{3}\partial^{\lambda}\epsilon^{\alpha\beta\delta(\mu}_{(1)}\partial_{\alpha}\Gamma_{\,\,\lambda,\beta\delta}^{\nu)}
  +\partial_{\lambda}\left[\partial^{(\lambda}\epsilon^{\alpha\beta\delta\mu\nu)}\partial_{\alpha}\mathcal{F}_{\beta\delta}\right] \nonumber\\
  &-&\frac{2}{3}\partial_{\lambda}\left[\epsilon^{\lambda\alpha\mu\nu}_{(1)}
  \partial_{\alpha}\mathcal{F}\right]+\frac{1}{6}\epsilon^{\alpha\beta\mu\nu}_{(1)}\partial_{\alpha}\partial_{\beta}\mathcal{F}
  +\partial^{(\alpha}\epsilon^{\beta\mu\nu)}_{(2)}\mathcal{F}_{\alpha\beta} + \frac{5}{3}\partial^{\alpha}\epsilon^{\beta\lambda\mu\nu)}_{(1)}\partial_{\lambda}\mathcal{F}_{\alpha\beta} \nonumber\\
  &-&\frac{5}{3}\partial_{\lambda}\left[\epsilon^{\lambda\alpha\beta(\mu}_{(1)}\partial_{\alpha}\mathcal{F}_{\beta}^{\nu)}\right] + \frac{1}{6}\Box\epsilon^{\alpha\beta\mu\nu}_{(1)}\mathcal{F}_{\alpha\beta}-\frac{1}{6}\partial^{\lambda}
  \epsilon^{\alpha\beta\mu\nu}_{(1)}\partial_{\lambda}\mathcal{F}_{\alpha\beta}-\frac{1}{2} \epsilon^{\alpha(\mu}_{(3)}\mathcal{F}^{\nu)}_{\alpha} .\nonumber \\
\end{eqnarray}
Substituting into this expression $\partial^{(\lambda}\epsilon^{\alpha\beta\delta\mu\nu)}$ with $\frac{1}{6}h^{(6)\lambda\alpha\beta\delta\mu\nu}$ , $\partial^{(\alpha}\epsilon^{\beta\mu\nu)}_{(2)}$ with $\frac{1}{4}h^{(4)\alpha\beta\mu\nu}$, and correspondingly $2\epsilon^{\alpha\beta\mu\nu}_{(1)}$ and $2\epsilon^{\alpha\beta}_{(3)}$ with their traces, we define a field redefinition for $h^{(2)\mu\nu}$
\begin{equation}
    h^{(2)\mu\nu} \rightarrow h^{(2)\mu\nu}+  R_{int}^{\mu\nu}(\Gamma,\mathcal{F},h^{(6)},h^{(4)}) ,
\end{equation}
using which we can drop the third line in (\ref{3.25}).
The second line in (\ref{3.25}) can be cancelled by the following deformation of the initial ansatz for the transformation (\ref{3.21})
\begin{eqnarray}
\delta^{1}_{\epsilon}h^{(2)}_{\alpha\beta}=\epsilon^{\mu\nu\rho\lambda\sigma}
\partial_{\mu}\partial_{\nu}\partial_{\rho}\Gamma_{\lambda\sigma,\alpha\beta}-\frac{4}{3}\partial^{\rho}\epsilon^{\,\,\,\,\mu\nu\lambda\sigma}_{\alpha}
\partial_{\lambda}\partial_{\sigma}\Gamma_{\beta\rho ,\mu\nu}+\frac{1}{3}\partial^{\rho}\partial^{\lambda}
  \epsilon^{\,\,\,\,\,\,\,\mu\nu\sigma}_{\alpha\beta}\partial_{\sigma}\Gamma_{\rho\lambda ,\mu\nu}.\quad\quad\label{3.30}
\end{eqnarray}

Thus we arrive at the promised result that the 2-2-6 interaction automatically includes also the 2-2-4 interaction constructed above, and the corresponding trilinear interaction Lagrangian is
\begin{eqnarray}
&& \mathcal{L}_{1}(h^{(2)}, h^{(4)}, h^{(6)})
=-\frac{1}{6}h^{(6)\alpha\beta\mu\nu\lambda\rho}\Psi^{(6)}_{(\Gamma)\alpha\beta\mu\nu\lambda\rho}
+\frac{1}{4}h^{(4)\alpha\beta\mu\nu}\Psi^{(4)}_{(\Gamma)\alpha\beta\mu\nu}\nonumber\\
&& =-\frac{1}{6}h^{(6)\alpha\beta\mu\nu\lambda\rho}\partial_{\alpha}\Gamma_{\beta\mu,}^{\quad\,\,\,\,\,\sigma\delta}
\partial_{\nu}\Gamma_{\lambda\rho,\sigma\delta}
+\frac{1}{6}h_{\alpha}^{(6)\alpha\mu\nu\lambda\rho}\partial_{\mu}\Gamma_{\nu}^{\ \kappa,\sigma\delta}\partial_{\lambda}\Gamma_{\rho\kappa,\sigma\delta}
\nonumber\\&&+\frac{1}{12}h_{\alpha}^{(6)\alpha\mu\nu\lambda\rho}\partial^{\kappa}\Gamma_{\mu\nu,}^{\quad\sigma\delta}
\partial_{\sigma}\Gamma_{\lambda\rho),\kappa\delta}
+\frac{1}{4}h^{(4)\alpha\beta\mu\nu}\Gamma_{\alpha\beta,\rho\sigma}\Gamma_{\mu\nu,}^{\ \ \ \rho\sigma}
-\frac{1}{6}h^{(4)\alpha\mu\nu}_{\alpha}\Gamma_{\mu}^{\ \rho,\sigma\lambda}\Gamma_{\nu\rho,\sigma\lambda} .\quad\quad\quad\quad
\end{eqnarray}
This formula together with the corrected gauge transformation (\ref{3.30}) solves completely Noether's equation (\ref{3.20}).

\subsection{2s-s-s interaction Lagrangian}

\quad
Now we turn to the generalization of the Noether procedure of the 2-2-4 case to the general s-s-2s interaction construction.
So we must propose a first order variation of the spin s field with respect to a spin 2s gauge transformation. Remembering that Fronsdal's higher spin gauge potential is double traceless, we must
make sure that the same holds for the variation. Expanding the general variation in powers of $a^{2}$
\begin{equation}\label{deltaexpansion}
    \delta h^{(s)}(a)=\delta h^{(s)}_{(1)}(a)+a^{2}\delta h^{(s-2)}(a)+(a^{2})^{2}\delta h^{(s-4)}(a)+ \dots ,
\end{equation}
we see that the double tracelessness condition $\Box^{2}_{a} \delta h^{(s)}(a)=0$ expresses the third and
higher terms of the expansion (\ref{deltaexpansion}) through the first two free parameters $\delta h^{(s)}_{(1)}(a)$ and $\delta h^{(s-2)}(a)$\footnote{For completeness we present here the solution for $\delta h^{(s-4)}(a)$ following from the double tracelessness condition
\begin{eqnarray}
&& \delta h^{(s-4)}(a)=-\frac{1}{8\alpha_{1}\alpha_{2}} \left[\Box^{2}_{a}\delta h^{(s)}_{(1)}(a)+4\alpha_{1}\Box_{a}\delta h^{(s-2)}(a)\right] ,\nonumber\\
  && \alpha_{k}=D+2s-(4+2k) ,\quad k\in \{1,2\} . \nonumber
\end{eqnarray} }.
>From the other hand Fronsdal's tensor is double traceless by definition and therefore all these $O(a^{4})$ terms are unimportant because they do not contribute to (\ref{lagrangianvariation}). This leaves us freedom in the choice of $\delta h^{(s-2)}(a)$. Substituting (\ref{deltaexpansion}) in (\ref{lagrangianvariation}) we discover that the following choice of  $\delta h^{(s-2)}(a)$
\begin{equation}\label{4.47}
    \delta h^{(s-2)}(a)=\frac{1}{2(D+2s-2)}\Box_{a}\delta h^{(s)}_{(1)}(a) ,
\end{equation}
reduces our variation (\ref{lagrangianvariation}) to
\begin{equation}\label{4.48}
     \delta_{(1)}\mathcal{L}_{0}(h^{(s)}(a))=-\mathcal{F}^{(s)}(a)*_{a}\delta h^{(s)}_{(1)}(a) .
\end{equation}
Then we propose the following spin 2s transformation of the spin s potential
\begin{equation}\label{4.49}
    \delta h^{(s)}_{(1)}(a)= \tilde{\mathcal{U}}(b, a, 2, s)\epsilon^{2s-1}(z;b)*_{b}\Gamma^{(s)}(z;b,a) ,
\end{equation}
where
\begin{equation}\label{4.50}
    \tilde{\mathcal{U}}(b, a, 2, s)=\frac{(-1)^{s}}{(s-1)!}\prod^{s}_{k=2}\left[(\nabla\partial_{b})-\frac{1}{k}(a\partial_{b})(\nabla\partial_{a})\right] ,
\end{equation}
is operator dual to
\begin{equation}\label{4.51}
    [(b\nabla)-\frac{1}{2}(a\nabla)(b\partial_{a})]\mathcal{U}(b, a, 3, s)=\prod^{s}_{k=2}[(b\nabla)-\frac{1}{k}(a\nabla)(b\partial_{a})] ,
\end{equation}
with respect to the $*_{a,b}$ contraction product. Taking into account (\ref{4.35}) and Bianchi identities (\ref{4.41}) we get
 \begin{eqnarray}
   &&  \delta_{(1)}\mathcal{L}_{0}(h^{(s)}(a))=\epsilon^{2s-1}(z;b)*_{b}\Gamma^{(s)}(z;b,a)*_{a} [(b\nabla)-\frac{1}{2}(a\nabla)(b\partial_{a})]\mathcal{U}(b, a, 3, s)\mathcal{F}^{(s)}(z;a) \nonumber\\
   && = \epsilon^{2s-1}(z;b)*_{b}\Gamma^{(s)}(z;b,a)*_{a} \frac{1}{s(s-1)}[(b\nabla)-\frac{1}{2}(a\nabla)(b\partial_{a})]\Box_{b}\Gamma^{(s)}(z;b,a)\nonumber\\
   && = \epsilon^{2s-1}(z;b)*_{b}\Gamma^{(s)}(z;b,a)*_{a}\frac{1}{s}(\nabla\partial_{b})\Gamma^{(s)}(z;b,a)\nonumber\\
   && = -(b\nabla) \epsilon^{2s-1}(b)*_{b}\Gamma^{(s)}(b,a)*_{a}\Gamma^{(s)}(b,a)
   -\epsilon^{2s-1}(b)*_{b}\nabla_{\mu}\Gamma^{(s)}(b,a)*_{a}\frac{1}{s}\partial^{\mu}_{b}\Gamma^{(s)}(b,a) .\quad\quad\quad\quad
 \end{eqnarray}
Then using a secondary Bianchi identity  (\ref{4.40}) and a primary one (\ref{Bianchi}) one can show that
\begin{eqnarray}
  && -\epsilon^{2s-1}(b)*_{b}\nabla_{\mu}\Gamma^{(s)}(b,a)*_{a}\frac{1}{s}\partial^{\mu}_{b}\Gamma^{(s)}(b,a)\nonumber\\&&\quad\quad\quad\quad=
  \frac{1}{2s(s+1)(2s-1)}(\nabla\partial_{b})\epsilon^{2s-1}(b)*_{b}\partial^{b}_{\mu}\Gamma^{(s)}(b,a)*_{a}\partial^{\mu}_{b}\Gamma^{(s)}(b,a).
  \quad\quad\quad\quad
\end{eqnarray}
Putting all together we see that the integrated first order interaction Lagrangian (with generalized Bell-Robinson current \cite{vanDam})
\begin{eqnarray}
  && \mathcal{L}_{1}(h^{(s)}(a),h^{(2s)}(b))=\frac{1}{2s}h^{(2s)}(z;b)*_{b}\Psi^{(2s)}_{(\Gamma)}(z;b) ,\\
  && \Psi^{(2s)}_{(\Gamma)}(z;b)= \Gamma^{(s)}(b,a)*_{a}\Gamma^{(s)}(b,a)-\frac{b^{2}}{2(s+1)}\partial^{b}_{\mu}\Gamma^{(s)}(b,a)*_{a}\partial^{\mu}_{b}\Gamma^{(s)}(b,a).
  \quad\quad\quad\quad
\end{eqnarray}
supplemented with transformation (\ref{4.49}) for $h^{(s)}(a)$ and the standard zero order for $h^{(2s)}(a)$
\begin{eqnarray}
  && \delta_{0} h^{(2s)}(z;b)=2s (b\nabla)\epsilon^{(2s-1)}(z;b) ,\\
  && \delta_{0} \Box_{b}h^{(2s)}(z;b)=4s (\nabla\partial_{b})\epsilon^{(2s-1)}(z;b) ,
\end{eqnarray}
completely solves Noether's equation
\begin{equation}\label{4.58s}
    \delta_{(1)}\mathcal{L}_{0}(h^{(s)}(a))+\delta_{0}\mathcal{L}_{1}(h^{(s)}(a),h^{(2s)}(b))=0 .
\end{equation}
Note that here just as in the 2-2-4 case we did not obtain an interaction with lower spins because all derivatives included in the ansatz were used for the lifting to the second curvature.


\section{Cubic selfinteraction for Higher Spin gauge fields}\label{self}

\setcounter{equation}{0}

\subsection{Higher spin gauge field selfinteraction: The Noether's procedure}\label{self1}
\quad  Here we present again
Fronsdal's Lagrangian
\begin{equation}\label{4.42(1)}
 \mathcal{L}_{0}(h^{(s)}(a))=-\frac{1}{2}h^{(s)}(a)*_{a}\mathcal{F}^{(s)}(a)
    +\frac{1}{8s(s-1)}\Box_{a}h^{(s)}(a)*_{a}\Box_{a}\mathcal{F}^{(s)}(a) ,
\end{equation}
where $\mathcal{F}^{(s)}(z;a)$ is the so-called Fronsdal tensor
\begin{eqnarray}
\mathcal{F}^{(s)}(z;a)=\Box h^{(s)}(z;a)-s(a\nabla)D^{(s-1)}(z;a) , \quad\label{4.32(1)}
\end{eqnarray}
and $D^{(s-1)}(z;a)$ is the so-called de Donder tensor or traceless divergence of the higher spin gauge field
\begin{eqnarray}\label{4.42(D)}
 && D^{(s-1)}(z;a) = Divh^{(s-1)}(z;a)
-\frac{s-1}{2}(a\nabla)Trh^{(s-2)}(z;a) ,\\
&& \Box_{a} D^{(s-1)}(z;a)=0 .
\end{eqnarray}
The initial gauge variation of a spin $s$ field that is of field order zero is
\begin{eqnarray}\label{4.5}
\delta_{(0)} h^{(s)}(z;a)=s (a\nabla)\epsilon^{(s-1)}(z;a) ,
\end{eqnarray}
with the traceless gauge parameter
\begin{eqnarray}
\Box_{a}\epsilon^{(s-1)}(z;a)=0 ,\label{4.6}
\end{eqnarray}
for the by definition double traceless gauge field
 \begin{eqnarray}
\Box_{a}^{2}h^{(s)}(z;a)=0 .
\end{eqnarray}
Therefore on this level we can see from (\ref{4.5}) and (\ref{4.6})  that a correct generalization of the Lorentz gauge condition in the case of $s>2$ could be only the so-called de Donder gauge condition
\begin{equation}\label{dd}
    D^{(s-1)}(z;a)=0 .
\end{equation}

The equation of motion following  from (\ref{4.42(1)}) is
\begin{equation}\label{4.45}
     \delta\mathcal{L}_{0}(h^{(s)}(a))=-(\mathcal{F}^{(s)}(a)-\frac{a^{2}}{4}\Box_{a}\mathcal{F}^{(s)}(a))*_{a}\delta h^{(s)}(a) ,
\end{equation}
and zero order gauge invariance (when $\delta h^{(s)}(a)=\delta_{(0)}h^{(s)}(a)$) can be  checked by substitution of (\ref{4.5}) into this variation and use of the duality relation (\ref{dual}) and identity (\ref{BianchiF}) taking into account tracelessness  of the gauge parameter (\ref{4.6}).

Now we turn to the formulation of Noether's general procedure for constructing the spin $s$ cubic selfinteraction.
Similar to  \cite{vanDam} Noether's equation in this case looks like\footnote{From now on we will admit integration everywhere where it is necessary (we work with a Lagrangian as with an action) and therefore we will neglect all $d$ dimensional space-time total derivatives when making a partial integration.}
\begin{equation}\label{4.58}
    \delta_{(1)}\mathcal{L}_{0}(h^{(s)}(a))+\delta_{0}\mathcal{L}_{1}(h^{(s)}(a))=0 ,
\end{equation}
where $\mathcal{L}_{1}(h^{(s)}(a))$ is a cubic interaction Lagrangian and $\delta_{(1)}h^{(s)}(a)$ is a gauge transformation that is of first order in the gauge field. Actually equation (\ref{4.58}) just expresses in the cubic order on the field the general gauge invariance
\begin{equation}\label{gge}
    \delta \mathcal{L}(h^{(s)}(a))=\frac{\delta \mathcal{L}(h^{(s)}(a)}{\delta h^{(s)}(a))}*_{a}\delta h^{(s)}(a)=0 ,
\end{equation}
where
\begin{eqnarray}
  \mathcal{L}(h^{(s)}(a)) &=& \mathcal{L}_{0}(h^{(s)}(a)) + \mathcal{L}_{1}(h^{(s)}(a)) + \dots , \\
  \delta h^{(s)}(a)  &=& \delta_{(0)} h^{(s)}(a)+ \delta_{(1)}h^{(s)}(a)+ \dots .
\end{eqnarray}
Combining (\ref{4.45}) and (\ref{4.58}) we obtain the following functional Noether's equation
\begin{equation}\label{4.481}
    \delta_{(0)}\mathcal{L}_{1}(h^{(s)}(a))=(\mathcal{F}^{(s)}(a)-\frac{a^{2}}{4}\Box_{a}\mathcal{F}^{(s)}(a))*_{a}\delta_{(1)} h^{(s)}(a) ,
\end{equation}
and \emph{we would like to present in this section the solution of the latter equation for the case $s=4$ and propose a generalization for any even $s$}.

First we investigate a first order variation of the spin $s$ gauge transformation. Remembering that Fronsdal's higher spin gauge potential  has scaling dimension $\Delta_{s}=s-2$ (zero for the $s=2$ graviton case) and ascribing the same dimensions to the free part of the Lagrangian that is quadratic in the fields and derivatives $\mathcal{L}_{0}(h^{(s)}(a))$ and to the interaction $\mathcal{L}_{1}(h^{(s)}(a))$ cubic in the fields, we arrive at the idea that \emph{the number of derivatives in the interaction should be $s$}. This type of interacting theories will behave in the same way as gravity. Then we can easily conclude from (\ref{4.58}) that the number of derivatives in the first order variation $\delta_{(1)}h^{(s)}(a)$ should be $s-1$. For $s=2$ this consideration is of course in full agrement with the linearized expansion of the Einstein-Hilbert action.

The next observation is connected with double tracelessness of Fronsdal's higher spin gauge potential. This means that we must
make sure that the same holds for the variation. Expanding the general variation in powers of $a^{2}$
\begin{equation}\label{4.46}
    \delta_{(1)} h^{(s)}(a)=\delta_{(1)} \tilde{h}^{(s)}(a)+a^{2}\delta_{(1)} h^{(s-2)}(a)+(a^{2})^{2}\delta h^{(s-4)}(a)+ \dots ,
\end{equation}
we see that the double tracelessness condition $\Box^{2}_{a} \delta h^{(s)}(a)=0$ expresses the third and
higher terms of the expansion (\ref{4.46}) through the first two free parameters $\delta_{(1)} h^{(s)}(a)$ and $\delta_{(1)} h^{(s-2)}(a)$\footnote{For completeness we present here the solution for $\delta h^{(s-4)}(a)$ following from the double tracelessness condition
\begin{eqnarray}
&& \delta h^{(s-4)}(a)=-\frac{1}{8\alpha_{1}\alpha_{2}} \left[\Box^{2}_{a}\delta h^{(s)}_{(1)}(a)+4\alpha_{1}\Box_{a}\delta h^{(s-2)}(a)\right] ,\nonumber\\
  && \alpha_{k}=D+2s-(4+2k) ,\quad k\in \{1,2\} . \nonumber
\end{eqnarray} }.
>From the other hand Fronsdal's tensor (and the r.h.s of (\ref{4.481})) is double traceless by definition and therefore all these $O(a^{4})$ terms are unimportant because they do not contribute to (\ref{4.481}). This leaves us freedom in the choice of initial $\delta_{(1)} h^{(s-2)}(a)$. Using this freedom we can shift the initial first order variation in the following way
\begin{equation}\label{4.47i}
    \delta_{(1)} h^{(s)}(a)\Rightarrow \delta_{(1)} h^{(s)}(a)+\frac{a^{2}}{2(D+2s-2)}\Box_{a}\delta h^{(s)}_{(1)}(a) ,
\end{equation}
and discover that (\ref{4.481}) reduces to
\begin{equation}\label{4.482}
     \delta_{(0)}\mathcal{L}_{1}(h^{(s)}(a))=\mathcal{F}^{(s)}(a)*_{a}\delta h^{(s)}_{(1)}(a) .
\end{equation}
Now to solve this equation we can formulate the following strategy:

1) First we can start from any  cubic ansatz with $s$ derivatives $\mathcal{L}_{1}(h^{(s)}(a))$ suitable in respect to the zero order variation (\ref{4.5}) and variate it inserting in the l.h.s. of (\ref{4.482})  .

2) Then we make a partial integration  and rearrange indices to extract an integrable part due to terms proportional to Fronsdal's tensor $\mathcal{F}^{(s)}(a)$ (or $Tr\mathcal{F}^{(s)}(a))$ in agreement with the r.h.s. of (\ref{4.482}).

3) Symmetrizing expressions in this way we classify terms as
\begin{itemize}
  \item integrable
  \item integrable and subjected to field redefinition (proportional to Fronsdal's tensor)
  \item non integrable but reducible by deformation of the initial ansatz for the gauge transformation (again proportional to Fronsdal's tensor)
\end{itemize}

Then if no other terms remain we can construct our interaction together with the corrected first order transformation.
Following this strategy we will consider the $s=2$ and $s=4$ cases in the next subsections in detail.  The exact and unique results after field redefinition and partial integration that are presented in the next two subsections are in full agreement with the prediction for general even spin $s$.
To formulate this prediction let us first introduce a classification of cubic monoms with $s$ derivatives. We will call leading terms all those monoms without traces and divergences or equivalently without $\bar{h}^{(s-2)}=Tr: h^{(s)}$  and $D^{(s-1)}$, where the derivatives are contracted only with gauge fields and not with other derivatives. This type of terms is interesting because any partial integration will map such term to the terms of the same type and create one additional term with a divergence, which we can map to $D$ dependent and trace dependent terms. Another important point of this class of monoms is that inside of this class we have the following important term involving the linearized Freedman-de Witt gauge invariant curvature \cite{deWit:pe, MR6}
\begin{eqnarray}
    &&\mathcal{L}^{initial}_{1}(h^{(s)}(a))=\frac{1}{2s}h^{(s)}(b)*_{b}\Gamma^{(s)}(b,a)*_a h^{(s)}(a),\label{impterm}\\
    &&\Gamma^{(s)}(z;b,a)=\sum_{k=0}^{s}\frac{(-1)^{k}}{k!}(b\nabla)^{s-k}(a\nabla)^{k}(b\partial_{a})^{k}h^{(s)}(z;a) .\label{curv}
\end{eqnarray}
This term we can use (and we used it in the case s=4) as an initial ansatz for the solution of (\ref{4.482}). Using (\ref{curvinv}) and (\ref{4.41}) we see that
\begin{equation}\label{variat}
    \delta_{(0)}\mathcal{L}^{initial}_{1}(h^{(s)}(a))=-\epsilon^{(s-1)}(z;b)(b\nabla)h^{(s)}(a)*_a *_b \Gamma^{(s)}(b,a) + O(\mathcal{F}^{(s)}) .
\end{equation}
It is easy to see from (\ref{curv}) that after variation in the r.h.s. of  (\ref{variat}) we get $s+1$ monoms linear on the gauge parameter $\epsilon^{(s-1)}(z;b)$ and quadratic in the gauge field, where some of them contain two factors $(b\nabla)$ of contracted derivatives. These terms we can separate as next level terms including the de Donder tensor $ D^{(s-1)}(z;b)$.  To prove this statement we note first that due to partial integration there is the following simple formula:
\begin{equation}\label{mumu}
    F(z)\nabla_{\mu}G(z)\nabla^{\mu}H(z)=\frac{1}{2}\left(\Box F(z) G(z) H(z) - F(z)\Box G(z) H(z)- F(z) G(z)\Box H(z)\right).
\end{equation}
The objects $ F(z), G(z), H(z)$ in our case are proportional to $h^{(s)}(z;a)$ or $\epsilon^{(s-1)}(z;a)$. Then using the definition of Fronsdal's operator (\ref{4.32(1)}) and from (\ref{4.42(D)}) and (\ref{4.5}) follows the transformation rule
\begin{equation}\label{trr}
    \delta_{(0)}D^{(s-1)}(z;a)=\Box \epsilon^{(s-1)}(z;a) .
\end{equation}
This implies that we can classify all terms with contracted derivatives (i.e. terms with Laplacians) as monoms containing $D^{(s-1)}(z;a)$ or $ \delta_{(0)}D^{(s-1)}(z;a)$ which therefore vanish in the de Donder gauge. Actually according to the r.h.s of (\ref{4.482}) we can during functional integration always replace any $\Box h^{(s)}(a)$ with $\mathcal{F}^{(s)}(a)+s(a\nabla)D^{(s-1)}(a)$ obtaining a contribution to $\delta_{(1)}$ and shifting this monom to the next level class comprising one more order of the de Donder tensor.

Operating in this way we can integrate Noether's equation (\ref{4.482}) (or equivalently express the r.h.s. of (\ref{variat}) as   $-\delta_{(0)}\mathcal{L}^{cubic}_{1}(h^{(s)})+O(\mathcal{F}^{(s)})$) using the initial ansatz (\ref{variat}) step by step: integrating first the leading terms without any de Donder tensor or trace, then integrate terms involving only traces but not $D^{(s-1)}(z;a)$. That is the solution in de Donder gauge. After that we can continue the integration and obtain terms linear on $D^{(s-1)}(z;a)$ , quadratic and so on. The procedure will be closed when we obtain a sufficient number of $D^{(s-1)}(z;a)$ to stop the production of terms with contracted derivatives and therefore the production of new level terms coming from formula (\ref{mumu}).

\emph{Collecting the leading terms and rearranging by partial integration derivatives in a cyclic way so that each derivative acting on a tensor gauge field is contracted with the preceding tensor} we finally come to the following prediction for the leading terms of the interaction for a general spin $s$ gauge field:
\begin{eqnarray}
&&\mathcal{L}^{leading}_{(1)}(h^{(s)}(z))=\frac{1}{3s(s!)^{3}}\sum_{\alpha+\beta+\gamma = s}\binom{s}{\alpha,\beta,\gamma}\int_{z_{1},z_{2},z_{3}} \delta(z-z_{1})\delta(z-z_{2})\delta(z-z_{3})\nonumber\\
&&\left[(\nabla_{1}\partial_{c})^{\gamma}(\nabla_{2}\partial_{a})^{\alpha}(\nabla_{3}\partial_{b})^{\beta}
(\partial_{a}\partial_{b})^{\gamma}(\partial_{b}\partial_{c})^{\alpha}(\partial_{c}\partial_{a})^{\beta}\right]h(a;z_{1})h(b;z_{2})h(c;z_{3}) ,\label{prediction}
\end{eqnarray}
where the relative coefficients between monoms are trinomial coefficients:
\begin{equation}\label{trinom}
    \binom{s}{\alpha,\beta,\gamma}=\frac{s!}{\alpha!\beta!\gamma!}, \quad s=\alpha+\beta+\gamma .
\end{equation}
Correspondingly the leading term of the first order gauge transformation should be
\begin{eqnarray}
&&\delta_{(1)}^{leading}h^{(s)}(c;z)=\frac{1}{s!(s-1)!}\sum_{\alpha+\beta+\gamma = s}(-1)^{\beta}\binom{s-1}{\alpha-1,\beta,\gamma}\int_{z_{1},z_{2}} \delta(z-z_{1})\delta(z-z_{2})\nonumber\\
&&\ \ \ \ \ \ \left[(c\nabla_{1})^{\gamma}(\nabla_{2}\partial_{a})^{\alpha-1}(\nabla_{1}\partial_{b})^{\beta}
(\partial_{a}\partial_{b})^{\gamma}(c\partial_{b})^{\alpha}(c\partial_{a})^{\beta}\right]\epsilon(a;z_{1})h(b;z_{2}) .
\end{eqnarray}
Splitting the trinomial into  two binomials we can rewrite this expression in a more elegant way
\begin{eqnarray}
  &&\delta_{(1)}^{leading}h^{(s)}(c;z)=\frac{1}{s!}\sum_{k=0}^{s-1}k!\binom{s-1}{k}\gamma^{(k)}_{(\epsilon^{(s-1)})}(c,b;a)*_{a,b} ,
(a\nabla)^{s-k-1}(c\partial_{b})^{s-k}h^{(s)}(b)\nonumber\\\label{pred1}
\end{eqnarray}
where
\begin{eqnarray}
&&\gamma^{(k)}_{(\epsilon^{(s-1)})}(c,b;a)\nonumber\\
&&\ \ \ \ =
    \frac{k!}{(s-1)!}\sum_{i=0}^{k}\frac{(-1)^{i}}{i!}(c\nabla)^{k-i}(b\nabla)^{i}(c\partial_{b})^{i}
    \left[(a\partial_{b})^{s-1-k}\epsilon^{(s-1)}(b)\right].\quad\label{gamma}
\end{eqnarray}
Comparing with (\ref{curv}) we see that
\begin{equation}\label{gamma1}
    \gamma^{(k)}_{(\epsilon^{(s-1)})}(c,b;a)=\Gamma^{(k)}(c,b;h^{(k)}_{a}(b)) ,
\end{equation}
where
\begin{equation}\label{gamma2}
 h^{(k)}_{a}(b)=\frac{k!}{(s-1)!}\left[(a\partial_{b})^{s-1-k}\epsilon^{(s-1)}(b)\right] ,
\end{equation}
and therefore the $ \gamma^{(k)}_{(\epsilon^{(s-1)})}(c,b;a)$ coefficients inherit in the $c,b$ index spaces all properties of the corresponding spin $k$ curvature described in details in Section \ref{techsetup}.
In the next two sections we show for the $s=2,4$ cases that fixing the leading terms by partial integration and field redefinition leads to the unique solution of Noether's equation (\ref{4.482}).

\subsection{Cubic selfinteraction and Noether's procedure, the spin two example}

\quad
Using our general basis for the spin 2 case
\begin{eqnarray}
&&h_{\mu\nu},\\
&&D_{\mu}=(\nabla h)_{\mu}-\frac{1}{2}\nabla_{\mu}h,\\
&&h=h^{\ \mu}_{\mu} ,
\end{eqnarray}
we can rewrite the free Fronsdal (linearized Einstein-Hilbert gravity) Lagrangian for the spin two gauge field in the following way:
\begin{eqnarray}
&&\mathcal{L}_{0}=-\frac{1}{2}h^{\mu\nu}(\Box h_{\mu\nu}
-2\nabla_{(\mu} D_{\nu)})+\frac{1}{4}h(\Box h-2(\nabla D)),\label{s2}\\
&& (\nabla D)=\nabla^{\mu}D_{\mu} .
\end{eqnarray}
This action is invariant with respect to the zero order gauge transformation
\begin{eqnarray}
\delta_{(0)}h_{\mu\nu}=2\nabla_{(\mu}\varepsilon_{\nu)}.
\end{eqnarray}

According to our strategy described in the previous section we obtain the following cubic interaction Lagrangian
\begin{eqnarray}
\mathcal{L}_{1}(h^{(2)})=&&\frac{1}{2}h^{\alpha\beta}\nabla_{\alpha}\nabla_{\beta}h_{\mu\nu}h^{\mu\nu}
                         +h^{\alpha\mu}\nabla_{\alpha}h^{\beta\nu}\nabla_{\beta}h_{\mu\nu}\nonumber\\
                         &&-\frac{1}{4}(\nabla D)h_{\mu\nu}h^{\mu\nu}
                         -\frac{1}{2}h^{\mu\nu}\nabla_{\mu}h D_{\nu},\label{intgravity}
\end{eqnarray}
supplemented with the Lie derivative form of the first order transformation law
\begin{eqnarray}
\delta_{(1)}h_{\mu\nu}=\varepsilon^{\rho}\nabla_{\rho}h_{\mu\nu}+2\nabla_{(\mu}\varepsilon^{\rho}h_{\nu)\rho} ,\label{deltagravity}
\end{eqnarray}
and the following field redefinition leading to this minimized form of Lagrangian (\ref{intgravity})
\begin{eqnarray}
h_{\mu\nu} \rightarrow h_{\mu\nu}+\frac{1}{4}(hh_{\mu\nu}-2h_{\mu}^{\ \rho}h_{\nu\rho}-\frac{1}{2(D-2)}h^{2}g_{\mu\nu}) ,\label{frgravity}
\end{eqnarray}
Note that the interaction Lagrangian in de Donder gauge
\begin{eqnarray}
D_{\mu}=0,
\end{eqnarray}
reduces to the first two leading terms of (\ref{intgravity}). This minimized form of the leading terms is equivalent to the expansion up to cubic terms of the Einstein-Hilbert action (see \cite{vanDam}) after partial integration and field redefinition, and is in full agreement with (\ref{prediction}) for $s=2$.

To see the same for the first order transformation law (\ref{deltagravity}) and (\ref{pred1}) we note that
the second term in the (\ref{deltagravity}) can be written in the form involving the vector curvature $\gamma^{(1)}_{\mu\nu}=2\nabla_{[\mu}\varepsilon_{\nu]}$ and the additional field redefinition
\begin{eqnarray}
(\nabla_{(\mu}\varepsilon^{\rho}-\nabla^{\rho}\varepsilon_{(\mu})h_{\nu)\rho}
+(\nabla_{(\mu}\varepsilon^{\rho}+\nabla^{\rho}\varepsilon_{(\mu})h_{\nu)\rho}\nonumber\\
=(\nabla_{(\mu}\varepsilon^{\rho}-\nabla^{\rho}\varepsilon_{(\mu})h_{\nu)\rho}
+\frac{1}{2}\delta_{\varepsilon}^{0}(h_{(\mu}^{\ \ \rho}h_{\nu)\rho}) .
\end{eqnarray}
Consequently the first order gauge variation becomes
\begin{eqnarray}
\delta_{(1)}h_{\mu\nu}=
\varepsilon^{\rho}\nabla_{\rho}h_{\mu\nu}
+\gamma^{(1)\,\rho}_{(\mu}h_{\nu)\rho} ,
\end{eqnarray}
and the field redefinition (\ref{frgravity}) reduces to
\begin{eqnarray}
h_{\mu\nu} \rightarrow h_{\mu\nu}+\frac{1}{4}(hh_{\mu\nu}-\frac{1}{2(D-2)}h^{2}g_{\mu\nu}) .\label{frspin2}
\end{eqnarray}

\subsection{The cubic selfinteraction for spin four}

\quad We  start this nontrivial case by introducing the free Fronsdal's Lagrangian for the spin four gauge field $h_{\alpha\beta\gamma\delta}$
\begin{eqnarray}
&&\mathcal{L}_{0}(h^{(4)})=-\frac{1}{2}h^{\alpha\beta\gamma\delta}\mathcal{F}_{\alpha\beta\gamma\delta}
                         +\frac{3}{2}\bar{h}^{\alpha\beta}\bar{\mathcal{F}}_{\alpha\beta} ,\label{f41}\\
&&\mathcal{F}_{\alpha\beta\gamma\delta}=\Box h_{\alpha\beta\gamma\delta}-4\nabla_{(\alpha}D_{\beta\gamma\delta)} ,\label{f42}\\
&&\bar{\mathcal{F}}_{\alpha\beta}=\mathcal{F}^{\gamma}_{\gamma\alpha\beta}=\Box \bar{h}^{}_{\alpha\beta}-2(\nabla D)_{\alpha\beta} ,\label{f43}
\end{eqnarray}
which is invariant under
\begin{eqnarray}
\delta_{(0)}h_{\alpha\beta\gamma\delta}=4\nabla_{(\alpha}\epsilon_{\beta\gamma\delta)} ,
\end{eqnarray}
where we defined the de Donder tensor and the trace of the gauge field by
\begin{eqnarray}
&D_{\alpha\beta\gamma}=(\nabla h)_{\alpha\beta\gamma}-\frac{3}{2}\nabla_{(\alpha}\bar{h}^{}_{\beta\gamma)},\\
&\bar{h}^{}_{\beta\gamma}=h^{\ \ \ \ \alpha}_{\beta\gamma\alpha},\\
&D_{\alpha\beta}^{\ \ \beta}=0, \ \ \bar{h}^{\ \beta}_{\beta}=0.
\end{eqnarray}

The spin four case is much more complicated than the spin two case and includes all difficulties and complexities of a general spin $s$ interaction but remains inside the domain of problems which one can handle analytically. To apply our strategy and integrate the corresponding Noether's equation completely we have to introduce the following table to classify terms and levels of the interaction Lagrangian.

\begin{equation}\label{table}
    \setlength{\unitlength}{0.254mm}
\begin{picture}(351,360)(120,-440)
        \allinethickness{0.254mm}\path(120,-80)(405,-80) 
        \allinethickness{0.254mm}\path(120,-80)(120,-440) 
        \allinethickness{0.254mm}\path(165,-80)(165,-440) 
        \allinethickness{0.254mm}\path(120,-120)(405,-120) 
        \allinethickness{0.254mm}\path(325,-80)(325,-360) 
        \allinethickness{0.254mm}\path(405,-80)(405,-280) 
        \allinethickness{0.254mm}\path(120,-200)(405,-200) 
        \allinethickness{0.254mm}\path(120,-280)(405,-280) 
        \allinethickness{0.254mm}\path(120,-360)(325,-360) 
        \allinethickness{0.254mm}\path(120,-440)(245,-440) 
        \allinethickness{0.254mm}\path(245,-80)(245,-440) 
        \allinethickness{0.254mm}\path(120,-80)(165,-120) 
        \put(200,-106){\shortstack{$0$}} 
        \put(275,-106){\shortstack{$1$}} 
        \put(355,-106){\shortstack{$2$}} 
        \put(135,-166){\shortstack{$0$}} 
        \put(135,-246){\shortstack{$1$}} 
        \put(135,-331){\shortstack{$2$}} 
        \put(135,-401){\shortstack{$3$}} 
        \put(145,-96){\shortstack{$D$}} 
        \put(135,-116){\shortstack{$\bar{h}$}} 
        \put(190,-166){\shortstack{$hhh$}} 
        \put(345,-166){\shortstack{$DDh$}} 
        \put(270,-166){\shortstack{$Dhh$}} 
        \put(190,-246){\shortstack{$\bar{h}hh$}} 
        \put(190,-331){\shortstack{$\bar{h}\bar{h}h$}} 
        \put(190,-401){\shortstack{$\bar{h}\bar{h}\bar{h}$}} 
        \put(265,-246){\shortstack{$\bar{h}Dh$}} 
        \put(340,-246){\shortstack{$\bar{h}DD$}} 
        \put(265,-331){\shortstack{$\bar{h}\bar{h}D$}} 
\end{picture}
\end{equation}
This table introduces some "coordinate system" for classification of our interaction
\begin{equation}\label{intlags}
    \mathcal{L}_{1}=\sum_{i,j=0,1,2,3 \atop  i+j\leq 3} \mathcal{L}^{\emph{int}}_{ij}(h^{(4)}) ,
\end{equation}
where
\begin{equation}\label{ijs}
    \mathcal{L}^{\emph{int}}_{ij}(h^{(4)}) \sim \nabla^{4-i} (D)^{i} (\bar{h}^{(4)})^{j} (h^{(4)})^{3-j-i} .
\end{equation}
In this notation the leading term described in the second section is $\mathcal{L}^{\emph{int}}_{00}(h^{(4)})$. On the other hand the first column of table (\ref{table}) is nothing else but the interaction Lagrangian in de Donder gauge $D_{\alpha\beta\gamma}=0$ and can be expressed as a sum
\begin{eqnarray}
\mathcal{L}^{\emph{int}}_{dD}(h^{(4)})=\sum_{j=0}^{3} \mathcal{L}^{\emph{int}}_{0j}(h^{(4)}) .
\end{eqnarray}
Integrating Noether's equation step by step (cell by cell in means of (\ref{table})) starting from the initial curvature ansatz (\ref{impterm}), we obtain after very long and tedious calculations the following cubic interaction Lagrangian:
\begin{eqnarray}
\mathcal{L}^{\emph{int}}_{00}(h^{(4)})=
&&\ \frac{1}{8}h^{\alpha\beta\gamma\delta}h^{\mu\nu\lambda\rho}\Gamma^{(4)}_{\alpha\beta\gamma\delta,\mu\nu\lambda\rho}
-\nabla^{\mu}h_{\alpha\beta\gamma\delta}\nabla^{\alpha}\nabla^{\beta}h^{\gamma\nu\lambda\rho}\nabla^{\delta}h_{\mu\nu\lambda\rho}\nonumber\\
&&+\frac{3}{4}\nabla^{\mu}h_{\alpha\beta\gamma\delta}\nabla^{\alpha}\nabla^{\nu}h^{\gamma\delta\lambda\rho}\nabla^{\beta}h_{\mu\nu\lambda\rho}\nonumber\\
&&+3\nabla^{\mu}\nabla_{\nu}h_{\alpha\beta\gamma\delta}h^{\alpha\nu\lambda\rho}\nabla^{\beta}\nabla_{\lambda}h_{\mu\rho}^{\ \ \ \gamma\delta},\label{00}
\end{eqnarray}
\begin{eqnarray}
\mathcal{L}^{\emph{int}}_{01}(h^{(4)})=
&&-\frac{3}{2}h_{\alpha\beta\gamma\delta}\nabla^{\alpha}\nabla^{\beta}h^{\gamma\nu\lambda\rho}\nabla^{\delta}\nabla_{\nu}\bar{h}^{}_{\lambda\rho}
-3h_{\alpha\beta\gamma\delta}h_{\nu\lambda\rho}^{\ \ \ \delta}\nabla^{\alpha}\nabla^{\beta}\nabla^{\nu}\nabla^{\lambda}\bar{h}^{\gamma\rho}\nonumber\\
&&+\frac{3}{2}\nabla_{\mu}h_{\alpha\beta\gamma\delta}\nabla^{\nu}\nabla^{\alpha}h^{\mu\beta\gamma\lambda}\nabla^{\delta}\bar{h}^{}_{\nu\lambda}
-\nabla^{\lambda}h^{\mu\alpha\beta\gamma}\nabla^{\rho}h^{\nu}_{\ \alpha\beta\gamma}\nabla_{\mu}\nabla_{\nu}\bar{h}^{}_{\lambda\rho}\nonumber\\
&&+\frac{1}{4}h^{\mu\alpha\beta\gamma}h^{\nu}_{\ \alpha\beta\gamma}\nabla_{\mu}\nabla_{\nu}(\nabla\nabla \bar{h}^{}),
\end{eqnarray}
\begin{eqnarray}
\mathcal{L}^{\emph{int}}_{02}(h^{(4)})=
&&-\frac{3}{2}h_{\alpha\beta\gamma\delta}\nabla^{\alpha}\nabla^{\beta}\nabla^{\mu}\bar{h}^{\gamma\nu}\nabla^{\delta}\bar{h}^{}_{\mu\nu}
+\frac{3}{2}h_{\alpha\beta\gamma\delta}\nabla^{\alpha}\nabla^{\mu}\bar{h}^{\beta\nu}\nabla^{\gamma}\nabla_{\nu}\bar{h}^{\ \delta}_{\mu}\nonumber\\
&&-\frac{3}{4}\nabla_{\mu}\nabla_{\nu}h_{\alpha\beta\gamma\delta}\nabla^{\alpha}\bar{h}^{\beta\nu}\nabla^{\gamma}\bar{h}^{\delta\mu}
-\frac{3}{4}h_{\alpha\beta\gamma\delta}\nabla^{\alpha}\nabla^{\beta}\bar{h}^{\gamma\delta}(\nabla\nabla \bar{h}^{})\nonumber\\
&&-3\nabla_{\mu}h_{\alpha\beta\gamma\delta}\nabla_{\nu}\nabla^{\alpha}\bar{h}^{\beta\gamma}\nabla^{\delta}\bar{h}^{}_{\mu\nu},
\end{eqnarray}
\begin{eqnarray}
\mathcal{L}^{\emph{int}}_{03}(h^{(4)})=
\frac{3}{4}\nabla_{\mu}\nabla_{\nu}\bar{h}^{}_{\alpha\beta}\nabla^{\alpha}\bar{h}^{\mu\lambda}\nabla^{\beta}\bar{h}^{\nu}_{\ \lambda}
-\frac{3}{4}\nabla_{\mu}\bar{h}^{\nu\lambda}\nabla_{\nu}\bar{h}^{\mu}_{\ \lambda}(\nabla\nabla \bar{h}^{}),
\end{eqnarray}
\begin{eqnarray}
\mathcal{L}^{\emph{int}}_{10}(h^{(4)})=
&&\ \ 3\nabla_{\alpha}\nabla_{\nu}D_{\lambda\rho\beta}h^{\alpha\beta\gamma\delta}\nabla_{\gamma}h^{\ \nu\lambda\rho}_{\delta}
+\frac{3}{2}\nabla^{\rho}D_{\alpha\beta\lambda}\nabla^{\mu}h^{\alpha\beta\gamma\delta}\nabla^{\lambda}h_{\rho\mu\gamma\delta}\nonumber\\
&&-2\nabla^{\delta}D_{\nu\lambda\rho}\nabla^{\nu}h_{\alpha\beta\gamma\delta}\nabla^{\lambda}h^{\rho\alpha\beta\gamma}
-\frac{3}{2}(\nabla D)^{\alpha\rho}\nabla^{\mu}h_{\alpha\beta\gamma\delta}\nabla^{\beta}h_{\rho\mu}^{\ \ \gamma\delta}\nonumber\\
&&+\frac{1}{4}(\nabla D)^{\mu\nu}\nabla_{\mu}h_{\alpha\beta\gamma\delta}\nabla_{\nu}h^{\alpha\beta\gamma\delta}
-\frac{1}{2}(\nabla D)^{\mu\nu}h_{\alpha\beta\gamma\delta}\nabla_{\mu}\nabla_{\nu}h^{\alpha\beta\gamma\delta}\nonumber\\
&&-\nabla_{\alpha}(\nabla D)^{\mu\nu}h^{\alpha\beta\gamma\delta}\nabla_{\mu}h_{\nu\beta\gamma\delta}
+\frac{3}{4}\nabla_{\alpha}(\nabla D)^{\mu\nu}h^{\alpha\beta\gamma\delta}\nabla_{\beta}h_{\mu\nu\gamma\delta},\qquad\,
\end{eqnarray}
\begin{eqnarray}
\mathcal{L}^{\emph{int}}_{11}(h^{(4)})=
&&-\frac{1}{2}\bar{h}^{\gamma\delta}\nabla_{\gamma}h_{\mu\nu\lambda\rho}\nabla_{\delta}\nabla^{\mu}D^{\nu\lambda\rho}
+\frac{1}{2}\bar{h}^{\gamma\delta}\nabla_{\gamma}\nabla_{\delta}h_{\mu\nu\lambda\rho}\nabla^{\mu}D^{\nu\lambda\rho}\nonumber\\
&&+\frac{3}{4}\nabla^{\mu}\bar{h}^{\gamma\delta}h_{\mu\nu\lambda\rho}\nabla_{\gamma}\nabla^{\nu}D_{\delta}^{\ \lambda\rho}
-\frac{3}{4}\nabla_{\mu}\bar{h}^{\gamma\delta}h^{\mu\nu\lambda\rho}\nabla_{\nu}\nabla_{\lambda}D_{\gamma\delta\rho}\nonumber\\
&&+\frac{9}{4}\nabla_{\mu}\bar{h}^{\gamma\delta}\nabla_{\rho}h_{\gamma\delta\nu\lambda}\nabla^{\lambda}D^{\mu\nu\rho}
+3\bar{h}^{\gamma\delta}\nabla_{\rho}h_{\gamma\mu\nu\lambda}\nabla^{\mu}\nabla^{\nu}D_{\delta}^{\ \lambda\rho}\nonumber\\
&&+\frac{3}{2}\bar{h}^{\gamma\delta}\nabla_{\rho}h_{\gamma\mu\nu\lambda}\nabla^{\mu}\nabla_{\delta}D^{\nu\lambda\rho}
-\frac{3}{2}\bar{h}^{\gamma\delta}\nabla_{\rho}\nabla_{\gamma}h_{\delta\mu\nu\lambda}\nabla^{\mu}D^{\nu\lambda\rho}\nonumber\\
&&-\frac{3}{4}(\nabla D)^{\gamma\delta}\nabla^{\mu}\bar{h}^{\nu\lambda}\nabla_{\gamma}h_{\delta\mu\nu\lambda}
-\frac{3}{4}\nabla^{\mu}(\nabla D)^{\gamma\delta}\bar{h}^{\nu\lambda}\nabla_{\lambda}h_{\gamma\delta\mu\nu}\nonumber\\
&&+6\nabla^{\mu}\nabla_{\nu}(\nabla D)^{\gamma\delta}\bar{h}^{}_{\gamma\lambda}h_{\delta\mu}^{\ \ \nu\lambda}
+\frac{1}{4}\bar{h}^{\gamma\delta}\Box D^{\mu\nu\rho}\nabla_{\gamma}h_{\delta\mu\nu\rho}\nonumber\\
&&-\frac{3}{8}\bar{h}^{\gamma\delta}\Box D^{\mu\nu\rho}\nabla_{\mu}h_{\gamma\delta\nu\rho},
\end{eqnarray}
\begin{eqnarray}
\mathcal{L}^{\emph{int}}_{12}(h^{(4)})=
&&\ \ \frac{3}{4}D^{\mu\nu\rho}(\nabla \bar{h}^{})^{\delta}\nabla_{\delta}\nabla_{\mu}\bar{h}^{}_{\nu\rho}
-\frac{9}{8}\nabla_{\gamma}\nabla_{\delta}D^{\mu\nu\rho}\bar{h}^{\gamma\delta}\nabla_{\mu}\bar{h}^{}_{\nu\rho}\nonumber\\
&&-3D^{\mu\nu\rho}\bar{h}^{\gamma\delta}\nabla_{\gamma}\nabla_{\mu}\nabla_{\nu}\bar{h}^{}_{\delta\rho}
-3\nabla_{\mu}D_{\nu\rho\gamma}\nabla^{\nu}\bar{h}^{\gamma\delta}\nabla_{\delta}\bar{h}^{\mu\rho}\nonumber\\
&&-\frac{3}{2}(\nabla D)^{\gamma\delta}\nabla_{\gamma}\bar{h}^{\mu\nu}\nabla_{\delta}\bar{h}^{}_{\mu\nu}
-\frac{3}{8}(\nabla D)^{\gamma\delta}\bar{h}^{\mu\nu}\nabla_{\gamma}\nabla_{\delta}\bar{h}^{}_{\mu\nu}\nonumber\\
&&+\frac{3}{2}\nabla^{\mu}(\nabla D)^{\gamma\delta}\nabla_{\gamma}\bar{h}^{}_{\mu\nu}\bar{h}_{\delta}^{\ \nu}
-3(\nabla D)^{\gamma\delta}\nabla_{\gamma}\nabla_{\mu}\bar{h}^{}_{\delta\nu}\bar{h}^{\mu\nu}\nonumber\\
&&-\frac{9}{4}(\nabla D)^{\gamma\delta}\nabla^{\mu}\bar{h}^{}_{\gamma\nu}\nabla^{\nu}\bar{h}^{}_{\delta\mu}
+\frac{3}{2}\nabla^{\mu}(\nabla D)^{\gamma\delta}\nabla^{\nu}\bar{h}^{}_{\gamma\delta}\bar{h}^{}_{\mu\nu}\nonumber\\
&&+\frac{3}{8}(\nabla D)^{\gamma\delta}\bar{h}^{}_{\gamma\delta}(\nabla\nabla \bar{h}^{})
+\frac{3}{8}\nabla^{\mu}\nabla^{\nu}(\nabla D)^{\gamma\delta}\bar{h}^{}_{\gamma\mu}\bar{h}^{}_{\delta\nu}\nonumber\\
&&-\frac{3}{2}\Box(\nabla D)^{\gamma\delta}\bar{h}^{}_{\gamma\mu}\bar{h}^{\ \mu}_{\delta},
\end{eqnarray}
\begin{eqnarray}
\mathcal{L}^{\emph{int}}_{20}(h^{(4)})=
&&\ \ 3D^{\alpha\beta\gamma}D^{\mu\nu\rho}\nabla_{\mu}\nabla_{\nu}h_{\rho\alpha\beta\gamma}
-\frac{9}{4}D^{\alpha\beta\gamma}D^{\mu\nu\rho}\nabla_{\alpha}\nabla_{\mu}h_{\beta\gamma\nu\rho}\nonumber\\
&&+3D^{\alpha\beta\gamma}\nabla^{\rho}D_{\gamma}^{\ \mu\nu}\nabla_{\alpha}h_{\beta\mu\nu\rho}
+\frac{1}{2}(\nabla D)^{\gamma\delta}D^{\mu\nu\rho}\nabla_{\gamma}h_{\delta\mu\nu\rho},
\end{eqnarray}
\begin{eqnarray}
\mathcal{L}^{\emph{int}}_{21}(h^{(4)})=
&&-\frac{3}{4}\bar{h}^{\gamma\delta}\nabla_{\gamma}D^{\mu\nu\rho}\nabla_{\delta}D_{\mu\nu\rho}
+\frac{1}{8}(\nabla\nabla \bar{h}^{})D^{\mu\nu\rho}D_{\mu\nu\rho}\nonumber\\
&&+\frac{3}{4}(\nabla \bar{h}^{})^{\delta}D^{\mu\nu\rho}\nabla_{\mu}D_{\delta\nu\rho}
+\frac{9}{4}\bar{h}^{\gamma\delta}\nabla_{\gamma}D^{\mu\nu\rho}\nabla_{\mu}D_{\delta\nu\rho}\nonumber\\
&&+3\bar{h}^{\gamma\delta}\nabla^{\mu}D_{\gamma}^{\ \nu\rho}\nabla_{\nu}D_{\delta\mu\rho}
+3\nabla^{\mu}\nabla_{\nu}\bar{h}^{\gamma\delta}D_{\gamma}^{\ \nu\rho}D_{\delta\mu\rho}\nonumber\\
&&-\frac{3}{4}\bar{h}^{\gamma\delta}(\nabla D)^{\mu\nu}\nabla_{\mu}D_{\nu\gamma\delta}
+6\bar{h}^{\gamma\delta}\nabla^{\mu}(\nabla D)_{\gamma}^{\ \nu}D_{\delta\mu\nu}\nonumber\\
&&+3\bar{h}^{\gamma\delta}(\nabla D)_{\gamma}^{\ \nu}(\nabla D)_{\delta\nu} .
\end{eqnarray}

Collecting factors coming with Fronsdal's equation of motion (Fronsdal's tensor) in Noether's equation we obtain
next to the free term $\delta_{(0)}h$ of the gauge transformation law for the spin four field the linear term
\begin{eqnarray}
\delta_{(1)}h_{\alpha\beta\gamma\delta}=
&&\epsilon^{\mu\nu\rho}\nabla_{\mu}\nabla_{\nu}\nabla_{\rho}h_{\alpha\beta\gamma\delta}\nonumber\\
&&+3(\nabla_{\alpha}\epsilon_{\rho}^{\ \mu\nu}-\nabla_{\rho}\epsilon_{\alpha}^{\ \mu\nu})
\nabla_{\mu}\nabla_{\nu}h_{\beta\gamma\delta}^{\ \ \ \ \rho}\nonumber\\
&&+3(\nabla_{\alpha}\nabla_{\beta}\epsilon_{\nu\rho}^{\ \ \ \mu}
-2\nabla_{\alpha}\nabla_{\nu}\epsilon_{\beta\rho}^{\ \ \ \mu}
+\nabla_{\nu}\nabla_{\rho}\epsilon_{\alpha\beta}^{\ \ \ \mu})\nabla_{\mu}h_{\gamma\delta}^{\ \ \ \nu\rho}\nonumber\\
&&+(\nabla_{\alpha}\nabla_{\beta}\nabla_{\gamma}\epsilon_{\mu\nu\rho}
-3\nabla_{\alpha}\nabla_{\beta}\nabla_{\mu}\epsilon_{\gamma\nu\rho}
+3\nabla_{\alpha}\nabla_{\mu}\nabla_{\nu}\epsilon_{\beta\gamma\rho}
-\nabla_{\mu}\nabla_{\nu}\nabla_{\rho}\epsilon_{\alpha\beta\gamma})h_{\delta}^{\ \ \mu\nu\rho}\nonumber\\
&&+(trace\  terms\  O(g_{\alpha\beta}))\nonumber\\
=&&\gamma^{(0)\mu\nu\rho}_{(\epsilon^{(3)})}\nabla_{\mu}\nabla_{\nu}\nabla_{\rho}h_{\alpha\beta\gamma\delta}
+3\gamma_{(\epsilon^{(3)})\alpha,\rho}^{(1)\ \ \ \ \ \mu\nu}\nabla_{\mu}\nabla_{\nu}h^{\ \ \ \ \rho}_{\beta\gamma\delta}\nonumber\\
&&+3\gamma_{(\epsilon^{(3)})\alpha\beta,\nu\rho}^{(2)\ \ \ \ \ \ \ \ \mu}\nabla_{\mu}h^{\ \ \ \nu\rho}_{\gamma\delta}
+\gamma_{(\epsilon^{(3)})\alpha\beta\gamma,\mu\nu\rho}^{(3)}h^{\ \mu\nu\rho}_{\delta}\nonumber\\
&&+(trace\  terms\  O(g_{\alpha\beta})),\label{translow}
\end{eqnarray}
where we assumed symmetrization of the indices $\alpha,\beta,\gamma,\delta$
and the spin four field redefinition
\begin{eqnarray}
h_{\alpha\beta\gamma\delta} \rightarrow h_{\alpha\beta\gamma\delta}
&&-\frac{9}{8}\nabla_{\mu}\nabla_{\nu}\bar{h}_{\alpha\beta}h_{\gamma\delta}^{\,\,\,\,\,\,\mu\nu}
-\frac{1}{4}(\nabla\nabla\bar{h})h_{\alpha\beta\gamma\delta}
-\frac{3}{4}\nabla_{\mu}\left[(\nabla\bar{h})^{\mu}h_{\alpha\beta\gamma\delta}\right]\nonumber\\
&&+\frac{1}{2}\bar{h}^{\mu\nu}\nabla_{\mu}\nabla_{\alpha}h_{\beta\gamma\delta\nu}
+\nabla_{\nu}(\nabla\bar{h})_{\alpha}h_{\beta\gamma\delta}^{\,\,\,\,\,\,\,\,\,\,\nu}
-\frac{3}{2}\nabla_{\mu}\bar{h}_{\nu\alpha}\nabla_{\beta}h_{\gamma\delta}^{\,\,\,\,\,\,\mu\nu}\nonumber\\
&&-\frac{3}{8}\bar{h}^{\mu\nu}\nabla_{\alpha}\nabla_{\beta}h_{\gamma\delta\mu\nu}
+\frac{1}{4}\nabla^{\mu}(\bar{h}_{\mu\alpha}D_{\beta\gamma\delta}-\frac{3}{2}\bar{h}_{\alpha\beta}D_{\gamma\delta\mu})\nonumber\\
&&+\frac{9}{2}\nabla^{\mu}\nabla_{\alpha}\bar{h}_{\beta\gamma}\bar{h}_{\delta\mu}
-\frac{21}{32}\nabla_{\nu}\bar{h}_{\alpha\beta}\nabla^{\nu}\bar{h}_{\gamma\delta}\nonumber\\
&&-\frac{3}{2}\nabla_{\alpha}\nabla_{\beta}\bar{h}_{\gamma}^{\,\,\,\mu}\bar{h}_{\delta\mu}
+\frac{15}{8}(\nabla\bar{h})_{\alpha}\nabla_{\beta}\bar{h}_{\gamma\delta}\nonumber\\
&&+(trace\  terms\  O(g_{\alpha\beta})),
\end{eqnarray}
where symmetrization over the indices $\alpha,\beta,\gamma,\delta$ is also understood.

Finally note that we did not obtain an $\mathcal{L}^{\emph{inter}}_{30} \sim (D)^{3}$ part of interaction (that's why we didn't draw corresponding cell in the first row of (\ref{table})) because we started the leading part $\mathcal{L}^{\emph{inter}}_{00}$ (\ref{00}) from the curvature term and fixed in this way partial integration freedom. After that as it was mentioned above all other terms of interaction could be constructed in a unique way up to some field redefinition. This particular way of derivative rearrangement (including partial integration of all other level terms)  does not lead to a $(D)^{3}$ term as opposed to other ways of rearranging the derivatives by means of the partial integration freedom.
On the other hand if we rearrange the derivatives as described in the subsection \ref{self1} we get leading part of the interaction $\mathcal{L}^{\emph{inter}}_{00}$ in complete agreement with our prediction (\ref{prediction}) for $s=4$. The same is true for the transformation law (\ref{translow}) and (\ref{pred1}).

\section{General cubic interaction $s_{1}-s_{2}-s_{3}$}\label{general}
\setcounter{equation}{0}
\subsection{Notations}
\quad

Here we present Fronsdal's Lagrangian:
\begin{equation}\label{1.42(1)}
 \mathcal{L}_{0}(h^{(s)}(a))=-\frac{1}{2}h^{(s)}(a)*_{a}\mathcal{F}^{(s)}(a)
    +\frac{1}{8s(s-1)}\Box_{a}h^{(s)}(a)*_{a}\Box_{a}\mathcal{F}^{(s)}(a) ,
\end{equation}
where $\mathcal{F}^{(s)}(z;a)$ is the Fronsdal tensor
\begin{eqnarray}
\mathcal{F}^{(s)}(z;a)=\Box h^{(s)}(z;a)-s(a\nabla)D^{(s-1)}(z;a) , \quad\label{0.32(1)}
\end{eqnarray}
and $D^{(s-1)}(z;a)$ is the deDonder tensor or traceless divergence of the higher spin gauge field
\begin{eqnarray}\label{0.42(D)}
 && D^{(s-1)}(z;a) = Divh^{(s-1)}(z;a)
-\frac{s-1}{2}(a\nabla)Trh^{(s-2)}(z;a) ,\\
&& \Box_{a} D^{(s-1)}(z;a)=0 .
\end{eqnarray}
The initial gauge variation of order zeroth in the spin $s$ field is
\begin{eqnarray}\label{0.5}
\delta_{(0)} h^{(s)}(z;a)=s (a\nabla)\epsilon^{(s-1)}(z;a) ,
\end{eqnarray}
with the traceless gauge parameter for the double traceless gauge field
\begin{eqnarray}
&&\Box_{a}\epsilon^{(s-1)}(z;a)=0 ,\label{0.6}\\
&&\Box_{a}^{2}h^{(s)}(z;a)=0 .
\end{eqnarray}
Therefore at this point we can see from (\ref{0.5}) and (\ref{0.6})  that the de Donder gauge condition \ref{dd}
is a correct generalization of the Lorentz gauge condition in the case of spin $s>2$.
Finally we note that in deDonder gauge (\ref{dd})  $\mathcal{F}^{(s)}(z;a)=\Box h^{(s)}(z;a)$  and the field $h^{(s)}$ decouples from it's trace in Fronsdal's Lagrangian (\ref{1.42(1)}).

\subsection{Noether's theorem in leading order: Trinomial coefficients}
\setcounter{equation}{0}
We consider three potentials $h^{(s_{1})}(z_{1};a), h^{(s_{2})}(z_{2};b), h^{(s_{3})}(z_{3};c)$ whose spins $s_{i}$ are assumed to be ordered
\begin{equation}
s_{1} \geq s_{2}\geq s_{3} .
\end{equation}
For the interaction we make the cyclic ansatz
\begin{eqnarray}
&&\mathcal{L}_{I}^{(0,0)}(h^{(s_{1})}(a),h^{(s_{2})}(b),h^{(s_{3})}(c))=\sum_{n_{i}} C_{n_{1},n_{2},n_{3}}^{s_{1},s_{2},s_{3}} \int dz_{1}dz_{2}dz_{3} \delta (z_{3}-z_{1}) \delta(z_{2}-z_{1})\nonumber\\
&&\hat{T}(Q_{12},Q_{23},Q_{31}|n_{1},n_{2},n_{3})h^{(s_{1})}(z_{1};a)h^{(s_{2})}(z_{2};b)h^{(s_{3})}(z_{3};c) ,\label{1.2}
\end{eqnarray}
where
\begin{eqnarray}
\hat{T}(Q_{12},Q_{23},Q_{31}|n_{1},n_{2},n_{3})= (\partial_{a}\partial_{b})^{Q_{12}}(\partial_{b}\partial_{c})^{Q_{23}} (\partial_{c}\partial_{a})^{Q_{31}}(\partial_{a}\nabla_{2})^{n_{1}}(\partial_{b}\nabla_{3})^{n_{2}}( \partial_{c}\nabla_{1})^{n_{3}} ,\nonumber\\\label{1.3}
\end{eqnarray}
and the notation $(0,0)$ as a superscript means that it is an ansatz for terms without $Divh^{(s_{i}-1)}=(\nabla_{i}\partial_{a_{i}})h^{(s_{i})}(a_{i})$ and $Trh^{(s_{i}-2)}=\frac{1}{s_{i}(s_{i}-1)}\Box_{a_{i}}h^{(s_{i})}(a_{i})$.
Denoting the number of derivatives by $\Delta$ we have
\begin{equation}
n_{1}+n_{2}+n_{3} = \Delta . \label{delta}
\end{equation}
 As balance equations we have
\begin{eqnarray}
n_{1}+Q_{12}+Q_{31} = s_{1} , \nonumber\\
n_{2}+Q_{23}+Q_{12} = s_{2} ,\nonumber\\
n_{3} + Q_{31} + Q_{23} = s_{3} .\label{1.6}
\end{eqnarray}
These equations are solved by
\begin{eqnarray}
Q_{12} = n_{3}-\nu_{3} ,\nonumber\\
Q_{23} = n_{1} - \nu_{1} , \nonumber\\
Q_{31} = n_{2} - \nu_{2} .
\end{eqnarray}
Since the l.h.s. cannot be negative, we have
\begin{equation}
n_{i} \geq  \nu_{i} .
\end{equation}
The $\nu_{i}$ are determined to be
\begin{equation}
\nu_{i} = 1/2 (\Delta +s_{i} -s_{j} -s_{k}), \quad i,j,k \quad  \textnormal{are all different.}
\end{equation}
These $\nu_{i}$ must also be nonnegative, since otherwise the natural limitation of the $n_{i}$ to nonnegative values
would imply a boundary value problem which has only a trivial solution.
It follows that the minimally possible $\Delta$ is expressed by Metsaev's  (see \cite{Metsaev} equ. (5.11)-(5.13)) formula (using the ordering of the $s_{i}$).
\begin{equation}
\Delta_{min} = \max{[s_{i} +s_{j} -s_{k}]} = s_{1}+ s_{2} -s_{3} .\label{2.25}
\end{equation}
 For example
\begin{equation}
\Delta_{min} = 6 \quad \textnormal{for}\quad s_{1}=s_{2} = 4, s_{3} =2 .
\end{equation}
This value and the ordering of the $s_{i}$ implies for the $\nu_{i}$
\begin{eqnarray}
\nu_{1}= s_{1}-s_{3} , \nonumber\\
\nu_{2} =s_{2}-s_{3} , \nonumber\\
\nu_{3} = 0 .
\end{eqnarray}
>From this result and the experience with the cubic selfinteraction for $s=4$ we can guess that the coefficient $C$ in the ansatz is  a \emph{trinomial}
\begin{equation}
C_{n_{1},n_{2},n_{3}}^{s_{1},s_{2},s_{3}} = const\quad {s_{3} \choose n_{1}-s_{1}+s_{3}, n_{2}-s_{2}+s_{3},n_{3}} , \label{cs}
\end{equation}
which entails
\begin{eqnarray}
\sum_{ij} Q_{ij} = \Delta - \sum_{i}\nu_{i} ,\nonumber\\
\sum_{i} \nu_{i} = 3/2 \Delta -1/2 \sum_{i} s_{i} ,\label{2.29}
\end{eqnarray}
and the expression (\ref{2.25}) for $\Delta_{min}$.

For the proof of this equation (\ref{cs}) we use Noether's theorem to derive recursion relations which are then solved. By variation w.r.t.
$h^{(s_{i})}$ we obtain three currents whose divergences must vanish on shell. We need only do the explicit variation once:
\begin{eqnarray}
J^{(3)}(z_{3};c) = \sum C_{n_{1},n_{2},n_{3}}^{s_{1},s_{2},s_{3}}\int dz_{1}dz_{2}\delta(z_{3}-z_{1})\delta(z_{3}-z_{2}) \nonumber \\
(\partial_{a}\partial_{b})^{Q_{12}} (\partial_{b} c)^{Q_{23}}
(c \partial)^{Q_{31}} (\partial_{a}\nabla_{2})^{n_{1}}(\partial_{b}\nabla_{3})^{n_{2}}(c\nabla_{1})^{n_{3}} \nonumber\\
h^{(s_{1})}(z_{1};a) h^{(s_{2})}(z_{2};b) ,
\end{eqnarray}
having the divergence
\begin{eqnarray}
(\partial_{c}\nabla_{3})J^{(3)}(z_{3};c) = \sum C_{n_{1},n_{2},n_{3}}^{s_{1},s_{2},s_{3}} \nonumber\\
\{n_{3} (\nabla_{1}\nabla_{3}) (\partial_{a}\partial_{b})^{Q_{12}}(\partial_{b}c)^{Q_{23}} (c\partial_{a})^{Q_{31}}(\partial_{a}\nabla_{2})^{n_{1}}
(\partial_{b}\nabla_{3})^{n_{2}}(c\nabla_{1})^{n_{3}-1} \nonumber\\
+Q_{23}(\partial_{a}\partial_{b})^{Q_{12}}(\partial_{b}c)^{Q_{23}-1} (c\partial_{a})^{Q_{31}}
(\partial_{a}\nabla_{2})^{n_{1}}(\partial_{b}\nabla_{3})^{n_{2}+1}(c\nabla_{1})^{n_{3}}\nonumber\\
+Q_{31}(\partial_{a}\partial_{b})^{Q_{12}}(\partial_{b}c)^{Q_{23}}(c\partial_{a})^{Q_{31}-1} (\partial_{a}\nabla_{2})^{n_{1}}(\partial_{a}\nabla_{3})
(\partial_{b}\nabla_{3})^{n_{2}}(c\nabla_{1})^{n_{3}}\}\nonumber\\
h^{(s_{1})}(z_{1};a)h^{(s_{2})}(z_{2};b) \mid z_{1}=z_{2}=z_{3} . \label{ddd}
\end{eqnarray}
This divergence (and the corresponding divergences of the currents $J^{(1,2)}$) must vanish on shell.

We shall develop now a recursive algorithm. First we study the terms not containing any deDonder expression
$D^{(s_{i}-1)}, i=1,2,3$:
\begin{equation}
D^{(s_i -1)} = \frac{1}{s_i}[(\partial_{a_{i}}\nabla_i) -1/2 (a_{i}\nabla_i)\Box_{a_{i}}] h^{(s_i)}(z_{i};a_{i}) , \quad a_{i}=a,b,c.
\end{equation}
We use that
\begin{equation}
(\nabla_{1}\nabla_{3}) = 1/2 [\Box_{2} - \Box_{1} -\Box_{3} ] ,\label{nablarule}
\end{equation}
and
\begin{eqnarray}
  \Box_i h^{(s_i)}(z_{i};a_{i}) &=& \mathcal{F}^{(s_{i})}(z_{i};a_{i})+ s_i (a_{i}\nabla_{i})D^{(s_i -1)} ,\label{1.18}\\
   \Box_i \epsilon^{(s_i-1)}(z_{i};a_{i})  &=& \delta^{(0)}_{i}D^{(s_i -1)} ,\label{1.19}
\end{eqnarray}
where $ \mathcal{F}^{(s_{i})}(z_{i};a_{i})$ is Fronsdal's gauge invariant equation of motion and can be dropped on shell. So the $n_{3}$-term of (\ref{ddd}) does not contribute to the leading order terms.
On the other hand the $Q_{23}$-term is purely leading order. The $Q_{31}$-term contains
\begin{equation}
(\partial_{a}\nabla_{3}) = - (\partial_{a}\nabla_{2}) -(\partial_{a}\nabla_{1}) .\label{1.20}
\end{equation}
Only the first term yields a leading order contribution, the next one is a divergence term.

In the leading order terms we renumber the powers $n_{1}\rightarrow n_{1}+1$ in the $Q_{23}$-term and $n_{2}\rightarrow n_{2}+1$ in the leading order $Q_{31}$ term. We get
\begin{eqnarray}
[(n_{1}+1-\nu_{1})C_{n_{1}+1,n_{2},n_{3}}^{s_{1},s_{2},s_{3}}- (n_{2}+1-\nu_{2})C_{n_{1},n_{2}+1,n_{3}}^{s_{1},s_{2},s_{3}}]\label{1.21}\\
(\partial_{a}\partial_{b})^{n_{3}-\nu_{3}}(\partial_{b}c)^{n_{1}-\nu_{1}}(c\partial_{a})^{n_{2}-\nu_{2}}
(\partial_{a}\nabla_{2})^{n_{1}+1}(\partial_{b}\nabla_{3})^{n_{2}+1}(c\nabla_{1})^{n_{3}} = 0 .\nonumber
\end{eqnarray}
It follows that the factor in the square bracket must vanish. Two analogous relations follow from the two other currents.
The solution of these three recursion relations is
\begin{equation}
C_{n_{1},n_{2},n_{3}}^{s_{1},s_{2},s_{3}} = const \quad {\sum n_{i}-\sum \nu_{i}\choose  n_{1}-\nu_{1},n_{2}-\nu_{2},n_{3}-\nu_{3}} ,\label{1.22}
\end{equation}
which is equivalent to (\ref{cs}) for $\Delta=\Delta_{min}$ and therefore $\nu_{3}=0$, and describes also all other $\Delta > \Delta_{min}$ cases.  Comparison with (\ref{2.25}), (\ref{2.29}) proves that in the $\Delta_{min}$ case we can present the trinomial coefficient also as
\begin{equation}
C_{Q_{12},Q_{23},Q_{31}}^{s_{1},s_{2},s_{3}} = const\quad {s_{min} \choose Q_{12},Q_{23},Q_{31}} .\label{1.23}
\end{equation}
We see that the number of contractions between indices of our three fields $Q_{12},Q_{23},Q_{31}$ define coefficients in our interaction completely.

Finally we want to make a remark concerning the case where two or all three of these fields are equal. Then we get only two or one current whose divergences vanish on shell. But in this case we have a symmetry which restores the result (\ref{cs}), (\ref{1.21}) and shows that this is correct in all cases.

\subsection{Cubic interactions for arbitrary spins: Complete solution of the Noether's procedure}

\quad
To derive the next terms of interaction containing one deDonder expression we turn to the Lagrangian formulation of the task and solve Noether's equation
\begin{equation}\label{2.24}
\sum^{3}_{i=1}\delta^{(1)}_{i}\mathcal{L}^{0}_{i}(h^{(s_{i})}(a)) ,
+\sum^{3}_{i=1}\delta^{(0)}_{i}\mathcal{L}_{I}(h^{(s_{1})}(a),h^{(s_{2})}(b),h^{(s_{3})}(c))=0 .
\end{equation}
where
\begin{eqnarray}
  \delta^{(0)}_{i}h^{(s_{i})}(a_{i}) &=& s_{i}(a_{i}\nabla_{i})\epsilon^{s_{i}-1}(z_{i};a_{i}) \\
  \mathcal{L}^{0}_{i}(h^{(s_{i})}(a))&=& -\frac{1}{2}h^{(s_{i})}(a_{i})*_{a_{i}}\mathcal{F}^{(s_{i})}(a_{i})
    +\frac{1}{8s_{i}(s_{i}-1)}\Box_{a_{i}}h^{(s_{i})}(a_{i})*_{a_{i}}\Box_{a_{i}}\mathcal{F}^{(s)}(a_{i})\nonumber\\
\end{eqnarray}
Shifting $\delta^{(1)}_{i}$ by a trace term in the same way as in previous section we obtain the following functional equation:
\begin{eqnarray}
 &&\sum^{3}_{i=1}\delta^{(0)}_{i}\mathcal{L}_{I}(h^{(s_{1})}(a),h^{(s_{2})}(b),h^{(s_{3})}(c))=0 + O(\mathcal{F}^{(s_{i})}(a_{i})) .\label{2.28}
\end{eqnarray}
We solve this equation starting from the ansatz (\ref{1.2}), (\ref{1.3}) and integrating level by level in means of its dependence on deDonder tensors and traces of higher spin gauge fields.

Actually we have to solve the following equation:
\begin{eqnarray}
   &&C^{\{s_i\}}_{\{n_i\}}\hat{T}(Q_{ij}|n_{i}) [(a\nabla_{1})\epsilon^{(s_{1}-1)}h^{(s_{2})}h^{(s_{3})}
   +h^{(s_{1})}(b\nabla_{2})\epsilon^{(s_{2}-1)}h^{(s_{3})}+h^{(s_{1})}h^{(s_{2})}(c\nabla_{3})\epsilon^{(s_{3}-1)}]\nonumber\\
   &&=0 +O(\mathcal{F}^{(s_{i})}(a_{i}), D^{(s_{i}-1)}(a_{i}), \Box_{a_{i}}h^{(s_{i})}(a_{i})) .
\end{eqnarray}
Taking into account that due to (\ref{1.6})
\begin{equation}\label{2.30}
  \hat{T}(Q_{ij}|n_{i})(a_{i}\nabla_{i})\epsilon^{(s_{i}-1)}(a_{i})=[\hat{T}(Q_{ij}|n_{i}),(a_{i}\nabla_{i})]\epsilon^{(s_{i}-1)}(a_{i}) ,
\end{equation}
we see that all necessary information for the recursion can be found calculating these commutators
\begin{eqnarray}
  [\hat{T}(Q_{ij}|n_{i}),(a\nabla_{1})]&=&Q_{31}\hat{T}(Q_{12},Q_{23},Q_{31}-1|n_{1}, n_{2}, n_{3}+1)\nonumber\\
  &&-Q_{12}\hat{T}(Q_{12}-1,Q_{23},Q_{31}|n_{1}, n_{2}+1, n_{3})\nonumber\\
  &&+n_{1}\hat{T}(Q_{12},Q_{23},Q_{31}|n_{1}-1, n_{2}, n_{3})(\nabla_{1}\nabla_{2})\quad\quad\nonumber\\
  &&-Q_{12}\hat{T}(Q_{12}-1,Q_{23},Q_{31}|n_{1}, n_{2}, n_{3})(\partial_{b}\nabla_{2}),\label{2.31}
\end{eqnarray}
\begin{eqnarray}
  [\hat{T}(Q_{ij}|n_{i}),(b\nabla_{2})]&=&Q_{12}\hat{T}(Q_{12}-1,Q_{23},Q_{31}|n_{1}+1, n_{2}, n_{3})\nonumber\\
  &&-Q_{23}\hat{T}(Q_{12},Q_{23}-1,Q_{31}|n_{1}, n_{2}, n_{3}+1)\nonumber\\
  &&+n_{2}\hat{T}(Q_{12},Q_{23},Q_{31}|n_{1}, n_{2}-1, n_{3})(\nabla_{2}\nabla_{3})\quad\quad\nonumber\\
  &&-Q_{23}\hat{T}(Q_{12},Q_{23}-1,Q_{31}|n_{1}, n_{2}, n_{3})(\partial_{c}\nabla_{3}),\label{2.32}
\end{eqnarray}
\begin{eqnarray}
  [\hat{T}(Q_{ij}|n_{i}),(c\nabla_{3})]&=&Q_{23}\hat{T}(Q_{12},Q_{23}-1,Q_{31}|n_{1}, n_{2}+1, n_{3})\nonumber\\
  &&-Q_{31}\hat{T}(Q_{12},Q_{23},Q_{31}-1|n_{1}+1, n_{2}, n_{3})\nonumber\\
  &&+n_{3}\hat{T}(Q_{12},Q_{23},Q_{31}|n_{1}, n_{2}, n_{3}-1)(\nabla_{3}\nabla_{1})\quad\quad\nonumber\\
  &&-Q_{31}\hat{T}(Q_{12},Q_{23},Q_{31}-1|n_{1}, n_{2}, n_{3})(\partial_{a}\nabla_{1}),\label{2.321}
\end{eqnarray}
where we used relations like (\ref{1.18}) and (\ref{1.20}).
In these commutators we can use also the following identities
\begin{eqnarray}
  \nabla_{1}\nabla_{2}&=&\frac{1}{2}(\Box_{3}-\Box_{2}-\Box_{1}),\nonumber\\
  \nabla_{2}\nabla_{3}&=&\frac{1}{2}(\Box_{1}-\Box_{2}-\Box_{3}),\nonumber\\
  \nabla_{3}\nabla_{1}&=&\frac{1}{2}(\Box_{2}-\Box_{3}-\Box_{1}).
\end{eqnarray}
Now we see immediately from the first two lines of (\ref{2.31})-(\ref{2.321}) that these contribute to (\ref{2.28}) as leading order terms and yield the same equations for the $C^{s_{i}}_{n_i}$ coefficients as (\ref{1.21})
\begin{eqnarray}
  && (Q_{31}+1)C_{n_{1},n_{2}+1,n_{3}}^{s_{1},s_{2},s_{3}}- (Q_{12}+1)C_{n_{1},n_{2},n_{3}+1}^{s_{1},s_{2},s_{3}}=0 ,\label{2.33}\\
  && (Q_{12}+1)C_{n_{1},n_{2},n_{3}+1}^{s_{1},s_{2},s_{3}}- (Q_{23}+1)C_{n_{1}+1,n_{2},n_{3}}^{s_{1},s_{2},s_{3}}=0 ,\label{2.34}\\
  && (Q_{23}+1)C_{n_{1}+1,n_{2},n_{3}}^{s_{1},s_{2},s_{3}}- (Q_{31}+1)C_{n_{1},n_{2}+1,n_{3}}^{s_{1},s_{2},s_{3}}=0 .\label{2.35}
\end{eqnarray}
with the solution (\ref{1.22}) or (\ref{1.23}).

To find the full interaction we follow the same strategy as in the case of $s=4$ selfinteraction and introduce the following classification for the higher order interaction terms in $D$ and $\bar{h}=Trh$ :
\begin{equation}\label{intlag}
    \mathcal{L}_{I}=\sum_{i,j=0,1,2,3 \atop  i+j\leq 3} \mathcal{L}_{I}^{(i,j)}(h^{(s_{k})}) ,
\end{equation}
where
\begin{equation}\label{ij}
    \mathcal{L}_{I}^{(i,j)}(h^{(s)}) \sim \nabla^{s-i} (D)^{i} (\bar{h}^{(s)})^{j} (h^{(s)})^{3-j-i} .
\end{equation}
In this notation the leading term described in the second section is $\mathcal{L}_{I}^{(0,0)}(h^{(s)})$.

To integrate Noether's equation next to the leading term we have to insert in (\ref{2.28}) the last two lines of (\ref{2.31})-(\ref{2.321}) and use two important relations (\ref{1.18}), (\ref{1.19}). Thus we arrive at the following $O(D)$ solution:
\begin{eqnarray}\label{l10}
  \mathcal{L}_{I}^{(1,0)}&=&\sum_{n_{i}} \int dzdz_{1}dz_{2}dz_{3} \delta(z_{1}-z)\delta (z_{3}-z)\delta(z_{2}-z)\nonumber\\
  \Big[&+&\frac{s_{1}n_{1}}{2}C_{n_{1},n_{2},n_{3}}^{s_{1},s_{2},s_{3}}\hat{T}(Q_{ij}|n_{1}-1,n_{2},n_{3})D^{(s_{1}-1)}h^{(s_{2})}h^{(s_{3})}\nonumber\\
  &+&\frac{s_{2}n_{2}}{2}C_{n_{1},n_{2},n_{3}}^{s_{1},s_{2},s_{3}}\hat{T}(Q_{ij}|n_{1},n_{2}-1,n_{3})h^{(s_{1})}D^{(s_{2}-1)}h^{(s_{3})}\nonumber\\
  &+&\frac{s_{3}n_{3}}{2}C_{n_{1},n_{2},n_{3}}^{s_{1},s_{2},s_{3}}\hat{T}(Q_{ij}|n_{1},n_{2},n_{3}-1)h^{(s_{1})}h^{(s_{2})}D^{(s_{3}-1)} \Big] .
\end{eqnarray}
The detailed proof of this formula can be found in the Appendix F where we describe also derivations of all other terms.

The next $O(D^{2})$ and $O(D^{3})$ level Lagrangians are
\begin{eqnarray}
 \mathcal{L}_{I}^{(2,0)}&=&\sum_{n_{i}} \int dzdz_{1}dz_{2}dz_{3} \delta(z_{1}-z)\delta (z_{3}-z)\delta(z_{2}-z)\nonumber\\
  \Big[&+&\frac{s_{3}n_{3}s_{1}n_{1}}{2}C_{n_{1},n_{2},n_{3}}^{s_{1},s_{2},s_{3}}\hat{T}(Q_{ij}|n_{1}-1,n_{2},n_{3}-1)D^{(s_{1}-1)}h^{(s_{2})}D^{(s_{3}-1)}\nonumber\\
  &+&\frac{s_{1}n_{1}s_{2}n_{2}}{2}C_{n_{1},n_{2},n_{3}}^{s_{1},s_{2},s_{3}}\hat{T}(Q_{ij}|n_{1}-1,n_{2}-1,n_{3})D^{(s_{1}-1)}D^{(s_{2}-1)}h^{(s_{3})}\nonumber\\
  &+&\frac{s_{2}n_{2}s_{3}n_{3}}{2}C_{n_{1},n_{2},n_{3}}^{s_{1},s_{2},s_{3}}\hat{T}(Q_{ij}|n_{1},n_{2}-1,n_{3}-1)h^{(s_{1})}D^{(s_{2}-1)}D^{(s_{3}-1)}                          \Big],\quad\quad\quad
\end{eqnarray}
and
\begin{eqnarray}
  \mathcal{L}_{I}^{(3,0)}&=&\sum_{n_{i}} \int dzdz_{1}dz_{2}dz_{3} \delta(z_{1}-z)\delta (z_{3}-z)\delta(z_{2}-z)\nonumber\\
  \Big[&+&\frac{s_{3}n_{3}s_{2}n_{2}s_{1}n_{1}}{2}C_{n_{1},n_{2},n_{3}}^{s_{1},s_{2},s_{3}}\hat{T}(Q_{ij}|n_{1}-1,n_{2}-1,n_{3}-1)D^{(s_{1}-1)}D^{(s_{2}-1)}D^{(s_{3}-1)} \Big] .\nonumber\\
\end{eqnarray}
The remaining terms in the Lagrangian contain at least one trace:
\begin{eqnarray}
\mathcal{L}_{I}^{(0,1)}&=&\mathcal{L}_{I}^{(0,2)}=0 ,\label{L01}\\
\mathcal{L}_{I}^{(0,3)}&=&\sum_{n_{i}} C_{n_{1},n_{2},n_{3}}^{s_{1},s_{2},s_{3}}\frac{Q_{12}Q_{23}Q_{31}}{8}\int dz_{1}dz_{2}dz_{3} \delta(z_{1}-z)\delta (z_{2}-z)\delta(z_{3}-z)\nonumber\\
  &&\Big[\hat{T}(Q_{12}-1,Q_{23}-1,Q_{31}-1|n_{1},n_{2},n_{3})\Box_{a}h^{(s_{1})}\Box_{b}h^{(s_{2})}\Box_{c}h^{(s_{3})}\Big] ,\quad\quad\quad\quad
\end{eqnarray}
\begin{eqnarray}
 \mathcal{L}_{I}^{(1,1)}&=&\sum_{n_{i}} C_{n_{1},n_{2},n_{3}}^{s_{1},s_{2},s_{3}}\int dz_{1}dz_{2}dz_{3} \delta(z_{1}-z)\delta (z_{2}-z)\delta(z_{3}-z)\nonumber\\
  \Big[&+&\frac{s_{1}Q_{12}n_{2}}{4}\hat{T}(Q_{12}-1,Q_{23},Q_{31}|n_{1},n_{2}-1,n_{3})
  D^{(s_{1}-1)}\Box_{b}h^{(s_{2})}h^{(s_{3})}\nonumber\\
  &+&\frac{s_{2}Q_{23}n_{3}}{4}\hat{T}(Q_{12},Q_{23}-1,Q_{31}|n_{1},n_{2},n_{3}-1)
  h^{(s_{1})}D^{(s_{2}-1)}\Box_{c}h^{(s_{3})}\nonumber\\
  &+&\frac{s_{3}Q_{31}n_{1}}{4}\hat{T}(Q_{12},Q_{23},Q_{31}-1|n_{1}-1,n_{2},n_{3})
  \Box_{a}h^{(s_{1})}h^{(s_{2})}D^{(s_{3}-1)} \Big] ,\quad\quad\quad\quad
\end{eqnarray}
\begin{eqnarray}
 \mathcal{L}_{I}^{(1,2)}&=&\sum_{n_{i}} C_{n_{1},n_{2},n_{3}}^{s_{1},s_{2},s_{3}}\int dz_{1}dz_{2}dz_{3} \delta(z_{1}-z)\delta (z_{2}-z)\delta(z_{3}-z)\nonumber\\
  \Big[&+&\frac{s_{1}Q_{12}Q_{23}n_{3}}{8}\hat{T}(Q_{12}-1,Q_{23}-1,Q_{31}|n_{1},n_{2},n_{3}-1)
  D^{(s_{1}-1)}\Box_{b}h^{(s_{2})}\Box_{c}h^{(s_{3})}\nonumber\\
  &+&\frac{s_{2}Q_{23}Q_{31}n_{1}}{8}\hat{T}(Q_{12},Q_{23}-1,Q_{31}-1|n_{1}-1,n_{2},n_{3})
  \Box_{a}h^{(s_{1})}D^{(s_{2}-1)}\Box_{c}h^{(s_{3})}\nonumber\\
  &+&\frac{s_{3}Q_{31}Q_{12}n_{2}}{8}\hat{T}(Q_{12}-1,Q_{23},Q_{31}-1|n_{1},n_{2}-1,n_{3})
  \Box_{a}h^{(s_{1})}\Box_{b}h^{(s_{2})}D^{(s_{3}-1)} \Big] ,\nonumber\\
\end{eqnarray}
\begin{eqnarray}
 \mathcal{L}_{I}^{(2,1)}&=&\sum_{n_{i}} C_{n_{1},n_{2},n_{3}}^{s_{1},s_{2},s_{3}}\int dz_{1}dz_{2}dz_{3} \delta(z_{1}-z)\delta (z_{2}-z)\delta(z_{3}-z)\nonumber\\
  \Big[&+&\frac{s_{2}s_{3}Q_{31}n_{1}n_{2}}{4}\hat{T}(Q_{12},Q_{23},Q_{31}-1|n_{1}-1,n_{2}-1,n_{3})
  \Box_{a}h^{(s_{1})}D^{(s_{2}-1)}D^{(s_{3}-1)}\nonumber\\
  &+&\frac{s_{1}s_{2}Q_{23}n_{3}n_{1}}{4}\hat{T}(Q_{12},Q_{23}-1,Q_{31}|n_{1}-1,n_{2},n_{3}-1)
  D^{(s_{1}-1)}D^{(s_{2}-1)}\Box_{c}h^{(s_{3})}\nonumber\\
  &+&\frac{s_{3}s_{1}Q_{12}n_{2}n_{3}}{4}\hat{T}(Q_{12}-1,Q_{23},Q_{31}|n_{1},n_{2}-1,n_{3}-1)
  D^{(s_{1}-1)}\Box_{b}h^{(s_{2})}D^{(s_{3}-1)} \Big] .\nonumber\\
\end{eqnarray}
So we integrated all cells of the following classification table corresponding to (\ref{intlag})
\begin{eqnarray}
\setlength{\unitlength}{0.254mm}
\begin{picture}(280,275)(125,-355)
        \allinethickness{0.254mm}\path(405,-80)(405,-175) 
        \allinethickness{0.254mm}\path(345,-235)(345,-80) 
        \allinethickness{0.254mm}\path(285,-295)(285,-80) 
        \allinethickness{0.254mm}\path(225,-355)(225,-80) 
        \allinethickness{0.254mm}\path(125,-80)(125,-355) 
        \allinethickness{0.254mm}\path(165,-80)(165,-355) 
        \allinethickness{0.254mm}\path(405,-80)(125,-80) 
        \allinethickness{0.254mm}\path(405,-175)(125,-175) 
        \allinethickness{0.254mm}\path(345,-235)(125,-235) 
        \allinethickness{0.254mm}\path(285,-295)(125,-295) 
        \allinethickness{0.254mm}\path(225,-355)(125,-355) 
        \allinethickness{0.254mm}\path(125,-115)(405,-115) 
        \allinethickness{0.254mm}\path(125,-80)(165,-115) 
        \put(130,-111){\shortstack{$\bar{h}$}} 
        \put(150,-96){\shortstack{$D$}} 
        \put(190,-106){\shortstack{$0$}} 
        \put(250,-106){\shortstack{$1$}} 
        \put(310,-106){\shortstack{$2$}} 
        \put(370,-106){\shortstack{$3$}} 
        \put(140,-151){\shortstack{$0$}} 
        \put(140,-211){\shortstack{$1$}} 
        \put(140,-271){\shortstack{$2$}} 
        \put(140,-331){\shortstack{$3$}} 
        \put(185,-151){\shortstack{$hhh$}} 
        \put(235,-151){\shortstack{$Dhh$}} 
        \put(295,-151){\shortstack{$DDh$}} 
        \put(355,-151){\shortstack{$DDD$}} 
        \put(185,-211){\shortstack{$\ 0$}} 
        \put(185,-271){\shortstack{$\ 0$}} 
        \put(185,-331){\shortstack{$\bar{h}\bar{h}\bar{h}$}} 
        \put(235,-211){\shortstack{$\bar{h}Dh$}} 
        \put(295,-211){\shortstack{$\bar{h}DD$}} 
        \put(235,-271){\shortstack{$\bar{h}\bar{h}D$}} 
\end{picture}
\end{eqnarray}
 and proved that \emph{after fixing the freedom of partial integration in the leading term (i.e. our cyclic ansatz)  all other terms of interaction can be integrated in a unique way when we avoid additional partial integration during recursions}.

 Summarizing we see that the interaction Lagrangian in deDonder gauge $D^{(s-1)}(z;a)=0$ can be expressed as a sum
\begin{eqnarray}
\mathcal{L}^{\emph{dD}}_{I}(h^{(s)})=\sum_{j=0}^{3} \mathcal{L}^{(0,j)}_{I}(h^{(s)}) .
\end{eqnarray}
and it is nothing else than the first column of this table. Therefore
(\ref{L01}) means that \emph{in deDonder gauge the traces of the HS fields decouple from the fields as they do in the free Lagrangian.}

\subsection{Discussion: Towards gauge transformations as open Lie algebras}

\quad
If all spins in the cubic interaction are equal $s$, we can derive the first order gauge transformation of $h^{(s)}$
from the r. h. s. of Noether's equation (\ref{2.28}) taken off shell
\begin{eqnarray}
&&[O(\mathcal{F}) \quad\textnormal{part of}\quad\delta^{0}_{\epsilon^{(s_{1}-1)}} \mathcal{L}_{I}]\nonumber\\
&=&\sum_{n_{i}} C_{n_{1},n_{2},n_{3}}^{s_{1},s_{2},s_{3}}\int dz_{1}dz_{2}dz_{3} \delta(z_{1}-z)\delta (z_{2}-z)\delta(z_{3}-z)\nonumber\\
  \Big[&-&\frac{s_{1}n_{1}}{2}\hat{T}(Q_{ij}|n_{1}-1,n_{2},n_{3})
  \epsilon^{(s_{1}-1)}\mathcal{F}^{(s_{2})}h^{(s_{3})}\nonumber\\
  &+&\frac{s_{1}Q_{12}n_{2}}{4}\hat{T}(Q_{12}-1,Q_{23},Q_{31}|n_{1},n_{2}-1,n_{3})
  \epsilon^{(s_{1}-1)}\Box_{b}\mathcal{F}^{(s_{2})}h^{(s_{3})}\nonumber\\
  &+&\frac{s_{3}s_{1}Q_{12}n_{2}n_{3}}{4}\hat{T}(Q_{12}-1,Q_{23},Q_{31}|n_{1},n_{2}-1,n_{3}-1)
  \epsilon^{(s_{1}-1)}\Box_{b}\mathcal{F}^{(s_{2})}D^{(s_{3}-1)} \nonumber\\
  &+&\frac{s_{1}Q_{12}Q_{23}n_{3}}{8}\hat{T}(Q_{12}-1,Q_{23}-1,Q_{31}|n_{1},n_{2},n_{3}-1)
  \epsilon^{(s_{1}-1)}\Box_{b}\mathcal{F}^{(s_{2})}\Box_{c}h^{(s_{3})}\nonumber\\
  &+&\frac{s_{1}n_{1}}{2}\hat{T}(Q_{ij}|n_{1}-1,n_{2},n_{3})
  \epsilon^{(s_{1}-1)}h^{(s_{2})}\mathcal{F}^{(s_{3})}\nonumber\\
  &+&\frac{s_{1}Q_{12}n_{2}}{4}\hat{T}(Q_{12}-1,Q_{23},Q_{31}|n_{1},n_{2}-1,n_{3})
  \epsilon^{(s_{1}-1)}\Box_{b}h^{(s_{2})}\mathcal{F}^{(s_{3})}\nonumber\\
  &+&\frac{s_{1}s_{2}n_{1}n_{2}}{4}\hat{T}(Q_{12},Q_{23},Q_{31}|n_{1}-1,n_{2}-1,n_{3})
  \epsilon^{(s_{1}-1)}D^{(s_{2}-1)}\mathcal{F}^{(s_{3})} \nonumber\\
  &-&\frac{s_{1}Q_{12}Q_{23}n_{3}}{8}\hat{T}(Q_{12}-1,Q_{23}-1,Q_{31}|n_{1},n_{2},n_{3}-1)
  \epsilon^{(s_{1}-1)}\Box_{b}h^{(s_{2})}\Box_{c}\mathcal{F}^{(s_{3})}\Big] .\nonumber\\\label{5.1}
\end{eqnarray}
  If we assume moreover that the gauge transformations form a Lie algebra of power series in some "coupling constant"
$g$, we can following along the ideas of Berends, Burger and Van Dam in their classical paper \cite{vanDam2}
derive conclusions on the higher order interactions. We sum up simple results:

 The arguments of these authors to show that such
power series algebra does not exist for $s=3$, cannot be generalized to even spins.

For a given gauge function $\epsilon^{(s-1)}(z;a)$ the gauge transformation is a substitution (classically) with expansion
\begin{equation}
h \rightarrow h + \delta_{\epsilon}h = h + \nabla \epsilon + \sum_{n\geq 1} g^{n}\Theta_{n}(h,h,...h;\epsilon) ,\label{5.2}
\end{equation}
with $\Theta_{n}$ depending on $\epsilon$ linearly and on $h$ in the n'th power. Moreover we assume that the commutator of two such transformations is given by
\begin{equation}
[\delta_{\epsilon},\delta_{\eta}]h  = \delta_{C(h;\epsilon,\eta)} h ,\label{5.3}
\end{equation}
with the expansion
\begin{equation}
C(h;\epsilon,\eta) = g\sum_{n\geq 0} g^{n}C_{n}(h, h,...h;\epsilon,\eta) ,\label{5.4}
\end{equation}
where each $C_{n}$ depends on $\epsilon$ and $\eta$ linearly and on $h$ in the n'th power. As substitutions gauge transformations are associative and their infinitesimals must satisfy the Jacobi identity. At order $g^{2}$ this is e.g.
\begin{equation}
\sum_{\eta,\epsilon,\zeta \textnormal{cyclic}}\{C_1(\nabla\zeta;\eta,\epsilon) +C_0(C_0(\eta,\epsilon),\zeta)\} = 0 .\label{5.5}
\end{equation}
The commutator can also be expanded
\begin{eqnarray}
[\delta_{\eta},\delta_{\epsilon}] &=& g(\Theta_1(\nabla\epsilon;\eta) -\Theta_1(\nabla\eta;\epsilon))\nonumber\\
&+& g^{2}\{[\Theta_1(\Theta_1(h;\epsilon);\eta) -\Theta_1(\Theta_1(h;\eta);\epsilon)]\nonumber\\
&+& [\Theta_2(\nabla\epsilon,h;\eta) -\Theta_2(\nabla\eta,h;\epsilon)]\nonumber\\
&+& [\Theta_2(h,\nabla\epsilon;\eta) -\Theta_2(h,\nabla\eta;\epsilon)]\}
+O(g^{3}) . \label{5.6}
\end{eqnarray}
Inserting this expansion into the definition of the functions $C_{n}$ we obtain
\begin{eqnarray}
\nabla C_0(\eta,\epsilon) &=& \Theta_1(\nabla\epsilon;\eta) - \Theta_1(\nabla\eta;\epsilon)  ,\label{5.7}\\
\nabla C_1(h; \eta,\epsilon) &=& \Theta_1(\Theta_1(h;\epsilon);\eta) -\Theta_1(\Theta_1(h;\eta);\epsilon)
 - \Theta_1(h; C_0(\eta,\epsilon)) \nonumber\\
 &+& \Theta_2(\nabla\epsilon,h;\eta) -\Theta_2(\nabla\eta,h;\epsilon) +\Theta_2(h, \nabla\epsilon; \eta) - \Theta_2(h, \nabla\eta;\epsilon) .\label{5.8}\nonumber\\
\end{eqnarray}
Assume that $\Theta_1(h;\epsilon)$ has been extracted from (\ref{5.1}) for the case of equal spins $s$. Then the order of derivations in $\Theta_1$ is
\begin{equation}
\natural\Theta_1(h; \epsilon) = s-1 .\label{5.9}
\end{equation}
Inserting this result into (\ref{5.5}), (\ref{5.7}) we obtain the number of derivations in $C_0, C_1$ as
\begin{equation}
\natural C_0(\eta,\epsilon) = s-1 \quad\textnormal{and}\quad \natural C_1(h;\eta,\epsilon) = 2s-3 .\label{5.10}
\end{equation}
This implies
\begin{equation}
\natural\Theta_2(h,h;\epsilon) = 2s-3 .\label{5.11}
\end{equation}
Consequently the quartic interaction must contain $2s-2$ derivatives. The argument can be continued to still higher interactions: For n'th order interactions the number is $(s-2)(n-2)+2$. This result is equivalent to introduction of a scale $L$ and dimensions
in the following way
\begin{equation}
[h] = L^{s-2}, \quad [\nabla] = \frac{1}{L} , \label{5.12}
\end{equation}
with a dimensionless coupling constant $g$, so that each term in the power series has the same dimension. Note that in the case of $\Delta=\Delta_{min}$ we obtained in the previous subsections of this section and in previous section the same dimensions for cubic selfinteractions  and a free  Fronsdal's action.

In \cite{vanDam2} the argument was presented that for spin $s=3$ a Lie algebra of gauge transformations in the form of power series
does not exist, the problem starting with the second power.  The argument was based on the term
\begin{equation}
(\partial_{a}\nabla_2)^{s-1}\epsilon^{(s-1)}(z_1;a) h^{(s)}(z_2;b) ,\label{5.13}
\end{equation}
which exists in $\Theta_1$. Such term is present in fact for any spin, as can be inspected from (\ref{5.1}). Namely, in the fifth term of the square bracket of (\ref{5.1}) (this is the unique localization) we get such expression for $n_1=s,n_2=n_3=0$. In equation (\ref{5.8}) in the first line we have thus $2s-2$ derivatives acting on the field $h$ in either term. In no other terms of (\ref{5.8}) such expression appears. Therefore they must cancel inside this line and they do cancel indeed for even spin only. There is in this case no obstruction of the power series algebra by these arguments. A deeper investigation of such algebras will follow in the future.


\chapter{Generating function of HSF cubic interactions}\label{GF}

\section{Generating function for the Free Lagrangian of all higher spin gauge fields}

\setcounter{equation}{0}
\quad We introduce a generating function for HS gauge fields by
\begin {equation}
\Phi(z;a) = \sum_{s=0}^{\infty} \frac{1}{s!}h^{(s)}(z;a) \label{sf}
\end{equation}
where we assume that the spin $s$ field has scaling dimension $s-2$, the $a_{i}$ vectors have dimension $-1$, and therefore all terms in the generating function for higher spin gauge fields (\ref{sf}) have the same dimension $-2$.

A zeroth order gauge transformation for this field reads as
\begin{eqnarray}
\delta^{0}_{\Lambda}\Phi(z;a) = (a\nabla)\Lambda(z;a)\label{gv},\\
\delta^{0}_{\Lambda}D_{a}\Phi(z;a) = \Box \Lambda(z;a),\\
\delta^{0}_{\Lambda}\Box_{a}\Phi(z;a) = 2(\nabla\partial_{a})\Lambda(z;a).
\end{eqnarray}
where
\begin{eqnarray}
\Lambda(z;a)=\sum_{s=1}^{\infty} \frac{1}{(s-1)!}\epsilon^{(s-1)}(z;a)\label{gp},
\end{eqnarray}
is the generating function of the gauge parameters and is dimensionless\footnote{The gauge parameter for spin $s$ field $\epsilon^{(s-1)}$ has scaling dimension $s-1$, therefore after contraction with $s-1$ $a$-s becomes dimensionless.}.

Fronsdal's constraint for the gauge parameter reads as
\begin{eqnarray}
\Box_{a}\Lambda(z;a)=0\label{FC},
\end{eqnarray}
For a spin $s$ field gauge variation we get as expected
\begin{eqnarray}
\delta_{\epsilon}^{0}h^{(s)}(z;a)=s(a\nabla)\epsilon^{(s-1)}(z;a)\label{gvs},
\end{eqnarray}

The second Fronsdal constraint of the gauge field reads in these notations
\begin{eqnarray}
\Box_{a}^{2}\Phi(z;a)=0\label{FC2},
\end{eqnarray}

We introduced the "de Donder" operator
\begin{equation}
D_{a_i} = (\partial_{a_i}\nabla_i)-\frac{1}{2}(a_i\nabla_i)\Box_{a_i} \label{D}
\end{equation}
This operator is "linear" in $\partial_{a_i}$.

Here we write the quadratic Lagrangian for free higher spin gauge fields in general form using the generating function for HS fields (\ref{sf}).
First we introduce Fronsdal's operator
\begin{eqnarray}
\mathcal{F}_{a_{i}}=\Box_{i}-(a_{i}\nabla_{i})(\nabla_{i}\partial_{a_{i}})+\frac{1}{2}(a_{i}\nabla_{i})^{2}\Box_{a_{i}},
\end{eqnarray}
or with the help of (\ref{D})
\begin{eqnarray}
\mathcal{F}_{a_{i}}=\Box_{i}-(a_{i}\nabla_{i})D_{a_{i}}.
\end{eqnarray}
The operator of the equation of motion can be written in the form
\begin{eqnarray}
\mathcal{G}_{a_{i}}=\mathcal{F}_{a_{i}}-\frac{a_{i}^{2}}{4}\Box_{a_{i}}\mathcal{F}_{a_{i}}
\end{eqnarray}
Now we can write the free Lagrangian for all gauge fields of any spin in a symmetric elegant form
\begin{eqnarray}
\mathcal{L}^{\emph{free}}(z)&=&\frac{\kappa}{2} exp[\lambda^{2}\partial_{a_{1}}\partial_{a_{2}}]\int_{z_{1}z_{2}} \delta (z_{1}-z) \delta(z_{2}-z)\nonumber\\
                            &&\{(\nabla_{1}\nabla_{2})-\lambda^{2}D_{a_1}D_{a_2}-\frac{\lambda^{4}}{4}(\nabla_{1}\nabla_{2})\Box_{a_1}\Box_{a_2}\}
                            \Phi(z_{1};a_{1})\Phi(z_{2};a_{2})\mid _{a_{1}=a_{2}=0}\quad\quad\quad\quad
\end{eqnarray}
where $\lambda$ has scaling dimension $-1$, therefore $\lambda^{2}$ compensates the dimension of the operator in the exponent.
We will see that all relative coupling constants of HS interactions can be expressed as powers of $\lambda$. The parameter $\kappa$ is a constant which makes the action dimensionless (analogous to Einstein's constant and simply connected with the latter). It has scaling dimension $6-d$, where $d$ is the space-time dimension. For Einstein's constant $\kappa_{E}$ we get
\begin{equation}
\kappa_{E}^{-2}=\kappa \lambda^{4}
\end{equation}
It is now obvious that in the free Lagrangian there is no mixing between gauge fields of different spin.
It can also be written in such forms
\begin{eqnarray}
\mathcal{L}^{\emph{free}}(z)&=&-\frac{1}{2}exp[\lambda^{2}\partial_{a_{1}}\partial_{a_{2}}]\int_{z_{1}} \delta(z_{1}-z)
                                (\mathcal{G}_{a_{1}})\Phi(z_{1};a_{1})\Phi(z;a_{2})\mid _{a_{1}=a_{2}=0}\nonumber\\
                            &=&-\frac{1}{2}exp[\lambda^{2}\partial_{a_{1}}\partial_{a_{2}}]\int_{z_{2}} \delta(z_{2}-z)
                                (\mathcal{G}_{a_{2}})\Phi(z;a_{1})\Phi(z_{2};a_{2})\mid _{a_{1}=a_{2}=0}\quad\quad\quad
\end{eqnarray}
These expressions reproduce Fronsdal's Lagrangians for all gauge fields with any spin.

\section{Generating Function for Cubic Interactions}
\setcounter{equation}{0}
\quad We are going to present a very beautiful and compact form of all HS gauge field interactions derived in the previous chapter.
First we rewrite the leading term of a general trilinear interaction of higher spin gauge fields with any spins $s_{1},s_{2},s_{3}$
\begin{eqnarray}
&&\mathcal{L}^{leading}_{(1)}(h^{(s_{1})}(z),h^{(s_{2})}(z),h^{(s_{3})}(z))\nonumber\\
&&=\sum_{\alpha+\beta+\gamma = n}\frac{1}{\alpha!\beta!\gamma!}\int_{z_{1},z_{2},z_{3}} \delta(z-z_{1})\delta(z-z_{2})\delta(z-z_{3})\nonumber\\
&&\left[(\nabla_{1}\partial_{c})^{s_{3}-n+\gamma}(\nabla_{2}\partial_{a})^{s_{1}-n+\alpha}(\nabla_{3}\partial_{b})^{s_{2}-n+\beta}
(\partial_{a}\partial_{b})^{\gamma}(\partial_{b}\partial_{c})^{\alpha}(\partial_{c}\partial_{a})^{\beta}\right]\nonumber\\
&&h^{(s_{1})}(a;z_{1})h^{(s_{2})}(b;z_{2})h^{(s_{3})}(c;z_{3}) ,\label{leading}
\end{eqnarray}
where the number of derivatives is
\begin{eqnarray}
\Delta=s_{1}+s_{2}+s_{3}-2n,\\
0\leq n\leq min(s_{1},s_{2},s_{3})
\end{eqnarray}
As we see, the minimal and maximal possible numbers of derivatives are
\begin{eqnarray}
\Delta_{min}&=&s_{1}+s_{2}+s_{3}-2min(s_{1},s_{2},s_{3}),\\
\Delta_{max}&=&s_{1}+s_{2}+s_{3}.
\end{eqnarray}
The case of $\Delta_{min}$ is important also because only in that case the interaction (\ref{leading}) has the same dimension as the lowest spin field free Lagrangian.

These interactions trivialize only if we have two equal spin values and the third value is odd. This we call the $\ell-s-s$ case, where $\ell$ is odd.
In that case we should have a multiplet of spin $s$ fields, with at least two charges to couple to the spin $\ell$ field. As example consider an odd spin self-interaction. In the case of $\ell-\ell-\ell$ odd spin self interaction, the number of possible charges in the multiplet should be at least 3.

The same Lagrangian can be written in the following way (due to a constant normalization factor $2^{\Delta}$)
\begin{eqnarray}
&&\mathcal{L}^{leading}_{(1)}(h^{(s_{1})}(z),h^{(s_{2})}(z),h^{(s_{3})}(z))\nonumber\\
&&=\sum_{\alpha+\beta+\gamma = n}\frac{1}{\alpha!\beta!\gamma!}\int_{z_{1},z_{2},z_{3}} \delta(z-z_{1})\delta(z-z_{2})\delta(z-z_{3})\nonumber\\
&&\left[(\nabla_{12}\partial_{c})^{s_{3}-n+\gamma}(\nabla_{23}\partial_{a})^{s_{1}-n+\alpha}(\nabla_{31}\partial_{b})^{s_{2}-n+\beta}
(\partial_{a}\partial_{b})^{\gamma}(\partial_{b}\partial_{c})^{\alpha}(\partial_{c}\partial_{a})^{\beta}\right]\nonumber\\
&&h^{(s_{1})}(a;z_{1})h^{(s_{2})}(b;z_{2})h^{(s_{3})}(c;z_{3}) ,\label{interaction}
\end{eqnarray}
where
\begin{eqnarray}
&&\nabla_{12}=\nabla_{1}-\nabla_{2},\\
&&\nabla_{23}=\nabla_{2}-\nabla_{3},\\
&&\nabla_{31}=\nabla_{3}-\nabla_{1}.
\end{eqnarray}
Now we can see that the following expression is a generating function for the leading term of all interactions of HS gauge fields.
\begin{eqnarray}
&\mathcal{A}^{00} = \int_{z_{1},z_{2},z_{3}} \delta(z-z_{1})\delta(z-z_{2})\delta(z-z_{3})exp W \nonumber \\
&\Phi_1(z_1;a_{1}+\frac{1}{2}\nabla_{23})\Phi_2(z_2;a_{2}+\frac{1}{2}\nabla_{31})\Phi_3(z_3;a_{3}+\frac{1}{2}\nabla_{12})\mid_{a_{1}=a_{2}=a_{3}=0}\label{GenF}
\end{eqnarray}
with
\begin{eqnarray}
W = \frac{\lambda^{2}}{2}[(\partial_{a_{1}}\partial_{a_{2}})(\partial_{a_{3}}\nabla_{12})+(\partial_{a_{2}}\partial_{a_{3}})(\partial_{a_{1}}\nabla_{23})
+(\partial_{a_{3}}\partial_{a_{1}})(\partial_{a_{2}}\nabla_{31})]\qquad \label{W1}
\end{eqnarray}
This can be written in another form
\begin{eqnarray}
\mathcal {A}^{00}(\Phi(z)) = \int_{z_{1},z_{2},z_{3}} \delta(z-z_{1,2,3}) exp \hat W \times\Phi(z_1;a_{1})\Phi(z_2;a_{2})\Phi(z_3;a_{3})\mid_{a_{1}=a_{2}=a_{3}=0}\quad\quad\label{exp}
\end{eqnarray}
where
\begin{eqnarray}
&\hat W = \frac{\lambda^{2}}{2}[(\partial_{a_{1}}\partial_{a_{2}})(\partial_{a_{3}}\nabla_{12})+(\partial_{a_{2}}\partial_{a_{3}})(\partial_{a_{1}}\nabla_{23})
+(\partial_{a_{3}}\partial_{a_{1}})(\partial_{a_{2}}\nabla_{31})]\nonumber \\
&+\frac{1}{2}[(\partial_{a_{3}}\nabla_{12})+(\partial_{a_{1}}\nabla_{23})+(\partial_{a_{2}}\nabla_{31})]\label{W},\\
&\int_{z_{1},z_{2},z_{3}} \delta(z-z_{1,2,3})=\int_{z_{1},z_{2},z_{3}} \delta(z-z_{1})\delta(z-z_{2})\delta(z-z_{3})
\end{eqnarray}
for brevity. Furthermore we will always assume this integration with delta functions, without writing it explicitly.
The operator in the second row of (\ref{W}) is a dimensionless operator, therefore it does not need any dimensional constant multiplier.

Now we can derive all other terms in the Lagrangian using the following important relation
\begin{eqnarray}
[exp\hat W, A]=exp\hat W [\hat W, A]+exp\hat W [\hat W, [\hat W, A]]+exp\hat W [\hat W, [\hat W, [\hat W, A]]]+...\quad\quad
\end{eqnarray}
for any operator $A$. And therefore
\begin{eqnarray}
&&[exp\hat W, (a_{1}\nabla_{1})]=exp\hat W [\hat W, (a_{1}\nabla_{1})],\\
&&[exp\hat W, (a_{2}\nabla_{2})]=exp\hat W [\hat W, (a_{2}\nabla_{2})],\\
&&[exp\hat W, (a_{3}\nabla_{3})]=exp\hat W [\hat W, (a_{3}\nabla_{3})].
\end{eqnarray}
The following commutators will be used many times while deriving trace and divergence terms
\begin{eqnarray}
&&[\hat W, (a_{1}\nabla_{1})]=-\frac{\lambda^{2}}{4}[(\partial_{a_{2}}\nabla_{2})(\partial_{a_{3}}\nabla_{12})
                                                    +(\partial_{a_{3}}\nabla_{3})(\partial_{a_{2}}\nabla_{31})]
                              +\frac{1}{2}[\lambda^{2}(\partial_{a_{2}}\partial_{a_{3}})+1]\nabla_{1}\nabla_{23},\qquad\quad\,\\
&&[\hat W, (a_{2}\nabla_{2})]=-\frac{\lambda^{2}}{4}[(\partial_{a_{3}}\nabla_{3})(\partial_{a_{1}}\nabla_{23})
                                                    +(\partial_{a_{1}}\nabla_{1})(\partial_{a_{3}}\nabla_{12})]
                              +\frac{1}{2}[\lambda^{2}(\partial_{a_{3}}\partial_{a_{1}})+1]\nabla_{2}\nabla_{31},\\
&&[\hat W, (a_{3}\nabla_{3})]=-\frac{\lambda^{2}}{4}[(\partial_{a_{1}}\nabla_{1})(\partial_{a_{2}}\nabla_{31})
                                                    +(\partial_{a_{2}}\nabla_{2})(\partial_{a_{1}}\nabla_{23})]
                              +\frac{1}{2}[\lambda^{2}(\partial_{a_{1}}\partial_{a_{2}})+1]\nabla_{3}\nabla_{12}.
\end{eqnarray}
Note that
\begin{eqnarray}
&&\nabla_{1}\nabla_{23}=\Box_{3}-\Box_{2},\\
&&\nabla_{2}\nabla_{31}=\Box_{1}-\Box_{3},\\
&&\nabla_{3}\nabla_{12}=\Box_{2}-\Box_{1},
\end{eqnarray}
which is obvious because\footnote{We always understand partial integrations to be performed, working with a Lagrangian as with an action.}
\begin{eqnarray}
\nabla_{1}+\nabla_{2}+\nabla_{3}=0.
\end{eqnarray}
We are working with the same type of diagram as in previous chapter.
\begin{eqnarray}
\setlength{\unitlength}{0.254mm}
\begin{picture}(280,275)(125,-355)
        \allinethickness{0.254mm}\path(405,-80)(405,-175) 
        \allinethickness{0.254mm}\path(345,-235)(345,-80) 
        \allinethickness{0.254mm}\path(285,-295)(285,-80) 
        \allinethickness{0.254mm}\path(225,-355)(225,-80) 
        \allinethickness{0.254mm}\path(125,-80)(125,-355) 
        \allinethickness{0.254mm}\path(165,-80)(165,-355) 
        \allinethickness{0.254mm}\path(405,-80)(125,-80) 
        \allinethickness{0.254mm}\path(405,-175)(125,-175) 
        \allinethickness{0.254mm}\path(345,-235)(125,-235) 
        \allinethickness{0.254mm}\path(285,-295)(125,-295) 
        \allinethickness{0.254mm}\path(225,-355)(125,-355) 
        \allinethickness{0.254mm}\path(125,-115)(405,-115) 
        \allinethickness{0.254mm}\path(125,-80)(165,-115) 
        \put(130,-110){\shortstack{$\Box_{a_{i}}$}} 
        \put(141,-94){\shortstack{$D_{a_{i}}$}} 
        \put(190,-106){\shortstack{$0$}} 
        \put(250,-106){\shortstack{$1$}} 
        \put(310,-106){\shortstack{$2$}} 
        \put(370,-106){\shortstack{$3$}} 
        \put(140,-151){\shortstack{$0$}} 
        \put(140,-211){\shortstack{$1$}} 
        \put(140,-271){\shortstack{$2$}} 
        \put(140,-331){\shortstack{$3$}} 
        \put(185,-151){\shortstack{$\mathcal {A}^{00}$}} 
        \put(235,-151){\shortstack{$\mathcal {A}^{10}$}} 
        \put(295,-151){\shortstack{$\mathcal {A}^{20}$}} 
        \put(355,-151){\shortstack{$\mathcal {A}^{30}$}} 
        \put(185,-211){\shortstack{$\mathcal {A}^{01}$}} 
        \put(185,-271){\shortstack{$\mathcal {A}^{02}$}} 
        \put(185,-331){\shortstack{$\mathcal {A}^{03}$}} 
        \put(235,-211){\shortstack{$\mathcal {A}^{11}$}} 
        \put(295,-211){\shortstack{$\mathcal {A}^{21}$}} 
        \put(235,-271){\shortstack{$\mathcal {A}^{12}$}} 
\end{picture}
\end{eqnarray}
Now we take a gauge variation of $\mathcal {A}^{00}$, and find generating functions for all other terms in the cubic Lagrangian.  A simple but elegant structure is exhibited by the first row of the diagram
\begin{eqnarray}
\mathcal {A}^{10}(\Phi(z))=\mathcal {A}^{30}(\Phi(&z&))=0,\\
\mathcal {A}^{20}(\Phi(z)) = \frac{1}{4} exp\hat W \{&&+[\lambda^{2}(\partial_{a_{1}}\partial_{a_{2}})+1][\lambda^{2}(\partial_{a_{2}}\partial_{a_{3}})+1]D_{a_{3}}D_{a_{1}}\nonumber\\
   &&+[\lambda^{2}(\partial_{a_{2}}\partial_{a_{3}})+1][\lambda^{2}(\partial_{a_{3}}\partial_{a_{1}})+1]D_{a_{1}}D_{a_{2}}\nonumber\\
   &&+[\lambda^{2}(\partial_{a_{3}}\partial_{a_{1}})+1][\lambda^{2}(\partial_{a_{1}}\partial_{a_{2}})+1]D_{a_{2}}D_{a_{3}}\}\nonumber\\
                                      &&\Phi(z_1;a_{1})\Phi(z_2;a_{2})\Phi(z_3;a_{3})\mid_{a_{1}=a_{2}=a_{3}=0}\qquad\qquad
\end{eqnarray}

Other terms are

\begin{eqnarray}
\mathcal {A}^{01}(\Phi(z))=0,&&\\
\mathcal {A}^{11}(\Phi(z)) = \frac{\lambda^{2}}{16}exp\hat W
\{&&+[\lambda^{2}(\partial_{a_{1}}\partial_{a_{2}})+1](\partial_{a_{1}}\nabla_{23})\Box_{a_{3}}D_{a_{2}}\nonumber\\
&&-[\lambda^{2}(\partial_{a_{1}}\partial_{a_{2}})+1](\partial_{a_{2}}\nabla_{31})\Box_{a_{3}}D_{a_{1}}\nonumber\\
&&+[\lambda^{2}(\partial_{a_{2}}\partial_{a_{3}})+1](\partial_{a_{2}}\nabla_{31})\Box_{a_{1}}D_{a_{3}}\nonumber\\
&&-[\lambda^{2}(\partial_{a_{2}}\partial_{a_{3}})+1](\partial_{a_{3}}\nabla_{12})\Box_{a_{1}}D_{a_{2}}\nonumber\\
&&+[\lambda^{2}(\partial_{a_{3}}\partial_{a_{1}})+1](\partial_{a_{3}}\nabla_{12})\Box_{a_{2}}D_{a_{1}}\nonumber\\
&&-[\lambda^{2}(\partial_{a_{3}}\partial_{a_{1}})+1](\partial_{a_{1}}\nabla_{23})\Box_{a_{2}}D_{a_{3}}\}\nonumber\\
&&\Phi(z_1;a_{1})\Phi(z_2;a_{2})\Phi(z_3;a_{3})\mid_{a_{1}=a_{2}=a_{3}=0}
\end{eqnarray}
and so on.

All these expressions can be written in a very elegant form. First we introduce Grassmann variables by
\begin{eqnarray}
\eta_{a_{1}}, \bar{\eta}_{a_{1}}, \eta_{a_{2}}, \bar{\eta}_{a_{2}}, \eta_{a_{3}}, \bar{\eta}_{a_{3}}.
\end{eqnarray}
Then we change expressions in the formula (\ref{exp}) in a following way
\begin{eqnarray}
&&(\partial_{a_{1}}\partial_{a_{2}}) \rightarrow (\partial_{a_{1}}\partial_{a_{2}})+\frac{1}{4}\eta_{a_{1}}\bar{\eta}_{a_{2}}\Box_{a_{2}}+\frac{1}{4}\eta_{a_{2}}\bar{\eta}_{a_{1}}\Box_{a_{1}},\\
&&(\partial_{a_{2}}\partial_{a_{3}}) \rightarrow (\partial_{a_{2}}\partial_{a_{3}})+\frac{1}{4}\eta_{a_{2}}\bar{\eta}_{a_{3}}\Box_{a_{3}}+\frac{1}{4}\eta_{a_{3}}\bar{\eta}_{a_{2}}\Box_{a_{2}},\\
&&(\partial_{a_{3}}\partial_{a_{1}}) \rightarrow (\partial_{a_{3}}\partial_{a_{1}})+\frac{1}{4}\eta_{a_{3}}\bar{\eta}_{a_{1}}\Box_{a_{1}}+\frac{1}{4}\eta_{a_{1}}\bar{\eta}_{a_{3}}\Box_{a_{3}},\\
&&(\partial_{a_{1}}\nabla_{23}) \rightarrow (\partial_{a_{1}}\nabla_{23})+\eta_{a_{1}}\bar{\eta}_{a_{2}}D_{a_{2}}-\eta_{a_{1}}\bar{\eta}_{a_{3}}D_{a_{3}}\\
&&(\partial_{a_{2}}\nabla_{31}) \rightarrow (\partial_{a_{2}}\nabla_{31})+\eta_{a_{2}}\bar{\eta}_{a_{3}}D_{a_{3}}-\eta_{a_{2}}\bar{\eta}_{a_{1}}D_{a_{1}}\\
&&(\partial_{a_{3}}\nabla_{12}) \rightarrow (\partial_{a_{3}}\nabla_{12})+\eta_{a_{3}}\bar{\eta}_{a_{1}}D_{a_{1}}-\eta_{a_{3}}\bar{\eta}_{a_{2}}D_{a_{2}}.
\end{eqnarray}
and can write
\begin{eqnarray}
\mathcal {A}(\Phi(z)) = \int d\eta_{a_{1}}d\bar{\eta}_{a_{1}}d\eta_{a_{2}}d\bar{\eta}_{a_{2}}d\eta_{a_{3}}d\bar{\eta}_{a_{3}}
                        (1+\eta_{a_{1}}\bar{\eta}_{a_{1}})(1+\eta_{a_{2}}\bar{\eta}_{a_{2}})(1+\eta_{a_{3}}\bar{\eta}_{a_{3}})\nonumber\\
                        exp \hat W \Phi(z_1;a_{1})\Phi(z_2;a_{2})\Phi(z_3;a_{3})\mid_{a_{1}=a_{2}=a_{3}=0},\qquad\label{exp1}
\end{eqnarray}
where
\begin{eqnarray}
\hat W &=& \frac{1}{2}[1+\lambda^{2}(\partial_{a_{1}}\partial_{a_{2}}+\frac{1}{4}\eta_{a_{1}}\bar{\eta}_{a_{2}}\Box_{a_{2}}
                        +\frac{1}{4}\eta_{a_{2}}\bar{\eta}_{a_{1}}\Box_{a_{1}})]
[\partial_{a_{3}}\nabla_{12}+\eta_{a_{3}}\bar{\eta}_{a_{1}}D_{a_{1}}-\eta_{a_{3}}\bar{\eta}_{a_{2}}D_{a_{2}}]\nonumber\\
&+&\frac{1}{2}[1+\lambda^{2}(\partial_{a_{2}}\partial_{a_{3}}+\frac{1}{4}\eta_{a_{2}}\bar{\eta}_{a_{3}}\Box_{a_{3}}
                        +\frac{1}{4}\eta_{a_{3}}\bar{\eta}_{a_{2}}\Box_{a_{2}})]
[\partial_{a_{1}}\nabla_{23}+\eta_{a_{1}}\bar{\eta}_{a_{2}}D_{a_{2}}-\eta_{a_{1}}\bar{\eta}_{a_{3}}D_{a_{3}}]\nonumber\\
&+&\frac{1}{2}[1+\lambda^{2}(\partial_{a_{3}}\partial_{a_{1}}+\frac{1}{4}\eta_{a_{3}}\bar{\eta}_{a_{1}}\Box_{a_{1}}
                        +\frac{1}{4}\eta_{a_{1}}\bar{\eta}_{a_{3}}\Box_{a_{3}})]
[\partial_{a_{2}}\nabla_{31}+\eta_{a_{2}}\bar{\eta}_{a_{3}}D_{a_{3}}-\eta_{a_{2}}\bar{\eta}_{a_{1}}D_{a_{1}}]\qquad\quad\,\,\label{SagnottiW}
\end{eqnarray}
This operator generates all terms in the cubic interaction of any three HS fields with any possible number of derivatives $\Delta$ in the range $\Delta_{min}\leq \Delta \leq \Delta_{max}$. Another possible form of the $\hat W$ operator is

\begin{eqnarray}
\hat W &=& [1+\lambda^{2}(\partial_{a_{1}}\partial_{a_{2}}+\frac{1}{2}\eta_{a_{1}}\bar{\eta}_{a_{2}}\Box_{a_{2}})]
[(\partial_{a_{3}}\nabla_{1})+\frac{1}{2}\eta_{a_{3}}\bar{\eta}_{a_{1}}D_{a_{1}}-\frac{1}{2}\eta_{a_{3}}\bar{\eta}_{a_{2}}D_{a_{2}}
+\frac{1}{2}\eta_{a_{3}}\bar{\eta}_{a_{3}}D_{a_{3}}]\nonumber\\
&+&[1+\lambda^{2}(\partial_{a_{2}}\partial_{a_{3}}+\frac{1}{2}\eta_{a_{2}}\bar{\eta}_{a_{3}}\Box_{a_{3}})]
[(\partial_{a_{1}}\nabla_{2})+\frac{1}{2}\eta_{a_{1}}\bar{\eta}_{a_{2}}D_{a_{2}}-\frac{1}{2}\eta_{a_{1}}\bar{\eta}_{a_{3}}D_{a_{3}}
+\frac{1}{2}\eta_{a_{1}}\bar{\eta}_{a_{1}}D_{a_{1}}]\nonumber\\
&+&[1+\lambda^{2}(\partial_{a_{3}}\partial_{a_{1}}+\frac{1}{2}\eta_{a_{3}}\bar{\eta}_{a_{1}}\Box_{a_{1}})]
[(\partial_{a_{2}}\nabla_{3})+\frac{1}{2}\eta_{a_{2}}\bar{\eta}_{a_{3}}D_{a_{3}}-\frac{1}{2}\eta_{a_{2}}\bar{\eta}_{a_{1}}D_{a_{1}}
+\frac{1}{2}\eta_{a_{2}}\bar{\eta}_{a_{2}}D_{a_{2}}]\qquad\quad\,\,\label{MMRW}
\end{eqnarray}
This case generates the Lagrangian derived in previous chapter. The leading term of that Lagrangian is (\ref{leading}). These two operators (\ref{SagnottiW}) and (\ref{MMRW}) generate two Lagrangians that differ from each other just by partial integration and field redefinition. All interactions of HS gauge fields with any number of derivatives are unique and are generated by both operators (\ref{SagnottiW}) and (\ref{MMRW}).

In the case of (\ref{SagnottiW}) we have
\begin{eqnarray}
\setlength{\unitlength}{0.254mm}
\begin{picture}(280,275)(125,-355)
        \allinethickness{0.254mm}\path(405,-80)(405,-175) 
        \allinethickness{0.254mm}\path(345,-235)(345,-80) 
        \allinethickness{0.254mm}\path(285,-295)(285,-80) 
        \allinethickness{0.254mm}\path(225,-355)(225,-80) 
        \allinethickness{0.254mm}\path(125,-80)(125,-355) 
        \allinethickness{0.254mm}\path(165,-80)(165,-355) 
        \allinethickness{0.254mm}\path(405,-80)(125,-80) 
        \allinethickness{0.254mm}\path(405,-175)(125,-175) 
        \allinethickness{0.254mm}\path(345,-235)(125,-235) 
        \allinethickness{0.254mm}\path(285,-295)(125,-295) 
        \allinethickness{0.254mm}\path(225,-355)(125,-355) 
        \allinethickness{0.254mm}\path(125,-115)(405,-115) 
        \allinethickness{0.254mm}\path(125,-80)(165,-115) 
        \put(130,-110){\shortstack{$\Box_{a_{i}}$}} 
        \put(141,-94){\shortstack{$D_{a_{i}}$}} 
        \put(190,-106){\shortstack{$0$}} 
        \put(250,-106){\shortstack{$1$}} 
        \put(310,-106){\shortstack{$2$}} 
        \put(370,-106){\shortstack{$3$}} 
        \put(140,-151){\shortstack{$0$}} 
        \put(140,-211){\shortstack{$1$}} 
        \put(140,-271){\shortstack{$2$}} 
        \put(140,-331){\shortstack{$3$}} 
        \put(185,-151){\shortstack{$\mathcal {A}^{00}$}} 
        \put(250,-151){\shortstack{$0$}} 
        \put(295,-151){\shortstack{$\mathcal {A}^{20}$}} 
        \put(370,-151){\shortstack{$0$}} 
        \put(190,-211){\shortstack{$0$}} 
        \put(185,-271){\shortstack{$\mathcal {A}^{02}$}} 
        \put(185,-331){\shortstack{$\mathcal {A}^{03}$}} 
        \put(235,-211){\shortstack{$\mathcal {A}^{11}$}} 
        \put(295,-211){\shortstack{$\mathcal {A}^{21}$}} 
        \put(250,-271){\shortstack{$\mathcal {A}^{12}$}} 
\end{picture}
\end{eqnarray}
In the case of (\ref{MMRW}) we have
\begin{eqnarray}
\setlength{\unitlength}{0.254mm}
\begin{picture}(280,275)(125,-355)
        \allinethickness{0.254mm}\path(405,-80)(405,-175) 
        \allinethickness{0.254mm}\path(345,-235)(345,-80) 
        \allinethickness{0.254mm}\path(285,-295)(285,-80) 
        \allinethickness{0.254mm}\path(225,-355)(225,-80) 
        \allinethickness{0.254mm}\path(125,-80)(125,-355) 
        \allinethickness{0.254mm}\path(165,-80)(165,-355) 
        \allinethickness{0.254mm}\path(405,-80)(125,-80) 
        \allinethickness{0.254mm}\path(405,-175)(125,-175) 
        \allinethickness{0.254mm}\path(345,-235)(125,-235) 
        \allinethickness{0.254mm}\path(285,-295)(125,-295) 
        \allinethickness{0.254mm}\path(225,-355)(125,-355) 
        \allinethickness{0.254mm}\path(125,-115)(405,-115) 
        \allinethickness{0.254mm}\path(125,-80)(165,-115) 
        \put(130,-110){\shortstack{$\Box_{a_{i}}$}} 
        \put(141,-94){\shortstack{$D_{a_{i}}$}} 
        \put(190,-106){\shortstack{$0$}} 
        \put(250,-106){\shortstack{$1$}} 
        \put(310,-106){\shortstack{$2$}} 
        \put(370,-106){\shortstack{$3$}} 
        \put(140,-151){\shortstack{$0$}} 
        \put(140,-211){\shortstack{$1$}} 
        \put(140,-271){\shortstack{$2$}} 
        \put(140,-331){\shortstack{$3$}} 
        \put(185,-151){\shortstack{$\mathcal {A}^{00}$}} 
        \put(235,-151){\shortstack{$\mathcal {A}^{10}$}} 
        \put(295,-151){\shortstack{$\mathcal {A}^{20}$}} 
        \put(355,-151){\shortstack{$\mathcal {A}^{30}$}} 
        \put(190,-211){\shortstack{$0$}} 
        \put(190,-271){\shortstack{$0$}} 
        \put(185,-331){\shortstack{$\mathcal {A}^{03}$}} 
        \put(235,-211){\shortstack{$\mathcal {A}^{11}$}} 
        \put(295,-211){\shortstack{$\mathcal {A}^{21}$}} 
        \put(235,-271){\shortstack{$\mathcal {A}^{12}$}} 
\end{picture}
\end{eqnarray}
Both forms of the same cubic Lagrangian are very useful for further investigations.


\chapter{Conclusions and outlook}

\quad In this Thesis I present first of all recent results in the theory of Higher Spin gauge field interactions. In addition a powerful method of constructing higher order conformal invariants is presented.

The cubic interaction problem in Higher Spin gauge field theory is solved in the most general case covering all possibilities. The other beauty of the result is the compactness and clearness of final formulas, that can be derived from a Generating Function, which reproduces \emph{all possible nontrivial interactions between massless Higher Spin gauge fields in a flat background spacetime}. However, it is not trivial to continue this construction to higher orders on the gauge field. The solution of the main problem of existence of a local (or nonlocal) Lagrangian for Higher Spin gauge fields is very close now: the problem of the quartic interaction is an urgent and real task which could shed light on the solution in all orders.

These results can be used also to test AdS/CFT correspondence between critical O(N) sigma model and Higher Spin gauge field theory on AdS  (after deformation to AdS from an even dimensional flat space).

Another important point here is that the interactions in flat space-time derived in this Thesis are independent of the space-time dimensions. At last we note that the structure of the Generating Function for these interactions leads to new connections with the massless regime of String Theory.


\chapter*{Appendix A}

\addcontentsline{toc}{chapter}{Appendix A}

\setcounter{equation}{0}
\renewcommand{\theequation}{A.\arabic{equation}}
\quad The Euclidian $AdS_{d+1}$ metric
\begin{equation}\label{eads}
    ds^{2}=g_{\mu \nu }(z)dz^{\mu }dz^{\nu
}=\frac{1}{(z^{0})^{2}}\delta _{\mu \nu }dz^{\mu }dz^{\nu }
\end{equation}
can be realized as an induced metric for the hypersphere defined by
the following embedding procedure in $d+2$ dimensional Minkowski
space
\begin{eqnarray}
  && X^{A}X^{B}\eta_{AB}=-X_{-1}^{2}+X_{0}^{2}+\sum^{d}_{i=1}X_{i}^{2}=-1 ,\\
  && X_{-1}(z)=\frac{1}{2}\left(\frac{1}{z_{0}}+\frac{z_{0}^{2}
  +\sum^{d}_{i=1}z^{2}_{i}}{z_{0}}\right) ,\\
  && X_{0}(z)=\frac{1}{2}\left(\frac{1}{z_{0}}-\frac{z_{0}^{2}
  +\sum^{d}_{i=1}z^{2}_{i}}{z_{0}}\right) ,\\
  && X_{i}(z)=\frac{z_{i}}{z_{0}} .
\end{eqnarray}
Using these embedding rules we can identify the variable
$\zeta(z,w)$ as an $SO(1,d+1)$ invariant scalar product
\begin{equation}\label{gd}
    -X^{A}(z)Y^{B}(w)\eta_{AB}=\frac{1}{2z_{0}w_{0}}\left(2z_{0}w_{0}
    +\sum^{d}_{\mu=0}(z-w)^{2}_{\mu}\right)=\zeta=u+1 ,
\end{equation}
and therefore can be realized by $cosh$ of a hyperbolic angle. Indeed we can
introduce another embedding
\begin{eqnarray}\label{hyp}
    &&X_{-1}(\eta,\omega_{\mu})=\cosh{\eta},\\
   && X_{\mu}(\eta,\omega_{\mu})=\sinh{\eta}\,\omega_{\mu}\quad,\quad\quad
    \sum^{d}_{\mu=0}\omega^{2}_{\mu}=1 ,\\
    &&ds^{2}=d\eta^{2}+\sinh^{2}{\eta}\, d\Omega_{d} .
\end{eqnarray}
In these coordinates the chordal distance $u$ between an arbitrary point
$X^{A}(\eta,\Omega_{\mu})$ and the pole of the hypersphere
$Y^{A}(\eta=0,\omega_{\mu})$ is simply
\begin{equation}\label{hd}
    \zeta= -X^{A}Y^{B}\eta_{AB}=\cosh{\eta} .
\end{equation}
Therefore the invariant measure is expressed as
\begin{equation}\label{invv}
    \sqrt{g}d\eta d\Omega_{d}=(\sinh\eta)^{d}d\eta
    d\Omega_{d}=[u(u+2)]^{\frac{d-1}{2}}du d\Omega_{d} .
\end{equation}
So we see that the integration measure for $d=3$ ($D=d+1=4$) will
cancel one order of $u^{-n}$ in short distance singularities and we
have to count the singularities starting from $u^{-2}$ which is  "logarithmically divergent"
in standard QFT terminology.

 In this article we use the following rules and relations for
$u(z,z')$, $I_{1a}$, $I_{2c}$ and the bitensorial basis
$\{I_{i}\}^{4}_{i=1}$
\begin{eqnarray}
  && \Box u=(d+1)(u+1) ,\quad \nabla_{\mu}\partial_{\nu}u=g_{\mu\nu}(u+1) ,
  \quad g^{\mu\nu}\partial_{\mu}u\partial_{\nu}u=u(u+2) ,\quad\quad\quad\label{start}\\
  &&   \partial_{\mu}\partial_{\nu'}u
  \nabla^{\mu}u=(u+1)\partial_{\nu'}u ,\quad
  \partial_{\mu}\partial_{\nu'}u \nabla^{\mu}\partial_{\mu'}u
  =g_{\mu'\nu'}+\partial_{\mu'}u\partial_{\nu'}u ,\\
&&\nabla_{\mu}\partial_{\nu}\partial_{\nu'}u \nabla^{\mu}u
  =\partial_{\nu}u\partial_{\nu'}u ,\quad
  \nabla_{\mu}\partial_{\nu}\partial_{\nu'}u
  =g_{\mu\nu}\partial_{\nu'}u ,\\
&&\frac{\partial}{\partial a^{\mu}}I_{1a}\frac{\partial}{\partial
a_{\mu}}I_{1a}=u(u+2) ,\quad \frac{\partial}{\partial
a^{\mu}}I_{1}\frac{\partial}{\partial
a_{\mu}}I_{1a}=(u+1) I_{2c} ,\\
&&\frac{\partial}{\partial a^{\mu}}I_{1}\frac{\partial}{\partial
a_{\mu}}I_{1}=c^{2}_{2}+ I_{2c}^{2} , \, \frac{\partial}{\partial
a^{\mu}}I_{1}\frac{\partial}{\partial a_{\mu}}I_{2}=(u+1) I_{2c}^{2}
,\,\Box_{a}I_{4}=2(d+1)c^{2}_{2} ,\\
 &&\frac{\partial}{\partial
a^{\mu}}I_{2}\frac{\partial}{\partial a_{\mu}}I_{2}=u(u+2)I_{2c}^{2}
,\quad
\Box_{a}I_{3}=2(d+1)I_{2c}^{2}+2c^{2}_{2}u(u+2) ,\\
&&\nabla^{\mu}\frac{\partial}{\partial a^{\mu}}I_{1}=(d+1)I_{2c}
,\,\nabla^{\mu}\frac{\partial}{\partial a^{\mu}}I_{2}=(d+2)(u+1)
I_{2c},\quad\nabla^{\mu} I_{1}\partial_{\mu}u=I_{2} ,\\
&&\nabla^{\mu}\frac{\partial}{\partial
a^{\mu}}I_{3}=4I_{1}I_{2c}+2(d+2)(u+1) c^{2}_{2}I_{1a}
,\quad\nabla^{\mu} I_{2}\partial_{\mu}u=2(u+1) I_{2} ,\\
&&\frac{\partial}{\partial a_{\mu}} I_{1}\partial_{\mu}u=(u+1)
I_{2c} ,\quad \frac{\partial}{\partial a_{\mu}}
I_{2}\partial_{\mu}u=u(u+2) I_{2c} ,\,\frac{\partial}{\partial
a_{\mu}}
I_{1}\nabla_{\mu} I_{1}=I_{1} I_{2c} ,\,\,\,\,\,\quad\quad\\
&&\frac{\partial}{\partial a_{\mu}} I_{1}\nabla_{\mu}
I_{2}=I_{2c}\left((u+1) I_{1}+I_{2}\right)+c^{2}_{2}I_{1a}
,\frac{\partial}{\partial a_{\mu}} I_{2}\nabla_{\mu}
I_{1}=I_{2c}I_{2} ,\\
&&\frac{\partial}{\partial a_{\mu}} I_{2}\nabla_{\mu} I_{2}=2(u+1)
I_{2c}I_{2} ,\quad \nabla^{\mu} I_{1}\nabla_{\mu}
I_{1}=a^{2}_{1}I_{2c}^{2} ,\quad \Box I_{1}=I_{1} ,\\
&&\nabla^{\mu} I_{1}\nabla_{\mu} I_{2}=I_{2}I_{1}+ a^{2}_{1}(u+1)
I_{2c}^{2} ,\quad \Box I_{2}=(d+2)I_{2}+2(u+1)
I_{1} ,\quad\\
&&\nabla^{\mu} I_{2}\nabla_{\mu} I_{2}=I_{2}^{2}+2(u+1)
I_{1}I_{2}+a^{2}_{1}I_{2c}^{2}(u+1)^{2}+c^{2}_{2}I_{1a}^{2}
,\\
&&a^{\mu}\nabla_{\mu}I_{1a}=a^{2}(u+1) ,\quad
a^{\mu}\nabla_{\mu}I_{2c}=I_{1},\quad
a^{\mu}\nabla_{\mu}I_{1}=a^{2}I_{2c},
\\&&a^{\mu}\nabla_{\mu}I_{2}=a^{2}(u+1) I_{2c}+I_{1a}I_{1},
.\label{end}
\end{eqnarray}
Using these relations we can derive ($
F'_{k}:=\frac{\partial}{\partial
  u}F_{k}(u)$)
\begin{itemize}
  \item Divergence map
\begin{eqnarray}
  && \nabla^{\mu}_{1}\frac{\partial}{\partial a^{\mu}}\Psi^{\ell}[F]=
  I_{2c}\Psi^{\ell-1}[Div_{\ell}F]+O(c^{2}_{2}) , \label{div}\\
  && (Div_{\ell}F)_{k}=(\ell-k)(u+1)
  F'_{k}+(k+1)u(u+2)F'_{k+1}\nonumber\\
  &&+(\ell-k)(\ell+d+k)F_{k}+(k+1)(\ell+d+k+1)(u+1) F_{k+1} .\label{dv}
  \end{eqnarray}
  \item Trace map
\begin{eqnarray}
    &&\Box_{a}\Psi^{\ell}[F]=I^{2}_{2c}\Psi^{\ell-2}
    [{Tr_{\ell}F}]+O(c^{2}_{2}) ,\label{tracemap}\\
&&(Tr_{\ell}F)_{k}=(\ell-k)(\ell-k-1)F_{k} +2(k+1)(\ell-k-1)(u+1)
F_{k+1}\nonumber\\&&\quad\quad\quad\quad\quad+(k+2)(k+1)u(u+2)F_{k+2} .\label{tr}
\end{eqnarray}
  \item Laplacian map
  \begin{eqnarray}
  && \Box_{1} \Psi^{\ell}[F]=\Psi^{\ell}[Lap_{\ell}F]+O(a^{2}_{1},c^{2}_{2}) , \label{lm1}\\
  &&(Lap_{\ell}F)_{k}=u(u+2)F''_{k}+(d+1+4k)(u+1)
  F'_{k}+[\ell+k(d+2\ell-k)]F_{k}\nonumber\\&&+2(u+1)(k+1)^{2}
  F_{k+1}+2(\ell-k+1)F'_{k-1},\label{lm2}\\
  &&\Box F_{k}(u)=u(u+2)F''_{k}+(d+1)(u+1) F'_{k}.\quad\label{lm3}\end{eqnarray}
  \item Gradient map
  \begin{eqnarray}
&&(a\cdot\nabla)_{1}\Psi^{\ell}[F]=I_{1a}\Psi^{\ell}[Grad_{\ell}F]+ O(a^{2}_{1}) ,\label{grad1}\\
  && (Grad_{\ell}F)_{k}=F'_{k}+(k+1)F_{k+1} .\label{grad2}
\end{eqnarray}

\end{itemize}

At the end we present all important commutation relations working in the space of symmetric rank $n$ tensors
\begin{eqnarray}
  &&[(\nabla\partial_{a}), \Box]f^{(n)}(z,a)= \left[2(a\nabla)\Box_{a}-(d+2n-2)(\nabla\partial_{a})\right]f^{(n)}(z,a) ;\quad \quad\label{A.6}\\
  &&[(\nabla\partial_{a}), (a\nabla)]f^{(n)}(z,a)= \Box f^{(n)}(z,a)
 + [\nabla_{\mu}, (a\nabla)]\partial_{a}^{\mu}f^{(n)}(z,a) ;\label{A.7}\\
&& [\nabla_{\mu}, (a\nabla)]\partial_{a}^{\mu}f^{(n)}(z,a)= \left[a^{2}\Box_{a}-n(d+n-1)\right]f^{(n)}(z,a) ;\label{A.8}\\
&&[\Box ,(a\nabla)]f^{(n)}(z,a)= \left[2a^{2}(\nabla\partial_{a})-(d+2n)(a\nabla)\right]f^{(n)}(z,a) ;\label{A.9}\\
&&\Box_{a}\left[a^{2}f^{(n)}(z,a)\right] = 2(d+2n+1)f^{(n)}(z,a)+a^{2}\Box_{a}f^{(n)}(z,a) .\label{A.10}
\end{eqnarray}


\chapter*{Appendix B}

\addcontentsline{toc}{chapter}{Appendix B}

\setcounter{equation}{0}
\renewcommand{\theequation}{B.\arabic{equation}}
\quad These two useful hypergeometric identities we learned from the book of H. Bateman  and A. Erdelyi ``Higher transcendental functions'' V.1, McGraw-Hill Book company Inc. 1953.
\begin{eqnarray}
   _{2}F_{1}(a,b,2b;z)&=& \left(1-\frac{z}{2}\right)^{-a} {}_{2}F_{1}\left(\frac{a}{2},\frac{a+1}{2},b+\frac{1}{2};\left(\frac{z}{2-z}\right)^{2}\right) ,\label{B1}\\
  {}_{2}F_{1}(a,b,c;z)&=&\frac{\Gamma(c)\Gamma(b-a)}{\Gamma(b)\Gamma(c-a)}(-z)^{-a} {}_{2}F_{1}(a,1-c+a,1-b+a;z^{-1})\nonumber\\
  &+&\frac{\Gamma(c)\Gamma(b-a)}{\Gamma(b)\Gamma(c-a)}(-z)^{-b} {}_{2}F_{1}(a,1-c+b,1-a+b;z^{-1}) .\label{B2}
\end{eqnarray}


\bigskip
\chapter*{Appendix C}

\addcontentsline{toc}{chapter}{Appendix C}

\setcounter{equation}{0}
\renewcommand{\theequation}{C.\arabic{equation}}
\quad Here we present the basic relations between different T-s, M-s and N-s which we use in section \ref{scalar}.
\begin{eqnarray}
&&T(n,k)=(-1)^{m}\sum_{i=0}^{m}\binom{m}{i}T(n-i,k+m-i),\\
&&T(n,k)=(-1)^{m}\sum_{i=0}^{m}\binom{m}{i}T(n-i,k-m),
\end{eqnarray}
and the same for M and N. There is another important relation
\begin{eqnarray}
&&T(n,k)=-M(n-1,k)-\frac{k(k-1)}{2L^{2}}N(n-1,k-1)\nonumber\\
&&-[\frac{(n-k-1)(2D+n-k-4)}{2L^{2}}+\frac{D(D-2)}{4L^{2}}]N(n-1,k)\nonumber\\
&&-\epsilon_{(\ell-n)}^{\mu_{1}...\mu_{n-1}}\nabla_{\mu_{1}}...\nabla_{\mu_{k}}\phi\nabla_{\mu_{k+1}}...\nabla_{\mu_{n-1}}
(\Box-\frac{D(D-2)}{4L^{2}})\phi,
\end{eqnarray}
and the 'symmetrization' relations
\begin{eqnarray}
&&M(2k+1,k)=M(2k+1,k+1)=-\frac{1}{2}M(2k,k),\\
&&N(2k+1,k)=N(2k+1,k+1)=-\frac{1}{2}N(2k,k),\\
&&T(2m,m)=\nabla^{(\alpha}\epsilon_{(\ell-2m)}^{\mu_{1}...\mu_{2m-1})}\nabla_{\mu_{1}}...\nabla_{\mu_{m-1}}\nabla_{\alpha}\phi
\nabla_{\mu_{m}}...\nabla_{\mu_{2m-1}}\phi\nonumber\\
&&+\frac{(m-1)(m-2)}{6L^{2}}N(2m-2,m-1)-\frac{(m-1)(m-2)}{12L^{2}}N(2m-4,m-2) ,\\
&&M(2m-2,m-1)=\epsilon_{(\ell-2m+1)\mu_{m}...\mu_{2m-2}}^{\mu_{1}...\mu_{m-1}}
\nabla_{(\mu_{1}}...\nabla_{\mu_{m-1}}\nabla_{\alpha)}\phi
\nabla^{(\mu_{m}}...\nabla^{\mu_{2m-2}}\nabla^{\alpha)}\phi\nonumber\\
&&+\frac{(m-1)(m-2)}{3L^{2}}N(2m-2,m-1,m-1)-\frac{(m-1)(m-2)}{6L^{2}}N(2m-4,m-2)\nonumber\\
\end{eqnarray}
We must mention here that these relations are satisfied up to full derivatives and therefore admit integration.


\chapter*{Appendix D}

\addcontentsline{toc}{chapter}{Appendix D}

\setcounter{equation}{0}
\renewcommand{\theequation}{D.\arabic{equation}}
\quad We use the following commutation relations in $AdS_{D}$
\begin{eqnarray}
&&\epsilon^{\mu_{1}\dots\mu_{\ell-1}}[\nabla^{\mu},\nabla_{\mu_{1}}.\dots\nabla_{\mu_{k}}]\phi=
\frac{k(k-1)}{2L^{2}}\epsilon^{\mu\mu_{2}...\mu_{\ell-1}}\nabla_{\mu_{2}}\dots\nabla_{\mu_{k}}\phi,\\
&&[\nabla_{\mu_{1}}\dots\nabla_{\mu_{k}},\nabla^{\mu}]
\epsilon^{\mu_{1}\dots\mu_{\ell-1}}=
\frac{2k(D+\ell-2)-k(k+1)}{2L^{2}}\epsilon_{(k-1)}^{\mu\mu_{k+1}...\mu_{\ell-1}},\\
&&\epsilon^{\mu_{1}\dots\mu_{\ell-1}}[\nabla_{\mu},\nabla_{\mu_{1}}\dots\nabla_{\mu_{k}}]\nabla^{\mu}\phi=
\frac{k(2D+k-3)}{2L^{2}}\epsilon^{\mu_{1}\mu_{2}\dots\mu_{\ell-1}}\nabla_{\mu_{1}}\dots\nabla_{\mu_{k}}\phi,\qquad\\
&&\epsilon^{\mu_{1}\dots\mu_{\ell-1}}[\nabla^{2},\nabla_{\mu_{1}}\dots\nabla_{\mu_{k}}]\phi=
\frac{k(D+k-2)}{L^{2}}\epsilon^{\mu_{1}\mu_{2}\dots\mu_{\ell-1}}\nabla_{\mu_{1}}\dots\nabla_{\mu_{k}}\phi,
\end{eqnarray}
where $\epsilon^{\mu_{1}...\mu_{\ell-1}}$ is the symmetric and traceless tensor.
Finally we list all necessary binomial identities
\begin{eqnarray}
&&\sum_{k=0}^{n-m}(-1)^{k}\binom{n}{k}=(-1)^{n-m}\binom{n-1}{m-1},\quad
\sum_{k=0}^{n-m}(-1)^{k}\binom{n}{m+k}=\binom{n-1}{m-1},\quad\quad\\
&&\binom{n}{k}=\binom{n-1}{k-1}+\binom{n-1}{k},\quad
\binom{\ell-m-1}{m-2}=\frac{m-1}{\ell-2m+1}\binom{\ell-m-1}{m-1}.\quad\qquad
\end{eqnarray}


\chapter*{Appendix E}

\addcontentsline{toc}{chapter}{Appendix E}

\setcounter{equation}{0}
\renewcommand{\theequation}{E.\arabic{equation}}
\quad Here we present  all  Weyl variations necessary for the derivation of (\ref{sys1})-(\ref{syslast})
\begin{eqnarray}
&&\delta_{\sigma}^{1}S_{0}=\Delta_{\ell}\int d^{D}z\sqrt{-g}
\{
\sum_{m=1}^{\frac{\ell}{2}-1}\binom{\ell-m-2}{m-1}
\nabla^{(\mu_{2m}}\sigma_{(\ell-2m-1)}^{\mu_{1}...\mu_{2m-1})}\Psi^{(2m)}_{\mu_{1}...\mu_{2m}}\nonumber\\
&&+\sum_{m=1}^{\frac{\ell}{2}-1}\frac{(-1)^{m-1}}{2}\binom{\ell-m-3}{m-1}
\Box\sigma_{(\ell-2m-2)}^{\mu_{1}...\mu_{2m}}\nabla_{\mu_{1}}...\nabla_{\mu_{m}}\phi
\nabla_{\mu_{m+1}}...\nabla_{\mu_{2m}}\phi\nonumber\\
&&+\sum_{m=1}^{\frac{\ell}{2}-1}(-1)^{m}\binom{\ell-m-3}{m-1}
\sigma_{(\ell-2m-2)}^{\mu_{1}...\mu_{2m}}\nabla_{\mu_{1}}...\nabla_{\mu_{m}}\nabla_{\alpha}\phi
\nabla_{\mu_{m+1}}...\nabla_{\mu_{2m}}\nabla^{\alpha}\phi\nonumber\\
&&+O(\frac{1}{L^{2}})\label{swl}
\}.
\end{eqnarray}

We don't have to calculate $O(\frac{1}{L^{2}})$ terms because they can be fixed from flat space considerations and gauge invariance of Fronsdal's operator in AdS.
The first term in (\ref{swl}) can be cancelled by an additional gauge transformation of all gauge fields with spin less than $\ell$.
To cancel other terms we calculate the variation of $\sum_{m=1}^{\ell/2}S_{1}^{\Psi^{(2m)}}(\phi,h^{(2m)})$:
\begin{eqnarray}
&&\delta_{\sigma}^{0}S_{1}^{\Psi^{(2m)}}(\phi,h^{(2)},...,h^{(2m)})\nonumber\\&&
=C_{\ell}^{m}\int d^{D}z\sqrt{-g}
\{-(m-1)[\nabla^{(\mu_{2m-2}}\sigma_{(\ell-2m+1)}^{\mu_{1}...\mu_{2m-3})}\Psi^{(2m-2)}_{\mu_{1}...\mu_{2m-2}}\nonumber\\
&&+\frac{(-1)^{m}}{2}\Box\sigma_{(\ell-2m)}^{\mu_{1}...\mu_{2m-2}}\nabla_{\mu_{1}}...\nabla_{\mu_{m-1}}\phi
\nabla_{\mu_{m}}...\nabla_{\mu_{2m-2}}\phi]\nonumber\\
&&+(-1)^{m}(1-\frac{D}{2})\sigma_{(\ell-2m)}^{\mu_{1}...\mu_{2m-2}}\nabla_{\mu_{1}}...\nabla_{\mu_{m-1}}\nabla_{\alpha}\phi
\nabla_{\mu_{m}}...\nabla_{\mu_{2m-2}}\nabla^{\alpha}\phi\}.
\end{eqnarray}
and  the variation of $\sum_{m=1}^{\ell/2}S_{1}^{r^{(2m)}}(\phi,h^{(2m)})$:
\begin{eqnarray}
&&\delta_{\sigma}^{0}S_{1}^{r^{(\ell)}}=\frac{1}{2}\sum_{m=1}^{\frac{\ell}{2}-1}\int d^{D}z\sqrt{-g}
\{[2m(2m-1)\xi_{\ell}^{m}-2(m-1)(D+4m-8)\xi_{\ell}^{m-1}]\times\nonumber\\
&&\times\nabla^{(\mu_{2m-2}}\sigma_{(\ell-2m+1)}^{\mu_{1}...\mu_{2m-3})}
\Psi^{(2m-2)}_{\mu_{1}...\mu_{2m-2}}\}\nonumber\\
&&-\frac{1}{2}\int d^{D}z\sqrt{-g}\{(\ell-2)(D+2\ell-8)\xi_{\ell}^{\ell/2-1}\nabla^{(\mu_{\ell-2}}
\sigma_{\ell(1)}^{\mu_{1}...\mu_{\ell-3})}\Psi^{(\ell-2)}_{\mu_{1}...\mu_{\ell-2}}
\}\nonumber\\
&&+\frac{1}{2}\sum_{m=1}^{\frac{\ell}{2}-1}\xi_{\ell}^{m}\int d^{D}z\sqrt{-g}\{2m(1-\frac{D}{2})\sigma_{(\ell-2m)}^{\mu_{1}...\mu_{2m-2}}\nabla_{\mu_{1}}...\nabla_{\mu_{m-1}}\nabla_{\alpha}\phi
\nabla_{\mu_{m}}...\nabla_{\mu_{2m-2}}\nabla^{\alpha}\phi
\}\nonumber\\&&+\frac{1}{2}\sum_{m=1}^{\frac{\ell}{2}-1}\int d^{D}z\sqrt{-g}\{[-m(m-1)\xi_{\ell}^{m}-(\ell-2m+2)(D+\ell+2m-5)\xi_{\ell}^{m-1}]\times\nonumber\\
&&\times\Box \sigma_{(\ell-2m)}^{\mu_{1}...\mu_{2m-2}}\nabla_{\mu_{1}}...\nabla_{\mu_{m-1}}\phi
\nabla_{\mu_{m}}...\nabla_{\mu_{2m-2}}\phi\}\nonumber\\
&&-\frac{1}{2}\int d^{D}z\sqrt{-g}\{2(D+2\ell-5)\xi_{\ell}^{\ell/2-1}\Box \sigma^{\mu_{1}...\mu_{\ell-2}}\nabla_{\mu_{1}}...\nabla_{\mu_{\ell/2-1}}\phi
\nabla_{\mu_{\ell/2}}...\nabla_{\mu_{\ell-2}}\phi\}\nonumber\\
&&+O(\frac{1}{L^{2}})
\end{eqnarray}
Then finally we get
\begin{eqnarray}
&&\delta_{\sigma}S^{WI}(\phi,h^{(2)},...,h^{(\ell)})=
\delta_{\sigma}^{1}S_{0}+\sum_{s=1}^{\ell/2}\delta_{\sigma}^{0}S_{1}^{\Psi^{(2s)}}
+\sum_{s=1}^{\ell/2}\delta_{\sigma}^{0}S_{1}^{r^{(\ell)}}\nonumber\\
&&=\sum_{m=1}^{\ell/2-1}\int d^{D}z\sqrt{-g}
\{\binom{\ell-m-2}{m-1}\Delta_{\ell}-mC_{\ell}^{m+1}[1+(D+4m-4)\xi_{2m+2}^{m}]\nonumber\\
&&+\frac{1}{2}\sum_{s=m+2}^{\ell/2}[(2m+2)(2m+1)\xi_{2s}^{m+1}-2m(D+4m-4)\xi_{2s}^{m}]\}\times\nonumber\\
&&\times\nabla^{(\mu_{2m}}\sigma_{(\ell-2m-1)}^{\mu_{1}...\mu_{2m-1})}\Psi^{(2m)}_{\mu_{1}...\mu_{2m}}\nonumber\\
&&+\int d^{D}z\sqrt{-g}\{\frac{(-1)^{\ell/2}}{2}(\Delta_{\ell}-\frac{\ell-2}{2})-(D+2\ell-5)\xi_{\ell}^{\ell/2-1}\}\times\nonumber\\
&&\times\Box \sigma^{\mu_{1}...\mu_{\ell-2}}\nabla_{\mu_{1}}...\nabla_{\mu_{\ell/2-1}}\phi\nabla_{\mu_{\ell/2}}...\nabla_{\mu_{\ell-2}}\phi\nonumber\\
&&+\sum_{m=1}^{\ell/2-1}\int d^{D}z\sqrt{-g}\{\frac{(-1)^{m}}{2}[\binom{\ell-m-2}{m-2}\Delta_{\ell}-(m-1)C_{\ell}^{m}]-C_{\ell}^{m}(D+4m-5)\xi_{2m}^{m-1}\nonumber\\
&&+\frac{1}{2}\sum_{s=m+1}^{\ell/2}C_{\ell}^{s}[-m(m-1)\xi_{2s}^{m}-(2s-2m+2)(D+2s+2m-5)\xi_{2s}^{m-1}]\}\times\nonumber\\
&&\times\Box\sigma_{(\ell-2m)}^{\mu_{1}...\mu_{2m-2}}\nabla_{\mu_{1}}...\nabla_{\mu_{m-1}}\phi
\nabla_{\mu_{m}}...\nabla_{\mu_{2m-2}}\phi\nonumber\\
&&+\int d^{D}z\sqrt{-g}\{(-1)^{\ell/2}(1-\frac{D}{2}-\Delta_{\ell})\sigma^{\mu_{1}...\mu_{\ell-2}}
\nabla_{\mu_{1}}...\nabla_{\mu_{\ell/2-1}}\nabla_{\alpha}\phi
\nabla_{\mu_{\ell/2}}...\nabla_{\mu_{\ell-2}}\nabla^{\alpha}\phi\}\nonumber\\
&&+\sum_{m=1}^{\ell/2-1}\int d^{D}z\sqrt{-g}\{(-1)^{m-1}\binom{\ell-m-2}{m-2}\Delta_{\ell}+(-1)^{m}(1-\frac{D}{2})C_{\ell}^{m}\nonumber\\
&&+(1-\frac{D}{2})\sum_{s=m+1}^{\ell/2}mC_{\ell}^{s}\xi_{2s}^{m}\}\sigma_{(\ell-2m)}^{\mu_{1}...\mu_{2m-2}}
\nabla_{\mu_{1}}...\nabla_{\mu_{m-1}}\nabla_{\alpha}\phi\nabla_{\mu_{m}}...\nabla_{\mu_{2m-2}}\nabla^{\alpha}\phi
\end{eqnarray}
>From this expression we can derive our system of equations (\ref{sys1})-(\ref{syslast}).


\chapter*{Appendix F\\ Proof of  $\mathcal{L}_{I}^{(i,j)}$}

\addcontentsline{toc}{chapter}{Appendix F}

\setcounter{equation}{0}
\renewcommand{\theequation}{F.\arabic{equation}}
\quad
The expression for $\mathcal{L}_{I}^{(1,0)}$ (\ref{l10}) is right since the following remaining group of terms vanishes:
\begin{eqnarray}
  && +\frac{s_{1}n_{1}s_{3}}{2}C_{(n_i)}^{(s_i)}[\hat{T}(Q_{ij}|n_{1}-1,n_{2},n_{3}),(c\nabla_{3})]
  (\epsilon^{(s_{1}-1)}h^{(s_{2})}D^{(s_{3}-1)}+D^{(s_{1}-1)}h^{(s_{2})}\epsilon^{(s_{3}-1)})\qquad\quad\label{F.1}\\
  && +\frac{s_{1}n_{1}s_{2}}{2}C_{(n_i)}^{(s_i)}[\hat{T}(Q_{ij}|n_{1}-1,n_{2},n_{3}),(b\nabla_{2})]
  (D^{(s_{1}-1)}\epsilon^{(s_{2}-1)}h^{(s_{3})}-\epsilon^{(s_{1}-1)}D^{(s_{2}-1)}h^{(s_{3})})\qquad\quad \label{F.2}\\
  && +\frac{s_{2}n_{2}s_{1}}{2}C_{(n_i)}^{(s_i)}[\hat{T}(Q_{ij}|n_{1},n_{2}-1,n_{3}),(a\nabla_{1})]
  (D^{(s_{1}-1)}\epsilon^{(s_{2}-1)}h^{(s_{3})}+\epsilon^{(s_{1}-1)}D^{(s_{2}-1)}h^{(s_{3})})\qquad\quad \label{F.3}\\
  &&+\frac{s_{2}n_{2}s_{3}}{2}C_{(n_i)}^{(s_i)}[\hat{T}(Q_{ij}|n_{1},n_{2}-1,n_{3}),(c\nabla_{3})]
  (h^{(s_{1})}D^{(s_{2}-1)}\epsilon^{(s_{3}-1)}-h^{(s_{1})}\epsilon^{(s_{2}-1)}D^{(s_{3}-1)})\qquad\quad \label{F.4}\\
  && +\frac{s_{3}n_{3}s_{2}}{2}C_{(n_i)}^{(s_i)}[\hat{T}(Q_{ij}|n_{1},n_{2},n_{3}-1),(b\nabla_{2})]
  (h^{(s_{1})}D^{(s_{2}-1)}\epsilon^{(s_{3}-1)}+h^{(s_{1})}\epsilon^{(s_{2}-1)}D^{(s_{3}-1)})\qquad\quad  \label{F.5}\\
  && +\frac{s_{3}n_{3}s_{1}}{2}C_{(n_i)}^{(s_i)}[\hat{T}(Q_{ij}|n_{1},n_{2},n_{3}-1),(a\nabla_{1})]
  (\epsilon^{(s_{1}-1)}h^{(s_{2})}D^{(s_{3}-1)}-D^{(s_{1}-1)}h^{(s_{2})}\epsilon^{(s_{3}-1)})\qquad\quad  \label{F.6}\\
  && - s_{1}Q_{12}s_{2}C_{(n_i)}^{(s_i)}\hat{T}(Q_{12}-1,Q_{23},Q_{31}|n_{i})\epsilon^{(s_{1}-1)}D^{(s_{2}-1)}h^{(s_{3})} \label{F.7}\\
  && - s_{2}Q_{23}s_{3}C_{(n_i)}^{(s_i)}\hat{T}(Q_{12},Q_{23}-1,Q_{31}|n_{i})h^{(s_{1})}\epsilon^{(s_{2}-1)}D^{(s_{3}-1)} \label{F.8}\\
  && - s_{3}Q_{31}s_{1}C_{(n_i)}^{(s_i)}\hat{T}(Q_{12},Q_{23},Q_{31}-1|n_{i})D^{(s_{1}-1)}h^{(s_{2})}\epsilon^{(s_{3}-1)} .\label{F.9}
\end{eqnarray}
Indeed calculating commutators in the leading order and using relation  (\ref{2.34}) we see that
\begin{eqnarray}
  && (\ref{F.1})+(\ref{F.6})\nonumber \\
  && =s_{1}s_{2}(Q_{23}+1)C^{s_{i}}_{n_{1}+1,n_{2},n_{3}}\hat{T}(Q_{12},Q_{23},Q_{31}|n_{1},n_{2}+1,n_{3})
  D^{(s_{1}-1)}h^{(s_{2})}\epsilon^{(s_{3}-1)} ,\quad\quad\qquad
\end{eqnarray}
which exactly cancels (\ref{F.9}) after a corresponding shift of $n_{2}$ and using relation (\ref{2.35}). In a similar way we can prove cancelation of the other two sets of three lines.

To prove formulas for $\mathcal{L}_{I}^{(2,0)}$ and $\mathcal{L}_{I}^{(3,0)}$  we should manage the commutators of T operators with a,b,c, gradients in the following expression
\begin{eqnarray}
&&\frac{s_{1}s_{2}s_{3}}{2} C_{(n_i)}^{(s_i)}\Big[\nonumber\\
  &&[n_{1}n_{3}\hat{T}(Q_{ij}|n_{1}-1,n_{2},n_{3}-1),(b\nabla_{2})]
  (D^{(s_{1}-1)}D^{(s_{2}-1)}\epsilon^{(s_{3}-1)}+D^{(s_{1}-1)}\epsilon^{(s_{2}-1)}D^{(s_{3}-1)})\qquad\quad\nonumber\\
  && [n_{2}n_{3}\hat{T}(Q_{ij}|n_{1},n_{2}-1,n_{3}-1),(a\nabla_{1})]
  (D^{(s_{1}-1)}\epsilon^{(s_{2}-1)}D^{(s_{3}-1)}+\epsilon^{(s_{1}-1)}D^{(s_{2}-1)}D^{(s_{3}-1)})\qquad\quad\nonumber\\
  && [n_{1}n_{2}\hat{T}(Q_{ij}|n_{1}-1,n_{2}-1,n_{3}),(c\nabla_{3})]
  (\epsilon^{(s_{1}-1)}D^{(s_{2}-1)}D^{(s_{3}-1)}+D^{(s_{1}-1)}D^{(s_{2}-1)}\epsilon^{(s_{3}-1)})\qquad\quad\nonumber\\
  && - n_{3}Q_{12}\hat{T}(Q_{12}-1,Q_{23},Q_{31}|n_{1},n_{2},n_{3}-1)(\epsilon^{(s_{1}-1)}D^{(s_{2}-1)}D^{(s_{3}-1)}
  -\epsilon^{(s_{1}-1)}D^{(s_{2}-1)}D^{(s_{3}-1)})\nonumber\\
  && - n_{2}Q_{31}\hat{T}(Q_{12},Q_{23},Q_{31}-1|n_{1},n_{2}-1,n_{3})(D^{(s_{1}-1)}D^{(s_{2}-1)}\epsilon^{(s_{3}-1)}
  -D^{(s_{1}-1)}\epsilon^{(s_{2}-1)}D^{(s_{3}-1)})\nonumber\\
  && - n_{1}Q_{23}\hat{T}(Q_{12},Q_{23}-1,Q_{31}|n_{1}-1,n_{2},n_{3})(D^{(s_{1}-1)}
  \epsilon^{(s_{2}-1)}D^{(s_{3}-1)}-\epsilon^{(s_{1}-1)}D^{(s_{2}-1)}D^{(s_{3}-1)})\Big] ,\nonumber\\ \label{F.11}
\end{eqnarray}
and use again (\ref{2.33})-(\ref{2.35}) to show that (\ref{F.11}) is zero.

The remaining terms are:
\begin{eqnarray}
  &&\frac{1}{2} C_{(n_i)}^{(s_i)}\Big[\nonumber\\
  && -s_{1}Q_{12}s_{2}(s_{2}-1)[\hat{T}(Q_{12}-1,Q_{23},Q_{31}|n_{i}),(b\nabla_{2})]\epsilon^{(s_{1}-1)}\bar{h}^{(s_{2}-2)}h^{(s_{3})}\nonumber\\
  &&-s_{2}Q_{23}s_{3}(s_{3}-1)[\hat{T}(Q_{12},Q_{23}-1,Q_{31}|n_{i}),(c\nabla_{3})]h^{(s_{1})}\epsilon^{(s_{2}-1)}\bar{h}^{(s_{3}-2)}\nonumber\\
  && -s_{3}Q_{31}s_{1}(s_{1}-1)[\hat{T}(Q_{12},Q_{23},Q_{31}-1|n_{i}),(a\nabla_{1})]\bar{h}^{(s_{1}-2)}h^{(s_{2})}\epsilon^{(s_{3}-1)}\Big] ,\quad\quad
\end{eqnarray}
and
\begin{eqnarray}
&&\frac{s_{1}s_{2}s_{3}}{4} C_{(n_i)}^{(s_i)}\Big[ \nonumber\\
  && - n_{3}Q_{12}(s_{2}-1)[\hat{T}(Q_{12}-1,Q_{23},Q_{31}|n_{1},n_{2},n_{3}-1),(b\nabla_{2})]\nonumber\\
  && (\epsilon^{(s_{1}-1)} \bar{h}^{(s_{2}-2)}D^{(s_{3}-1)}-D^{(s_{1}-1)}\bar{h}^{(s_{2}-2)}\epsilon^{(s_{3}-1)})\nonumber\\
  &&- n_{1}Q_{23}(s_{3}-1)[\hat{T}(Q_{12},Q_{23}-1,Q_{31}|n_{1}-1,n_{2},n_{3}),(c\nabla_{3})]\nonumber\\
  &&(D^{(s_{1}-1)}\epsilon^{(s_{2}-1)}\bar{h}^{(s_{3}-2)}
  -\epsilon^{(s_{1}-1)}D^{(s_{2}-1)}\bar{h}^{(s_{3}-2)}) \nonumber\\
  &&- n_{2}Q_{31}(s_{1}-1)[\hat{T}(Q_{12},Q_{23},Q_{31}-1|n_{1},n_{2}-1,n_{3}),(a\nabla_{1})]\nonumber\\
  &&(\bar{h}^{(s_{1}-2)}D^{(s_{2}-1)}\epsilon^{(s_{3}-1)}
  -\bar{h}^{(s_{1}-2)}\epsilon^{(s_{2}-1)}D^{(s_{3}-1)})\Big] ,\label{A.13}
 \end{eqnarray}
 \begin{eqnarray}
  && -\frac{s_{3}s_{1}(s_{1}-1)n_{1}}{4}C_{(n_i)}^{(s_i)}Q_{31}\hat{T}(Q_{12},Q_{23},Q_{31}-1|n_{1}-1,n_{2},n_{3})\nonumber\\
  &&(\delta\bar{h}^{(s_{1}-2)}h^{(s_{2})}D^{(s_{3}-1)} +2 (\nabla D)^{(s_{1}-2)}h^{(s_{2})}\epsilon^{(s_{3}-1)}) \nonumber\\
  && -\frac{s_{1}s_{2}(s_{2}-1)n_{2}}{4}C_{(n_i)}^{(s_i)}Q_{12}\hat{T}(Q_{12}-1,Q_{23},Q_{31}|n_{1},n_{2}-1,n_{3})\nonumber\\
  &&(D^{(s_{1}-1)}\delta\bar{h}^{(s_{2}-2)}h^{(s_{3})} +2\epsilon^{(s_{1}-1)}(\nabla D)^{(s_{2}-2)}h^{(s_{3})}) \nonumber\\
  && -\frac{s_{2}s_{3}(s_{3}-1)n_{3}}{4}C_{(n_i)}^{(s_i)}Q_{23}\hat{T}(Q_{12},Q_{23}-1,Q_{31}|n_{1},n_{2},n_{3}-1)\nonumber\\
  &&(h^{(s_{1})}D^{(s_{2}-1)}\delta\bar{h}^{(s_{3}-2)} +2h^{(s_{1})}\epsilon^{(s_{2}-1)}(\nabla D)^{(s_{3}-2)}) , \label{A.14}
\end{eqnarray}
The last $DD\bar{h}$ terms coming from our calculation are:
\begin{eqnarray}
&&\frac{s_{1}s_{2}s_{3}}{4} C_{(n_i)}^{(s_i)}\Big[\nonumber\\
  && - (s_{3}-1)n_{1}n_{3}Q_{23}\hat{T}(Q_{12},Q_{23}-1,Q_{31}|n_{1}-1,n_{2},n_{3}-1)\nonumber\\
  &&(D^{(s_{1}-1)}D^{(s_{2}-1)}\delta\bar{h}^{(s_{3}-2)}
  +2D^{(s_{1}-1)}\epsilon^{(s_{2}-1)}(\nabla D)^{(s_{3}-2)})\nonumber\\
  &&- (s_{2}-1)n_{2}n_{3}Q_{12}\hat{T}(Q_{12}-1,Q_{23},Q_{31}|n_{1},n_{2}-1,n_{3}-1)\nonumber\\
  &&(D^{(s_{1}-1)}\delta\bar{h}^{(s_{2}-2)}D^{(s_{3}-1)}
  +2\epsilon^{(s_{1}-1)}(\nabla D)^{(s_{2}-2)}D^{(s_{3}-1)}) \nonumber\\
  &&- (s_{1}-1)n_{1}n_{2}Q_{31}\hat{T}(Q_{12},Q_{23},Q_{31}-1|n_{1}-1,n_{2}-1,n_{3})\nonumber\\
  &&(\delta\bar{h}^{(s_{1}-2)}D^{(s_{2}-1)}D^{(s_{3}-1)}
  +2(\nabla D)^{(s_{3}-2)}D^{(s_{2}-1)}\epsilon^{(s_{3}-1)})\Big] .\label{A.15}
 \end{eqnarray}
\quad These terms can be used in the same fashion for proving the remaining part of $\mathcal{L}_{I}^{(i,j)}$ to contain traces.


\end{document}